\documentclass[review]{elsarticle}

\usepackage[colorlinks,citecolor=blue,linktoc=all,linkcolor=cyan]{hyperref}
\usepackage{graphicx}

\usepackage[T1]{fontenc}
\usepackage{dsfont}               
\usepackage{mathrsfs}             
\usepackage{slashed}              
\usepackage{amsmath}
\usepackage{amssymb}
\usepackage{amsbsy}
\usepackage{amsfonts}

\usepackage{verbatim}

\usepackage{bm}
\usepackage{longtable}

\numberwithin{equation}{section}
\numberwithin{table}{section}
\numberwithin{figure}{section}

\journal{Progress in Particle and Nuclear Physics}

\usepackage{titlesec}
\usepackage{sectsty}
\titleformat{\section}{\normalfont\Large\bfseries}{\thesection}{1em}{}
\titleformat{\subsection}{\normalfont\large\bfseries}{\thesubsection}{1em}{}
\titleformat{\subsubsection}{\normalfont\normalsize\bfseries}{\thesubsubsection}{1em}{}

\begin{document}
	
	\begin{frontmatter}
		
		\title{Foundations and Applications of Quantum Kinetic Theory}

		\author[address1,address2,address3]{Yoshimasa Hidaka}
		\ead{hidaka@post.kek.jp}
		\author[mymainaddress]{Shi Pu} 
		\ead{shipu@ustc.edu.cn}
		\author[mymainaddress]{Qun Wang\corref{mycorrespondingauthor}}
		\cortext[mycorrespondingauthor]{Corresponding author}
		\ead{qunwang@ustc.edu.cn}
		\author[mysecondaryaddress]{Di-Lun Yang}
		\ead{dilunyang@gmail.com}
		

		\address[address1]{KEK Theory Center, Tsukuba 305-0801, Japan}
		\address[address2]{Graduate University for Advanced Studies (Sokendai),
        Tsukuba 305-0801, Japan}
        \address[address3]{RIKEN iTHEMS, RIKEN, Wako 351-0198, Japan}
        
		\address[mymainaddress]{Department of Modern Physics, University of Science and Technology of China, Hefei, Anhui 230026, China}
		\address[mysecondaryaddress]{Institute of Physics, Academia Sinica, Taipei, 11529, Taiwan}
		
		\begin{abstract}
			Many novel quantum phenomena emerge in non-equilibrium relativistic quantum matter under extreme conditions such as strong magnetic fields and rotations. The quantum kinetic theory based on Wigner functions in quantum field theory provides a powerful and effective microscopic description of these quantum phenomena. In this article we review some of recent advances in the quantum kinetic theory and its applications in describing these quantum phenomena. 
		\end{abstract}
		
		\begin{keyword}
			Wigner function \sep quantum kinetic theory \sep chiral kinetic theory \sep global polarization effect \sep spin Boltzmann equation 
			
		\end{keyword}
		
	\end{frontmatter}
	
	\newpage
	
	\thispagestyle{empty}
	\tableofcontents
	

	\newpage
	
	\section{Introduction} \label{sec:introduction}

Non-equilibrium relativistic quantum matter under extreme
conditions 
incorporates the interplay between a variety of physical mechanisms such as the chiral anomaly, spin-orbit interaction, and other quantum effects with the underlying quantum field theory, which result in profound transport phenomena.

One playground to study non-equilibrium phenomena under extreme conditions
in high energy frontiers is relativistic heavy-ion collisions,
in which two nuclei collide with each other to excite quarks and gluons
from the vacuum. These quarks and gluons undergo strong interaction and
form a hot and dense matter known as the quark-gluon plasma (QGP)
\cite{Rischke:2003mt,Gyulassy:2004vg,Gyulassy:2004zy,Shuryak:2004cy,Csernai:2006zz,Akiba:2015jwa,Busza:2018rrf}.
The QGP is found to be strongly coupled and more like a ``perfect liquid''
than an ideal gas. 
The existence of such a strongly coupled QGP (sQGP) is supported by measurements of collective flows \cite{STAR:2003wqp,STAR:2003xyj,STAR:2002hbo,Sorensen:2003kp} with hydrodynamical descriptions \cite{Kolb:2000sd,Kolb:2003dz,Hama:2004rr,Huovinen:2006jp,Ollitrault:2007du,Teaney:2003kp,Lacey:2006bc,Gale:2013da}.

Strong electromagnetic (EM) fields of the order $10^{18}$ Gauss \cite{Bloczynski:2012en,Deng:2012pc,Tuchin:2013ie,Tuchin:2013apa,Roy:2015coa,Li:2016tel,Holliday:2016lbx,Stewart:2017zsu,Siddique:2021smf,Chen:2021nxs}
can be generated in high-energy heavy-ion collisions. Though the EM
fields have peak values of a few $m_{\pi}^{2}$, where $m_{\pi}$
is the pion mass; they decay very fast in a vacuum \cite{Kharzeev:2007jp}.
To describe the coupling between charged media and EM fields, one
can employ the relativistic magneto-hydrodynamics, which are hydrodynamical
conservation equations coupled with Maxwell's equations. Several analytic
solutions of the relativistic magneto-hydrodynamics \cite{Pu:2016ayh,Roy:2015kma,Pu:2016bxy,Pu:2016rdq,Roy:2017xtz,Siddique:2019gqh,Wang:2020qpx}
have shown that the magnetic fields decay as $1/\tau$ in the limit
of infinite electric conductivity with $\tau$ being the proper time,
implying that the EM field can live much longer in a medium than in
vacuum. Numerical simulations of ideal magneto-hydrodynamics and relativistic
Boltzmann equations coupled to Maxwell's equations have been carried
out in Refs.~\cite{Inghirami:2016iru,Inghirami:2019mkc} and Ref.~\cite{Yan:2021zjc}, 
respectively. Recently, the ordinary magneto-hydrodynamics
has been extended to the second order via the moment expansion \cite{Denicol:2018rbw,Denicol:2019iyh}.


Many novel quantum phenomena emerge in such strong EM fields. The
chiral magnetic effect (CME) is one of them, in which a charge current
is induced along the magnetic field \cite{Vilenkin:1980fu,Nielsen:1983rb,Alekseev:1998ds,Kharzeev:2004ey,Kharzeev:2007jp,Fukushima:2008xe}.
In the QGP, the chiral symmetry is restored, and light quarks can be
considered as massless fermions (or chiral fermions). The spin orientation is always aligned parallel or anti-parallel to the 
momentum direction for a massless fermion having the positive (right-handed) or negative (left-handed) helicity, 
respectively. Equivalently, we can also introduce
the chirality for fermions. For a massless fermion, its chirality
is the same as its helicity, while for a massless anti-fermion, its
chirality is opposite to its helicity. The massless fermions can be
polarized by strong magnetic fields generated in heavy-ion collisions.
Due to their chiralities, these polarized fermions will move along
the magnetic fields. Meanwhile, topological fluctuations
of Quantum Chromodynamics (QCD) can lead to chirality imbalance. A locally fluctuating axial charge density could be produced by sphaleron transition at finite temperature \cite{Manton:1983nd,Klinkhamer:1984di,McLerran:1990de,Arnold:1996dy,Moore:2010jd} or by overpopulated gluons in the initial stage of the collision \cite{Mace:2016svc,Tanji:2016dka}. Such an axial charge density may also induce turbulent (chromo) electromagnetic fields via the chiral plasma instability \cite{Joyce:1997uy,Akamatsu:2013pjd,Mace:2019cqo}.
A net chirality can lead to a net charge current along the magnetic,
fields 
\begin{eqnarray}
\mathbf{J} & = & \frac{e^{2}}{2\pi^{2}}\mu_{5}\mathbf{B},\label{eq:CME}
\end{eqnarray}
where $\mu_{5}$ is the chiral chemical potential to quantify the
chirality imbalance. One can accordingly utilize the CME to probe local parity violation in QCD matter. 
As a dual effect of the CME, the chiral separation effect (CSE) generates an axial current along the magnetic fields \cite{Son:2004tq,Metlitski:2005pr}.
In the presence of a constant magnetic field, the interplay between CME and CSE yields the 
magnetic waves \cite{Kharzeev:2010gd,Burnier:2011bf,Shovkovy:2018tks}. Analogously, the rotational chiral matter can induce the currents along the vortical field, known as the chiral vortical effect (CVE) \cite{Vilenkin:1979ui,Erdmenger:2008rm,Banerjee:2008th,Son:2009tf,Landsteiner:2011cp,Pu:2010as,Sadofyev:2010is,Gao:2012ix}, and results in the chiral vortical waves \cite{Jiang:2015cva}.   
Besides, the electric fields can also lead to the chiral separation
effect through interactions \cite{Huang:2013iia,Pu:2014cwa,Jiang:2014ura,Pu:2014fva}.
There are also novel nonlinear EM responses in chiral
systems \cite{Pu:2014fva,Chen:2016xtg,Gorbar:2016qfh,Gorbar:2016sey,Gorbar:2016ygi,Gorbar:2017cwv}.
More discussions and references can be found in recent reviews \cite{Kharzeev:2012ph,Kharzeev:2015znc,Liao:2014ava,Miransky:2015ava,Huang:2015oca,Fukushima:2018grm,Bzdak:2019pkr,Zhao:2019hta,Gao:2020vbh,Shovkovy:2021yyw}.

There have been longstanding search for the CME in heavy-ion experiments at Relativistic Heavy Ion Collider (RHIC)
\cite{STAR:2009wot,STAR:2009tro,Wang:2012qs,STAR:2013ksd,STAR:2013zgu,STAR:2014uiw,Tribedy:2017hwn,STAR:2019xzd}
and Large Hadron Collider (LHC) \cite{ALICE:2012nhw,CMS:2016wfo,CMS:2017lrw} (also see recent
reviews in Refs.~\cite{Kharzeev:2015znc,Li:2020dwr}). 
Based on the charge separation along the magnetic field, the measurements of the difference between two-particle correlators with the same-sign and opposite-sign \cite{STAR:2009wot,STAR:2009tro} were believed to be a signal for the CME \cite{Voloshin:2004vk,Bloczynski:2012en,Deng:2016knn,Zhao:2019crj}. 
However, later studies \cite{CMS:2016wfo,CMS:2017lrw} show that the background
contributions from, e.g., the transverse momentum conservation \cite{Pratt:2010zn,Bzdak:2012ia},
local charge conservation \cite{Schlichting:2010qia} and non-flow
correlations \cite{Xu:2017zcn,Xu:2020sln}, are dominant in the two-particle
correlators \cite{STAR:2009wot}. It is hence challenging to disentangle the CME signal and the backgrounds. Several other correlators or observables,
such as a modified two-particle correlator \cite{STAR:2013ksd},
the charged multiplicity asymmetry correlator \cite{STAR:2013zgu},
the multi-particle charge-sensitive correlator \cite{Ajitanand:2010rc,Magdy:2017yje},
and the signed balance functions \cite{Tang:2019pbl}, were also
proposed for searching the CME. To separate the signal from backgrounds in experiments, the isobar collisions were proposed \cite{Voloshin:2010ut}.
Nevertheless, although the recent STAR measurements of isobar collisions \cite{STAR:2021mii} 
do not observe the CME signature that satisfies the predefined criteria, CME 
could still be potentially pinned down with the available correlators or observables.
On the other hand, the CME has
been observed in Dirac \cite{Li:2014bha,CMEScience,Feng:PhysRevB2015,Li:NC2015,Li:NC2016} and Weyl semimetals \cite{Arnold:2015vvs,Huang:2015eia,Zhang:NC2016, Hirschberger:NM2016, Wang:PhysRevB2016, Du:SCP2016, Li:PNAS2018} 
and is potentially applicable to quantum computing \cite{Kharzeev:2019ceh}.

Despite the difficulty in measuring the CME in heavy-ion collisions, it is tempting to study non-equilibrium transport in chiral matter.
One macroscopic theory to describe chiral transport phenomena related
to EM fields near equilibrium is the anomalous magneto-hydrodynamics \cite{Son:2009tf,Lublinsky:2009wr,Pu:2010as,Sadofyev:2010pr,Kharzeev:2010gr,Jensen:2012jy,Jensen:2012jh,Pu:2012wn,Kalaydzhyan:2016dyr,Hattori:2017usa,Ammon:2020rvg,Speranza:2021bxf}.
The analytic solutions to anomalous magneto-hydrodynamics in Bjorken
flow have been given \cite{Siddique:2019gqh,Wang:2020qpx}. The simulations
based on models of anomalous-viscous fluid dynamics have been applied
to the studies of chiral transport phenomena \cite{Jiang:2016wve,Shi:2017cpu,Shi:2017ucn,Shi:2018sah,Shi:2019wzi}.
On the other hand, the chiral kinetic theory (CKT) is the microscopic theory to describe
the chiral transport phenomena in terms of the distribution
functions of massless fermions in phase space. The CKT has been built from the Hamiltonian approaches
\cite{Son:2012wh,Son:2012zy}, path integrals \cite{Stephanov:2012ki,Chen:2013iga,Chen:2014cla,Chen:2015gta},
Wigner function approaches \cite{Gao:2012ix,Chen:2012ca,Son:2012zy,Hidaka:2016yjf,Huang:2018wdl}, world-line
formalism \cite{Mueller:2017arw,Mueller:2017lzw} and effective theories
\cite{Son:2012zy,Manuel:2013zaa,Manuel:2014dza,Carignano:2018gqt,Carignano:2019zsh,Lin:2018aon,Lin:2019ytz,Manuel:2021oah}.
In the CKT, both CME and chiral anomaly are connected to the Berry
phase \cite{Berry:1984jv}, which is a topological phase introduced in
condensed matter physics \cite{Xiao:2005qw,Xiao:2009rm,Stone:2013sga,Stone:2015kla}.
Recently, further applications  \cite{Chen:2016xtg,Gorbar:2016qfh,Gorbar:2017toh,Hidaka:2017auj,Huang:2017tsq, Rybalka:2018uzh,Dayi:2017xrr,Ebihara:2017suq,Hidaka:2018ekt,Fukushima:2018osn,Yang:2018lew,Dayi:2019hod,Lin:2019fqo,Lin:2021sjw,Yang:2021eoz,Chen:2021azy,Fang:2022ttm} and the numerical simulations for the CKT have been developed in the context of heavy-ion collisions \cite{Sun:2016nig,Sun:2016mvh,Sun:2017xhx,Sun:2018idn,Sun:2018bjl,Zhou:2018rkh,Zhou:2019jag,Liu:2019krs}.
There are also many other studies of chiral phenomena related to EM
fields in 
the perturbative quantum field theory or other
effective theories \cite{Fukushima:2008xe,Lublinsky:2009wr,Charbonneau:2009ax,Golkar:2012kb,Hou:2012xg,Jensen:2013vta,Basar:2013qia,Satow:2014lva,Wu:2016dam,Feng:2017dom,Lin:2018aon,Wu:2016dam,Hou:2017szz,Feng:2018tpb,Lin:2018nxj,Feng:2018tpb,Horvath:2019dvl,Horvath:2019dvl,Dong:2020zci},
and in non-perturbative frameworks, such as the world-line formalism
\cite{Fukushima:2010vw,Copinger:2018ftr,Copinger:2020nyx}, the Anti-de Sitter/conformal field theory (AdS/CFT) correspondence \cite{Erdmenger:2008rm,Banerjee:2008th,Torabian:2009qk,Rebhan:2009vc,Sahoo:2009yq,Yee:2009vw,Gynther:2010ed,Rebhan:2010ax,Gorsky:2010xu,Kalaydzhyan:2011vx,Hoyos:2011us,Kharzeev:2011rw,Landsteiner:2011cp,Landsteiner:2011iq,Gahramanov:2012wz,Ballon-Bayona:2012qnu,Zakharov:2012vv,Landsteiner:2012kd,Jensen:2012kj,Hoyos:2013qwa,Lin:2013sga,Landsteiner:2016led,Bu:2016oba,Bu:2016vum,Ammon:2016fru, Landsteiner:2017lwm,Bu:2018psl,Bu:2019mow,Fernandez-Pendas:2019rkh,Morales-Tejera:2020xuv,Yin:2021zhs,Ghosh:2021naw, Grieninger:2021rxd}, and lattice simulations \cite{Yamamoto:2011gk,Muller:2016jod,Mace:2016shq,Mace:2016svc,Mace:2019cqo}.


The CKT has been widely applied to a variety of physical systems besides
heavy-ion collisions. The CME has been studied in Weyl and Dirac
semimetals \cite{Nielsen:1983rb,Wan:2010fyf,Burkov:2011ene,Xu:2011dn,Gorbar:2017cwv}
through negative magneto-resistance \cite{Son:2012bg,Li:2014bha}.
A CKT for quasi-particles was constructed \cite{Gorbar:2016ygi}
to address the non-anomalous consistent current \cite{Landsteiner:2016led},
which incorporates pseudo-electromagnetic fields induced by a time-dependent
chiral shift or mechanical strains \cite{Cortijo:2016wnf}. It was
also employed to study the chiral magnetic plasmons as novel collective
excitations \cite{Gorbar:2016ygi,Gorbar:2016sey,Gorbar:2021ebc}.
In astrophysics, the chiral effects on lepton transport may affect
the evolution of core-collapse supernovae \cite{Yamamoto:2015gzz,Yamamoto:2016xtu,Masada:2018swb} and
magnetars \cite{Ohnishi:2014uea,Onishi:2020rqr} via the chiral
plasma instability \cite{Joyce:1997uy,Akamatsu:2013pjd} and the pulsar
kicks \cite{Kaminski:2014jda,Yamamoto:2021hjs} from the chiral transport of neutrinos. To study non-equilibrium
chiral effects in these astrophysical systems, the chiral radiation
transport theory of neutrinos has been recently developed \cite{Yamamoto:2020zrs,Yamamoto:2021hjs}
based on the generalized CKT in curved spacetime \cite{Liu:2018xip,Hayata:2020sqz}.


In addition to the strong EM fields, large orbital angular momenta (OAM) are
generated in heavy-ion collisions. It is straightforward to estimate
the global OAM of two colliding Au nuclei at $\sqrt{s_{NN}}=200$
GeV and impact parameter $b=10$ fm as large as $J_{0}\sim10^{5}\hbar$.
Part of large OAM is converted to the polarization of particles with
spins through spin-orbit couplings \cite{Liang:2004ph,Liang:2004xn,Gao:2007bc}.
The spin-vorticity coupling can be generated from spin-orbit coupling
in a microscopic description \cite{Zhang:2019xya}. Recently, STAR
experiments measured the global polarization of $\Lambda$ and $\overline{\Lambda}$
hyperons indicating an average angular velocity or vorticity of the
QGP as $\omega\sim10^{22}\,\mathrm{s}^{-1}$ \cite{STAR:2017ckg},
which reveal that the QGP is the most vortical fluid ever observed in nature. The
properties of the vorticity in heavy-ion collisions have been extensively
studied in various models, such as hydrodynamic models \cite{Betz:2007kg,Csernai:2013bqa,Becattini:2013vja,Becattini:2015ska,Pang:2016igs},
models of A Multi-Phase Transport (AMPT) model and Heavy Ion Jet Interaction Generator (HIJING) \cite{Jiang:2016woz,Deng:2016gyh,Li:2017slc,Wei:2018zfb}
and Ultra-relativistic Quantum Molecular Dynamics (UrQMD) model \cite{Wei:2018zfb}. The properties of relativistic spinning
particles can be described by quantum statistical models \cite{Becattini:2007nd,Becattini:2007sr,Becattini:2013fla,Becattini:2016gvu}
(also see recent reviews in Refs.~\cite{Wang:2017jpl,Becattini:2020ngo,Becattini:2020sww,Gao:2020vbh}).
The data on the global polarization of $\Lambda$ and $\overline{\Lambda}$
hyperons in collisions at intermediate energies can be well described
by various phenomenological models \cite{Karpenko:2016jyx,Li:2017slc,Xie:2017upb,Sun:2017xhx,Shi:2017wpk,Wei:2018zfb,Wu:2019eyi,Wu:2020yiz,Fu:2020oxj,Ryu:2021lnx,ExHIC-P:2020tcv}.
At low collision energies, there are a few numerical studies from
hydrodynamical models \cite{Guo:2021udq,Ivanov:2020udj} and transport
models \cite{Deng:2020ygd,Deng:2021miw} that can partially describe
the data \cite{STAR:2021beb,Kornas:2020qzi}.

On the other hand, the STAR collaboration has also measured the local
spin polarization, which is the polarization of $\Lambda$ and $\overline{\Lambda}$
hyperons along the beam direction in Au+Au collisions at $\sqrt{s_{NN}}=$
200 GeV \cite{STAR:2019erd}. However, the theoretical model calculations
based on the freezeout formula in terms of thermal vorticity contradict
the data of the longitudinal spin polarization as a function of the
azimuthal angle \cite{Becattini:2017gcx,Xia:2018tes}. This inconsistency
cannot be explained by feed-down effects \cite{Becattini:2019ntv,Xia:2019fjf,Li:2021jvn}.
There have been some phenomenological models in an attempt to describe
the data \cite{Liu:2019krs,Wu:2019eyi,Voloshin:2017kqp,Fu:2021pok,Becattini:2021iol}
but with external assumptions. It has been realized that the freezeout
formula in terms of thermal vorticity based on the equilibrium
condition (the Killing equation for the thermal velocity) may not
be valid in a hydrodynamical simulation of the QGP, implying a possible
off-equilibrium contribution to the local polarization \cite{Wu:2019eyi}.
It has also been realized that the temperature gradient plays an
important role in the local spin polarization \cite{Wu:2019eyi}.
Recently an off-equilibrium (or more precisely local-equilibrium) effect from the shear tensor is
found to provide an additional contribution to the local spin polarization
\cite{Liu:2020dxg,Liu:2021uhn,Becattini:2021suc} (see also Ref.~\cite{Hidaka:2017auj} for the generic corrections of massless fermions and Refs.~\cite{Liu:2021nyg,Florkowski:2021xvy} for related studies).
Whereby the numerical simulations yield the results in quantitative agreement with experimental data by tuning
appropriate parameters \cite{Fu:2021pok,Becattini:2021iol}. However, there are a few problems that remain unsettled
such as the sensitivity to the equation of state and the temperature
gradient in hydrodynamical simulations \cite{Yi:2021ryh, Florkowski:2021xvy,Sun:2021nsg}.
Very recently,  
helicity polarization defined as the local spin polarization projected
to the momentum direction of polarized hadrons has been proposed and studied
in Refs.~\cite{Becattini:2020xbh,Gao:2021rom,Yi:2021unq}, which could be implemented to probe local parity violation
characterized by an axial chemical potential in quark matter at finite temperature. Helicity polarization is also found to be useful for probing the local strength of vorticity in rotational QGP \cite{Yi:2021unq}.
In general, it is essential to understand dynamical spin polarization in heavy-ion collisions.

The relativistic spin hydrodynamics is a macroscopic theory for tracking the
evolution of spin degrees of freedom, in which the conservation equation of angular
momentum is added to those of energy-momentum and conserved charges
in conventional hydrodynamics. The spin hydrodynamics can be constructed
from the entropy principle \cite{Hattori:2019lfp,Fukushima:2020qta,Fukushima:2020ucl,Li:2020eon,She:2021lhe},
the effective Lagrangian theory \cite{Montenegro:2017lvf,Montenegro:2017rbu},
kinetic approaches \cite{Florkowski:2017ruc,Florkowski:2018myy,Florkowski:2018fap,Yang:2018lew,Bhadury:2020puc,Shi:2020qrx},
and quantum field theory \cite{Becattini:2018duy,Gallegos:2021bzp,Hongo:2021ona}.
The ideal spin hydrodynamics can be constructed from spin-dependent
distribution functions \cite{Florkowski:2017dyn,Florkowski:2017ruc}
with further applications \cite{Florkowski:2018ahw,Florkowski:2018myy,Florkowski:2019qdp,Florkowski:2019voj,Becattini:2018duy,Bhadury:2020puc},
see Ref.~\cite{Florkowski:2018fap} for a recent review on spin hydrodynamics.
With a gradient expansion, the dissipative spin hydrodynamics can
be derived in a canonical \cite{Hattori:2019lfp,Bhadury:2020cop,Shi:2020htn,Shi:2020qrx,Hongo:2021ona}
and a Belinfante form \cite{Fukushima:2020qta,Fukushima:2020ucl}. 
The spin hydrodynamics has also been applied to background expanding
systems \cite{Florkowski:2019qdp,Singh:2020rht,Florkowski:2021wvk} and analytic solutions
can be found in a Bjorken flow \cite{Wang:2021ngp} and Gubser flow \cite{Wang:2021wqq}. Recent reviews
on the spin polarization in a hot and dense matter in various perspectives
can be found in Refs.~\cite{Wang:2017jpl,Florkowski:2018fap,Becattini:2020ngo,Becattini:2020sww,Gao:2020vbh,Liu:2020ymh}.


As a microscopic description, the quantum kinetic theory (QKT) or quantum transport theory for
massive fermions is a natural extension of CKT for massless fermions.
The QKT for massive fermions has been developed from the covariant
Wigner-function approach based on quantum field theory \cite{Gao:2019znl,Weickgenannt:2019dks,Weickgenannt:2020aaf,Hattori:2019ahi,Yang:2020hri,Liu:2020flb,Weickgenannt:2021cuo,Sheng:2021kfc,Wang:2021qnt,Huang:2020wrr,Wang:2020dws,Wang:2019moi,Wang:2022yli}.
The covariant Wigner function also plays an important role in quantum
statistical model for relativistic particles with spins \cite{Becattini:2007nd,Becattini:2007sr,Becattini:2013fla}.
The QKT provides a unified framework to study various chiral transport
and spin polarization effects in fermion systems. In particular, it directly gives rise to the spectrum of spin polarization. In this article,
we will review the main developments of the QKT based on the covariant
Wigner function in quantum field theory.


The structure of this article is as follows. 
We start with a brief introduction to Wigner functions in quantum mechanics 
and the closed-time-path formalism in Sec.~\ref{sec:Wigner-function-and}. 
In Sec.~\ref{sec:Master-equation}, we give a general derivation of the master equations 
for Wigner functions of massless fermions in background electromagnetic fields. 
In Sec.~\ref{sec:CCKE}, we derive a set of master equations for chiral components of Wigner functions 
or chiral Wigner functions, then we solve the master equations in equilibrium in constant fields. 
The chiral Wigner functions are found order by order in the Planck constant. 
In Sec.~\ref{sec:dwf-theorem}, we discuss the theorem for disentanglement of chiral Wigner functions, 
derive the kinetic theory for on-shell chiral fermions and
 discuss chiral Wigner functions in a general Lorentz frame as well as
the decomposition of 
the CME and CVE currents. In Sec.~\ref{sec:CKT_effective}, we briefly introduce 
the CKT derived from effective theories in comparison with the Wigner function approach. 
The collision kernels of the CKT are discussed in Sec.~\ref{sec:CKT_collisions}. 
We derive the QKT for massive fermions in Sec.~\ref{sec:massive fermion with collisions} 
and collision terms in Sec.~\ref{sec:QKT_collision}. In Sec.~\ref{sec:spin-Boltzmann}, 
we give a brief introduction to spin Boltzmann equations and non-local collisions in terms of spin-dependent distributions. Summary and outlook of the article are given in the final section or Sec.~\ref{sec:conclusion}.

Unless explicitly stated, we will absorb the fermion's
electric charge into EM fields such as the field potential,
the field strength tensor, the electric and magnetic field, i.e.,
$QA^{\mu}\rightarrow A^{\mu}$, $QF^{\mu\nu}\rightarrow F^{\mu\nu}$,
$QE^{\mu}\rightarrow E^{\mu}$, $QB^{\mu}\rightarrow B^{\mu}$. The
Minkowski metric tensor is denoted as $\eta^{\alpha\beta}=\mathrm{diag}(1,-1,-1,-1)$. 
The sign convention for the Levi-Civita symbol is $\epsilon^{0123}=-\epsilon_{0123}=1$. 
A coordinate four-vector is denoted as $x^{\mu}=(x^{0},\mathbf{x})$
or $x_{\mu}=(x_{0},-\mathbf{x})$ with its spatial components being
denoted as a three-vector $\mathbf{x}$ or $-\mathbf{x}$ in boldface.
Sometime the index of $x^{\mu}$ is suppressed and denoted as $x$
for notational simplicity. Similarly a momentum is denoted as $k^{\mu}=(k^{0},\mathbf{k})$
or $p^{\mu}=(p^{0},\mathbf{p})$, or sometime just $k$ or $p$ for
simplicity. In some parts, we keep the (reduced) Planck constant $\hbar$ 
explicitly in order to make an expansion in it, while in other parts 
we simply suppress it, i.e. $\hbar$ set to 1, if it is irrelevant to 
the physics in discussion. For convenience, we define $A^{(\mu}B^{\nu)}=A^\mu B^\nu + B^\mu A^\nu$ and $A^{[\mu}B^{\nu]}=A^\mu B^\nu - B^\mu A^\nu$. 

\newpage

\section{Two ingredients of quantum kinetic theory\label{sec:Wigner-function-and}}
In this section, we give a brief introduction to two ingredients of quantum kinetic theory: the Wigner function and the Closed-Time-Path (CTP) formalism. 

\subsection{Wigner functions in quantum mechanics}

The Wigner distribution function or Wigner function (WF) in short
was firstly introduced by Eugene Wigner \cite{Wigner:1932eb} in 1932
in his seminal paper entitled ``On the Quantum Correction for
Thermodynamic Equilibrium'' as a quantum analog of the phase space
distribution function. So quantum mechanics can be re-formulated in
a statistical language. Later on Groenewold \cite{Groenewold:1946kp}
and Moyal \cite{Moyal:1949sk} developed the idea and established
the WF as an apparatus for an independent formulation of quantum mechanics
in phase space. Nowadays, the WF has profound influence on many aspects
of physics: statistical mechanics, condensed-matter physics, gravitational-wave
detection, nuclear and particle physics, communication theory, see,
e.g., Refs.~\cite{Hillery:1983ms,Zachos:2005gri} for reviews.

In classical mechanics the dynamics of a moving particle can be described
in terms of the classical distribution $f(t,\mathbf{x},\mathbf{p})$
in phase space, a space of the position and momentum. Note that $f(t,\mathbf{x},\mathbf{p})$
is well-defined for a classical particle since its position $\mathbf{x}$
and momentum $\mathbf{p}$ can be determined simultaneously. The average
of any phase-space function $A(\mathbf{x},\mathbf{p})$ is given by
\begin{equation}
\left\langle A\right\rangle =\int d^{3}\mathbf{x}\int\frac{d^{3}\mathbf{p}}{(2\pi\hbar)^{3}}A(\mathbf{x},\mathbf{p})f(t,\mathbf{x},\mathbf{p}).\label{eq:av-a-class}
\end{equation}
In quantum mechanics, however, it is non-trivial to define a particle's
phase-space distribution due to the uncertainty principle: $\mathbf{x}$
and $\mathbf{p}$ cannot be determined at the same time. The particle's
movement is described by the time evolution of its wave function, which
follows the Schr\"odinger equation. The wave function can be represented
either in coordinate space $\psi(t,\mathbf{x})$ or in momentum space
$\widetilde{\psi}(t,\mathbf{p})$ which are connected by Fourier transform
but not in both. The modular square of the wave function gives the
probability of the particle at a position $\mathbf{x}$ {[}for $\psi(t,\mathbf{x})${]}
or with a momentum $\mathbf{p}$ {[}for $\widetilde{\psi}(t,\mathbf{p})${]}.
Another effective description of quantum mechanics is
through the density operator $\hat{\rho}$, which was proposed by
Landau and von Neumann. The average of the operator $\hat{A}$ as
a function of coordinate and momentum operators is defined as 
\begin{equation}
\left\langle \widehat{A}\right\rangle =\mathrm{Tr}\left(\widehat{A}\widehat{\rho}\right).\label{eq:av-op-a}
\end{equation}
It can be shown that $\left\langle \widehat{A}\right\rangle $ in
above formula can be written in a similar form to Eq.~(\ref{eq:av-a-class}),
\begin{equation}
\left\langle \widehat{A}\right\rangle =\int d^{3}\mathbf{x}\int\frac{d^{3}\mathbf{p}}{(2\pi\hbar)^{3}}A(\mathbf{x},\mathbf{p})W(\mathbf{x},\mathbf{p}),\label{eq:av-a-aw-xp}
\end{equation}
where $W(\mathbf{x},\mathbf{p})$ is the WF and defined through Weyl
transformation \cite{Weyl:1927vd} as 
\begin{equation}
W(\mathbf{x},\mathbf{p})=\int d^{3}\mathbf{y}\exp\left(\frac{i}{\hbar}\mathbf{p}\cdot\mathbf{y}\right)\left\langle \mathbf{x}-\frac{\mathbf{y}}{2}\right|\widehat{\rho}\left|\mathbf{x}+\frac{\mathbf{y}}{2}\right\rangle ,\label{eq:wf-def}
\end{equation}
and $A(\mathbf{x},\mathbf{p})$ is defined in the same way as 
\begin{equation}
A(\mathbf{x},\mathbf{p})=\int d^{3}\mathbf{y}\exp\left(\frac{i}{\hbar}\mathbf{p}\cdot\mathbf{y}\right)\left\langle \mathbf{x}-\frac{\mathbf{y}}{2}\right|\widehat{A}\left|\mathbf{x}+\frac{\mathbf{y}}{2}\right\rangle .\label{eq:class-a-def}
\end{equation}
From the above equation, we have 
\begin{equation}
\int d^{3}\mathbf{x}\int\frac{d^{3}\mathbf{p}}{(2\pi\hbar)^{3}}A(\mathbf{x},\mathbf{p})=\int d^{3}\mathbf{x}\left\langle \mathbf{x}\right|\widehat{A}\left|\mathbf{x}\right\rangle =\mathrm{Tr}\left(\widehat{A}\right).
\end{equation}
Similarly one can verify Eq.~(\ref{eq:av-a-aw-xp}) using expressions
of $W(\mathbf{x},\mathbf{p})$ and $A(\mathbf{x},\mathbf{p})$ in
Eqs.~(\ref{eq:wf-def}) and (\ref{eq:class-a-def}).

Note that Eq.~(\ref{eq:av-a-aw-xp}) in quantum mechanics is similar
to the classical formula (\ref{eq:av-a-class}). This is a remarkable
feature that a quantum ensemble average can be carried out by a phase
space integration. The WF (\ref{eq:wf-def}) is then a quantum counterpart
of the classical phase-space distribution: it is a quantum object
since it depends on $\hbar$ and may allow negative values. Another
nice feature of the WF is that $\mathbf{x}$ and $\mathbf{p}$ are
commutable as operators since $\mathbf{p}$ is conjugate to $\mathbf{y}$,
the distance of two points, while $\mathbf{x}$ is the center position
of two points.

\subsection{Two-point correlations in Closed-Time-Path formalism} 
\label{sec:CTP}
The Closed-Time-Path (CTP) formalism was originally developed by Schwinger and Keldysh for non-equilibrium
dynamics \cite{Martin:1959jp,Keldysh:1964ud}, see, e.g., Refs.~\cite{Chou:1984es,Blaizot:2001nr,Berges:2004yj},
for reviews of this topic.

\begin{figure}
\begin{center}
  \includegraphics{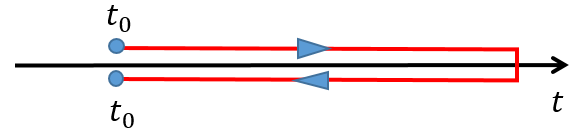}   
  \caption{\label{fig:ctp}The Closed-Time Path.}
\end{center}
\end{figure}

We see in Eq.~(\ref{eq:wf-def}) that the WF is defined through a
two-point correlation function. In non-equilibrium quantum field theory,
the two-point correlation function is defined through the density
operator at the initial time,
\begin{equation}
G(x_{1},x_{2})=\left\langle T\phi(x_{1})\phi(x_{2})\right\rangle \equiv\frac{1}{\mathrm{Tr}\rho(t_{0})}\mathrm{Tr}[\rho(t_{0})T\phi(x_{1})\phi(x_{2})],\label{eq:green-non-eq}
\end{equation}
where $\phi$ is a scalar field operator (can be replaced by other
fields, we will suppress the hat for all operators from now on) in
Heisenberg's picture, $\rho(t_{0})$ is the density operator at the
initial time $t_{0}$, and the trace is taken over initial states
which can be expanded in eigenstates of free Hamiltonian $H_{0}$.
The interaction $H_{I}$ is is switched on at a later time. We can rewrite
Eq.~(\ref{eq:green-non-eq}) in terms of fields in the interaction
picture as 
\begin{eqnarray}
G(x_{1},x_{2}) & = & \frac{1}{\mathrm{Tr}\rho(t_{0})}\mathrm{Tr}[\rho(t_{0})TU^{\dagger}(x_{1}^{0},t_{0})\phi_{I}(x_{1})U(x_{1}^{0},t_{0}) 
 \times U^{\dagger}(x_{2}^{0},t_{0})\phi_{I}(x_{2})U(x_{2}^{0},t_{0})]\nonumber \\
 & = & \frac{1}{\mathrm{Tr}\rho(t_{0})}\mathrm{Tr}[\rho(t_{0})TU^{\dagger}(x_{1}^{0},t_{0})\phi_{I}(x_{1})U(x_{1}^{0},x_{2}^{0})\phi_{I}(x_{2})U(x_{2}^{0},t_{0})]\nonumber \\
 & = & \frac{1}{\mathrm{Tr}\rho(t_{0})}\mathrm{Tr}[\rho(t_{0})T_{P}\phi_{I}(x_{1})\phi_{I}(x_{2})U_{CTP}(t_{0})]\nonumber \\
 & \equiv & \left\langle T_{P}\phi(x_{1})\phi(x_{2})\right\rangle \nonumber \\
 & = & \theta_{P}(x_{1}-x_{2})\left\langle \phi(x_{1})\phi(x_{2})\right\rangle +\theta_{P}(x_{2}-x_{1})\left\langle \phi(x_{2})\phi(x_{1})\right\rangle ,\label{eq:g-non-eq}
\end{eqnarray}
where $U(t,t_{0})$ is the unitary operator connecting the interaction
and Heisenberg's picture,
\begin{eqnarray}
U(t,t_{0}) & = & e^{iH_{0}(t-t_{0})}e^{-iH(t-t_{0})}\nonumber \\
 & = & T\left[\exp\left(-i\int_{t_{0}}^{t}dt^{\prime}H_{I}(t^{\prime})\right)\right],
\end{eqnarray}
with $T$ denoting the time ordering operator and $H_{I}$ being the
interaction Hamiltonian in the interaction picture. In the third equality
of Eq.~(\ref{eq:g-non-eq}), $U_{CTP}(t_{0})$ is the unitary evolution
operator defined on the CTP starting from $t_{0}$ to $+\infty$ (positive
time branch) and back to $t_{0}$ (negative time branch) as shown
in Fig. \ref{fig:ctp}, 
\begin{align}
U_{CTP}(t_{0}) & \equiv T_{P}\left[\exp\left(-i\int_{CTP}dtH_{I}(t)\right)\right]\nonumber \\
 & =T_{P}\left[\exp\left(-i\int_{t_{0}}^{\infty}dt_{+}H_{I}(t_{+})+i\int_{t_{0}}^{\infty}dt_{-}H_{I}(t_{-})\right)\right].\label{eq:u-ctp}
\end{align}
In Eqs.~(\ref{eq:g-non-eq}) and (\ref{eq:u-ctp}), $T_{P}$ is the
ordering operator and on the CTP, and $\theta_{P}(x_{1}-x_{2})$ is
the step function on the CTP, which is 1 or 0 for the case that $x_{1}^{0}$
is later or earlier than $x_{2}^{0}$ on the CTP, respectively (for
example, the time in the negative time branch is later than the one
in the positive time branch). From Eq.~(\ref{eq:u-ctp}) we see that
the interaction terms leading to vertices in Feynman rules on the
negative time branch have got a minus sign relative to those on the
positive time branch.

The difference between $G(x_{1},x_{2})$ in vacuum and non-equilibrium
is that there is no evolution of an in-state to a state at $t_{0}$
and a state at $t_{0}$ to an out-state. Only states at $t_{0}$ are
involved, which leads to the CTP.

In Eq.~(\ref{eq:g-non-eq}), $x_{1}^{0}$ and $x_{2}^{0}$ can be
on the positive or negative time branch, so there are four kinds of
two-point correlation functions that can be put into a matrix form,
\begin{equation}
G(x_{1},x_{2})=\left(\begin{array}{cc}
G^{++}(x_{1},x_{2}) & G^{+-}(x_{1},x_{2})\\
G^{-+}(x_{1},x_{2}) & G^{--}(x_{1},x_{2})
\end{array}\right)=\left(\begin{array}{cc}
G^{F}(x_{1},x_{2}) & \pm G^{<}(x_{1},x_{2})\\
G^{>}(x_{1},x_{2}) & G^{\overline{F}}(x_{1},x_{2})
\end{array}\right),\label{eq:two-point-funct}
\end{equation}
where $G^{ij}(x_{1},x_{2})$ (with $i,j=+,-$) means that the first
time argument $x_{1}^{0}$ lives on the time branch $i$ and the second
time argument $x_{2}^{0}$ lives on the time branch $j$. For $G^{++}$,
both $x_{1}^{0}$ and $x_{2}^{0}$ are on the positive time branch.
Then, the ordering on the CTP is just the standard time ordering of
quantum field theory, so that $G^{++}$ is simply the Feynman propagator
$G^{F}$. On the other hand, $G^{-+}$ means that $x_{1}^{0}$ lives
on the negative and $x_{2}^{0}$ on the positive time branch, respectively,
such that $x_{1}^{0}$ is later than $x_{2}^{0}$ in accordance with
the ordering on the CTP. Consequently, this two-point function is
denoted as $G^{>}$. Analogously, we have $G^{+-}\equiv\pm G^{<}$
with the upper sign for bosons and lower sign for fermions. Note that
the minus sign for fermions arises from the interchange of two fermion
fields for $G^{+-}$, but the sign is introduced by definition so
that $G^{<}$ does not have such a sign. Equivalently one can also
absorb such a sign into the definition of $G^{<}$ as was also done
in some literature \cite{Schonhofen:1994zf,Sheng:2021kfc}. When both
$x_{1}^{0}$ and $x_{2}^{0}$ live on the negative time branch, $x_{1}^{0}>x_{2}^{0}$
actually means that, on the CTP, $x_{2}^{0}$ is later than $x_{1}^{0}$,
so the ordering on the CTP is actually equivalent to the standard
anti-time ordering, and hence we denote $G^{--}$ as $G^{\overline{F}}$.

For spin-1/2 fermions, the definitions of the various two-point correlation
functions are 
\begin{eqnarray}
G_{\alpha\beta}^{F}(x_{1},x_{2}) & = & \left\langle T\psi_{\alpha}(x_{1})\overline{\psi}_{\beta}(x_{2})\right\rangle ,\nonumber \\
G_{\alpha\beta}^{\overline{F}}(x_{1},x_{2}) & = & \left\langle T_{A}\psi_{\alpha}(x_{1})\overline{\psi}_{\beta}(x_{2})\right\rangle ,\nonumber \\
G_{\alpha\beta}^{<}(x_{1},x_{2}) & = & \left\langle \overline{\psi}_{\beta}(x_{2})\psi_{\alpha}(x_{1})\right\rangle ,\nonumber \\
G_{\alpha\beta}^{>}(x_{1},x_{2}) & = & \left\langle \psi_{\alpha}(x_{1})\overline{\psi}_{\beta}(x_{2})\right\rangle ,\label{eq:def-green-func}
\end{eqnarray}
where $\psi_{\alpha}(x_{1})$ and $\overline{\psi}_{\beta}(x_{2})=(\psi^{\dagger}\gamma_{0})_{\beta}$
denote the Dirac spinor quantum fields with $\alpha,\beta=1,2,3,4$
being spinor indices, $T$ and $T_{A}$ denote the time-ordering and
anti-time-ordering operators, respectively, and angular brackets denote
averages weighted by $\rho(t_{0})$. 
Not all of
the four types of two-point functions appearing in Eq.~(\ref{eq:def-green-func})
are independent, for example they satisfy 
\begin{equation}
G^{F}+G^{\overline{F}}=G^{>}-G^{<},\label{eq:g-less-g-larger-gff}
\end{equation}
which is a direct consequence of the anticommutation relations for
fermion field operators. Equivalently we can use the following two-point
functions, which are linear combinations of those in Eq.~(\ref{eq:def-green-func}),
\begin{eqnarray}
-iG^{R} & = & G^{F}+G^{<}=-G^{\overline{F}}+G^{>},\nonumber \\
-iG^{A} & = & G^{F}-G^{>}=-G^{\overline{F}}-G^{<},\nonumber \\
-iG^{C} & = & G^{F}+G^{\overline{F}}=G^{>}-G^{<},
\end{eqnarray}
where $G^{R}$ and $G^{A}$ are the retarded and advanced two-point
functions, respectively, whose explicit forms are given by 
\begin{eqnarray}
G^{R}(x_{1},x_{2}) & = & i\Theta(t_{1}-t_{2})\left[G^{>}(x_{1},x_{2})+G^{<}(x_{1},x_{2})\right],\nonumber \\
G^{A}(x_{1},x_{2}) & = & -i\Theta(t_{2}-t_{1})\left[G^{>}(x_{1},x_{2})+G^{<}(x_{1},x_{2})\right].\label{eq:ga-explicit}
\end{eqnarray}
We can actually express all two-point functions in terms of $G^{>}$
and $G^{<}$ with the help of the step functions $\Theta(t_{1}-t_{2})$
and $\Theta(t_{2}-t_{1})$.

\newpage
\section{Wigner functions for chiral fermions in background electromagnetic fields \label{sec:Master-equation}}

Let us start from the Lagrangian for massless fermions in a background
electromagnetic field, 
\begin{eqnarray}
\mathcal{L}  =  \overline{\psi}i\gamma\cdot D\psi
  =  \chi_{R}^{\dagger}i\sigma\cdot D\chi_{R}+\chi_{L}^{\dagger}i\overline{\sigma}\cdot D\chi_{L},\label{eq:chiral-lagrangian}
\end{eqnarray}
where we have used the Weyl representation so that the fields are
$\psi=(\chi_{L},\chi_{R})^{T}$ and $\overline{\psi}=\psi^{\dagger}\gamma^{0}=(\chi_{R}^{\dagger},\chi_{L}^{\dagger})$,
with $\chi_{R,L}$ being right-handed and left-handed Pauli spinors,
$\sigma^{\mu}=(1,\boldsymbol{\sigma})$, $\overline{\sigma}^{\mu}=(1,-\boldsymbol{\sigma})$,
and $D_{\mu}=\hbar\partial_{\mu}+iA_{\mu}$ is the covariant derivative.
We see that the Dirac equations of motion for right- and left-handed
fermions are decoupled. So one can describe massless fermions by two-dimension
Pauli spinors instead of four-dimension Dirac spinors.

For convenience, we focus on the right-handed fermion, the corresponding
equations for the left-handed fermion can be easily obtained. The
Dirac equation for the right-handed fermion in terms of Pauli spinors
reads
\begin{equation}
i\sigma\cdot D\chi_{R}=0,\label{eq:R_EOM_01}
\end{equation}
and its conjugation one, 
\begin{equation}
-\chi_{R}^{\dagger}i\sigma\cdot\overleftarrow{D^{\dagger}}=0,\label{eq:R_EOM_02}
\end{equation}
For notational simplicity, we suppress the subscript $R$ in the rest
part of this section.

We will follow the standard method to derive the kinetic theory from
a Lagrangian in Ref.~\cite{Blaizot:2001nr}. Our strategy is as follows.
First, from the equation of motion (\ref{eq:R_EOM_01}) and its conjugate
one (\ref{eq:R_EOM_02}), we will derive the equation of motion for
the two-point Green function. Then we will generalize the two-point
function to the gauge-invariant one. Finally, we apply the gradient
expansion to all equations and take the Wigner transformation. The
equation of motion for the gauge-invariant two-point function can
also be derived directly from the Dirac equation for massive fermions
in Ref.~\cite{Elze:1986ii,Elze:1986qd,Vasak:1987um}.

In accordance with the definition of the two-point function on the
CTP, we define $S^{>}$ and $S^{<}$ for right-handed fermions as
\begin{eqnarray}
S_{ab}^{>}(x_{1},x_{2}) & = & \left\langle \chi_{a}(x_{1})\chi_{b}^{\dagger}(x_{2})\right\rangle ,\nonumber \\
S_{ab}^{<}(x_{1},x_{2}) & = & \left\langle \chi_{b}^{\dagger}(x_{2})\chi_{a}(x_{1})\right\rangle ,
\end{eqnarray}
where $a,b=1,2$ denote the components of the Pauli spinors $\chi$
and $\chi^{\dagger}$. Note that sometimes $S^{<}(x_{1},x_{2})$ can
include a minus sign by definition as we have mentioned in the paragraph
of Eq.~(\ref{eq:two-point-funct}). Using Eqs.~(\ref{eq:R_EOM_01})
and (\ref{eq:R_EOM_02}), $S^{>}$ and $S^{<}$ satisfy 
\begin{eqnarray}
i\sigma\cdot D_{x_{1}}S(x_{1},x_{2}) & = & 0,\nonumber \\
-S(x_{1},x_{2})i\sigma\cdot\overleftarrow{D_{x_{2}}^{\dagger}} & = & 0,\label{eq:Master_eq_pre_01}
\end{eqnarray}
where we have suppressed the subscripts of $S$ for notational simplicity.
Without loss of generality, in the rest part of the section, we assume
$S=S^{<}$. Equivalently one can also set $S=S^{>}$ and follow the
same steps to derive the corresponding equations for $S^{>}$.

The two-point Green function $S(x_{1},x_{2})$ is not gauge invariant.
We can define the gauge invariant two-point function by including
gauge links, 
\begin{equation}
\widetilde{S}(x,y)=U\left(x,x_{1}\right)S(x_{1},x_{2})U\left(x_{2},x\right),\label{eq:s-invariant-op}
\end{equation}
where we have used the center and distance of two space-time points,
\begin{equation}
x=\frac{1}{2}(x_{1}+x_{2}),\;y=x_{1}-x_{2},\label{eq:X-y-def}
\end{equation}
and $U$ is the gauge link from $x_{2}$ to $x_{1}$, 
\begin{equation}
U(x_{1},x_{2})=\mathcal{P}\exp\left[-i\frac{1}{\hbar}\int_{x_{2}}^{x_{1}}dz\cdot A(z)\right],\label{eq:gauge_link}
\end{equation}
with $\mathcal{P}$ is the path ordering operator. Here we have chosen
the straight line from $x_{2}$ to $x_{1}$ as the integral path.
This path corresponds to the interpretation of $p_{\mathrm{c}}$,
conjugate to $y$, in the Wigner function as the canonical momentum
\cite{Elze:1986ii,Elze:1986qd,Vasak:1987um}. If electromagnetic fields
are classical background ones, $U(x_{1},x_{2})$ is just a phase factor
without the operator $\mathcal{P}$, and Eq.~(\ref{eq:s-invariant-op})
becomes 
\begin{equation}
\widetilde{S}(x_{1},x_{2})=S(x_{1},x_{2})U(x_{2},x_{1})=U(x_{2},x_{1})S(x_{1},x_{2}).
\end{equation}

Then the left-hand-side of Eq.~(\ref{eq:Master_eq_pre_01}) can be
rewritten as 
\begin{eqnarray}
i\sigma\cdot D_{x_{1}}S(x_{1},x_{2}) & = & i\sigma\cdot[\hbar\partial_{x_{1}}+ieA(x_{1})]S(x_{1},x_{2})\nonumber \\
 & = & i\sigma\cdot D_{x_{1}}U\left(x_{1},x_{2}\right)\widetilde{S}(x_{1},x_{2})\nonumber \\
 & = & i\sigma\cdot\left[D_{x_{1}}U\left(x_{1},x_{2}\right)\right]\widetilde{S}(x_{1},x_{2})+U\left(x_{1},x_{2}\right)i\hbar\sigma\cdot\partial_{x_{1}}\widetilde{S}(x_{1},x_{2}), \label{eq:kinetic-part} 
\end{eqnarray}
where we have used $U^{\dagger}\left(x_{2},x_{1}\right)=U\left(x_{1},x_{2}\right)$.
It is convenient to use $x$ and $y$ in Eq.~(\ref{eq:X-y-def}),
so the derivatives have the form,
\begin{equation}
\partial_{x_{1}}=\frac{1}{2}\partial_{x}+\partial_{y},\;\partial_{x_{2}}=\frac{1}{2}\partial_{x}-\partial_{y}.
\end{equation}
For the first term of Eq.~(\ref{eq:kinetic-part}), we use Eq.~(3.16)
of Ref.~\cite{Elze:1986qd},
\begin{eqnarray}
D_{x_{1},\mu}U\left(x_{1},x_{2}\right) & = & -U\left(x_{1},x_{2}\right)i(x_{1}^{\nu}-x_{2}^{\nu})\int_{0}^{1}ds\,sF_{\mu\nu}[z(s)]\nonumber \\
 & = & -U\left(x_{1},x_{2}\right)iy^{\nu}\int_{0}^{1}ds\,sF_{\mu\nu}[z(s)],\label{eq:elze-formula}
\end{eqnarray}
where $z(s)=x+(s-1/2)y$ with $z(0)=x_{2}$ and $z(1)=x_{1}$. Here
we adopted the notation for the field strength tensor $F_{\mu\nu}(z)=F_{\mu\nu}[z(s)]$ as a function
of space-time.

With Eqs.~(\ref{eq:kinetic-part}) and (\ref{eq:elze-formula}),
Eq.~(\ref{eq:Master_eq_pre_01}) is expressed in the gauge-invariant
two-point function,
\begin{equation}
\sigma^{\mu}\left\{ \frac{1}{2}i\hbar\frac{\partial}{\partial x^{\mu}}+i\hbar\frac{\partial}{\partial y^{\mu}}+y^{\nu}\int_{0}^{1}ds\,sF_{\mu\nu}[z(s)]\right\} \widetilde{S}(x,y)=0.\label{eq:kin-gauge-inv-1}
\end{equation}
We can expand $F_{\mu\nu}[z(s)]$ in powers of $y$, so the integral
of $F_{\mu\nu}$ becomes 
\begin{eqnarray}
\int_{0}^{1}ds\,sF_{\mu\nu}[z(s)] & = & \sum_{n=0}\frac{1}{n!}\int_{0}^{1}ds\,s\left(s-\frac{1}{2}\right)^{n}\left(y\cdot\partial_{x}\right)^{n}F_{\mu\nu}(x)\nonumber \\
 & = & \sum_{n=0}\frac{\left[(-1)^{n}+3+2n\right]}{4(n+2)!}\left(\frac{1}{2}y\cdot\partial_{x}\right)^{n}F_{\mu\nu}(x).
\end{eqnarray}
Equation (\ref{eq:kin-gauge-inv-1}) can be put into momentum space
conjugate to $y$ by Fourier transformation as 
\begin{equation}
\sigma\cdot\left(\frac{1}{2}i\hbar\nabla+\Pi\right)\widetilde{S}(x,p)=0,\label{eq:master-full}
\end{equation}
where the operators $\nabla$ and $\Pi$ are defined as 
\begin{eqnarray}
\nabla_{\mu} & = & \partial_{\mu}^{x}-j_{0}(\Delta)F_{\mu\nu}(x)\partial_{p}^{\nu},\nonumber \\
\Pi_{\mu} & = & p_{\mu}-\frac{1}{2}\hbar j_{1}(\Delta)F_{\mu\nu}(x)\partial_{p}^{\nu},\label{eq:operator}
\end{eqnarray}
where $\Delta=(1/2)\hbar\partial_{x}\cdot\partial_{p}$ is an operator
with $\partial_{x}$ acting only on $F_{\mu\nu}(x)$, $j_{0}(z)=\sin z/z$
and $j_{1}(z)=(\sin z-z\cos z)/z^{2}$ are spherical Bessel functions,
we have used the same symbol $\widetilde{S}$ to denote both $\widetilde{S}(x,y)$
and $\widetilde{S}(x,p)$ that are related through the Fourier (Wigner)
transformation,
\begin{eqnarray}
\widetilde{S}(x,p) & = & \int d^{4}y\exp\left(\frac{i}{\hbar}p\cdot y\right)\widetilde{S}(x,y),\nonumber \\
\widetilde{S}(x,y) & = & \int\frac{d^{4}p}{(2\pi\hbar)^{4}}\exp\left(-\frac{i}{\hbar}p\cdot y\right)\widetilde{S}(x,p).\label{eq:fourier-y}
\end{eqnarray}
Note that the above definition of the Wigner transformation differs
from the one in Refs.~\cite{Vasak:1987um,Gao:2012ix,Gao:2015zka,Gao:2018jsi,Gao:2018wmr,Weickgenannt:2019dks,Yang:2020mtz}
in that the factor $(2\pi\hbar)^{-4}$ appears in the coordinate integration
in $\widetilde{S}(x,p)$ instead of in $\widetilde{S}(x,y)$.

Equation (\ref{eq:master-full}) is the master equation, and our starting
point, which has a similar form to Eq.~(4.18) of Ref.~\cite{Vasak:1987um},
but it is a two-dimension equation with Pauli matrices while the latter
is a four-dimension equation with Dirac matrices. To be consistent
with the symbols used in some literature, in the rest part of this
section, we resume the use of $S(x,p)$ for $\widetilde{S}(x,p)$,
i.e., $S(x,p)\equiv S^{<}(x,p)\equiv\widetilde{S}(x,p)$. Then
Eq.~(\ref{eq:master-full}) is rewritten as 
\begin{equation}
\sigma\cdot\left(\frac{1}{2}i\hbar\nabla+\Pi\right)S(x,p)=0.\label{eq:master}
\end{equation}
We can also derive the conjugate
form of Eq.~(\ref{eq:master}) as 
\begin{equation}
\left(-\frac{1}{2}i\hbar\nabla+\Pi\right)S(x,p)\cdot\sigma=0.\label{eq:master-conjugate}
\end{equation}
Note that $S(x,p)$ is the Wigner function for right-handed (left-handed)
fermions whose equations of motion are Eqs.~(\ref{eq:master}) and
(\ref{eq:master-conjugate}). Alternatively we can rewrite Eqs.~(\ref{eq:master})
and (\ref{eq:master-conjugate}) in different forms by taking their
sum and difference, 
\begin{eqnarray}
\left[\sigma^{\mu},\hbar\nabla_{\mu}S(x,p)\right] & = & 2i\left\{ \sigma^{\mu},\Pi_{\mu}S(x,p)\right\} ,\nonumber \\
\left\{ \sigma^{\mu},\hbar\nabla_{\mu}S(x,p)\right\}  & = & 2i\left[\sigma^{\mu},\Pi_{\mu}S(x,p)\right].\label{eq:master-diff-sum}
\end{eqnarray}
The operators $\nabla$ and $\Pi$ can be expanded in even powers
of $\hbar$ as 
\begin{eqnarray}
\nabla^{\mu} & = & \sum_{n=0}^{\infty}\hbar^{2n}\nabla_{(2n)}^{\mu},\nonumber \\
\Pi^{\mu} & = & \sum_{n=0}^{\infty}\hbar^{2n}\Pi_{(2n)}^{\mu}.\label{eq:expansion-op}
\end{eqnarray}
For $n=0$, we have 
\begin{equation}
\nabla_{\mu}^{(0)}=\partial_{\mu}^{x}-F_{\mu\nu}\partial_{p}^{\nu},\;\;\Pi_{\mu}^{(0)}=p_{\mu}.
\end{equation}
For $n>0$, we have 
\begin{eqnarray}
\nabla_{(2n)}^{\mu} & = & (-1)^{n+1}\frac{1}{(2n+1)!}\left(\frac{1}{2}\hbar\partial_{x}\cdot\partial_{p}\right)^{2n}F_{\mu\nu}(x)\partial_{p}^{\nu},\nonumber \\
\Pi_{(2n)}^{\mu} & = & (-1)^{n}\frac{n}{(2n+1)!}\left(\frac{1}{2}\hbar\partial_{x}\cdot\partial_{p}\right)^{2n-1}F_{\mu\nu}(x)\partial_{p}^{\nu}. \label{eq:Pi_operator}
\end{eqnarray}
In constant background fields with $\partial_{x}F_{\mu\nu}=0$, the
operators $\nabla$ and $\Pi$ are just $\nabla_{\mu}^{(0)}$ and
$\Pi_{\mu}^{(0)}$ respectively,
\begin{equation}
\nabla_{\mu}^{(0)}=\partial_{\mu}^{x}-F_{\mu\nu}\partial_{p}^{\nu},\;\;\Pi_{\mu}^{(0)}=p_{\mu},\label{eq:constant-field-op}
\end{equation}
and Eqs.~(\ref{eq:master-simp}) and (\ref{eq:master-conjugate})
have simple forms 
\begin{eqnarray}
\sigma^{\mu}\left(\frac{1}{2}i\hbar\nabla_{\mu}^{(0)}+p_{\mu}\right)S(x,p) & = & 0,\label{eq:master-simp}\\
\left(-\frac{1}{2}i\hbar\nabla_{\mu}^{(0)}+p_{\mu}\right)S(x,p)\sigma^{\mu} & = & 0.\label{eq:master-simp-conj}
\end{eqnarray}
For notational simplicity, sometimes we will denote $\nabla_{\mu}^{(0)}$
and $\Pi_{\mu}^{(0)}$ simply as $\nabla$ and $\Pi$ respectively
if without ambiguity.

Note that we have suppressed the index ``$R$'' in $S(x,p)$ for
right-handed fermions. To obtain the equations of motion for left-handed
fermions, we simply replace $\sigma^{\mu}$ with $\overline{\sigma}^{\mu}$
in Eqs.~(\ref{eq:master})-(\ref{eq:master-diff-sum}), and Eqs.~(\ref{eq:master-simp})
and (\ref{eq:master-simp-conj}).

\newpage
\section{Master equations for chiral Wigner functions and their solutions \label{sec:CCKE}}

In this section, we will derive the master equations for chiral components of Wigner functions or chiral Wigner functions. Then we solve the master equations in equilibrium and constant fields. The chiral Wigner functions are found order by order in the Planck constant. 

\subsection{Master equations}

As we have seen in Sec.~\ref{sec:Master-equation} that the equation
of motion and the dynamics for chiral fermions can be simplified in
terms of Pauli spinors. As two-dimension matrices, right-handed and
left-handed components of Wigner functions (gauge-invariant two-point
functions) can be expanded in $\overline{\sigma}^{\mu}$ and $\sigma^{\mu}$
respectively,
\begin{eqnarray}
S_{R}(x,p) & = & \overline{\sigma}^{\mu}\mathscr{J}_{\mu}^{+},\nonumber \\
S_{L}(x,p) & = & \sigma^{\mu}\mathscr{J}_{\mu}^{-},\label{eq:wr-wl}
\end{eqnarray}
where $\mathscr{J}_{\mu}^{s}$ with $s=\pm$ represent right/left-handed
components of Wigner functions and can be extracted by taking following
traces,
\begin{equation}
\mathscr{J}_{\mu}^{+}=\frac{1}{2}\textrm{Tr }(\sigma_{\mu}S_{R}),\;\;\mathscr{J}_{\mu}^{-}=\frac{1}{2}\textrm{Tr }(\overline{\sigma}_{\mu}S_{L}),\label{eq:wrwl-1}
\end{equation}
where we have used $\textrm{Tr }(\sigma^{\mu}\overline{\sigma}^{\nu})=2\eta^{\mu\nu}$.
From the right-handed and left-handed components, one can obtain the
vector and axial-vector components,
\begin{equation}
\mathcal{V}_{\mu}=\mathscr{J}_{\mu}^{+}+\mathscr{J}_{\mu}^{-},\;\;\mathcal{A}_{\mu}=\mathscr{J}_{\mu}^{+}-\mathscr{J}_{\mu}^{-}.
\end{equation}

Inserting Eq.~(\ref{eq:wr-wl}) into Eq.~(\ref{eq:master-simp})
for right-handed fermions and into the corresponding equation for
left-handed fermions, we obtain 
\begin{eqnarray}
(\sigma^{\mu}\overline{\sigma}^{\nu})\left[\Pi_{\mu}+\frac{1}{2}i\hbar\nabla_{\mu}\right]\mathscr{J}_{\nu}^{+} & = & 0,\nonumber \\
(\overline{\sigma}^{\nu}\sigma^{\mu})\left[-\Pi_{\mu}+\frac{1}{2}i\hbar\nabla_{\mu}\right]\mathscr{J}_{\nu}^{+} & = & 0,\nonumber \\
(\overline{\sigma}^{\mu}\sigma^{\nu})\left[\Pi_{\mu}+\frac{1}{2}i\hbar\nabla_{\mu}\right]\mathscr{J}_{\nu}^{-} & = & 0,\nonumber \\
(\sigma^{\nu}\overline{\sigma}^{\mu})\left[-\Pi_{\mu}+\frac{1}{2}i\hbar\nabla_{\mu}\right]\mathscr{J}_{\nu}^{-} & = & 0,\label{eq:emo-sigma}
\end{eqnarray}
where $\nabla$ and $\Pi$ are just $\nabla_{\mu}^{(0)}$ and $\Pi_{\mu}^{(0)}$
in Eq.~(\ref{eq:constant-field-op}) respectively, i.e., the index
'(0)' is suppressed for notational simplicity. We can use the following
formula,
\begin{eqnarray}
\sigma^{\mu}\overline{\sigma}^{\nu} & = & \eta^{\mu\nu}-\frac{1}{2}i\epsilon^{\mu\nu\lambda\rho}\sigma_{\lambda}\overline{\sigma}_{\rho},\nonumber \\
\overline{\sigma}^{\nu}\sigma^{\mu} & = & \eta^{\mu\nu}+\frac{1}{2}i\epsilon^{\mu\nu\lambda\rho}\sigma_{\lambda}\overline{\sigma}_{\rho},\label{eq:sigma-mu-sigma-nu}
\end{eqnarray}
to simplify Eq.~(\ref{eq:emo-sigma}) and obtain equations for right-handed
and left-handed components of Wigner functions,
\begin{eqnarray}
\Pi^{\mu}\mathscr{J}_{\mu}^{s}(x,p) & = & 0,\label{eq:wig-eq-constrain-1}\\
\nabla^{\mu}\mathscr{J}_{\mu}^{s}(x,p) & = & 0,\label{eq:wig-eq-constrain-2}\\
2s\left(\Pi^{\mu}\mathscr{J}_{s}^{\nu}-\Pi^{\nu}\mathscr{J}_{s}^{\mu}\right) & = & -\hbar\epsilon^{\mu\nu\rho\sigma}\nabla_{\rho}\mathscr{J}_{\sigma}^{s},\label{eq:wig-eq}
\end{eqnarray}
where $s=\pm$ is the chirality index for right-handed ($+$) and
left-handed ($-$) fermions. Equations (\ref{eq:wig-eq-constrain-1})
and (\ref{eq:wig-eq-constrain-2}) can be obtained by taking traces
of Eq.~(\ref{eq:emo-sigma}), involving $\eta^{\mu\nu}$ terms according
to Eq.~(\ref{eq:sigma-mu-sigma-nu}). Equation (\ref{eq:wig-eq})
is the result of $\epsilon^{\mu\nu\lambda\rho}\sigma_{\lambda}\overline{\sigma}_{\rho}$
terms according to Eq.~(\ref{eq:sigma-mu-sigma-nu}), for the right-handed
sector, such term in the first line of Eq.~(\ref{eq:emo-sigma})
can be put into the form,
\begin{eqnarray}
0 & = & -\frac{1}{2}i\epsilon^{\mu\nu\lambda\rho}\sigma_{\lambda}\overline{\sigma}_{\rho}\left(\Pi_{\mu}\mathscr{J}_{\nu}^{+}+\frac{1}{2}i\hbar\nabla_{\mu}\mathscr{J}_{\nu}^{+}\right)\nonumber \\
 & = & \frac{1}{2}\sigma_{\lambda}\overline{\sigma}_{\rho}\left(\Pi^{\lambda}\mathscr{J}_{+}^{\rho}-\Pi^{\rho}\mathscr{J}_{+}^{\lambda}+\frac{1}{2}\hbar\epsilon^{\lambda\rho\mu\nu}\nabla_{\mu}\mathscr{J}_{\nu}^{+}\right),
\end{eqnarray}
where we have used Eq.~(\ref{eq:sigma-mu-sigma-nu}) to rewrite $\sigma_{\lambda}\overline{\sigma}_{\rho}$
in the first term $\epsilon^{\mu\nu\lambda\rho}\sigma_{\lambda}\overline{\sigma}_{\rho}\Pi_{\mu}\mathscr{J}_{\nu}^{+}$.
In the same way, we can also derive the equation for $\mathscr{J}_{\mu}^{-}$
of left-handed fermions.

We note that Eqs.~(\ref{eq:wig-eq-constrain-1})-(\ref{eq:wig-eq})
can be obtained from Eqs.~(5.7)-(5.21) of Ref.~\cite{Vasak:1987um}
when taking the massless limit. They were revisited and first solved
in the context of chiral effects in a constant background field~\cite{Gao:2012ix}.

\subsection{Recursive relation and semi-classical expansion in equilibrium and constant fields}

\label{sec:solutions-wigner}The current can be obtained from chiral
components of the Wigner function by an integration over the momentum,
\begin{equation}
J^{\mu}(x)=2\int\frac{d^{4}p}{(2\pi)^{4}}\mathscr{J}^{\mu}(x,p),\label{eq:current_def}
\end{equation}
where we have suppressed the chirality index $s=\pm$ for fermions
and we will keep doing so in cases of no ambiguity. Note that the
factor $1/(2\pi)^{4}$ in the definition of $J^{\mu}(x)$ which is
different from that in Refs.~\cite{Gao:2012ix,Gao:2015zka,Gao:2018jsi,Gao:2018wmr,Yang:2020mtz}
due to the convention of the Wigner function in Eq.~(\ref{eq:fourier-y}),
and factor 2 is introduced by convention that is consistent to
the definition of the charge density in quantum field theory.
Note that compared with Eq.~(\ref{eq:fourier-y}),
we took $\hbar=1$ in Eq.~(\ref{eq:current_def}) that does not contribute to the 
derivative expansion. Hereafter, the $\hbar$ that appears with $2\pi$ 
in the momentum integral is set to unity, 
while the $\hbar$ in the derivative expansion is kept.

Equations (\ref{eq:wig-eq-constrain-1})-(\ref{eq:wig-eq}) are still
very complicated and hard to solve generally. Since we assume a system
close to local equilibrium under a constant external field $F^{\mu\nu}$,
therefore, $\mathscr{J}_{\mu}(x,p)$ depend on $x$ through the fluid
velocity $u^{\mu}(x)$, the temperature $T(x)$, the chemical potential
$\mu(x)$ and the chiral chemical potential $\mu_{5}(x)$. The analytic
form of the Wigner function can be expressed in terms of \{$p^{\rho}$,
$F^{\rho\sigma}$, $\beta u^{\rho}$, $\beta\mu$, $\beta\mu_{5}$\}
with $\beta=1/T(x)$ being the inverse temperature. Note that the
chiral number density $n_{5}=n_{R}-n_{L}$ is not generally conserved due to explicit breaking by mass and quantum anomaly, so $\mu_{5}$
is not well defined. However, we can still introduce $\mu_{5}$ as an effective chemical potential if the time scale of chirality flipping is much larger than the time scale that we are interested in.
In QCD at high temperature, the chirality flipping rate is estimated from the sphaleron rate as $\Gamma_\mathrm{sphal}\sim \alpha_s^5 T^4$~\cite{McLerran:1990de,Arnold:1996dy, Son:2002sd, Moore:2010jd,Iatrakis:2015fma, Hou:2017szz}, where $\alpha_s$ is the strong coupling constant.
The time scale $T^3/\Gamma_\mathrm{sphal}$ is much larger than the typical time scale $1/(\alpha_s T)$ as the inverse of the damping rate \cite{Blaizot:1999xk} described by kinetic theory, so that $\mu_5$ is applicable. 
The chiral chemical potential in $\pi^0$ channel also works because 
the divergence of the chiral number current contains only electromagnetic fields,
i.e., there is no gluon contribution from sphaleron or instanton.
Throughout this review, we assume that $\mu_{5}$ vary slowly
and can be considered as a constant of time.

We further assume that the space-time derivative $\partial_{x}$ and
the field strength tensor $F_{\mu\nu}$ are small quantities at the
same order and can be used as power expansion parameters. 
The expansion in terms of $\partial_{x}$ and $F^{\mu\nu}$ are equivalent to $\hbar$ expansion in the context of Wigner functions in background EM fields since $\hbar$ always comes with $\partial_x$ that acts on $F^{\mu\nu}$ as shown in Eqs. (\ref{eq:operator}) and (\ref{eq:Pi_operator}).
Then $\mathscr{J}^{\mu}(x,p)$ can be expanded as 
\begin{equation}
\mathscr{J}^{\mu}(x,p)=\mathscr{J}_{(0)}^{\mu}(x,p)+\hbar\mathscr{J}_{(1)}^{\mu}(x,p)+\hbar^{2}\mathscr{J}_{(2)}^{\mu}(x,p)+\cdots,\label{eq:expansion-wig}
\end{equation}
where the indices $(n)$ with $n=0,1,2,\cdots$ denote the powers
of $(\partial_{x})^{n}$ and $(F_{\mu\nu})^{n}$. Note that $\mathscr{J}_{(n)}^{\mu}$
are related to $\mathscr{J}_{(n-1)}^{\mu}$ via Eq.~(\ref{eq:wig-eq})
for $n\geq1$, so Eq.~(\ref{eq:wig-eq}) serves as a recursive relation
which can be used to solve $\mathscr{J}_{(n)}^{\mu}$ order by order.

For a constant external field $F^{\mu\nu}$, we have $\nabla_{\mu}=\partial_{\mu}^{x}-F_{\mu\nu}\partial_{p}^{\nu}$
and $\Pi_{\mu}=p_{\mu}$ in Eqs.~(\ref{eq:wig-eq-constrain-1})-(\ref{eq:wig-eq}).
Contracting $p_{\lambda}$ with Eq.~(\ref{eq:wig-eq}) and using
Eq.~(\ref{eq:wig-eq-constrain-1}), we obtain 
\begin{equation}
p^{2}\mathscr{J}_{\mu}^{(n)}=\frac{s}{2}\epsilon_{\mu\nu\rho\sigma}p^{\nu}\nabla^{\rho}\mathscr{J}_{(n-1)}^{\sigma}.\label{eq:recursive-n-n-1}
\end{equation}
The general form of $\mathscr{J}_{(n)}^{\mu}$ is 
\begin{equation}
\mathscr{J}_{(n)}^{\mu}=p^{\mu}f_{(n)}\delta(p^{2})+X_{(n)}^{\mu}\delta(p^{2})+\frac{s}{2p^{2}}\epsilon^{\mu\nu\rho\sigma}p_{\nu}\nabla_{\rho}\mathscr{J}_{\sigma}^{(n-1)},\label{eq:general_sol_01}
\end{equation}
where $X_{(n)}^{\mu}$ satisfies from the constraint (\ref{eq:wig-eq-constrain-1}),
\begin{equation}
p^{\mu}X_{\mu}^{(n)}=0.
\end{equation}
Here $f_{(n)}$ and $X_{\mu}^{(n)}$ in Eq.~(\ref{eq:general_sol_01})
are functions of $x$ and $p$. We will see shortly that $f_{(0)}$
is actually the distribution function by mapping it to the ideal Fermi
gas. It seems that the last term in Eq.~(\ref{eq:general_sol_01})
is proportional to $1/p^{2}$, so it might diverge when $p^{2}$ is
small, and then the whole power counting scheme would fail. However
$1/p^{2}$ plays the same role as $\delta(p^{2})$, the mass shell
condition in the other two terms of Eq.~(\ref{eq:general_sol_01}). These
singularities in the Wigner function can be removed when obtaining
physical observables as space-time functions by taking momentum integrals.
One can also write Eq.~(\ref{eq:general_sol_01}) in another form
without $1/p^{2}$, see Ref.~\cite{Hidaka:2016yjf}.

\subsection{Zeroth order results}

At the leading or zeroth order $O(\hbar^{0})$, the right-hand side
of Eq.~(\ref{eq:wig-eq}) is vanishing. The solution to Eq.~(\ref{eq:wig-eq-constrain-1})
can be written in the form,
\begin{equation}
\mathscr{J}_{(0)}^{\rho}(x,p)=p^{\rho}f_{(0)}(x,p)\delta(p^{2}),\label{eq:j-0}
\end{equation}
where $f_{(0)}\equiv f_{s}^{(0)}$ denote phase space distributions
of massless fermions at the zeroth order, 
\begin{eqnarray}
f_{(0)}(x,p) & = & 2\pi\left\{ \Theta(p_{0})f_{\mathrm{FD}}(p_{0}-\mu_{s})\right. 
\left.+\Theta(-p_{0})\left[f_{\mathrm{FD}}(-p_{0}+\mu_{s})-1\right]\right\} .\label{eq:dist}
\end{eqnarray}
Here $p_{0}\equiv u\cdot p$, $f_{\mathrm{FD}}(y)\equiv1/[\exp(\beta y)+1]$
is the Fermi-Dirac distribution function, $\mu_{s}=\mu+s\mu_{5}$
is the chemical potential with $s=\pm1$ denoting the chirality, and
$\Theta(p_{0})$ is the step function with $\Theta(p_{0})=1$ for
$p_{0}>0$ and $\Theta(p_{0})=0$ for $p_{0}<0$. 
Actually one cannot determine the explicit form of the function $f_{\mathrm{FD}}(y)$ in Eq. (\ref{eq:dist}) since it can be in any other form provided $f_{(0)}$ satisfies global equilibrium conditions. 
The reason why we choose it to be the Fermi-Dirac distribution is that it is the distribution in global equilibrium for fermions which is consistent with global equilibrium conditions in Eqs. (\ref{eq:static-eq-01})-(\ref{eq:static-eq-03}).
Once we fix $f_{(0)}(x,p)$
to be the distribution function (\ref{eq:dist}), $X_{(0)}^{\mu}$
can be set to zero. Note that $f_{(0)}(x,p)$ in Eq.~(\ref{eq:dist})
differs from the same distribution in Refs.~\cite{Gao:2012ix,Gao:2015zka,Gao:2018jsi,Gao:2018wmr}
by the prefactor due to the convention of the Wigner function in Eq.~(\ref{eq:fourier-y})
and the factor 2 in Eq.~(\ref{eq:current_def}).
In Eq. (\ref{eq:dist}), we have included the vacuum term '$-1$'. The vacuum term does not contribute as an infinite constant at the first order in $\hbar$, e.g. in Eq. (\ref{eq:sol-first-order}), because there are derivatives acting on $f^{(0)}$. But for some quantities such as the energy-momentum tensor the vacuum contribution should be present, see subsection \ref{sec:EMT_2nd}. The vacuum term is essential to deriving the equation for the Wigner function from the equation of motion for the field. 

The zeroth-order solution (\ref{eq:j-0}) must satisfy Eq.~(\ref{eq:wig-eq-constrain-2}),
we obtain 
\begin{eqnarray}
\nabla_{\rho}\mathscr{J}_{(0)}^{\rho} & = & \delta(p^{2})p^{\rho}\nabla_{\rho}f_{(0)}\nonumber \\
 & = & \delta(p^{2})f_{(0)}^{\prime}\left[\frac{1}{2}p^{\rho}p^{\sigma}\left(\partial_{\rho}\beta_{\sigma}+\partial_{\sigma}\beta_{\rho}\right)
-p^{\rho}\partial_{\rho}\left(\overline{\mu}+s\overline{\mu}_{5}\right)-F_{\rho\sigma}p^{\rho}\beta^{\sigma}\right]=0,\label{eq:constraint-eq}
\end{eqnarray}
where we have used $f_{(0)}^{\prime}\equiv\partial f_{(0)}/\partial(\beta\cdot p)$,
$\beta^{\rho}\equiv\beta u^{\rho}$, $\overline{\mu}_{s}\equiv\beta\mu_{s}$,
$\overline{\mu}\equiv\beta\mu$, and $\overline{\mu}_{5}\equiv\beta\mu_{5}$.
In order for Eq.~(\ref{eq:constraint-eq}) to hold, the following
conditions must be fulfilled \cite{Gao:2012ix,Gao:2018jsi},
\begin{eqnarray}
\partial_{\rho}\beta_{\sigma}+\partial_{\sigma}\beta_{\rho} & = & 0,\label{eq:static-eq-01}\\
\partial_{\rho}\overline{\mu}+F_{\rho\sigma}\beta^{\sigma} & = & 0,\label{eq:static-eq-02}\\
\partial_{\rho}\overline{\mu}_{5} & = & 0.\label{eq:static-eq-03}
\end{eqnarray}
These conditions are global equilibrium conditions for a system under
static and uniform vorticity and electromagnetic field. The solutions
to Eqs.~(\ref{eq:static-eq-01})-(\ref{eq:static-eq-03}) are 
\begin{eqnarray}
\beta_{\mu} & = & -\Omega_{\mu\nu}x^{\nu} + c_\mu,\nonumber \\
\overline{\mu} & = & -\frac{1}{2}F^{\mu\lambda}x_{\lambda}\Omega_{\mu\nu}x^{\nu}+c,\nonumber \\
\overline{\mu}_{5} & = & c_{5},
\end{eqnarray}
where $c_\mu,c,c_5$ are constants and $\Omega_{\mu\nu}$ is the thermal vorticity tensor,
\begin{equation}
\Omega_{\mu\nu}=\frac{1}{2}\left(\partial_{\mu}\beta_{\nu}-\partial_{\nu}\beta_{\mu}\right).
\end{equation}
Equations (\ref{eq:static-eq-01}) and (\ref{eq:static-eq-02}) also
implies the integrability condition,
\begin{equation}
F_{\;\;\lambda}^{\mu}\Omega^{\nu\lambda}=F_{\;\;\lambda}^{\nu}\Omega^{\mu\lambda},\label{eq:int-cond}
\end{equation}
which can be proved via the identity $\partial_{\lambda}\partial_{\rho}\overline{\mu}=\partial_{\rho}\partial_{\lambda}\overline{\mu}$
and applying Eqs.~(\ref{eq:static-eq-02}) and (\ref{eq:static-eq-01}).

\subsection{First-order results}

We can obtain the first order solution by inserting the zeroth order
one (\ref{eq:j-0}) into the recursive relation (\ref{eq:general_sol_01}),
\begin{eqnarray}
\mathscr{J}_{(1)}^{\mu} & = & p^{\mu}f_{(1)}\delta(p^{2})+X_{(1)}^{\mu}\delta(p^{2})+\frac{s}{2p^{2}}\epsilon^{\mu\nu\rho\sigma}p_{\nu}\nabla_{\rho}\mathscr{J}_{\sigma}^{(0)}\nonumber \\
 & = & X_{(1)}^{\mu}\delta(p^{2})+s\widetilde{F}^{\mu\nu}p_{\nu}f_{(0)}\delta^{\prime}(p^{2}),\label{eq:j1-x1-f}
\end{eqnarray}
where we have chosen $f_{(1)}=0$, $\widetilde{F}^{\mu\nu}=(1/2)\epsilon^{\mu\nu\rho\sigma}F_{\rho\sigma}$
and $\delta^{\prime}(x)=-(1/x)\delta(x)$. The function $X_{(1)}^{\mu}(x,p)$
can be determined by inserting $\mathscr{J}_{(1)}^{\mu}$ into Eq.
(\ref{eq:wig-eq}),
\begin{eqnarray}
(p^{\mu}X_{(1)}^{\nu}-p^{\nu}X_{(1)}^{\mu})\delta(p^{2}) & = & -\frac{s}{2}\epsilon^{\mu\nu\rho\lambda}p_{\lambda}\delta(p^{2})\nabla_{\rho}f_{(0)}\nonumber \\
 & = & -\frac{s}{2}\epsilon^{\mu\nu\rho\lambda}f_{(0)}^{\prime}\delta(p^{2})p_{\lambda}p^{\sigma}\Omega_{\rho\sigma}\nonumber \\
 & = & -\frac{s}{2}\left[p^{\mu}\widetilde{\Omega}^{\nu\lambda}p_{\lambda}-p^{\nu}\widetilde{\Omega}^{\mu\lambda}p_{\lambda}\right]f_{(0)}^{\prime}\delta(p^{2}),\label{eq:x-mu}
\end{eqnarray}
where we have used equilibrium conditions (\ref{eq:static-eq-01})-(\ref{eq:static-eq-03})
and Schouten identity,
\begin{equation}
\epsilon^{\mu\nu\rho\lambda}p^{\sigma}+\epsilon^{\nu\rho\lambda\sigma}p^{\mu}+\epsilon^{\rho\lambda\sigma\mu}p^{\nu}+\epsilon^{\lambda\sigma\mu\nu}p^{\rho}+\epsilon^{\sigma\mu\nu\rho}p^{\lambda}=0.
\end{equation}
One can directly read out $X_{(1)}^{\mu}$ from Eq.~(\ref{eq:x-mu}),
\begin{equation}
X_{(1)}^{\mu}=-\frac{s}{2}\widetilde{\Omega}^{\mu\lambda}p_{\lambda}f_{(0)}^{\prime},\label{eq:sol-x-mu-1}
\end{equation}
where $\widetilde{\Omega}^{\mu\nu}=(1/2)\epsilon^{\mu\nu\rho\sigma}\Omega_{\rho\sigma}$.
So the first order solution to chiral components of Wigner functions
can be obtain by putting Eq.~(\ref{eq:sol-x-mu-1}) into Eq.~(\ref{eq:j1-x1-f}),
\begin{equation}
\mathscr{J}_{(1)}^{\mu}=-\frac{s}{2}\widetilde{\Omega}^{\mu\lambda}p_{\lambda}f_{(0)}^{\prime}\delta(p^{2})+s\widetilde{F}^{\mu\nu}p_{\nu}f_{(0)}\delta^{\prime}(p^{2}).\label{eq:sol-first-order}
\end{equation}
One can verify that the solution (\ref{eq:sol-first-order}) satisfies
Eq.~(\ref{eq:wig-eq-constrain-2}) under equilibrium conditions (\ref{eq:static-eq-01})-(\ref{eq:static-eq-03}).

The result in Eq.~(\ref{eq:sol-first-order}) is the first order
solution to the Wigner function for chiral fermions under the equilibrium
conditions (\ref{eq:static-eq-01})-(\ref{eq:static-eq-03}) and it
was first derived in Ref.~\cite{Gao:2012ix}. It has one term with
the electromagnetic field and one term with the vorticity field which
will give rise to the chiral magnetic effect (CME) and chiral vortical
effect (CVE) respectively.

\subsection{Second-order results}

The second order solution can also be obtained from the first order
one by the iterative equation (\ref{eq:general_sol_01}) with the
the first order solution (\ref{eq:sol-first-order}), 
\begin{eqnarray}
\mathscr{J}_{\mu}^{(2)} & = & X_{\mu}^{(2)}\delta(p^{2})+\frac{s}{2p^{2}}\epsilon_{\mu\nu\rho\sigma}p^{\nu}\nabla^{\rho}\mathscr{J}_{(1)}^{\sigma}\nonumber \\
 & = & \frac{1}{4p^{2}}\left(p_{\mu}\Omega_{\gamma\beta}p^{\beta}-p^{2}\Omega_{\gamma\mu}\right)\Omega^{\gamma\lambda}p_{\lambda}f_{(0)}^{\prime\prime}\delta(p^{2})\nonumber \\
 &  & +\frac{1}{(p^2)^{2}}\left(p_{\mu}F_{\gamma\beta}p^{\beta}-p^{2}F_{\gamma\mu}\right)\Omega^{\gamma\lambda}p_{\lambda}f_{(0)}^{\prime}\delta(p^{2})\nonumber \\
 &  & +\frac{2}{(p^2)^{3}}\left(p_{\mu}F_{\gamma\beta}p^{\beta}-p^{2}F_{\gamma\mu}\right)F^{\gamma\lambda}p_{\lambda}f_{(0)}\delta(p^{2}),\label{eq:second-order-solution}
\end{eqnarray}
where we have set $X_{\mu}^{(2)}=0$ since $X_{\mu}^{(2)}$ can be
constrained by inserting above $\mathscr{J}_{\mu}^{(2)}$ and $\mathscr{J}_{(1)}^{\mu}$
in Eq.~(\ref{eq:sol-first-order}) into Eq.~(\ref{eq:wig-eq}) at
$O(\hbar^{2})$ as 
\begin{eqnarray}
\left(p_{\mu}X_{\nu}^{(2)}-p_{\nu}X_{\mu}^{(2)}\right)\delta(p^{2}) & = & 0,\label{X-2}
\end{eqnarray}
leading to $X_{\mu}^{(2)}=0$. In deriving (\ref{X-2}) we have used
equilibrium conditions (\ref{eq:static-eq-01})-(\ref{eq:static-eq-03}).
Note that we see in Eq.~(\ref{eq:second-order-solution}) that the
number of derivatives of $f_{(0)}$ is the same as the number of the
vorticity tensors in each term of $\mathscr{J}_{\mu}^{(2)}$. The
second-order contribution (\ref{eq:second-order-solution}) was first
derived in Ref.~\cite{Yang:2020mtz}.

\subsection{Vector and axial currents up to second order}

With chiral components of Wigner functions in phase space, one can
obtain the chiral current densities $J_{s}^{\mu}(x)$ ($s=\pm$ for
right-handed and left-handed respectively) by integrating over four-momentum
for $\mathscr{J}_{s}^{\mu}(x,p)$ as done in Eq.~(\ref{eq:current_def}).
Here we have recovered the chirality index $s=\pm$. The vector (fermion
number or charge) and axial charge (charge) current densities are
linear combination of chiral ones,
\begin{eqnarray}
J^{\mu} & = & J_{+}^{\mu}+J_{-}^{\mu}=J_{(0)}^{\mu}+\hbar J_{(1)}^{\mu}+\hbar^{2}J_{(2)}^{\mu}+\cdots,\nonumber \\
J_{5}^{\mu} & = & J_{+}^{\mu}-J_{-}^{\mu}=J_{5,(0)}^{\mu}+\hbar J_{5,(1)}^{\mu}+\hbar^{2}J_{5,(2)}^{\mu}+\cdots,\label{eq:vector-axial-vector-current}
\end{eqnarray}
where we have expanded $J^{\mu}$ and $J_{5}^{\mu}$ in powers of
$\hbar$. Up to the second order the charge and axial charge currents
are obtained from chiral components of Wigner functions. The results
for the charge (vector) current of a fermion gas are given by 
\begin{eqnarray}
J_{(0)}^{\mu} & = & \rho u^{\mu},\nonumber \\
J_{(1)}^{\mu} & = & \xi\omega^{\mu}+\xi_{B}B^{\mu},\nonumber \\
J_{(2)}^{\mu} & = & -\frac{\mu}{2\pi^{2}}(\varepsilon^{2}+\omega^{2})u^{\mu}-\frac{1}{4\pi^{2}}(\varepsilon\cdot E+\omega\cdot B)u^{\mu}\nonumber \\
 &  & -\frac{C}{12\pi^{2}}(E^{2}+B^{2})u^{\mu}-\frac{1}{4\pi^{2}}\epsilon^{\mu\nu\rho\sigma}u_{\nu}E_{\rho}\omega_{\sigma}\nonumber \\
 &  & -\frac{C}{6\pi^{2}}\epsilon^{\mu\nu\rho\sigma}u_{\nu}E_{\rho}B_{\sigma},\label{eq:2nd-porder-current-1}
\end{eqnarray}
while the results for the axial charge (vector) current are given
by 
\begin{eqnarray}
J_{5,(0)}^{\mu} & = & \rho_{5}u^{\mu},\nonumber \\
J_{5,(1)}^{\mu} & = & \xi_{5}\omega^{\mu}+\xi_{B5}B^{\mu},\nonumber \\
J_{5,(2)}^{\mu} & = & -\frac{\mu_{5}}{2\pi^{2}}(\varepsilon^{2}+\omega^{2})u^{\mu}-\frac{C_{5}}{12\pi^{2}}(E^{2}+B^{2})u^{\mu} 
-\frac{C_{5}}{6\pi^{2}}\epsilon^{\mu\nu\rho\sigma}u_{\nu}E_{\rho}B_{\sigma},\label{eq:2nd-order-current-2}
\end{eqnarray}
where coefficients are listed in Table \ref{tab:coeff-j-j5}. In Eqs.~(\ref{eq:2nd-porder-current-1}) and (\ref{eq:2nd-order-current-2})
the currents are expressed in terms of electric and magnetic components
of field strength and vorticity tensors. As we have mentioned, to
find the solution to the Wigner function, the field strength tensor
$F^{\mu\nu}$ is assumed to be a constant in the lab frame. In the
local comoving frame of a fluid cell, $F^{\mu\nu}$ can be decomposed
into the electric and magnetic parts,
\begin{equation}
F_{\mu\nu}=E_{\mu}u_{\nu}-E_{\nu}u_{\mu}+\epsilon_{\mu\nu\rho\sigma}u^{\rho}B^{\sigma},
\end{equation}
where the electric and magnetic field are given by 
\begin{equation}
E_{\sigma}=F_{\sigma\rho}u^{\rho},\;B_{\sigma}=\frac{1}{2}\epsilon_{\sigma\mu\nu\rho}u^{\mu}F^{\nu\rho},\label{eq:em-field-u}
\end{equation}
which depend on $x$ via $u(x)$. Similarly we can decompose the vorticity
tensor into the electric and magnetic parts,
\begin{equation}
T\Omega_{\mu\nu}=\varepsilon_{\mu}u_{\nu}-\varepsilon_{\nu}u_{\mu}+\epsilon_{\mu\nu\rho\sigma}u^{\rho}\omega^{\sigma},\label{eq:vorticity-tensor}
\end{equation}
where the electric and magnetic part of the tensor can be extracted,
\begin{eqnarray}
\varepsilon^{\mu} & = & T\Omega^{\mu\nu}u_{\nu},\nonumber \\
\omega^{\mu} & = & T\widetilde{\Omega}^{\mu\nu}u_{\nu}=\frac{1}{2}T\epsilon^{\mu\nu\alpha\beta}u_{\nu}\partial_{\alpha}^{x}(\beta u_{\beta})\nonumber \\
 & = & \frac{1}{2}\epsilon^{\mu\nu\alpha\beta}u_{\nu}\partial_{\alpha}^{x}u_{\beta}.\label{eq:vorticity-decomp}
\end{eqnarray}
Note that the magnetic part is the vorticity vector $\omega^{\mu}$.
The first order currents $J_{(1)}^{\mu}$ and $J_{5,(1)}^{\mu}$ were
first obtained from the Wigner function in Ref.~\cite{Gao:2012ix},
which show that the currents can be generated by the magnetic and
vorticity field, i.e., the chiral magnetic and vortical effects.

As we see in the second-order vector current in Eq.~(\ref{eq:2nd-porder-current-1}),
the first three terms indicate that the charge density is modified
by quadratic terms $\varepsilon^{2}$, $\omega^{2}$, $E^{2}$, $B^{2}$,
$\varepsilon\cdot E$ and $\omega\cdot B$, while the last two terms
of Eq.~(\ref{eq:2nd-porder-current-1}) are the Hall currents induced
along the direction orthogonal to both $E^{\mu}$ and $\omega^{\nu}$
or that orthogonal to both $E^{\mu}$ and $B^{\nu}$ in the comoving
frame of the fluid cell. There is no Hall current induced by $\varepsilon^{\mu}$
and $\omega^{\nu}$. Note that the mixed Hall current $\epsilon^{\mu\nu\rho\sigma}u_{\nu}E_{\rho}\omega_{\sigma}$
is actually identical to $\epsilon^{\mu\nu\rho\sigma}u_{\nu}\varepsilon_{\rho}B_{\sigma}$
due to the integrability condition,
\begin{eqnarray}
\epsilon_{\mu\nu\rho\sigma}E^{\rho}\omega^{\sigma} & = & \epsilon_{\mu\nu\rho\sigma}\varepsilon^{\rho}B^{\sigma},\nonumber \\
\epsilon_{\mu\nu\rho\sigma}E^{\rho}\varepsilon^{\sigma} & = & -\epsilon_{\mu\nu\rho\sigma}\omega^{\rho}B^{\sigma}.\label{eq:integrability-1}
\end{eqnarray}

For the second-order axial current, the first and second terms in
Eq.~(\ref{eq:2nd-order-current-2}) indicate that the axial charge
density gets modified by quadratic terms $\varepsilon^{2}$, $\omega^{2}$,
$E^{2}$ and $B^{2}$ but not from mixed terms $\varepsilon\cdot E$
and $\omega\cdot B$ due to the symmetry, which is different from the
charge density. The last term in Eq.~(\ref{eq:2nd-order-current-2})
is the axial Hall current generated by $E^{\mu}$ and $B^{\nu}$ only,
but not by $E^{\mu}$ and $\omega^{\nu}$ or $\varepsilon^{\mu}$
and $B^{\nu}$, which is also different from the vector current. Like
the charge current, there is also no axial Hall current from $\varepsilon^{\mu}$
and $\omega^{\nu}$.

\begin{table}
\caption{Coefficients in the vector and axial vector current densities. Here
$C_{\pm}$ are defined as an integral in Eq.~(4.22) of Ref.~\cite{Yang:2020mtz},
$\zeta(x)$ is the Riemann zeta function and $\zeta^{\prime}(x)=d\zeta(x)/dx$.
The last two rows are the asymptotic values of $C$ and $C_{5}$ at
two limits $\overline{\mu}_{s}\equiv\mu_{s}/T\ll1$ (small chemical
potential) and $\overline{\mu}_{s}\gg1$ (large chemical potential).
Note that we have absorbed the fermion charge into the fields by convention.
In normal convention, $\xi_{B}$ and $\xi_{B5}$ have an additional
factor, the electric charge $Q$ of the fermion, while $C$ and $C_{5}$
have an additional factor $Q^{2}$. \label{tab:coeff-j-j5}}

\centering{}%
\begin{tabular}{|c|c|c|c|}
\hline 
 & $J^{\mu}$ &  & $J_{5}^{\mu}$\tabularnewline
\hline 
$\rho$ & $\frac{\mu}{3\pi^{2}}\left(\pi^{2}T^{2}+\mu^{2}+3\mu_{5}^{2}\right)$ & $\rho_{5}$  & $\frac{\mu_{5}}{3\pi^{2}}\left(\pi^{2}T^{2}+3\mu^{2}+\mu_{5}^{2}\right)$\tabularnewline
\hline 
$\xi$  & $\mu\mu_{5}/\pi^{2}$  & $\xi_{5}$  & $\frac{1}{6\pi^{2}}\left[\pi^{2}T^{2}+3(\mu^{2}+\mu_{5}^{2})\right]$\tabularnewline
\hline 
$\xi_{B}$  & $\mu_{5}/(2\pi^{2})$  & $\xi_{B5}$  & $\mu/(2\pi^{2})$\tabularnewline
\hline 
$C$  & $\frac{1}{2}(C_{+}+C_{-})$  & $C_{5}$  & $\frac{1}{2}(C_{+}-C_{-})$\tabularnewline
\hline 
$C$  & $-14\zeta^{\prime}(-2)\mu/T^{2},\;(\overline{\mu}_{s}\ll1)$  & $C_{5}$  & $-14\zeta^{\prime}(-2)\mu_{5}/T^{2},\;(\overline{\mu}_{s}\ll1)$\tabularnewline
\hline 
$C$  & $\mu/(\mu^{2}-\mu_{5}^{2}),\;(\overline{\mu}_{s}\gg1)$  & $C_{5}$  & $-\mu_{5}/(\mu^{2}-\mu_{5}^{2}),\;(\overline{\mu}_{s}\gg1)$\tabularnewline
\hline 
\end{tabular}
\end{table}

\subsection{Energy-momentum tensors up to second order} \label{sec:EMT_2nd}

The energy-momentum tensor can be obtained by the vector component
of the Wigner function as 
\begin{equation}
T^{\mu\nu}=2\int\frac{d^{4}p}{(2\pi)^{4}}\mathcal{V}^{\mu}p^{\nu}=2\int\frac{d^{4}p}{(2\pi)^{4}}(\mathscr{J}_{+}^{\mu}+\mathscr{J}_{-}^{\mu})p^{\nu},
\end{equation}
where the factor $(2\pi)^{-4}$ is absent in the definition of the
energy-momentum tensor in Refs.~\cite{Gao:2012ix,Gao:2015zka,Gao:2018jsi,Gao:2018wmr,Yang:2020mtz}
due to the convention of the Wigner function in Eq.~(\ref{eq:fourier-y}),
and the factor 2, as in Eq.~(\ref{eq:current_def}), is put to be
consistent with the definition of the charge density in quantum field
theory. Up to the second order, the energy-momentum tensor reads 
\begin{eqnarray}
T_{(0)}^{\mu\nu} & = & \epsilon u^{\mu}u^{\nu}-\frac{1}{3}\epsilon\Theta^{\mu\nu},\nonumber \\
T_{(1)}^{\mu\nu} & = & \rho_{5}\left(u^{\mu}\omega^{\nu}+u^{\nu}\omega^{\mu}\right)+\frac{\xi}{2}\left(u^{\mu}B^{\nu}+u^{\nu}B^{\mu}+\epsilon^{\mu\nu\alpha\beta}u_{\alpha}E_{\beta}\right)\nonumber \\
 &  & -\frac{1}{2}\rho_{5}\left(u^{\mu}\omega^{\nu}-u^{\nu}\omega^{\mu}-\epsilon^{\mu\nu\alpha\beta}u_{\alpha}\varepsilon_{\beta}\right),\nonumber \\
T_{(2)}^{\mu\nu} & = & T_{(2),\textrm{vv}}^{\mu\nu}+T_{(2),\textrm{ve}}^{\mu\nu}+T_{(2),\textrm{ee}}^{\mu\nu},\label{eq:t0-t1-t2}
\end{eqnarray}
where $\Theta^{\mu\nu}\equiv \eta ^{\mu\nu}-u^\mu u^\nu$ is a projector and 
$\epsilon$ is the energy density of a fermion gas given by 
\begin{equation}
\epsilon=\frac{T^{4}}{4\pi^{2}}\left[\frac{7}{15}\pi^{4}+2\pi^{2}\frac{\mu^{2}+\mu_{5}^{2}}{T^{2}}+\frac{\mu^{4}}{T^{4}}+6\frac{\mu^{2}\mu_{5}^{2}}{T^{4}}+\frac{\mu_{5}^{4}}{T^{4}}\right].
\end{equation}
We see in Eq.~(\ref{eq:t0-t1-t2}) that there are three parts in the
second-order contribution $T_{(2)}^{\mu\nu}$: the coupling terms
of vorticity-vorticity, vorticity-electromagnetic-field, and electromagnetic-field-electromagnetic-field,
which are denoted as ``vv'', ``ve'' and ``ee'' respectively.
These terms are given by 
\begin{eqnarray}
T_{(2),\textrm{vv}}^{\mu\nu} & = & -\frac{1}{2}\xi_{5}\left[3u^{\mu}u^{\nu}(\omega^{2}+\varepsilon^{2})-\Theta^{\mu\nu}(\omega^{2}+\varepsilon^{2})\right.\nonumber \\
 &  & \left.-2(u^{\mu}\epsilon^{\nu\alpha\beta\gamma}+u^{\nu}\epsilon^{\mu\alpha\beta\gamma})u_{\alpha}\varepsilon_{\beta}\omega_{\gamma}-2(u^{\mu}\epsilon^{\nu\alpha\beta\gamma}-u^{\nu}\epsilon^{\mu\alpha\beta\gamma})u_{\alpha}\varepsilon_{\beta}\omega_{\gamma}\right],\label{eq:t2-vv}\\
T_{(2),\textrm{ve}}^{\mu\nu} & = & -\frac{1}{2}\xi_{B5}\left[u^{\mu}u^{\nu}(\omega\cdot B+\varepsilon\cdot E)-(\omega^{\mu}B^{\nu}+E^{\mu}\varepsilon^{\nu})\right.\nonumber \\
 &  & \left.-(u^{\mu}\epsilon^{\nu\alpha\beta\gamma}+u^{\nu}\epsilon^{\mu\alpha\beta\gamma})u_{\alpha}E_{\beta}\omega_{\gamma}-2(u^{\mu}\epsilon^{\nu\alpha\beta\gamma}-u^{\nu}\epsilon^{\mu\alpha\beta\gamma})u_{\alpha}E_{\beta}\omega_{\gamma}\right],\label{eq:t2-ve}\\
T_{(2),\textrm{ee}}^{\mu\nu} & = & -\frac{1}{12}(\kappa_{+}^{\epsilon}+\kappa_{-}^{\epsilon})\left(\frac{1}{4}\eta^{\mu\nu}F_{\gamma\beta}F^{\gamma\beta}-F^{\gamma\mu}F_{\gamma}^{\;\;\nu}\right)\nonumber \\
 &  & +\frac{1}{24\pi^{2}}\left[u^{\mu}u^{\nu}E^{2}-\Theta^{\mu\nu}\left(E^{2}+2B^{2}\right)\right.\nonumber \\
 &  & +4\left(E^{\mu}E^{\nu}+B^{\mu}B^{\nu}\right)+3\left(u^{\mu}\epsilon^{\nu\alpha\beta\gamma}+u^{\nu}\epsilon^{\mu\alpha\beta\gamma}\right)u_{\alpha}E_{\beta}B_{\gamma}\nonumber \\
 &  & \left.+3\left(u^{\mu}\epsilon^{\nu\alpha\beta\gamma}-u^{\nu}\epsilon^{\mu\alpha\beta\gamma}\right)u_{\alpha}E_{\beta}B_{\gamma}\right],\label{eq:t2-ee}
\end{eqnarray}
where $\kappa_{s=\pm}^{\epsilon}$ is defined by 
\begin{eqnarray}
\kappa_{s}^{\epsilon} & = & \frac{4\pi^{(3-\epsilon)/2}T^{-\epsilon}}{\Gamma\left((3-\epsilon)/2\right)(2\pi)^{3-\epsilon}}\int_{0}^{\infty}\frac{dy}{y^{1+\epsilon}}\left[\frac{1}{e^{(y-\overline{\mu}_{s})}+1}+\frac{1}{e^{(y+\overline{\mu}_{s})}+1}-1\right].\label{kappa-s-epsilon}
\end{eqnarray}
Here $\epsilon$ is a small positive number in dimension regularization.
The coefficient of the first term of $T_{(2),\textrm{ee}}^{\mu\nu}$
contains logarithmic ultraviolet divergence at the limit $\epsilon\rightarrow0$,
which can be seen by an expansion of $\kappa_{s}^{\epsilon}$ in $\epsilon$
as 
\begin{eqnarray}
\kappa_{s}^{\epsilon} & = & -\frac{1}{\pi^{2}}\left[\frac{1}{\epsilon}+\ln2+\frac{1}{2}\ln\pi+\frac{1}{2}\psi\left(\frac{3}{2}\right)-\ln T+\hat{\kappa}_{s}\right],\label{kappa}
\end{eqnarray}
where $\psi(x)$ is the digamma function and $\hat{\kappa}_{s}$ is
given by 
\begin{eqnarray}
\hat{\kappa}_{s} & = & \int_{0}^{\infty}dy\,\ln y\ \frac{d}{dy}\left[\frac{1}{e^{(y-\overline{\mu}_{s})}+1}+\frac{1}{e^{(y+\overline{\mu}_{s})}+1}\right].\label{hatkappa}
\end{eqnarray}
The derivation of the second-order energy-momentum tensors in Eqs.~(\ref{eq:t2-vv})-(\ref{eq:t2-ee})
is given in Ref.~\cite{Yang:2020mtz}.

It is straightforward to verify with Eqs.~(\ref{eq:t0-t1-t2})-(\ref{eq:t2-ee})
that the trace of the energy-momentum tensor up to the second order
is free of trace anomaly,
\begin{equation}
\eta_{\mu\nu}T_{(0)}^{\mu\nu}=\eta_{\mu\nu}T_{(1)}^{\mu\nu}=\eta_{\mu\nu}T_{(2)}^{\mu\nu}=0.
\end{equation}
Note that when taking the trace of $T_{(2),\textrm{ee}}^{\mu\nu}$
we have used $\eta_{\mu\nu}\eta^{\mu\nu}=4-\epsilon$. The above result
is due to the fact that the electromagnetic field is assumed to be
a classical background field. An interesting observation is that the
divergent term in $T_{(2),\textrm{ee}}^{\mu\nu}$ is exactly the contribution
from the quantum correction of the electromagnetic field \cite{Peskin:1995ev}
but with a wrong sign. Adding the quantum correction, a renormalization
procedure, we arrive at the finite result,
\begin{eqnarray}
T_{(2),\textrm{ee,R}}^{\mu\nu} & = & \frac{1}{6\pi^{2}}\left(\hat{\kappa}_{+}+\hat{\kappa}_{-}+\ln\frac{\Lambda}{T}\right)\left(\frac{1}{4}\eta^{\mu\nu}F_{\gamma\beta}F^{\gamma\beta}-F^{\gamma\mu}{F_{\gamma}}^{\nu}\right)\nonumber \\
 &  & +\frac{1}{24\pi^{2}}\left[u^{\mu}u^{\nu}E^{2}-\Theta^{\mu\nu}\left(E^{2}+2B^{2}\right)\right.\nonumber \\
 &  & +4\left(E^{\mu}E^{\nu}+B^{\mu}B^{\nu}\right)+3\left(u^{\mu}\epsilon^{\nu\alpha\beta\gamma}+u^{\nu}\epsilon^{\mu\alpha\beta\gamma}\right)u_{\alpha}E_{\beta}B_{\gamma}\nonumber \\
 &  & \left.+3\left(u^{\mu}\epsilon^{\nu\alpha\beta\gamma}-u^{\nu}\epsilon^{\mu\alpha\beta\gamma}\right)u_{\alpha}E_{\beta}B_{\gamma}\right],\label{eq:renorm-t2-ee}
\end{eqnarray}
where $\Lambda$ is a renormalization scale in quantum electromagnetic
field into which all possible remaining constant terms in Eq.~(\ref{kappa})
have been absorbed. After removing the divergence in $T_{(2),\textrm{ee}}^{\mu\nu}$,
one can verify the trace anomaly,
\begin{equation}
\eta_{\mu\nu}T_{(2)}^{\mu\nu}=\eta_{\mu\nu}T_{(2),\textrm{ee,R}}^{\mu\nu}=\frac{1}{24\pi^{2}}F_{\mu\nu}F^{\mu\nu},
\end{equation}
which originates from the quantum correction of the electromagnetic
field.

\subsection{Conservation laws}

With $J^{\mu}$ and $J_{5}^{\mu}$ in Eq.~(\ref{eq:2nd-porder-current-1})
and (\ref{eq:2nd-order-current-2}), and $T^{\mu\nu}$ in Eqs.~(\ref{eq:t0-t1-t2})-(\ref{eq:t2-ee}),
we can derive conservation laws for these quantities in constant and
homogeneous electromagnetic fields under the conditions, (\ref{eq:static-eq-01})-(\ref{eq:static-eq-03}),
(\ref{eq:int-cond}) and (\ref{eq:integrability-1}) as \cite{Yang:2020mtz}
\begin{eqnarray}
\partial^{\mu}J_{\mu} & = & 0,\nonumber \\
\partial^{\mu}J_{\mu}^{5} & = & -\frac{1}{2\pi^{2}}E\cdot B,\nonumber \\
\partial^{\mu}T_{\mu\nu} & = & F_{\nu\mu}J^{\mu}.
\end{eqnarray}
The second-order correction to the axial current does not contribute
to the chiral anomaly as expected. The term with $\ln\Lambda/T$ in
Eq.~(\ref{eq:renorm-t2-ee}) is essential to conserve the energy-momentum
when there is vorticity.

\newpage
\section{Disentanglement of chiral Wigner functions and chiral kinetic equations}

\label{sec:dwf-theorem}The system of chiral fermions in an electromagnetic
background field can be described by the system of kinetic equations
(\ref{eq:wig-eq-constrain-1})-(\ref{eq:wig-eq}) for the Wigner function.
We see in these equations that the multiple components of the Wigner
function are entangled, while in the Boltzmann-like equation, the single-component
distribution is involved. It has been shown in Ref.~\cite{Gao:2018wmr}
that the Wigner function can be disentangled in the way that the system
of kinetic equations for the multiple components can be reduced to
one chiral kinetic equation (CKE) for a single-component distribution
function. The above result has been proved to any order of $\hbar$
\cite{Gao:2018wmr} and is called the theorem for the disentanglement
of chiral Wigner function or the DWF theorem. Note that no assumptions
such as equilibrium conditions for distribution functions and constant
electromagnetic fields are made in this section. So the DWF theorem
holds for general distribution functions and the electromagnetic background
field that can vary with space-time.

\subsection{DWF theorem}

\label{subsec:DWF-theorem}To understand the DWF theorem, it is more
transparent to rewrite Eqs.~(\ref{eq:wig-eq-constrain-1})-(\ref{eq:wig-eq})
in time and spatial components,
\begin{eqnarray}
\Pi^{0}\mathscr{J}^{0}-\boldsymbol{\Pi}\cdot\pmb{\mathscr{J}} & = & 0,\label{eq:constraint-1}\\
\nabla^{0}\mathscr{J}^{0}+\boldsymbol{\nabla}\cdot\pmb{\mathscr{J}} & = & 0,\label{eq:evo-1}\\
2s(\boldsymbol{\Pi}\times\pmb{\mathscr{J}}) & = & \hbar\left[\nabla^{0}\pmb{\mathscr{J}}+\boldsymbol{\nabla}\mathscr{J}^{0}\right],\label{eq:evo-2}\\
2s\left(\boldsymbol{\Pi}\mathscr{J}^{0}-\Pi^{0}\pmb{\mathscr{J}}\right) & = & -\hbar\boldsymbol{\nabla}\times\pmb{\mathscr{J}},\label{eq:constraint-2}
\end{eqnarray}
where we have used $\nabla^{\mu}=(\nabla^{0},-\boldsymbol{\nabla})$,
$\Pi^{\mu}=(\Pi^{0},\boldsymbol{\Pi})$, and $\mathscr{J}^{\mu}=(\mathscr{J}^{0},\pmb{\mathscr{J}})$.
As we have mentioned in Eqs.~(\ref{eq:expansion-op}) and (\ref{eq:expansion-wig})
that $\nabla^{\mu}$, $\Pi^{\mu}$ and $\mathscr{J}^{\mu}$ can be
expanded in powers of $\hbar$. Equations (\ref{eq:evo-1}) and (\ref{eq:evo-2})
having time derivatives through with $\nabla^{0}$ are evolution equations,
while Eqs.~(\ref{eq:constraint-1}) and (\ref{eq:constraint-2})
without time derivatives are constraint equations.

One can make an expansion of Eqs.~(\ref{eq:constraint-1})-(\ref{eq:constraint-2})
in powers of $\hbar$ through expansions of $\nabla^{\mu}$, $\Pi^{\mu}$
and $\mathscr{J}^{\mu}$. The evolution equation (\ref{eq:evo-1})
at $O(\hbar^{n})$ for $n=0,1,2$ reads 
\begin{eqnarray}
\nabla_{(0)}^{0}\mathscr{J}_{(0)}^{0}+\boldsymbol{\nabla}_{(0)}\cdot\pmb{\mathscr{J}}_{(0)} & = & 0\,,\nonumber \\
\nabla_{(0)}^{0}\mathscr{J}_{(1)}^{0}+\boldsymbol{\nabla}_{(0)}\cdot\pmb{\mathscr{J}}_{(1)} & = & 0\,,\nonumber \\
\nabla_{(0)}^{0}\mathscr{J}_{(2)}^{0}+\boldsymbol{\nabla}_{(0)}\cdot\pmb{\mathscr{J}}_{(2)} & = & -\nabla_{(2)}^{0}\mathscr{J}_{(0)}^{0}-\boldsymbol{\nabla}_{(2)}\cdot\pmb{\mathscr{J}}_{(0)}\,.\label{eq:evolution1}
\end{eqnarray}
The evolution equation (\ref{eq:evo-2}) at $O(\hbar^{n})$ for $n=-1,0,1,2$
reads 
\begin{eqnarray}
0 & = & 2s(\mathbf{p}\times\pmb{\mathscr{J}}_{(0)}),\nonumber \\
\nabla_{(0)}^{0}\pmb{\mathscr{J}}_{(0)}+\boldsymbol{\nabla}_{(0)}\mathscr{J}_{(0)}^{0} & = & 2s(\mathbf{p}\times\pmb{\mathscr{J}}_{(1)}),\nonumber \\
\nabla_{(0)}^{0}\pmb{\mathscr{J}}_{(1)}+\boldsymbol{\nabla}_{(0)}\mathscr{J}_{(1)}^{0} & = & 2s(\mathbf{p}\times\pmb{\mathscr{J}}_{(2)}+\boldsymbol{\Pi}_{(2)}\times\pmb{\mathscr{J}}_{(0)}),\nonumber \\
\nabla_{(0)}^{0}\pmb{\mathscr{J}}_{(2)}+\boldsymbol{\nabla}_{(0)}\mathscr{J}_{(2)}^{0}+\nabla_{(2)}^{0}\pmb{\mathscr{J}}_{(0)}+\boldsymbol{\nabla}_{(2)}\mathscr{J}_{(0)}^{0} & = & 2s(\mathbf{p}\times\pmb{\mathscr{J}}_{(3)}+\boldsymbol{\Pi}_{(2)}\times\pmb{\mathscr{J}}_{(1)})\,.\nonumber \\
\label{eq:evolution2}
\end{eqnarray}
We see in Eq.~(\ref{eq:evolution2}) that the first line is degenerated
to a constraint condition for $\pmb{\mathscr{J}}^{(0)}$ and that
$\mathbf{p}\times\pmb{\mathscr{J}}^{(n+1)}$ appears in the $n$-th
order equation. The constraint equation (\ref{eq:constraint-1}) at
$O(\hbar^{n})$ for $n=0,1,2$ read 
\begin{eqnarray}
p_{0}\mathscr{J}_{(0)}^{0}-\mathbf{p}\cdot\pmb{\mathscr{J}}_{(0)} & = & 0,\nonumber \\
p_{0}\mathscr{J}_{(1)}^{0}-\mathbf{p}\cdot\pmb{\mathscr{J}}_{(1)} & = & 0,\nonumber \\
p_{0}\mathscr{J}_{(2)}^{0}-\mathbf{p}\cdot\pmb{\mathscr{J}}_{(2)}+\Pi_{(2)}^{0}\mathscr{J}_{(0)}^{0}-\boldsymbol{\Pi}_{(2)}\cdot\pmb{\mathscr{J}}_{(0)} & = & 0\,,\label{eq:constraints1}
\end{eqnarray}
which provide mass-shell conditions at each order. The constraint
equation (\ref{eq:constraint-2}) at $O(\hbar^{n})$ for $n=-1,0,1,2$
read 
\begin{eqnarray}
0 & = & 2s(p_{0}\pmb{\mathscr{J}}_{(0)}-\mathbf{p}\mathscr{J}_{(0)}^{0}),\nonumber \\
\boldsymbol{\nabla}_{(0)}\times\pmb{\mathscr{J}}_{(0)} & = & 2s(p_{0}\pmb{\mathscr{J}}_{(1)}-\mathbf{p}\mathscr{J}_{(1)}^{0}),\nonumber \\
\boldsymbol{\nabla}_{(0)}\times\pmb{\mathscr{J}}_{(1)} & = & 2s(p_{0}\pmb{\mathscr{J}}_{(2)}-\mathbf{p}\mathscr{J}_{(2)}^{0}+\Pi_{(2)}^{0}\pmb{\mathscr{J}}_{(0)}-\boldsymbol{\Pi}_{(2)}\mathscr{J}_{(0)}^{0}),\nonumber \\
\boldsymbol{\nabla}_{(0)}\times\pmb{\mathscr{J}}_{(2)}+\boldsymbol{\nabla}_{(2)}\times\pmb{\mathscr{J}}_{(0)} & = & 2s(p_{0}\pmb{\mathscr{J}}_{(3)}-\mathbf{p}\mathscr{J}_{(3)}^{0}+\Pi_{(2)}^{0}\pmb{\mathscr{J}}_{(1)}-\boldsymbol{\Pi}_{(2)}\mathscr{J}_{(1)}^{0})\,.
\label{eq:constraints2}
\end{eqnarray}
We see in each equation of Eq.~(\ref{eq:constraints2}) there is
a term $p_{0}\pmb{\mathscr{J}}_{(n)}$ ($n=0,1,2,3$) and other terms
involving either $\pmb{\mathscr{J}}_{(m)}$ with $m<n$ or $\mathscr{J}_{(m)}^{0}$
with $m\leq n$. So we can express $\pmb{\mathscr{J}}$ in terms of
$\mathscr{J}_{0}$ order by order as 
\begin{eqnarray}
\pmb{\mathscr{J}}_{(0)} & = & \frac{\mathbf{p}}{p_{0}}\mathscr{J}_{(0)}^{0},\nonumber \\
\pmb{\mathscr{J}}_{(1)} & = & \frac{\mathbf{p}}{p_{0}}\mathscr{J}_{(1)}^{0}+\frac{s}{2p_{0}}\boldsymbol{\nabla}_{(0)}\times\pmb{\mathscr{J}}_{(0)},\nonumber \\
\pmb{\mathscr{J}}_{(2)} & = & \frac{\mathbf{p}}{p_{0}}\mathscr{J}_{(2)}^{0}+\frac{s}{2p_{0}}\boldsymbol{\nabla}_{(0)}\times\pmb{\mathscr{J}}_{(1)}\nonumber \\
 &  & -\frac{1}{p_{0}}\Pi_{(2)}^{0}\pmb{\mathscr{J}}_{(0)}+\frac{1}{p_{0}}\boldsymbol{\Pi}_{(2)}\mathscr{J}_{(0)}^{0},\nonumber \\
\pmb{\mathscr{J}}_{(3)} & = & \frac{\mathbf{p}}{p_{0}}\mathscr{J}_{(3)}^{0}+\frac{s}{2p_{0}}\boldsymbol{\nabla}_{(0)}\times\pmb{\mathscr{J}}_{(2)}-\frac{1}{p_{0}}\Pi_{(2)}^{0}\pmb{\mathscr{J}}_{(1)}\nonumber \\
 &  & +\frac{1}{p_{0}}\boldsymbol{\Pi}_{(2)}\mathscr{J}_{(1)}^{0}+\frac{s}{2p_{0}}\boldsymbol{\nabla}_{(2)}\times\pmb{\mathscr{J}}_{(0)}\,.\label{eq:j0-j}
\end{eqnarray}
In the first line of Eq.~(\ref{eq:j0-j}) we see that $\pmb{\mathscr{J}}_{(0)}$
is determined by $\mathscr{J}_{(0)}^{0}$. As shown in the second
of Eq.~(\ref{eq:j0-j}), $\pmb{\mathscr{J}}_{(1)}$ depends on $\mathscr{J}_{0}^{(0)}$
through $\pmb{\mathscr{J}}^{(0)}$ and $\mathscr{J}_{(1)}^{0}$. Similarly
$\pmb{\mathscr{J}}^{(2)}$ is determined by $(\mathscr{J}_{0}^{(0)},\mathscr{J}_{0}^{(1)},\mathscr{J}_{0}^{(2)})$,
and $\pmb{\mathscr{J}}^{(3)}$ is determined by $(\mathscr{J}_{0}^{(0)},\mathscr{J}_{0}^{(1)},\mathscr{J}_{0}^{(2)},\mathscr{J}_{0}^{(3)})$.
So $\pmb{\mathscr{J}}^{(n)}$ is determined by $\mathscr{J}_{0}$
up to the order $n$.

Let us look at the evolution equation. At the zeroth order, one evolution
equation is the first line of Eq.~(\ref{eq:evolution1}). Another
evolution equation is from the second equation of Eq.~(\ref{eq:evolution2})
by inserting the second equation of Eq.~(\ref{eq:j0-j}), 
\begin{equation}
\nabla_{(0)}^{0}\pmb{\mathscr{J}}_{(0)}+\boldsymbol{\nabla}_{(0)}\mathscr{J}_{(0)}^{0}=\frac{1}{p_{0}}\mathbf{p}\times\left(\boldsymbol{\nabla}_{(0)}\times\pmb{\mathscr{J}}_{(0)}\right).\label{eq:second-evo-0th}
\end{equation}
One can show that Eq.~(\ref{eq:second-evo-0th}) is consistent with
the first line of Eq.~(\ref{eq:evolution1}) and therefore redundant.
At the first order, one evolution equation is the second line of Eq.~(\ref{eq:evolution1}).
Another evolution equation can be obtained by inserting the third
equation of Eq.~(\ref{eq:j0-j}) into the third equation of Eq.~(\ref{eq:evolution2}).
One can prove that these two evolution equations are consistent with
each other, so at least one equation is redundant. The same result
can be proved to any order of $\hbar$ \cite{Gao:2018wmr}.

We can solve $\mathscr{J}_{(i)}^{0}$ order by order through mass-shell
conditions, which can be done by substituting Eq.~(\ref{eq:j0-j})
into Eq.~(\ref{eq:constraints1}). At the zeroth order, we obtain
from the first equation of Eq.~(\ref{eq:constraints1}) by using
the first equation of Eq.~(\ref{eq:j0-j}), 
\begin{equation}
\mathscr{J}_{(0)}^{0}=p_{0}f_{(0)}\delta(p^{2}),\label{eq:0th-j0}
\end{equation}
where $f_{(0)}$ is an arbitrary scalar function of $x$ and $p$
without singularity at $p^{2}=0$ and can be determined by the first
evolution equation of Eq.~(\ref{eq:evolution1}). At the first order,
we insert the second equation of Eq.~(\ref{eq:j0-j}) into the second
equation of Eq.~(\ref{eq:constraints1}) and obtain 
\begin{equation}
\mathscr{J}_{(1)}^{0}=p_{0}f_{(1)}\delta(p^{2})+s(\mathbf{p}\cdot\mathbf{B})f_{(0)}\delta^{\prime}(p^{2}),\label{eq:j0-(1)}
\end{equation}
where $f_{(1)}$ is an arbitrary scalar function of $x$ and $p$
without singularity at $p^{2}=0$ and can be determined by the second
evolution equation of Eq.~(\ref{eq:evolution1}). Here we have used
the derivative of a delta function $\delta^{\prime}(y)\equiv d\delta(y)/dy=-\delta(y)/y$.
Similarly one can derive $\mathscr{J}_{(2)}^{0}$ as 
\begin{eqnarray}
\mathscr{J}_{(2)}^{0} & = & p_{0}f_{(2)}\delta(p^{2})+s(\mathbf{p}\cdot\mathbf{B})f_{(1)}\delta^{\prime}(p^{2})+\frac{(\mathbf{p}\cdot\mathbf{B})^{2}}{2p_{0}}f_{(0)}\delta^{\prime\prime}(p^{2})\nonumber \\
 &  & +\frac{1}{4p^{2}}\mathbf{p}\cdot\left\{ \boldsymbol{\nabla}_{(0)}\times\left[\frac{1}{p_{0}}\boldsymbol{\nabla}_{(0)}\times\left(\mathbf{p}f_{(0)}\delta(p^{2})\right)\right]\right\} -\frac{p_{0}}{p^{2}}\Pi_{\mu}^{(2)}p^{\mu}f_{(0)}\delta(p^{2})\nonumber \\
 &  & +\frac{1}{p^{2}}\mathbf{p}\cdot\left(\boldsymbol{\Pi}_{(2)}p_{0}-\Pi_{(2)}^{0}\mathbf{p}\right)f_{(0)}\delta(p^{2})\,,\label{eq:j0-(2)}
\end{eqnarray}
where $f_{(2)}$ is an arbitrary scalar function of $x$ and $p$
without singularity at $p^{2}=0$ and can be determined by the third
evolution equation of Eq.~(\ref{eq:evolution1}).

We can combine $\mathscr{J}_{(0)}^{0}$, $\mathscr{J}_{(1)}^{0}$
and the first three terms of $\mathscr{J}_{(2)}^{0}$ to obtain $\mathscr{J}_{0}$
up to the second order,
\begin{equation}
\mathscr{J}^{0}\approx p_{0}f(x,p)\delta\left(p^{2}+s\hbar\frac{\mathbf{p}\cdot\mathbf{B}}{p_{0}}\right),
\end{equation}
where $f(x,p)\equiv f_{(0)}(x,p)+\hbar f_{(1)}(x,p)+\hbar^{2}f_{(2)}(x,p)$
is the off-shell distribution function up to $O(\hbar^{2})$. We see
that to the first order the energy poles have been shifted to 
\begin{equation}
E_{p}^{(\pm)}=\pm E_{p}(1\mp s\hbar\mathbf{B}\cdot\boldsymbol{\Omega}_{p}),
\end{equation}
where $E_{p}\equiv E_{\mathbf{p}}=|\mathbf{p}|$ is the energy of the free fermion and
$\boldsymbol{\Omega}_{p}\equiv\boldsymbol{\Omega}_{\mathbf{p}}=\mathbf{p}/(2|\mathbf{p}|^{3})$ is
the Berry curvature in momentum space. The energy correction can be
regarded as the magnetic moment energy of chiral fermions \cite{Son:2012zy,Manuel:2013zaa,Chen:2014cla,Gao:2015zka}.

\subsection{Chiral kinetic equation for on-shell fermions}

\label{subsec:Chiral-kinetic-equation}In the former subsection we
have shown that all components of the chiral Wigner function can be
expressed by its time component in the background electromagnetic
field order by order in $\hbar$. In this subsection, we will derive
the CKE in terms of the single-component distribution function from
the evolution equations in Eq.~(\ref{eq:evolution1}).

We start from the zeroth order equation, the first equation of Eq.~(\ref{eq:evolution1}),
\begin{eqnarray}
\nabla_{\mu}^{(0)}\mathscr{J}_{(0)}^{\mu} & = & \left(\partial_{\mu}^{x}-F_{\mu\nu}\partial_{p}^{\nu}\right)\mathscr{J}_{(0)}^{\mu}\nonumber \\
 & = & \left(\partial_{\mu}^{x}-F_{\mu\nu}\partial_{p}^{\nu}\right)[p^{\mu}f_{(0)}\delta(p^{2})]\nonumber \\
 & = & \left(\partial_{t}+\mathbf{E}\cdot\nabla_{p}\right)[p_{0}f_{(0)}\delta(p^{2})]\nonumber \\
 &  & +\left(\nabla_{x}+\mathbf{E}\partial_{p_{0}}+\mathbf{B}\times\nabla_{p}\right)\cdot[\mathbf{p}f_{(0)}\delta(p^{2})]\nonumber \\
 & = & 0,\label{eq:0th-cke-off-shell}
\end{eqnarray}
where we have used $\mathscr{J}_{(0)}^{\mu}=p^{\mu}f_{(0)}\delta(p^{2})$
from Eqs.~(\ref{eq:j0-j}) and (\ref{eq:0th-j0}). The chiral kinetic
equation for on-shell fermions or anti-fermions can be obtained by
an integration of Eq.~(\ref{eq:0th-cke-off-shell}) over $p_{0}$
in the range $[0,\infty)$ or $(-\infty,0]$ respectively. Then for
on-shell fermions, the chiral kinetic equation reads 
\begin{equation}
\left(\partial_{t}+\mathbf{v}\cdot\nabla_{x}\right)f_{(0)}+(\mathbf{E}+\mathbf{v}\times\mathbf{B})\cdot\nabla_{p}f_{(0)}=0,
\end{equation}
where $\mathbf{v}=\mathbf{p}/E_{p}$ is the velocity of the fermion.
In deriving the above equation, the term with $\partial_{p_{0}}$
in Eq.~(\ref{eq:0th-cke-off-shell}) gives a vanishing result in
the integration over $p_{0}$ from the boundary condition.

Before the integration over $p_{0}$, the first order equation is
provided by the second equation of Eq.~(\ref{eq:evolution1}). We
can add the zeroth and first order equations, i.e. the first two equations
of Eq.~(\ref{eq:evolution1}), and obtain the evolution equation
up to $O(\hbar)$, 
\begin{equation}
p_{\mu}\nabla_{(0)}^{\mu}\left[f\delta(\widetilde{p}^{2})\right]+\hbar\frac{s}{2}\boldsymbol{\nabla}_{(0)}\cdot\left\{ \frac{1}{p_{0}}\boldsymbol{\nabla}_{(0)}\times\left[\mathbf{p}f_{(0)}\delta(p^{2})\right]\right\} =0,\label{eq:1st-cke-off}
\end{equation}
where $\widetilde{p}^{2}\equiv p^{2}+s\hbar(\mathbf{p}\cdot\mathbf{B})/p_{0}$
and $f\equiv f_{(0)}(x,p)+\hbar f_{(1)}(x,p)$ is the off-shell distribution
function up to $O(\hbar)$. After the integration of the above equation
over $p_{0}$ in the range $[0,\infty)$, we obtain the CKE for on-shell
fermions (not antifermions) with non-vanishing momenta up to $O(\hbar)$
\cite{Son:2012zy,Manuel:2013zaa,Manuel:2014dza,Hidaka:2016yjf,Huang:2018wdl},
\begin{eqnarray}
\left(1+\hbar s\boldsymbol{\Omega}_{p}\cdot\mathbf{B}\right)\partial_{t}f(x,E_{p},\mathbf{p})\nonumber \\
+\left[\mathbf{v}+\hbar s(\mathbf{E}\times\boldsymbol{\Omega}_{p})+\hbar s\frac{1}{2|\mathbf{p}|^{2}}\mathbf{B}\right]\cdot\boldsymbol{\nabla}_{x}f(x,E_{p},\mathbf{p})\nonumber \\
+\left[\widetilde{\mathbf{E}}+\mathbf{v}\times\mathbf{B}+\hbar s(\mathbf{E}\cdot\mathbf{B})\boldsymbol{\Omega}_{p}\right]\cdot\boldsymbol{\nabla}_{p}f(x,E_{p},\mathbf{p}) & = & 0,\label{eq:cke-1}
\end{eqnarray}
where $\widetilde{\mathbf{E}}\equiv\mathbf{E}-\boldsymbol{\nabla}_{x}E_{p}^{(+)}$
is the effective electric field, $\mathbf{v}\equiv\boldsymbol{\nabla}_{p}E_{p}^{(+)}$
is the effective velocity, and $f(x,E_{p},\mathbf{p})$ is the on-shell
distribution function up to $O(\hbar)$.

Note that in the integration over $p_{0}$ in the range $[0,\infty)$
there are some infrared singular terms from the derivative in $p_{0}$
that cannot be dropped causally. It turns out that two additional
terms in the above CKE appear, which are singular at $|\mathbf{p}|=0$
but were previously neglected \cite{Gao:2018wmr}. These new terms
do not contribute to the CKE at non-vanishing momentum, but it is
present in the anomalous conservation equation obtained by an integration
over $p_{0}$ in the full range $(-\infty,\infty)$ \cite{Gao:2018wmr}.
This may imply that the chiral anomaly may arise from another source
other than the well-known Berry phase in three-momentum, which seems
to be consistent with the observation of Ref.~\cite{Mueller:2017lzw}.

\subsection{Wigner functions in a general Lorentz frame}

\label{subsec:WF_frame}In Secs~\ref{subsec:DWF-theorem} and
\ref{subsec:Chiral-kinetic-equation}, we work in a specific Lorentz
frame. It is straightforward to introduce a time-like frame vector
$n^{\mu}$ to rewrite all formulas in a general Lorentz frame, where
$n^{\mu}$ behaves like a four-velocity satisfying $n^{2}=1$. For
simplicity, we assume $n^{\mu}$ is a constant vector. Then we can
decompose any vector $X^{\mu}$ into the component parallel to $n^{\mu}$
and that perpendicular to $n^{\mu}$, $X^{\mu}=(X\cdot n)n^{\mu}+\overline{X}^{\mu}$
with $\overline{X}\cdot n=0$. For $n^{\mu}=(1,0,0,0)$ we have the
normal decomposition: $X\cdot n=X_{0}$ and $\overline{X}^{\mu}=(0,\mathbf{X})$.
In such a decomposition, we can rewrite Eqs.~(\ref{eq:evolution1})-(\ref{eq:constraints2})
in a general Lorentz frame with the frame vector $n^{\mu}$. Corresponding
to Eqs.~(\ref{eq:j0-j}), (\ref{eq:0th-j0}), and (\ref{eq:j0-(1)}),
the zeroth and first-order solutions to the chiral Wigner function
in a general Lorentz frame read 
\begin{eqnarray}
\mathscr{J}_{(0)}^{\mu} & = & p^{\mu}f_{(0)}\delta(p^{2}),\nonumber \\
\mathscr{J}_{(1)}^{\mu} & = & p^{\mu}\frac{n\cdot\mathscr{J}_{(1)}}{n\cdot p}-\frac{s}{2(n\cdot p)}\epsilon^{\mu\nu\rho\sigma}n_{\nu}\nabla_{\sigma}\mathscr{J}_{(0)\rho}\nonumber \\
 & = & p^{\mu}f_{(1)}\delta(p^{2})-sp^{\mu}\frac{1}{n\cdot p}\widetilde{F}_{\lambda\nu}n^{\nu}p^{\lambda}f_{(0)}\delta^{\prime}(p^{2})\nonumber \\
 &  & -\frac{s}{2(n\cdot p)}\epsilon^{\mu\nu\rho\sigma}n_{\nu}\nabla_{\sigma}\mathscr{J}_{(0)\rho},\label{eq:j0-mu-j1-mu}
\end{eqnarray}
where we have used 
\begin{equation}
\frac{n\cdot\mathscr{J}_{(1)}}{n\cdot p}=f_{(1)}\delta(p^{2})-s\frac{1}{n\cdot p}\widetilde{F}_{\lambda\nu}n^{\nu}p^{\lambda}f_{(0)}\delta^{\prime}(p^{2}).\label{eq:n-j1-div-n-p}
\end{equation}
Then the chiral Wigner function up to the first order in $\hbar$
has the form,
\begin{eqnarray}
\mathscr{J}^{\mu} & = & \mathscr{J}_{(0)}^{\mu}+\hbar\mathscr{J}_{(1)}^{\mu}\nonumber \\
 & \approx & p^{\mu}f\delta\left(p^{2}-\hbar s\frac{1}{n\cdot p}\widetilde{F}_{\lambda\nu}n^{\nu}p^{\lambda}\right)\nonumber \\
 &  & -\hbar s\frac{1}{n\cdot p}\widetilde{F}^{\mu\lambda}n_{\lambda}f_{(0)}\delta(p^{2})\nonumber \\
 &  & -\hbar s\frac{1}{2(n\cdot p)}\epsilon^{\mu\nu\rho\sigma}n_{\nu}p_{\rho}\nabla_{\sigma}[f_{(0)}\delta(p^{2})],
\end{eqnarray}
where $f\equiv f_{(0)}+\hbar f_{(1)}$.

One has a freedom to choose another frame with a different velocity
$n_{\mu}^{\prime}$, then $\mathscr{J}_{(0)}^{\mu}$ and $\mathscr{J}_{(1)}^{\mu}$
in Eq.~(\ref{eq:j0-mu-j1-mu}) have the form,
\begin{eqnarray}
\mathscr{J}_{(0)}^{\prime\mu} & = & p^{\mu}f_{(0)}\delta(p^{2}),\nonumber \\
\mathscr{J}_{(1)}^{\prime\mu} & = & p^{\mu}f_{(1)}^{\prime}\delta(p^{2})-sp^{\mu}\frac{1}{n^{\prime}\cdot p}\widetilde{F}_{\lambda\nu}n^{\prime\nu}p^{\lambda}f_{(0)}\delta^{\prime}(p^{2})\nonumber \\
 &  & -\frac{s}{2(n^{\prime}\cdot p)}\epsilon^{\mu\nu\rho\sigma}n_{\nu}^{\prime}\nabla_{\sigma}\mathscr{J}_{(0)\rho},\label{eq:j0-mu-j1-mu-1}
\end{eqnarray}
where $f_{(1)}^{\prime}$ is the first-order distribution function
corresponding to the frame $n_{\mu}^{\prime}$ (the prime does not
denote the derivative). One can verify both $\mathscr{J}_{(0)}^{\mu}$
and $\mathscr{J}_{(1)}^{\mu}$ are independent of the choice of the
frame vector $n^{\mu}$, 
\begin{eqnarray}
\mathscr{J}_{(0)}^{\prime\mu}-\mathscr{J}_{(0)}^{\mu} & = & 0,\nonumber \\
\mathscr{J}_{(1)}^{\prime\mu}-\mathscr{J}_{(1)}^{\mu} & = & 0.\label{eq:frame-dep}
\end{eqnarray}
The first equation of Eq.~(\ref{eq:frame-dep}) is obvious since
$\mathscr{J}_{(0)}^{\mu}$ does not have an explicit dependence on
$n^{\mu}$, which also means that $f_{(0)}$ does not have the frame
dependence. It takes more effort to prove the second equation of Eq.~(\ref{eq:frame-dep})
\cite{Gao:2018wmr} which implies that $f_{(1)}$ is not invariant
with respect to the frame change. The change of $n\cdot\mathscr{J}_{(1)}/(n\cdot p)$
has the form,
\begin{eqnarray}
\delta\left(\frac{n\cdot\mathscr{J}_{(1)}}{n\cdot p}\right) & \equiv & \frac{n^{\prime}\cdot\mathscr{J}_{(1)}}{n^{\prime}\cdot p}-\frac{n\cdot\mathscr{J}_{(1)}}{n\cdot p}\nonumber \\
 & = & \frac{n_{\lambda}n_{\nu}^{\prime}\left(p^{\lambda}\mathscr{J}_{(1)}^{\nu}-p^{\nu}\mathscr{J}_{(1)}^{\lambda}\right)}{(n\cdot p)(n^{\prime}\cdot p)}\nonumber \\
 & = & -s\frac{\epsilon^{\lambda\nu\rho\sigma}n_{\lambda}n_{\nu}^{\prime}\nabla_{\rho}\mathscr{J}_{(0)\sigma}}{2(n\cdot p)(n^{\prime}\cdot p)},\label{eq:diff-n-j1-div-n-p}
\end{eqnarray}
where we have used Eq.~(\ref{eq:wig-eq}) at the first order. On
the other hand the change of $n\cdot\mathscr{J}_{(1)}/(n\cdot p)$
has two contributions according to Eq.~(\ref{eq:n-j1-div-n-p}),
\begin{equation}
\delta\left(\frac{n\cdot\mathscr{J}_{(1)}}{n\cdot p}\right)=\delta f_{(1)}\delta(p^{2})-s\left(\frac{\widetilde{F}_{\lambda\nu}n^{\prime\nu}}{n^{\prime}\cdot p}-\frac{\widetilde{F}_{\lambda\nu}n^{\nu}}{n\cdot p}\right)p^{\lambda}f_{(0)}\delta^{\prime}(p^{2}),\label{eq:diff-n-j1-div-n-p-1}
\end{equation}
where we have defined $\delta f_{(1)}\equiv f_{(1)}^{\prime}-f_{(1)}$.
From Eqs.~(\ref{eq:diff-n-j1-div-n-p}) and (\ref{eq:diff-n-j1-div-n-p-1}),
we obtain the variation of the distribution function due to the change
of the frame as 
\begin{eqnarray}
\delta f_{(1)}\delta(p^{2}) & = & s\left(\frac{\widetilde{F}_{\lambda\nu}n^{\prime\nu}}{n^{\prime}\cdot p}-\frac{\widetilde{F}_{\lambda\nu}n^{\nu}}{n\cdot p}\right)p^{\lambda}f_{(0)}\delta^{\prime}(p^{2})\nonumber \\
 &  & -s\frac{\epsilon^{\lambda\nu\rho\sigma}n_{\lambda}n_{\nu}^{\prime}\nabla_{\rho}[p_{\sigma}f_{(0)}\delta(p^{2})]}{2(n\cdot p)(n^{\prime}\cdot p)}\nonumber \\
 & = & -s\frac{\epsilon^{\lambda\nu\rho\sigma}n_{\lambda}n_{\nu}^{\prime}p_{\sigma}}{2(n\cdot p)(n^{\prime}\cdot p)}\delta(p^{2})\nabla_{\rho}f_{(0)},\label{eq:delta-f1}
\end{eqnarray}
where only the term proportional to $\nabla_{\rho}f_{(0)}$ survives
while other terms cancel in the second equality. Equation (\ref{eq:delta-f1})
is related to the side-jump effect \cite{Chen:2015gta,Hidaka:2016yjf}.

\subsection{Decomposition of CME and CVE current}

\label{sec:frame-decomposition}In Sec.~\ref{sec:solutions-wigner}
we derive the chiral Wigner function in the covariant form up to the second
order in $\hbar$ under the equilibrium conditions (\ref{eq:static-eq-01})-(\ref{eq:static-eq-03}).
The results are given in Eqs.~(\ref{eq:j-0}), (\ref{eq:sol-first-order}),
and (\ref{eq:second-order-solution}). When carrying out an integration
over the four-momentum for the chiral Wigner function, we obtain the
chiral current in Eq.~(\ref{eq:current_def}) and then the charge
(vector) and axial charge (vector) currents in Eq.~(\ref{eq:vector-axial-vector-current}).
The results for the charge and axial charge currents in Eqs.~(\ref{eq:2nd-porder-current-1})
and (\ref{eq:2nd-order-current-2}) show that they all have explicit
Lorentz covariance.

The first-order charge current $J_{(1)}^{\mu}=\xi\omega^{\mu}+\xi_{B}B^{\mu}$
in Eq.~(\ref{eq:2nd-porder-current-1}) gives the CME and CVE. In
the semiclassical expansion without assuming equilibrium conditions
and constant electromagnetic fields, one can still solve the vector
component $\mathscr{J}^{\mu}$ of the Wigner function order by order
in $\hbar$ as we have shown in Sec.~\ref{sec:dwf-theorem}. In
this formalism, to any order of $\hbar$, only the time component
of $\mathscr{J}^{\mu}$ is independent, while spatial components can
be derived from the time component. There is a freedom to choose any
reference frame in which the time component is defined. So the explicit
Lorentz covariance seems to be lost.

In this section, we will show that under equilibrium conditions (\ref{eq:static-eq-01})-(\ref{eq:static-eq-03}),
the CVE current can be decomposed into two parts, which can be identified
as the normal current and magnetization current. Each part depends
on the reference frame, but the sum of two parts does give the total
CVE current in a Lorentz covariant form without the frame dependence
\cite{Gao:2018jsi}. Similar properties also exist for the CME current.

We start from the application of the equilibrium conditions (\ref{eq:static-eq-01})-(\ref{eq:static-eq-03})
to $\delta f_{(1)}\delta(p^{2})$ in Eq.~(\ref{eq:delta-f1}) 
\begin{eqnarray}
\delta f_{(1)}\delta(p^{2}) & = & -s\frac{\epsilon^{\lambda\nu\rho\sigma}n_{\lambda}n_{\nu}^{\prime}p_{\sigma}}{2(n\cdot p)(n^{\prime}\cdot p)}\delta(p^{2})\nabla_{\rho}f_{(0)}\nonumber \\
 & = & -s\frac{p^{\mu}\epsilon^{\lambda\nu\rho\sigma}n_{\lambda}n_{\nu}^{\prime}p_{\sigma}\Omega_{\rho\mu}}{2\left(n^{\prime}\cdot p\right)\left(n\cdot p\right)}\frac{df_{(0)}}{d(\beta\cdot p)}\delta(p^{2})\nonumber \\
 & = & -s\frac{n_{\alpha}^{\prime}p_{\gamma}\widetilde{\Omega}^{\alpha\gamma}}{2\left(n^{\prime}\cdot p\right)}\frac{df_{(0)}}{d(\beta\cdot p)}\delta(p^{2})+s\frac{n_{\alpha}p_{\gamma}\widetilde{\Omega}^{\alpha\gamma}}{2\left(n\cdot p\right)}\frac{df_{(0)}}{d(\beta\cdot p)}\delta(p^{2}),\label{eq:delta-f1-simp}
\end{eqnarray}
where we have used following identity in equilibrium,
\begin{eqnarray}
\nabla_{\rho}f_{(0)} & = & \frac{df_{(0)}}{d(\beta\cdot p)}\frac{1}{2}(\partial_{\rho}\beta_{\mu}-\partial_{\mu}\beta_{\rho})p^{\mu}\nonumber \\
 & = & \frac{df_{(0)}}{d(\beta\cdot p)}\Omega_{\rho\mu}p^{\mu}.
\end{eqnarray}
The derivation of the above identity can be found in intermediate steps
of Eq.~(\ref{eq:constraint-eq}). Obviously $\delta f_{(1)}$ in
Eq.~(\ref{eq:delta-f1-simp}) depends on the frame, but we single
out the contribution with an explicit frame dependence from that independent
of the frame as 
\begin{eqnarray}
f_{(1)} & = & \widetilde{f}_{(1)}-s\frac{n_{\alpha}p_{\gamma}\widetilde{\Omega}^{\alpha\gamma}}{2\left(n\cdot p\right)}\frac{df_{(0)}}{d(\beta\cdot p)},\label{f1}
\end{eqnarray}
where $\widetilde{f}_{(1)}$ does not depend on the reference vector
$n^{\mu}$, i.e. $\delta\widetilde{f}_{(1)}=0$. A simple choice of
a specific solution is $\widetilde{f}_{(1)}=0$.

Substituting Eq.~(\ref{f1}) into Eq.~(\ref{eq:j0-mu-j1-mu}) yields
\begin{eqnarray}
\mathscr{J}_{(1)}^{\mu} & = & \mathscr{J}_{(1)B}^{\mu}+\mathscr{J}_{(1)\omega}^{\mu},
\end{eqnarray}
where 
\begin{eqnarray}
\mathscr{J}_{(1)B}^{\mu} & = & \frac{s}{n\cdot p}p^{\mu}n_{\nu}\widetilde{F}^{\nu\lambda}p_{\lambda}f_{(0)}\delta^{\prime}\left(p^{2}\right)\nonumber \\
 &  & +\frac{s}{2n\cdot p}\epsilon^{\mu\nu\rho\sigma}n_{\nu}F_{\sigma\lambda}f_{(0)}\partial_{p}^{\lambda}\left[p_{\rho}\delta\left(p^{2}\right)\right],\label{eq:frame-decomp-j1-b}\\
\mathscr{J}_{(1)\omega}^{\mu} & = & -p^{\mu}\frac{s}{2\left(n\cdot p\right)}n_{\alpha}p_{\gamma}\widetilde{\Omega}^{\alpha\gamma}\frac{df_{(0)}}{d(\beta\cdot p)}\delta(p^{2})\nonumber \\
 &  & -\frac{s}{2n\cdot p}p^{\lambda}\epsilon^{\mu\nu\rho\sigma}n_{\nu}p_{\rho}\Omega_{\sigma\lambda}\frac{df_{(0)}}{d(\beta\cdot p)}\delta(p^{2}).\label{eq:frame-decomp-j1-omega}
\end{eqnarray}
We see that $\mathscr{J}_{(1)B}^{\mu}$ and $\mathscr{J}_{(1)\omega}^{\mu}$
have superficial dependence on the frame vector $n^{\mu}$, and the
first term of $\mathscr{J}_{(1)B}^{\mu}$ in Eq.~(\ref{eq:frame-decomp-j1-b})
is proportional to $p^{\mu}$ while the second one is perpendicular
to it, so is $\mathscr{J}_{(1)\omega}^{\mu}$ in Eq.~(\ref{eq:frame-decomp-j1-omega}).
However one can prove that $\mathscr{J}_{(1)B}^{\mu}$ in Eq.~(\ref{eq:frame-decomp-j1-b})
and $\mathscr{J}_{(1)\omega}^{\mu}$ in Eq.~(\ref{eq:frame-decomp-j1-omega})
do give the covariant result in Eq.~(\ref{eq:sol-first-order}),
which are frame independent \cite{Gao:2018jsi}.

In order to show the physical meaning of $\mathscr{J}_{(1)B}^{\mu}$
and $\mathscr{J}_{(1)\omega}^{\mu}$ in Eqs.~(\ref{eq:frame-decomp-j1-b})
and (\ref{eq:frame-decomp-j1-omega}), let us evaluate the current
by assuming that $n^{\mu}$ is the fluid velocity $u^{\mu}$. For
the CME current, it is easy to rewrite the currents corresponding
to two terms of $\mathscr{J}_{(1)B}^{\mu}$ in Eq.~(\ref{eq:frame-decomp-j1-b})
which we denote as $J_{s,B}^{\mu}(1\mathrm{st})$ and $J_{s,B}^{\mu}(2\mathrm{nd})$,
\begin{eqnarray}
J_{s,B}^{\mu}(1\mathrm{st}) & = & \hbar su_{\nu}\widetilde{F}^{\nu\lambda}\int\frac{d^{4}p}{(2\pi)^{4}}\frac{1}{p_{0}}p^{\mu}p_{\lambda}f_{(0)}\delta^{\prime}\left(p^{2}\right)\nonumber \\
 & = & -\hbar sB^{\lambda}\int\frac{d^{4}p}{(2\pi)^{4}}\frac{1}{p_{0}^{3}}\overline{p}^{\mu}\overline{p}_{\lambda}f_{(0)}\delta\left(p^{2}\right)\nonumber \\
 &  & +\hbar\frac{1}{2}sB^{\lambda}\int\frac{d^{4}p}{(2\pi)^{4}}\frac{1}{p_{0}^{2}}\overline{p}^{\mu}\overline{p}_{\lambda}\frac{df_{(0)}}{dp_{0}}\delta\left(p^{2}\right),\label{eq:jsb-1}
\end{eqnarray}
\begin{eqnarray}
J_{s,B}^{\mu}(2\mathrm{nd}) & = & \hbar\frac{s}{2}\epsilon^{\mu\nu\rho\sigma}u_{\nu}F_{\sigma\lambda}\int\frac{d^{4}p}{(2\pi)^{4}}\frac{1}{p_{0}}f_{(0)}\partial_{p}^{\lambda}\left[p_{\rho}\delta\left(p^{2}\right)\right]\nonumber \\
 & = & -\hbar sB^{\mu}\int\frac{d^{4}p}{(2\pi)^{4}}\frac{1}{p_{0}}f_{(0)}\delta\left(p^{2}\right)\nonumber \\
 &  & +\hbar s\epsilon^{\mu\nu\rho\sigma}u_{\nu}F_{\sigma\lambda}\int\frac{d^{4}p}{(2\pi)^{4}}\frac{1}{p_{0}^{3}}\overline{p}_{\rho}\overline{p}^{\lambda}f_{(0)}\delta\left(p^{2}\right)\nonumber \\
 &  & -\hbar\frac{s}{2}\epsilon^{\mu\nu\rho\sigma}u_{\nu}F_{\sigma\lambda}\int\frac{d^{4}p}{(2\pi)^{4}}\frac{1}{p_{0}^{2}}\overline{p}_{\rho}\overline{p}^{\lambda}\frac{df_{(0)}}{dp_{0}}\delta\left(p^{2}\right),\label{eq:jsb-2}
\end{eqnarray}
where $p_{0}\equiv u\cdot p$, $\overline{p}^{\mu}\equiv p^{\mu}-p_{0}u^{\mu}$,
$\xi_{B,s}=(\xi_{B}+s\xi_{B5})/2$ is the CME coefficient corresponding
to the chirality $s$, and $f_{(0)}$ is given by Eq.~(\ref{eq:dist})
but without the unity in the anti-particle sector proportional to
$\Theta(-p_{0})$. Note that we have put back $\hbar$ in $J_{s,B}^{\mu}(1\mathrm{st})$
and $J_{s,B}^{\mu}(2\mathrm{nd})$. It can be proved that $J_{s,B}^{\mu}(2\mathrm{nd})=0$,
i.e. the current from the second term gives vanishing result. Now
we analyze the two terms of $J_{s,B}^{\mu}(1\mathrm{st})$ in Eq.~(\ref{eq:jsb-1})
which we denote as $J_{s,B}^{\mu}(a)$ and $J_{s,B}^{\mu}(b)$, 
\begin{eqnarray}
J_{s,B}^{\mu}(a) & = & \frac{1}{2}\hbar sB^{\lambda}\int\frac{d^{4}p}{(2\pi)^{4}}\frac{1}{p_{0}^{2}}\overline{p}^{\mu}\overline{p}_{\lambda}\frac{df_{(0)}}{dp_{0}}\delta\left(p^{2}\right)\nonumber \\
 & = & -\frac{1}{2}\hbar s\beta\int\frac{d^{3}\mathbf{p}}{(2\pi)^{3}}\frac{1}{E_{p}^{3}}\overline{p}^{\mu}(\overline{p}\cdot B)\nonumber \\
 &  & \times\left[f_{\mathrm{FD}}^{+}(1-f_{\mathrm{FD}}^{+})-f_{\mathrm{FD}}^{-}(1-f_{\mathrm{FD}}^{-})\right]\nonumber \\
 & = & \frac{1}{3}\xi_{Bs}B^{\mu},\label{eq:jsb-a}\\
J_{s,B}^{\mu}(b) & = & -\hbar sB^{\lambda}\int\frac{d^{4}p}{(2\pi)^{4}}\frac{1}{p_{0}^{3}}\overline{p}^{\mu}\overline{p}_{\lambda}f_{(0)}\delta\left(p^{2}\right)\nonumber \\
 & = & -\hbar s\int\frac{d^{3}\mathbf{p}}{(2\pi)^{3}}\frac{1}{E_{p}^{4}}\overline{p}^{\mu}(\overline{p}\cdot B)\left(f_{\mathrm{FD}}^{+}-f_{\mathrm{FD}}^{-}\right)\nonumber \\
 & = & \frac{2}{3}\xi_{Bs}B^{\mu},\label{eq:jsb-b}
\end{eqnarray}
where $E_{p}=|\mathbf{p}|$ and $f_{\mathrm{FD}}^{\pm}\equiv f_{\mathrm{FD}}(E_{p}\mp\mu_{s})$.
We see that $J_{s,B}^{\mu}(a)$ and $J_{s,B}^{\mu}(b)$ give one-third
and two-thirds of the CME, respectively. In order to go further to
show the physical meaning of $J_{s,B}^{\mu}(a)$ and $J_{s,B}^{\mu}(b)$,
we can rewrite them in the local rest frame with $u^{\mu}=(1,0,0,0)$,
\begin{eqnarray}
\mathbf{J}_{s,B}(a) & = & \hbar s\beta\int\frac{d^{3}\mathbf{p}}{(2\pi)^{3}}\frac{1}{2|\mathbf{p}|^{3}}\mathbf{p}(\mathbf{p}\cdot\mathbf{B})\nonumber \\
 &  & \times\left[f_{\mathrm{FD}}^{+}(1-f_{\mathrm{FD}}^{+})-f_{\mathrm{FD}}^{-}(1-f_{\mathrm{FD}}^{-})\right]\nonumber \\
 & \approx & \int\frac{d^{3}\mathbf{p}}{(2\pi)^{3}}\frac{\mathbf{p}}{|\mathbf{p}|}\left[f_{\mathrm{FD}}\left(|\mathbf{p}|-\mu_{s}-s\hbar\frac{\mathbf{p}\cdot\mathbf{B}}{2|\mathbf{p}|^{2}}\right)\right.\nonumber \\
 &  & \left.+f_{\mathrm{FD}}\left(|\mathbf{p}|+\mu_{s}+s\hbar\frac{\mathbf{p}\cdot\mathbf{B}}{2|\mathbf{p}|^{2}}\right)\right],\nonumber \\
\mathbf{J}_{s,B}(b) & = & \hbar s\beta\int\frac{d^{3}\mathbf{p}}{(2\pi)^{3}}\frac{1}{|\mathbf{p}|^{4}}\mathbf{p}(\mathbf{p}\cdot\mathbf{B})\nonumber \\
 &  & \times\left[f_{\mathrm{FD}}(E_{p}-\mu_{s})-f_{\mathrm{FD}}(E_{p}+\mu_{s})\right].
\end{eqnarray}
We see that the current $\mathbf{J}_{s,B}(a)$ comes from the three-momentum
integral of the velocity $\mathbf{v}=\mathbf{p}/|\mathbf{p}|$
on the Fermi-Dirac distribution, which includes the energy correction
from the magnetic moment in the magnetic field with the magnetic moment
being $\hbar sQ\mathbf{p}/(2|\mathbf{p}|^{2})$ for fermions/anti-fermions
with the electric charge $Q=\pm$ respectively. According to Ref.
\cite{Kharzeev:2016sut}, $\mathbf{J}_{s,B}(b)$ is from the modification
of the quasi-particle velocity due to the Berry curvature.

In a similar way, we can analyze the CVE current from $\mathscr{J}_{(1)\omega}^{\mu}$
in Eq.~(\ref{eq:frame-decomp-j1-omega}), in which the two terms
are denoted as $J_{s,\omega}^{\mu}(a)$ and $J_{s,\omega}^{\mu}(b)$.
Their results are 
\begin{eqnarray}
J_{s,\omega}^{\mu}(a) & = & -\frac{1}{2}\hbar su_{\alpha}\widetilde{\Omega}^{\alpha\gamma}\int\frac{d^{4}p}{(2\pi)^{4}}\frac{1}{p_{0}}p^{\mu}p_{\gamma}\frac{df_{s}}{d(\beta\cdot p)}\delta(p^{2})\nonumber \\
 & = & -\frac{1}{2}\hbar s\beta\omega^{\gamma}\int\frac{d^{3}\mathbf{p}}{(2\pi)^{3}}\frac{\overline{p}^{\mu}\overline{p}_{\gamma}}{E_{p}^{2}}\left[f_{\mathrm{FD}}^{+}(1-f_{\mathrm{FD}}^{+})+f_{\mathrm{FD}}^{-}(1-f_{\mathrm{FD}}^{-})\right]\nonumber \\
 & = & \frac{1}{3}\xi_{s}\omega^{\mu},\nonumber \\
J_{s,\omega}^{\mu}(b) & = & -\frac{1}{2}\hbar s\epsilon^{\mu\nu\rho\sigma}u_{\nu}\int\frac{d^{4}p}{(2\pi)^{4}}\frac{1}{p_{0}}p_{\rho}(\partial_{\sigma}^{x}f_{s})\delta(p^{2})\nonumber \\
 & = & \frac{2}{3}\xi_{s}\omega^{\mu}.\label{eq:j1-j2}
\end{eqnarray}
We see that $J_{s,\omega}^{\mu}(a)$ and $J_{s,\omega}^{\mu}(b)$
contribute to the full CVE current by 1/3 and 2/3, respectively. By
choosing the local rest frame $u^{\mu}=(1,0,0,0)$ at one space-time
point but with $\partial_{\mu}u_{\nu}\neq0$ in its vicinity, we obtain
the three-dimension form for the two terms,
\begin{eqnarray}
\mathbf{J}_{s,\omega}(a) & = & \frac{1}{2}\hbar s\beta\int\frac{d^{3}\mathbf{p}}{(2\pi)^{3}}\frac{\mathbf{p}}{|\mathbf{p}|^{2}}(\mathbf{p}\cdot\boldsymbol{\omega})\left[f_{\mathrm{FD}}^{+}(1-f_{\mathrm{FD}}^{+})+f_{\mathrm{FD}}^{-}(1-f_{\mathrm{FD}}^{-})\right]\nonumber \\
 & \approx & \int\frac{d^{3}p}{(2\pi)^{3}}\frac{\mathbf{p}}{|\mathbf{p}|}\left[f_{\mathrm{FD}}\left(|\mathbf{p}|-\mu_{s}-s\hbar\frac{\mathbf{p}\cdot\boldsymbol{\omega}}{2|\mathbf{p}|}\right)\right.\nonumber 
 \left.+f_{\mathrm{FD}}\left(|\mathbf{p}|+\mu_{s}-s\hbar\frac{\mathbf{p}\cdot\boldsymbol{\omega}}{2|\mathbf{p}|}\right)\right],\nonumber \\
\mathbf{J}_{s,\omega}(b) & = & \lim_{|\mathbf{v}|=0}\nabla\times\int\frac{d^{3}\mathbf{p}}{(2\pi)^{3}}\left(\frac{s\mathbf{p}}{2|\mathbf{p}|^{2}}\hbar\right)\nonumber \\
 &  & \times\left[f_{\mathrm{FD}}(\gamma|\mathbf{p}|-\gamma\mathbf{v}\cdot\mathbf{p}-\mu_{s})+f_{\mathrm{FD}}(\gamma|\mathbf{p}|-\gamma\mathbf{v}\cdot\mathbf{p}+\mu_{s})\right],
\end{eqnarray}
where in obtaining $\mathbf{J}_{s,\omega}(b)$ we have taken the limit
at $|\mathbf{v}|=0$ with $u^{\mu}=(\gamma,\gamma\mathbf{v})$ and
$\gamma=1/\sqrt{1-|\mathbf{v}|^{2}}$. We can see that $\mathbf{J}_{s,\omega}(a)$
comes from the momentum integration of the fermion's velocity $\mathbf{v}_{p}$
with the Fermi-Dirac distribution function in which the fermion's
energy is modified by the spin-vorticity coupling, while $\mathbf{J}_{s,\omega}(b)$
is the ``magnetization'' current due to the ``magnetic'' moment
of chiral fermions in vorticity fields \cite{Chen:2014cla}. Note
that the energy correction from the spin-vorticity coupling has the
same sign for fermions and anti-fermions.

\newpage
\section{Chiral kinetic theory from effective theories}\label{sec:CKT_effective}

In this section, we briefly introduce other effective theories related to the chiral transport phenomena. We first introduce the concept of Berry phase and Berry curvature in quantum mechanics. We then derive the kinetic theory with quantum corrections proportional to Berry curvature, which is the 3-dimensional chiral kinetic theory, by the path integrals of the effective Hamiltonian of chiral fermions. Later, we discuss the symplectic form for the phase space in the presence of electromagnetic fields coupled to the Berry curvature and compute the modified Poisson bracket to derive the chiral kinetic theory. After that, we consider the Lorentz transformation to the effective action of chiral fermions and obtain the quantum corrections to both space and momentum under Lorentz transformation, i.e., named side jump effects. We also introduce the high-density effective theory and comment on other related methods.
For simple notations, we choose $\hbar=1$ in this section. 

\subsection{Berry phase}

\label{subsec:Berry}In this subsection, we briefly introduce the
idea of the Berry phase \cite{Berry:1984jv,Xiao:2009rm}.

We consider a Hamiltonian $H(t)$ whose time dependence is through
a set of parameters $\mathbf{R}$,

\begin{equation}
H=H(\mathbf{R}(t)).
\end{equation}
As we will show later, $\mathbf{R}$ can be the momentum operator
$\mathbf{p}$. We consider an adiabatic evolution of the system: $\mathbf{R}(t)$
moves so slowly along a path $C$ that the instantaneous orthonormal
basis can be defined at any time $t$,

\begin{equation}
H(\mathbf{R}(t))\left|n(\mathbf{R}(t))\right\rangle =\varepsilon_{n}(\mathbf{R}(t))\left|n(\mathbf{R}(t))\right\rangle ,
\end{equation}
where $\left|n(\mathbf{R}(t))\right\rangle $ and $\varepsilon_{n}(\mathbf{R}(t))$
denote the eigenstate and eigenvalue of $H(\mathbf{R}(t))$ at $t$
respectively. Different from normal eigenvectors of a time independent
Hamiltonian, the extra phase factor of $\left|n(\mathbf{R}(t))\right\rangle $
can be physical and cannot be simply ignored. The wave function has
the form,
\begin{equation}
\left|\psi_{n}(t)\right\rangle =e^{i\gamma_{n}(t)}e^{-ih(t)}\left|n(\mathbf{R}(t))\right\rangle ,\label{eq:wave_function_01}
\end{equation}
where $e^{-ih(t)}=\exp\left[-i\int_{0}^{t}dt^{\prime}\varepsilon_{n}(\mathbf{R}(t^{\prime}))\right]$
is usually called the dynamical phase factor and $\gamma_{n}(t)$
is the extra phase factor. Inserting the wave function (\ref{eq:wave_function_01})
into the Schr\"odinger equation $i\partial_{t}\psi=H\psi$ and multiplying
it from the left by $\left\langle n(\mathbf{R}(t))\right|$ we obtain
\begin{equation}
\gamma_{n}=\int_{0}^{t}dt^{\prime}\frac{d\mathbf{R}(t^{\prime})}{dt^{\prime}}\cdot\left\langle n(\mathbf{R}(t^{\prime}))\right|\frac{\partial}{\partial\mathbf{R}}\left|n(\mathbf{R}(t^{\prime}))\right\rangle \equiv\int_{c}d\mathbf{R}\cdot\mathbf{a}_{n}(\mathbf{R}),\label{eq:Berry_phase_01}
\end{equation}
where we have defined the Berry connection or potential as 
\begin{equation}
\mathbf{a}_{n}(\mathbf{R})=i\left\langle n(\mathbf{R}(t^{\prime}))\right|\frac{\partial}{\partial\mathbf{R}}\left|n(\mathbf{R}(t^{\prime}))\right\rangle .\label{eq:Berry_connection_01}
\end{equation}

Similar to the vector potential $\mathbf{A}$ for electromagnetic
fields, the Berry connection $\mathbf{a}_{n}$ is not gauge invariant.
Under the transformation $\left|n(\mathbf{R})\right\rangle \rightarrow e^{i\xi(\mathbf{R})}\left|n(\mathbf{R})\right\rangle $,
Berry connection transforms as $\mathbf{a}_{n}\rightarrow\mathbf{a}_{n}-\nabla_{\mathbf{R}}\xi(\mathbf{R})$
and the phase factor changes as $\gamma_{n}\rightarrow\gamma_{n}+\xi(\mathbf{R}(0))-\xi(\mathbf{R}(t))$.
Therefore, one can always choose a suitable $\xi(\mathbf{R}(0))-\xi(\mathbf{R}(t))$
to make $\gamma_{n}$ vanish. However, if one chooses a closed path,
i.e. $\xi(\mathbf{R}(0))=\xi(\mathbf{R}(t))$ or $\xi(\mathbf{R}(0))=\xi(\mathbf{R}(t))+2\pi l$
with $l$ being an integer, then the phase factor,
\begin{equation}
\gamma_{c}=\ointop_{C}d\mathbf{R}\cdot\mathbf{a}_{n}(\mathbf{R}),
\end{equation}
cannot be removed. Such a gauge-invariant phase is called the Berry
phase after Michael Berry \cite{Berry:1984jv}.

If the parameter space for $\mathbf{R}$ has three dimensions, one
can also rewrite $\gamma_{c}$ as 
\begin{equation}
\gamma_{c}=\int_{S}d\mathbf{S}\cdot\boldsymbol{\Omega}_{n}(\mathbf{R}),
\end{equation}
where $S$ is the area surrounded by the\textbf{ }closed loop C and
\begin{equation}
\boldsymbol{\Omega}_{n}(\mathbf{R})=\nabla_{\mathbf{R}}\times\mathbf{a}_{n}(\mathbf{R)},
\end{equation}
is called the Berry curvature.

\subsection{Chiral kinetic theory from path integrals \label{subsec:Berry-massless-path}}

In this subsection, we construct the chiral kinetic theory for massless
fermions incorporating the Berry phase through path integrals \cite{Stephanov:2012ki,Chen:2013iga,Chen:2014cla}.

We consider a spin-$1/2$ massless particle moving in a background
electromagnetic field with the vector potential \textbf{$A^{\mu}=(\phi(\mathbf{x}),\mathbf{A}(\mathbf{x}))$}.
The Hamiltonian for the right-handed fermions can be expressed as
\begin{eqnarray}
H & = & \boldsymbol{\sigma}\cdot\left[\mathbf{p}_{c}-e\mathbf{A}(\mathbf{x})\right]+e\phi(\mathbf{x}),\label{eq:chiral-f-h}
\end{eqnarray}
where $\mathbf{p}_{c}=\mathbf{p}+e\mathbf{A}(\mathbf{x})$ is the
canonical momentum with $\mathbf{p}$ being the mechanical one, and
$\boldsymbol{\sigma}$ are Pauli matrices $\boldsymbol{\sigma}=(\sigma_{1},\sigma_{2},\sigma_{3})$.
Note that the Hamiltonian $H$ is a $2\times2$ matrix.

Now, we compute the quantum correction to the equation of motion for
particles via the path integral quantization. The transition matrix
element in the path integral is 
\begin{eqnarray}
K_{\text{fi}} & = & \langle\mathbf{x}_{\text{f}}|e^{-iH(t_{\text{f}}-t_{\text{i}})}|\mathbf{x}_{\text{i}}\rangle\nonumber \\
 & = & \int[D\mathbf{x}][D\mathbf{p}_{c}]P\exp\left[i\int_{t_{\text{i}}}^{t_{\text{f}}}dt(\mathbf{p}_{c}\cdot\dot{\mathbf{x}}-H)\right],\label{eq:transition-am}
\end{eqnarray}
where $t_{\mathrm{i}}$ and $t_{\mathrm{f}}$ denote the initial and
final time respectively, $\mathbf{x}(t_{\mathrm{i}})=\mathbf{x}_{\text{i}}$
and $\mathbf{x}(t_{\text{f}})=\mathbf{x}_{\text{f}}$ denote the initial
and final position of the path respectively, $[D\mathbf{x}][D\mathbf{p}_{c}]$
denotes the path integral elements in phase space, $\dot{\mathbf{x}}\equiv d\mathbf{x}/dt$,
and $P$ is the path ordering operator.

We note that there are two eigenstates of $H$ in Eq.~(\ref{eq:chiral-f-h}),
\begin{equation}
\sigma\cdot\mathbf{p}\chi_{\pm}(\mathbf{p})=\pm|\mathbf{p}|\chi_{\pm}(\mathbf{p}),\qquad\chi_{\pm}^{\dagger}\chi_{\pm}=1,
\end{equation}
Here, the lower index $\pm$ denotes the particle (upper) and anti-particle
(lower) respectively. Our propose is to derive the effective action
for right-handed particle or anti-particle only. Therefore, we need
to diagonalize $H$ by using a unitary matrix $U_{\mathbf{p}}$ with
$\mathbf{p}=\mathbf{p}_{c}-e\mathbf{A}$, 
\begin{eqnarray}
U_{\mathbf{p}}^{\dagger}HU_{\mathbf{p}} & = & \left[\begin{array}{cc}
|\mathbf{p}|+e\phi(\mathbf{x}) & 0\\
0 & -|\mathbf{p}|+e\phi(\mathbf{x})
\end{array}\right]\nonumber \\
 & = & \sigma_{3}\epsilon(\mathbf{p}_{c}-e\mathbf{A})+e\phi(\mathbf{x}),\label{eq:diag-h}
\end{eqnarray}
where $\epsilon(\mathbf{p})\equiv|\mathbf{p}|$ is the particle's
energy and 
\begin{equation}
U_{\mathbf{p}}=(\chi_{+},\chi_{-})=\left(\begin{array}{cc}
e^{-i\phi}\cos\frac{\theta}{2} & -e^{-i\phi}\sin\frac{\theta}{2}\\
\sin\frac{\theta}{2} & \cos\frac{\theta}{2}
\end{array}\right),\quad U_{\mathbf{p}}^{\dagger}=\left(\begin{array}{c}
\chi_{+}^{\dagger}\\
\chi_{-}^{\dagger}
\end{array}\right),\label{eq:uu-daggar}
\end{equation}
where $\theta$ and $\phi$ are the polar and azimuthal angle of $\mathbf{p}$
respectively. Note that when using $\mathbf{p}_{c}$ to label $U_{\mathbf{p}}$
as we will do shortly we denote $U_{\mathbf{p}}$ as $U(\mathbf{x},\mathbf{p}_{c})$
since $\mathbf{p}$ contains $\mathbf{A}(\mathbf{x})$.

Inserting Eqs.~(\ref{eq:diag-h}, \ref{eq:uu-daggar}) into Eq.~(\ref{eq:transition-am})
yields
\begin{eqnarray}
K_{\text{fi}} & = & \lim_{N\rightarrow\infty}\int\left[\prod_{j=1}^{N}d\mathbf{x}_{j}d\mathbf{p}_{j}^{c}\right]d\mathbf{x}_{0}\langle\mathbf{x}_{\text{f}}|\mathbf{x}_{N}\rangle\nonumber \\
 &  & \times\left(\prod_{j=1}^{N}\langle\mathbf{x}_{j}|e^{-3iH\Delta t}|\mathbf{p}_{j}^{c}\rangle\langle\mathbf{p}_{j}^{c}|e^{-3iH\Delta t}|\mathbf{x}_{j-1}\rangle\right)\langle\mathbf{x}_{0}|\mathbf{x}_{\text{i}}\rangle.\label{eq:am1}
\end{eqnarray}
In the $N\rightarrow\infty$ limit, the amplitude (\ref{eq:am1})
can be put into the form,
\begin{eqnarray}
K_{\text{fi}} & = & \int[D\mathbf{x}][D\mathbf{p}_{c}]U(\mathbf{x}_{\mathrm{f}},\mathbf{p}_{\mathrm{f}}^{c})P\exp\left\{ i\int_{t_{\text{i}}}^{t_{\text{f}}}dt\left[\mathbf{p}\cdot\dot{\mathbf{x}}+e\mathbf{A}(\mathbf{x})\cdot\mathbf{x}\right.\right.\nonumber \\
 &  & \left.\left.-\sigma_{3}\epsilon(\mathbf{p})-e\phi(\mathbf{x})-\boldsymbol{\mathcal{A}}(\mathbf{p})\cdot\dot{\mathbf{p}}\right]\right\} U^{\dagger}(\mathbf{x}_{\mathrm{i}},\mathbf{p}_{\mathrm{i}}^{c}),\label{eq:am2}
\end{eqnarray}
where we have used the following approximation,
\begin{eqnarray}
U^{\dagger}(\mathbf{x}_{j},\mathbf{p}_{j}^{c})U(\mathbf{x}_{j},\mathbf{p}_{j-1}^{c}) & \approx & \exp\left[-i\boldsymbol{\mathcal{A}}(\mathbf{x}_{j},\mathbf{p}_{j-1}^{c})\cdot(\mathbf{p}_{j}^{c}-\mathbf{p}_{j-1}^{c})\right],\nonumber \\
U^{\dagger}(\mathbf{x}_{j+1},\mathbf{p}_{j}^{c})U(\mathbf{x}_{j},\mathbf{p}_{j}^{c}) & \approx & \exp\left[i\boldsymbol{\mathcal{A}}(\mathbf{x}_{j},\mathbf{p}_{j}^{c})\cdot(e\mathbf{A}(\mathbf{x}_{j+1})-e\mathbf{A}(\mathbf{x}_{j}))\right],\label{eq:uu+}
\end{eqnarray}
and defined the Berry potential,
\begin{eqnarray}
\boldsymbol{\mathcal{A}}(\mathbf{p}) & \equiv-iU_{\mathbf{p}}^{\dagger}\boldsymbol{\nabla}_{\mathbf{p}}U_{\mathbf{p}}= & -\frac{1}{2|\mathbf{p}|}\left(\begin{array}{cc}
\mathbf{e}_{\phi}\cot\frac{\theta}{2} & \mathbf{e}_{\phi}-i\mathbf{e}_{\theta}\\
\mathbf{e}_{\phi}+i\mathbf{e}_{\theta} & \mathbf{e}_{\phi}\tan\frac{\theta}{2}
\end{array}\right).\label{eq:berry-m0}
\end{eqnarray}
With the amplitude in a general form,
\begin{equation}
K_{\mathrm{fi}}=\int[D\mathbf{x}][D\mathbf{p}]\exp(iS),
\end{equation}
one can read out the effective action for the right-handed particle
from Eq.~(\ref{eq:am2}),
\begin{eqnarray}
S & = & \int dt\left[\mathbf{p}\cdot\dot{\mathbf{x}}+e\mathbf{A}(\mathbf{x})\cdot\mathbf{x}-\sigma_{3}\epsilon(\mathbf{p})-e\phi(\mathbf{x})-\boldsymbol{\mathcal{A}}(\mathbf{p})\cdot\dot{\mathbf{p}}\right].\label{eq:action}
\end{eqnarray}
Here we assume that there is no transition between the particle and
anti-particle states during the adiabatic evolution, so one can neglect
the off-diagonal terms in Eq.~(\ref{eq:action}). More details can
be found in Ref.~\cite{Stephanov:2012ki,Chen:2014cla} or in the appendix
of Ref.~\cite{Chen:2013iga}.

Then the diagonal terms of Eq.~(\ref{eq:action}) have the form \cite{Stephanov:2012ki},
\begin{equation}
S_{\pm}=\int dt\left[\mathbf{p}\cdot\dot{\mathbf{x}}+e\mathbf{A}(\mathbf{x})\cdot\mathbf{x}\mp\epsilon(\mathbf{p})-e\phi(\mathbf{x})-\mathbf{a}_{\pm}(\mathbf{p})\cdot\dot{\mathbf{p}}\right],\label{eq:action-11}
\end{equation}
where $\mathbf{a}_{\pm}(\mathbf{p})=\boldsymbol{\mathcal{A}}_{11/22}(\mathbf{p})$
is the Berry connection or potential for particles (upper) and anti-particles
(lower). More details can be found in Ref.~\cite{Chen:2013iga}. 
In general, there are quantum corrections to particle's energy $\epsilon(\mathbf{p})$ 
due to the Zeeman effects \cite{Son:2012zy,Chen:2013iga}. More general expression for $\epsilon(\mathbf{p})$  can be found in Sec.
\ref{subsec:symplectic form} and \ref{subsec:Other-effective-theories}. Later, we will show that this quantum corrections are essential to get the Lorentz invariance in Sec. \ref{subsec:Side-jump,-Lorentz}. 

Using the Euler-Lagrangian equations, we obtain the equations of motion
for $\mathbf{x}$ and $\mathbf{p}$,
\begin{eqnarray}
\sqrt{G}\mathbf{\dot{x}} & = & \hat{\mathbf{p}}+\mathbf{E}\times\boldsymbol{\Omega}_{\mathbf{p}}+\mathbf{B}(\widehat{\mathbf{p}}\cdot\boldsymbol{\Omega}_{\mathbf{p}}),\nonumber \\
\sqrt{G}\dot{\mathbf{p}} & = & \mathbf{E}+\widehat{\mathbf{p}}\times\mathbf{B}+\boldsymbol{\Omega}_{\mathbf{p}}(\mathbf{E}\cdot\mathbf{B}),\label{eq:EOM_01}
\end{eqnarray}
where $\widehat{\mathbf{p}}\equiv\mathbf{p}/|\mathbf{p}|$, $\sqrt{G}=1+\boldsymbol{\Omega}_{\mathbf{p}}\cdot\mathbf{B}$
with 
\begin{equation}
\boldsymbol{\Omega}_{\mathbf{p}}=\pm\frac{\mathbf{p}}{2|\mathbf{p}|^{3}},\label{eq:Berry_curv_01}
\end{equation}
being the Berry curvature for the right-handed particle (upper sign)
and anti-particle (lower sign). The first term for $\sqrt{G}\mathbf{\dot{x}}$
is the standard velocity of a single particle and $\mathbf{E}+\widehat{\mathbf{p}}\times\mathbf{B}$
in $\sqrt{G}\dot{\mathbf{p}}$ is the Lorentz force. The new terms
proportional to $\boldsymbol{\Omega}_{\mathbf{p}}$ are quantum corrections
to the effective velocity $\mathbf{\dot{x}}$ and force $\dot{\mathbf{p}}$.
An important feature for the Berry curvature is
\begin{equation}
\boldsymbol{\nabla}_{\mathbf{p}}\cdot\boldsymbol{\Omega}_{\mathbf{p}}=2\pi\delta^{3}(\mathbf{p}),\label{eq:Berry_monopole}
\end{equation}
which we will show later leads to the chiral anomaly.

Once one has the equations of motion (\ref{eq:EOM_01}) for single
particles, it is straightforward to derive the kinetic equation by
introducing the distribution function $f(t,\mathbf{x},\mathbf{p})$
representing the probability to find particles (the particle number
density) in phase space.

In a non-interacting system of fermions without collisions, the total
time derivative of the distribution function should vanish, so we have
\begin{equation}
\left[\sqrt{G}\partial_{t}+\sqrt{G}\dot{\mathbf{x}}\cdot\nabla_{\mathbf{x}}+\sqrt{G}\dot{\mathbf{p}}\cdot\nabla_{\mathbf{p}}\right]f(t,\mathbf{x},\mathbf{p})=0.\label{eq:CKT_non_interaction_01}
\end{equation}
Here we have multiplied $\sqrt{G}$ to the original equation. The
above equation is usually called the chiral kinetic equation for massless
fermions.

Integrating Eq.~(\ref{eq:CKT_non_interaction_01}) over $\mathbf{p}$,
we obtain the continuity equation for the particle number of right-handed
fermions (chiral charge), 
\begin{equation}
\partial_{t}\rho+\boldsymbol{\nabla}\cdot\mathbf{J}=\frac{1}{2\pi}(\mathbf{E}\cdot\mathbf{B})f(\mathbf{p}=0),\label{eq:anomaly_02}
\end{equation}
where $\rho$ and $\mathbf{J}$ are the particle number density and
current density, 
\begin{eqnarray}
\rho & = & \int\frac{d^{3}\mathbf{p}}{(2\pi)^{3}}\sqrt{G}f(t,\mathbf{x},\mathbf{p}),\nonumber \\
\mathbf{J} & = & \int\frac{d^{3}\mathbf{p}}{(2\pi)^{3}}\sqrt{G}\dot{\mathbf{x}}f(t,\mathbf{x},\mathbf{p}),\label{eq:current_CKE_01}
\end{eqnarray}
and, we have also used the Maxwell equations and
\begin{equation}
\partial_{t}\sqrt{G}+\nabla_{\mathbf{x}}\cdot(\sqrt{G}\dot{\mathbf{x}})+\nabla_{\mathbf{p}}\cdot(\sqrt{G}\dot{\mathbf{p}})=2\pi(\mathbf{E}\cdot\mathbf{B})\delta^{3}(\mathbf{p}).\label{eq:var_phase_space_01}
\end{equation}
Equation (\ref{eq:anomaly_02}) means that the total chiral charge
is no longer conserved but can be changed by a source $2\pi(\mathbf{E}\cdot\mathbf{B})\delta^{3}(\mathbf{p})$
coming from the Berry curvature (\ref{eq:Berry_monopole}). It is
remarkable that the Berry curvature plays a crucial role to reproduce
the chiral anomaly.

We assume that $f(t,\mathbf{x},\mathbf{p})$ is the Fermi-Dirac distribution
function, then we can compute the current and the right-hand side
of Eq.~(\ref{eq:anomaly_02}). Inserting Eq.~(\ref{eq:EOM_01}) into
the current (\ref{eq:current_CKE_01}), it gives the chiral magnetic
effect for right-handed fermions, 
\begin{equation}
\mathbf{J}_{R}=\frac{1}{2\pi^{2}}\mu_{R}\mathbf{B},
\end{equation}
and the continuity equation for the chiral charge with chiral anomaly,
\begin{equation}
\partial_{t}\rho_{R}+\boldsymbol{\nabla}\cdot\mathbf{J}_{R}=\frac{1}{4\pi^{2}}(\mathbf{E}\cdot\mathbf{B}),\label{eq:nR_01}
\end{equation}
where $\mu_{R}$ is the chemical potential for right-handed fermions.

Although the chiral anomaly is a topological effect and should be intact by the interaction, it seems that the correct chiral anomaly
coefficient depends on the value $f(\mathbf{p}=\mathbf{0})$ as shown in Eq.~(\ref{eq:anomaly_02}). One may wonder what happens if $f(\mathbf{p}=\mathbf{0})\neq1$
in off-equilibrium systems. Such a problem has been clarified in Refs.
\cite{Gao:2019zhk,Gao:2020pfu} in the covariant Wigner function approach.
It has been shown that the chiral anomaly is actually from the Dirac
sea or the vacuum contribution in the Wigner function without normal-ordering,
and this contribution modifies the chiral kinetic equation for antiparticles
\cite{Gao:2019zhk,Gao:2020pfu}.

The path integral approaches for the non-Abelian Berry phases are discussed in
Ref. \cite{Chen:2013iga} and also see Ref. \cite{Pu:2017apt} for the lattice QCD simulations of non-Abelian Berry curvatures. 

\subsection{Hamiltonian approach and symplectic form} \label{subsec:symplectic form}

\label{subsec:Hamiltonian-approaches}In this subsection, we follow
Ref.~\cite{Son:2012wh,Chen:2012ca} to illustrate the Hamiltonian
approach from the action (\ref{eq:action-11}).

In a general case, the particle might be dressed, e.g., in a condensed
matter system, we, therefore, extend the action (\ref{eq:action-11})
to \cite{Xiao:2005qw,Duval:2005vn}

\begin{equation}
S=\int dt\left[\mathbf{p}\cdot\dot{\mathbf{x}}+e\mathbf{A}(\mathbf{x})\cdot\mathbf{x}-\mathbf{a}_\mathbf{p}\cdot\dot{\mathbf{p}}-H(\mathbf{x},\mathbf{p})\right],\label{eq:action_02}
\end{equation}
where $H(\mathbf{x},\mathbf{p})$ contains $\epsilon(\mathbf{p})$
and $\phi(\mathbf{x})$. In general the particle's energy can depend
on both $\mathbf{x}$ and $\mathbf{p}$. For simplicity, here we only
consider the action for particles (similar for anti-particles).

We introduce a $6$ dimensional parameter $\xi^{a}=(\mathbf{x},\mathbf{p})$
in phase space and the action (\ref{eq:action_02}) can be cast into
\begin{equation}
S=\int dt\left[-w_{a}\dot{\xi}^{a}-H(\xi)\right],\label{eq:action_03}
\end{equation}
where $w_{a}=(-\dot{\mathbf{p}}-\dot{\mathbf{A}},\mathbf{a}_\mathbf{p})$ and lower index $a=1,2,...,6$.
The equations of motion or the Euler equations give 
\begin{eqnarray}
\omega_{ab}\dot{\xi}^{b} & = & -\frac{\partial H}{\partial\xi^{a}},\label{eq:equation_motion_01}
\end{eqnarray}
with 
\begin{eqnarray}
\omega_{ab} & \equiv & \partial_{a} w_{b}-\partial_{b}w_{a}=\left(\begin{array}{cc}
-\epsilon_{ijk}\Omega_{k} & -\delta_{jk}\\
\delta_{jk} & \epsilon_{ijk}B_{k}
\end{array}\right)\nonumber \\
 & = & \left(\begin{array}{cccccc}
0 & \Omega_{3} & -\Omega_{2} & -1 & 0 & 0\\
-\Omega_{3} & 0 & \Omega_{1} & 0 & -1 & 0\\
\Omega_{2} & -\Omega_{1} & 0 & 0 & 0 & -1\\
1 & 0 & 0 & 0 & -B_{3} & B_{2}\\
0 & 1 & 0 & B_{3} & 0 & -B_{1}\\
0 & 0 & 1 & -B_{2} & B_{1} & 0
\end{array}\right),
\end{eqnarray}
where $\boldsymbol{\Omega}$ is the Berry curvature in Eq.~(\ref{eq:Berry_curv_01}).

The dynamics of the system by the action (\ref{eq:action_03}) is
defined in phase space and can be written in a symplectic form \cite{Duval:2005vn}.
Similar to the metric tensor $\eta_{\mu\nu}$, $\omega_{ab}$ is called
the symplectic matrix and it is anti-symmetric with its inverse $\omega^{ab}=(\omega^{-1})^{ab}$
satisfying 
\begin{equation}
\omega^{ab}\omega_{bc}=\delta_{c}^{a}.
\end{equation}
The invariant volume in this symplectic form is given by 
\begin{equation}
dV=\frac{1}{d!}\omega^{d}=\sqrt{\det\omega_{ab}}\prod_{\alpha=1}^{2d}d\xi^{\alpha},\label{eq:volume_01}
\end{equation}
where 
\begin{equation}
\det\omega_{ab}=(1+\boldsymbol{\Omega}_{\mathbf{p}}\cdot\mathbf{B})^{2}.
\end{equation}
Therefore the invariant phase space volume becomes $\sqrt{G}d^{3}\mathbf{x}d^{3}\mathbf{p}$.
Similar to $\eta_{\mu\nu}$, we also have
\begin{equation}
\partial_{b}(\sqrt{\det(\omega_{ab})}\;\omega^{ab})=0.\label{eq:identity_omega_sympetical_01}
\end{equation}
We also introduce the Poisson bracket in a symplectic form related
to $\omega_{ab}$,
\begin{equation}
\{f,g\}_{\omega}=\omega^{ab}\partial_{a}f\partial_{b}g.\label{eq:Possion_bracket_01}
\end{equation}
Equation (\ref{eq:equation_motion_01}) can be put into the form,
\begin{equation}
\dot{\xi}^{a}=\{H,\xi^{a}\}_{\omega}=\{\xi^{b},\xi^{a}\}_{\omega}\frac{\partial H}{\partial\xi^{b}}.
\end{equation}
The following Poisson brackets are given by \cite{Duval:2005vn,Son:2012wh}
\begin{eqnarray}
\{\mathbf{p}_{i},\mathbf{p}_{j}\}_{\omega} & = & -\frac{\epsilon_{ijk}B_{k}}{1+\boldsymbol{\Omega}_{\mathbf{p}}\cdot\mathbf{B}},\nonumber \\
\{\mathbf{x}_{i},\mathbf{x}_{j}\}_{\omega} & = & \frac{\epsilon_{ijk}\Omega_{k}}{1+\boldsymbol{\Omega}_{\mathbf{p}}\cdot\mathbf{B}},\nonumber \\
\{\mathbf{p}_{i},\mathbf{x}_{j}\}_{\omega} & = & \frac{\delta_{ij}+\Omega_{i}B_{j}}{1+\boldsymbol{\Omega}_{\mathbf{p}}\cdot\mathbf{B}},\label{eq:relation_commutator_01}
\end{eqnarray}
which are quite different from conventional Poisson brackets for $x$
and $p$ in classical mechanics. The structure of the phase space
in the presence of the Berry curvature becomes more complicated.

Our aim is to compute the time evolution of the occupation number
density operator $n_{\mathbf{p}}$ through the Heisenberg equation,
\begin{equation}
\partial_{t}n_{\mathbf{p}}(\mathbf{x})=i[H,n_{\mathbf{p}}(\mathbf{x})].\label{eq:Heisenberg_01}
\end{equation}
In a usual quantum system, we can assume that two operators $\widehat{A}_{1}$
and $\widehat{A}_{2}$ are linear in $n_{\mathbf{p}}$, 
\begin{equation}
\widehat{A}_{i}=\int\frac{d^{3}\mathbf{p}d^{3}\mathbf{x}}{(2\pi)^{3}}A_{i}(\mathbf{x},\mathbf{p})n_{\mathbf{p}}(\mathbf{x}),\label{eq:A_exp_np}
\end{equation}
where $A_{i}(\mathbf{x},\mathbf{p})$ for $i=1,2$ are phase space
functions. Then their commutator is given by 
\begin{equation}
[\widehat{A}_{1},\widehat{A}_{2}]=-i\int\frac{d^{3}\mathbf{p}d^{3}\mathbf{x}}{(2\pi)^{3}}
\left\{ 
\frac{\partial A_{1}}{\partial\mathbf{p}}\cdot\frac{\partial A_{2}}{\partial\mathbf{x}}-\frac{\partial A_{1}}{\partial\mathbf{x}}\cdot\frac{\partial A_{2}}{\partial\mathbf{p}}
\right\}
n_{\mathbf{p}}(\mathbf{x}),\label{eq:commutator_01}
\end{equation}
where 
we have used 
\begin{equation}
[n_{\mathbf{p}}(\mathbf{x}),n_{\mathbf{q}}(\mathbf{y})]=-i(2\pi)^{3}[\nabla_{\mathbf{p}}\delta(p-q)\cdot\nabla_{\mathbf{x}}\delta(x-y)][n_{\mathbf{p}}(\mathbf{y})-n_{\mathbf{q}}(\mathbf{x})].
\end{equation}

In the system with action (\ref{eq:action_02}), we need to revisit
the above calculation. In the symplectic manifold, the operators $\widehat{A}_{i}$
now become 
\begin{equation}
\widehat{A}_{i}=\int d\xi\sqrt{\det\omega_{ab}}A_{i}(\xi)n(\xi),
\end{equation}
where $n(\xi)$ is the occupation number density operator in phase
space of $\xi$. 
The bracket term $\{...\}$ in the left handed sided of Eq.~(\ref{eq:commutator_01}),
which is the classical Poisson brackets,
is changed to the one in Eq.~(\ref{eq:Possion_bracket_01}). It is
natural to define the commutator (\ref{eq:commutator_01}) as \cite{Son:2012wh}
\begin{equation}
[\widehat{A}_{1},\widehat{A}_{2}]=-i\int d\xi\sqrt{\det\omega_{ab}}\{A_{1},A_{2}\}_{\omega}n_{\mathbf{p}}(\mathbf{x}).\label{eq:commutator_02}
\end{equation}
Usually, the above description is implemented to the low energy or
low-temperature systems, e.g., Fermi liquid at the low temperature,
in which the physics near the Fermi surface is relevant. Integrating
by part and using Eq.~(\ref{eq:identity_omega_sympetical_01}), one
can rewrite Eq.~(\ref{eq:commutator_02}) as 
\begin{equation}
[\widehat{A}_{1},\widehat{A}_{2}]=-\frac{i}{2}\int d\xi\sqrt{\det\omega_{cd}}\omega^{ba}(A_{1}\partial_{b}A_{2}-A_{2}\partial_{b}A_{1})\partial_{a}n(\xi).\label{eq:commutator_03}
\end{equation}
If $\widehat{A}_{i}$ are general functions of $n(\xi)$, one can
also generalize the above commutator as 
\begin{equation}
[\widehat{A}_{1},\widehat{A}_{2}]=-\frac{i}{2}\int d\xi\sqrt{\det\omega_{cd}}\omega^{ba}\left(\frac{\delta A_{1}}{\delta n(\xi)}\partial_{b}\frac{\delta A_{2}}{\delta n(\xi)}-\frac{\delta A_{2}}{\delta n(\xi)}\partial_{b}\frac{\delta A_{1}}{\delta n(\xi)}\right)\partial_{a}n(\xi).\label{eq:commutator_04}
\end{equation}

Now we implement Eq.~(\ref{eq:commutator_03}) to compute the commutator
of two operators for the particle number density,
\begin{eqnarray}
[\rho(\mathbf{y}),\rho(\mathbf{z})] & = & -\frac{i}{2}\int\frac{d^{3}\mathbf{p}d^{3}\mathbf{x}}{(2\pi)^{3}}(1+\boldsymbol{\Omega}_{\mathbf{p}}\cdot\mathbf{B})\nonumber \\
 &  & \times[\delta(\mathbf{x}-\mathbf{y})\partial_{i}^{x}\delta(\mathbf{x}-\mathbf{z})-\delta(\mathbf{x}-\mathbf{z})\partial_{i}^{x}\delta(\mathbf{x}-\mathbf{y})]\nonumber \\
 &  & \times\left[\{x^{i},x^{j}\}_{\omega}\frac{\partial n_{\mathbf{p}}(\mathbf{x})}{\partial x^{j}}+\{x^{i},p^{j}\}_{\omega}\frac{\partial n_{\mathbf{p}}(\mathbf{x})}{\partial p^{j}}\right],\label{eq:comm-rho-x-rho-y}
\end{eqnarray}
where we have used the relation for the particle number density operator,
\begin{eqnarray}
\rho(\mathbf{x}) & = & \int\frac{d^{3}\mathbf{p}}{(2\pi)^{3}}n_{\mathbf{p}}(\mathbf{x})\sqrt{\det\omega_{ab}}\nonumber \\
 & = & \int\frac{d^{3}\mathbf{p}d^{3}\mathbf{x}^{\prime}}{(2\pi)^{3}}(1+\boldsymbol{\Omega}_{\mathbf{p}}\cdot\mathbf{B})\delta(\mathbf{x}-\mathbf{x}^{\prime})n_{\mathbf{p}}(\mathbf{x}^{\prime}).
\end{eqnarray}
Using the relations (\ref{eq:relation_commutator_01}), we can compute
each term in Eq.~(\ref{eq:comm-rho-x-rho-y}). For the terms proportional
to $\{x^{i},x^{j}\}_{\omega}$, we find 
\begin{eqnarray}
I_{1} & = & -i\int\frac{d^{3}\mathbf{p}d^{3}\mathbf{x}}{(2\pi)^{3}}(1+\boldsymbol{\Omega}_{\mathbf{p}}\cdot\mathbf{B})\delta(\mathbf{x}-\mathbf{y})\partial_{i}^{x}\delta(\mathbf{x}-\mathbf{z})\{x^{i},x^{j}\}_{\omega}\frac{\partial n_{\mathbf{p}}(\mathbf{x})}{\partial x^{j}}\nonumber \\
 & = & -i\left\{ \nabla_{\mathbf{y}}\times\boldsymbol{\sigma}(\mathbf{y})\right\} \cdot\nabla_{\mathbf{y}}\delta(\mathbf{y}-\mathbf{z}).
\end{eqnarray}
where 
\begin{equation}
\sigma_{i}(\mathbf{x})\equiv-\int\frac{d^{3}\mathbf{p}}{(2\pi)^{3}}p_{i}\Omega_{k}\frac{\partial n_{\mathbf{p}}(\mathbf{x})}{\partial p^{k}}.
\end{equation}
While the terms proportional to $\{x^{i},p^{j}\}_{\omega}$ can be
calculated as
\begin{eqnarray}
I_{2} & = & -i\int\frac{d^{3}\mathbf{p}d^{3}\mathbf{x}}{(2\pi)^{3}}(1+\boldsymbol{\Omega}_{\mathbf{p}}\cdot\mathbf{B})\delta(\mathbf{x}-\mathbf{y})\partial_{i}^{x}\delta(\mathbf{x}-\mathbf{z})\{x^{i},p^{j}\}_{\omega}\frac{\partial n_{\mathbf{p}}(\mathbf{x})}{\partial p^{j}}\nonumber \\
 & = & -i[\mathbf{B}\cdot\nabla_{\mathbf{y}}\delta(\mathbf{y}-\mathbf{z})]\int\frac{d^{3}\mathbf{p}}{(2\pi)^{3}}(\nabla_{\mathbf{p}}\cdot\boldsymbol{\Omega}_{\mathbf{p}})n_{\mathbf{p}}(\mathbf{y}).
\end{eqnarray}
In reaching the above result for $I_{2}$, we have assumed that the
number density is just a Fermi sphere $n_{\mathbf{p}}^{\mathrm{F.S.}}(\mathbf{x})\propto\theta(|\mathbf{p}|-p_{F})$
with $p_{F}$ being the Fermi momentum, which is a good approximation
at low temperatures and high densities. Then we obtain 
\begin{equation}
I_{2}=-\frac{\mathcal{K}}{8\pi^{2}}\mathbf{B}\cdot\nabla_{\mathbf{y}}\delta(\mathbf{y}-\mathbf{z}),
\end{equation}
where $\mathcal{K}$ is the monopole charge 
\begin{equation}
\mathcal{K}\equiv\frac{1}{2\pi}\int d\mathbf{S}\cdot\boldsymbol{\Omega_{\mathbf{p}}}.
\end{equation}
The commutator can then be put into a compact form,
\begin{equation}
[\rho(\mathbf{y}),\rho(\mathbf{z})]=-i\left[\nabla_{\mathbf{y}}\times\boldsymbol{\sigma}(\mathbf{y})+\frac{\mathcal{K}}{4\pi^{2}}\mathbf{B}\right]\cdot\nabla_{\mathbf{y}}\delta(\mathbf{y}-\mathbf{z}).\label{eq:n_n_commutator_01}
\end{equation}

We now compute the evolution equation for $\rho(\mathbf{x})$. The
Hamiltonian $H$ in the action (\ref{eq:action_02}) can be written
as 
\begin{equation}
H=\int\frac{d^{3}\mathbf{p}d^{3}\mathbf{x}}{(2\pi)^{3}}\epsilon_{\mathbf{p}}\delta n_{\mathbf{p}}+\int d^{3}\mathbf{x}\phi(\mathbf{x})\rho(\mathbf{x}),
\end{equation}
where $\epsilon_{\mathbf{p}}$ is the energy of a single quasi-particle,
$\delta n_{\mathbf{p}}=n_{\mathbf{p}}-n_{\mathbf{p}}^{\textrm{F.S.}}$
is the deviation from the ground state $n_{\mathbf{p}}^{\textrm{F.S.}}$,
and $\phi(\mathbf{x})$ is the electric potential of the electromagnetic
field appearing in Eq.~(\ref{eq:chiral-f-h}). In general, the interaction
can change $\epsilon_{\mathbf{p}}$, one can also introduce an extra
term,
\begin{equation}
\frac{1}{2}\int\frac{d^{3}\mathbf{p}d^{3}\mathbf{q}d^{3}\mathbf{x}}{(2\pi)^{6}}g(\mathbf{p},\mathbf{q})\delta n_{\mathbf{p}}\delta n_{\mathbf{q}},
\end{equation}
to account for the modification from the interaction with the Landau
parameter $g(\mathbf{p},\mathbf{q})$. Using Eqs.~(\ref{eq:commutator_04}) and 
(\ref{eq:n_n_commutator_01}), we find 
\begin{eqnarray}
\partial_{t}\rho(\mathbf{x}) & = & i[H,\rho(\mathbf{x})]=-\nabla\cdot\mathbf{j}+\frac{\mathcal{K}}{4\pi^{2}}\mathbf{B}\cdot\mathbf{E},\label{eq:partial_n_02}
\end{eqnarray}
where we have used $\mathbf{E}=-\nabla\phi(\mathbf{x})$.

Equation (\ref{eq:partial_n_02}) is the same as Eq.~(\ref{eq:nR_01}).
The current is 
\begin{equation}
\mathbf{J}(\mathbf{x})=\int\frac{d^{3}\mathbf{p}}{(2\pi)^{3}}\left[\frac{\partial\epsilon_{\mathbf{p}}}{\partial\mathbf{p}}n_{\mathbf{p}}-\left(\boldsymbol{\Omega}_{\mathbf{p}}\cdot\frac{\partial n_{\mathbf{p}}}{\partial\mathbf{p}}\right)\epsilon_{\mathbf{p}}\mathbf{B}-\epsilon_{\mathbf{p}}\boldsymbol{\Omega}_{\mathbf{p}}\times\frac{\partial n_{\mathbf{p}}}{\partial\mathbf{x}}\right]+\mathbf{E}\times\boldsymbol{\sigma}(\mathbf{x}),\label{eq:current_03}
\end{equation}
where $\partial\epsilon_{\mathbf{p}}/\partial\mathbf{p}$ is the particle's
velocity, the second term proportional to $\mathbf{B}$ leads to the
chiral magnetic currents, and the last term $\mathbf{E}\times\boldsymbol{\sigma}(\mathbf{x})$
corresponds to $\mathbf{E}\times\boldsymbol{\Omega}_{\mathbf{p}}$
in Eq.~(\ref{eq:EOM_01}). We see that the term $\boldsymbol{\Omega}_{\mathbf{p}}\times(\partial n_{\mathbf{p}}/\partial\mathbf{x})$
implies that an inhomogeneous distribution function can also lead
to a current. Recently, such a term provides new sources for the local
spin polarization, e.g., the polarization induced by the shear viscous
term and gradient of the chemical potential over temperature \cite{Hidaka:2018ekt,Liu:2021uhn,Fu:2021pok,Becattini:2021suc,Becattini:2021iol,Ryu:2021lnx}.

\subsection{Lorentz transformation and side jump \label{subsec:Side-jump,-Lorentz}}

We consider the Lorentz covariance of the chiral kinetic equation
in this subsection. Restoring the Planck constant, the action (\ref{eq:action-11})
for the particle is 
\begin{equation}
S=\int\left[(\mathbf{p}+\mathbf{A})\cdot d\mathbf{x}-\left(|\mathbf{p}|-\hbar\frac{\mathbf{B}\cdot\mathbf{p}}{2|\mathbf{p}|^{2}}+\phi\right)dt-\hbar\mathbf{a}_{\mathbf{p}}\cdot d\mathbf{p}\right],\label{eq:action_Zeeman_01}
\end{equation}
where $\mathbf{B}\cdot\mathbf{p}/(2|\mathbf{p}|^{2})$ is the correction
to the particle's energy from the magnetic fields, i.e. Zeeman effect
for massless fermions.

At $O(\hbar^{0})$, it is straightforward to verify that the action
is invariant under the standard infinitesimal Lorentz transformation
\cite{Chen:2014cla},
\begin{eqnarray}
\delta\mathbf{x} & = & \boldsymbol{\beta}_{u}t,\nonumber \\
\delta t & = & \boldsymbol{\beta}_{u}\cdot\mathbf{x},\nonumber \\
\delta\mathbf{p} & = & \boldsymbol{\beta}_{u}|\mathbf{p}|,\nonumber \\
\delta\mathbf{B} & = & \boldsymbol{\beta}_{u}\times\mathbf{E},\nonumber \\
\delta\mathbf{E} & = & -\boldsymbol{\beta}_{u}\times\mathbf{B},\label{eq:Lorentz_01}
\end{eqnarray}
where $\boldsymbol{\beta}_{u}$ is the boost velocity. At $O(\hbar^{1})$,
under the transformation (\ref{eq:Lorentz_01}) we have 
\begin{eqnarray}
\delta\left(\frac{\mathbf{B}\cdot\mathbf{p}}{2|\mathbf{p}|^{2}}dt\right) & = & \left[-\frac{(\boldsymbol{\beta}_{u}\times\mathbf{E})\cdot\mathbf{p}}{2|\mathbf{p}|^{2}}-\frac{\mathbf{B}\cdot\boldsymbol{\beta}_{u}}{2|\mathbf{p}|}-\frac{(\mathbf{B}\cdot\mathbf{p})(\boldsymbol{\beta}_{u}\cdot\mathbf{p})}{|\mathbf{p}|^{3}}\right]dt\nonumber \\
 &  & -\frac{\mathbf{B}\cdot\mathbf{p}}{2|\mathbf{p}|^{2}}(\boldsymbol{\beta}_{u}\cdot d\mathbf{x}),\nonumber \\
\delta(\mathbf{a}_{\mathbf{p}}\cdot d\mathbf{p}) & = & |\mathbf{p}|(\boldsymbol{\beta}_{u}\times\boldsymbol{\Omega}_{\mathbf{p}})\cdot d\mathbf{p}-d(\mathbf{a}_{\mathbf{p}}\cdot\boldsymbol{\beta}_{u}|\mathbf{p}|),
\end{eqnarray}
which lead to the variation of the action,
\begin{eqnarray}
\delta S & = & -\hbar\int\left[\frac{(\boldsymbol{\beta}_{u}\times\mathbf{p})}{2|\mathbf{p}|^{2}}\cdot(d\mathbf{p}-\mathbf{E}-\widehat{\mathbf{p}}\times\mathbf{B})-\frac{(\mathbf{B}\cdot\mathbf{p})(\boldsymbol{\beta}_{u}\cdot\mathbf{p})}{2|\mathbf{p}|^{3}}\right]dt\nonumber \\
 &  & +\hbar\int\left[\frac{\mathbf{B}\cdot\mathbf{p}}{2|\mathbf{p}|^{2}}(\boldsymbol{\beta}_{u}\cdot d\mathbf{x})dt\right].\label{eq:var_S_h_01}
\end{eqnarray}
Here we have dropped the total derivative term $+\hbar\int d\left(\mathbf{a}_{\mathbf{p}}\cdot\boldsymbol{\beta}|\mathbf{p}|\right)$.
It is obvious that $\delta S$ is non-vanishing in general. The
standard Lorentz symmetry seems to be broken.

There are two ways to solve this problem. One way is to modify the
standard infinitesimal Lorentz transformation (\ref{eq:Lorentz_01}),
e.g., add quantum corrections to Eq.~(\ref{eq:Lorentz_01}) to cancel
$\delta S$ in Eq.~(\ref{eq:var_S_h_01}). The second way is more
rigorous. We emphasize that in quantum field theory the Lorentz transformation
is well-defined and should not be modified. In the Wigner function
approach, we will demonstrate that there are quantum corrections to
the distribution function under the Lorentz transformation \cite{Hidaka:2016yjf}.
We will discuss it later in this subsection.

In the first way, we assume the modified Lorentz transformation to
the order of $\hbar$ \cite{Chen:2014cla},
\begin{equation}
\delta\mathbf{x}=\boldsymbol{\beta}_{u}t+\hbar\mathbf{G},\;\delta\mathbf{p}=\boldsymbol{\beta}_{u}|\mathbf{p}|+\hbar\mathbf{F},
\end{equation}
then we have 
\begin{eqnarray}
 &  & \delta[(\mathbf{p}+\mathbf{A})\cdot d\mathbf{x}-(|\mathbf{p}|+\phi)dt]\nonumber \\
 & = & -\mathbf{F}\cdot d\mathbf{x}-\mathbf{p}\cdot d\mathbf{G}+\frac{\mathbf{F}\cdot\mathbf{p}}{2|\mathbf{p}|}dt+(\mathbf{G}\times\mathbf{B})\cdot d\mathbf{x}+(\mathbf{G}\cdot\mathbf{E})dt.
\end{eqnarray}
Compared with Eq.~(\ref{eq:var_S_h_01}), we obtain 
\begin{eqnarray}
\delta\mathbf{x} & = & \boldsymbol{\beta}_{u}t+\hbar\frac{\boldsymbol{\beta}_{u}\times\widehat{\mathbf{p}}}{2|\mathbf{p}|},\nonumber \\
\delta\mathbf{p} & = & \boldsymbol{\beta}_{u}|\mathbf{p}|-\hbar\frac{\mathbf{B}\cdot\mathbf{p}}{2|\mathbf{p}|^{2}}+\hbar\frac{\boldsymbol{\beta}_{u}\times\widehat{\mathbf{p}}}{2|\mathbf{p}|}\times\mathbf{B},\label{eq:Lorentz_trans_01}
\end{eqnarray}
which is called the side-jump \cite{Skagerstam:1992er}.

\begin{figure}
\begin{centering}
\includegraphics[scale=0.55]{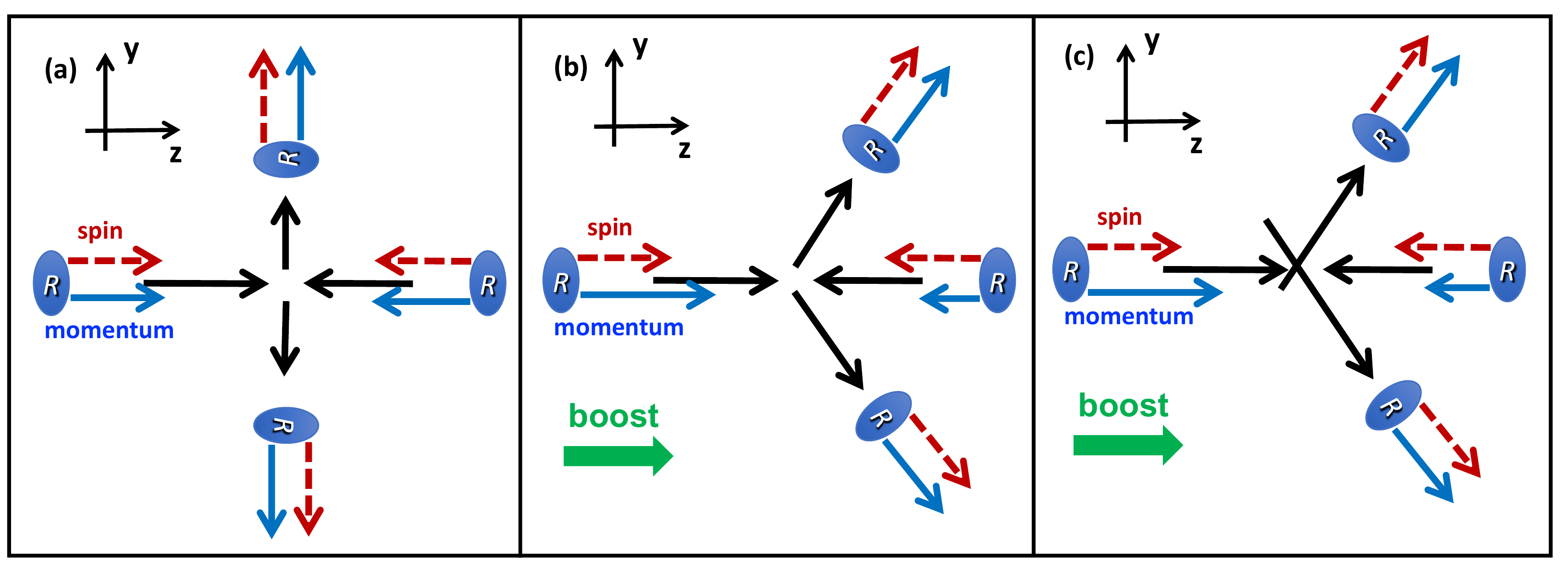}
\par\end{centering}
\caption{Illustration of the side jump and total angular momentum conservation
for massless fermions. (a) Two right-handed fermions collide with
each other. The initial and final momenta of fermions are assumed
to be along $z$ and $y$ directions, respectively. The orbital angular
momentum and spin are conserved independently in this special case.
(b) We boost the system along the $z$ direction. The spins in the
final state do not cancel. (c) To keep the total angular momentum
conserved, there are quantum corrections to the position of the final
state. \label{fig:side-jump}}
\end{figure}

The origin of the side jump effect comes from the total angular momentum
conservation. To illustrate it, we can consider a collision of two
right-handed fermions in the absence of background electromagnetic fields
as shown in Fig. \ref{fig:side-jump}. In Fig. \ref{fig:side-jump}(a),
we simply assume that the initial and final fermions move along the
$z$ and $y$ direction, respectively. We see that that the (classical)
orbital angular momentum and spin are conserved separately. Then we
make a Lorentz boost along the $z$ direction as shown in Fig. \ref{fig:side-jump}(b).
The total orbital angular momentum and spin in the initial state are
vanishing separately. In the final state, the total orbital angular
momentum is not modified. Due to the change of final momenta, the
total spin in the final state becomes nonzero. In order to make the
total angular momentum conserve, the only way in this scenario is
to introduce additional corrections to the orbital angular momentum
to cancel such a change of the total spin. That is the reason for
the extra quantum correction to $\delta\mathbf{x}$ in Eq.~(\ref{eq:Lorentz_trans_01}).

The above discussion can also be extended to the finite Lorentz transformation
\cite{Chen:2015gta}. For massless fermions, the total angular momentum
tensor is in a canonical form,
\begin{equation}
J^{\mu\nu}=x^{\mu}p^{\nu}-x^{\nu}p^{\mu}+S^{\mu\nu},\label{eq:TAM_01}
\end{equation}
where $S^{\mu\nu}$ is usually called the canonical spin. In general,
one can also introduce another decomposition of $J^{\mu\nu}$ through
the pseudo-gauge transformation. One can always make the following transformation
for $x^{\mu}$ and $S^{\mu\nu}$,
\begin{eqnarray}
x^{\mu} & \rightarrow & x^{\prime\mu}=x^{\mu}+\delta x^{\mu},\nonumber \\
S^{\mu\nu} & \rightarrow & S^{\prime\mu\nu}=S^{\mu\nu}+\delta x^{\nu}p^{\mu}-\delta x^{\mu}p^{\nu},\label{eq:spin_trans_01}
\end{eqnarray}
that keeps $J^{\mu\nu}$ invariant. One can find more detailed discussion
in recent reviews \cite{Gao:2020vbh,Fukushima:2020qta} and related
works for dissipative spin hydrodynamics \cite{Hattori:2019lfp,Fukushima:2020ucl,Li:2020eon,Wang:2021ngp,Hongo:2021ona}.

As a spin tensor, $S^{\mu\nu}$ has 6 degrees of freedom, while the
spin of the massless fermion is always parallel or anti-parallel to
its momentum. To remove non-physical degrees of freedom, one needs
to impose an additional condition, Frenkel-Weyssenhoff condition,
\begin{equation}
p_{\mu}S^{\mu\nu}=0.\label{eq:pS_01}
\end{equation}
We find that Eq.~(\ref{eq:pS_01}) only requires $p\cdot\delta x=0$
under the transformation (\ref{eq:spin_trans_01}), which means that
the condition (\ref{eq:pS_01}) is insufficient to fix $S^{\mu\nu}$.
Another condition is needed, and one can choose 
\begin{equation}
n^{\mu}S_{\mu\nu}=0,\label{eq:nS_02}
\end{equation}
where $n^{\mu}$ is the frame vector or velocity of an arbitrary frame.
These two conditions, (\ref{eq:pS_01}) and (\ref{eq:nS_02}), indicate the
following form of $S^{\mu\nu}$,
\begin{equation}
S_{(n)}^{\mu\nu}=\lambda\frac{\epsilon^{\mu\nu\alpha\beta}p_{\alpha}n_{\beta}}{2(p\cdot n)},
\label{eq:spin}
\end{equation}
where $\lambda=\pm$ denotes the helicity for the right- (upper sign)
and left-handed (lower sign) fermions, and the subscript $(n)$ stands
for the frame vector $n^{\mu}$. There are two common choices for
$n^{\mu}$: the local rest frame $(1,\boldsymbol{0})$ and the fluid
velocity $u^{\mu}$.

Now we consider the finite Lorentz transformation for the coordinate
$x^{\mu}$. Under the transformation from the frame $n^{\mu}$ to
$n^{\prime\mu}$, using Eq.~(\ref{eq:spin_trans_01}), we obtain 
\begin{equation}
S_{(n^{\prime})}^{\prime\mu\nu}-S_{(n)}^{\mu\nu}=\delta x_{(nn^{\prime})}^{\nu}p^{\mu}-\delta x_{(nn^{\prime})}^{\mu}p^{\nu},
\end{equation}
which leads to \cite{Duval:2014cfa,Stone:2015kla,Chen:2015gta}
\begin{equation}
\delta x_{(nn^{\prime})}^{\mu}=\lambda\frac{\epsilon^{\mu\alpha\beta\gamma}p_{\alpha}n_{\beta}n_{\gamma}^{\prime}}{2(p\cdot n)(p\cdot n^{\prime})}.\label{eq:delta_x_02}
\end{equation}
Here the subscript $(nn^{\prime})$ indicates the frame dependence
of $\delta x^{\mu}$. In deriving the above equation, we have imposed
another condition $\delta x_{(nn^{\prime})}\cdot n=\delta x_{(nn^{\prime})}\cdot n^{\prime}=0$.
Equation (\ref{eq:delta_x_02}) is an extension of Eq.~(\ref{eq:Lorentz_trans_01}).

In relativistic kinetic theory, the distribution function $f(t,\mathbf{x},\mathbf{p})$
is not a scalar under Lorentz transformation as expected. Due to the
quantum corrections to both $x$ and $p$, we need to look at $f(x,p)$
in two different frames $n$ and $n^{\prime}$. After a rigorous derivation
of both $\delta x^{\mu}$ and $\delta p^{\mu}$ in finite Lorentz
transformation in quantum field theory, we find 
\begin{eqnarray}
x^{\mu} & \rightarrow & x^{\prime\mu}=x^{\mu}+\hbar N_{(nn^{\prime})}^{\mu},\nonumber \\
p^{\mu} & \rightarrow & p^{\prime\mu}=p^{\mu}+\hbar N_{\nu}^{(nn^{\prime})}F^{\mu\nu},
\end{eqnarray}
where $N_{(nn^{\prime})}^{\mu}$ is just $\delta x_{(nn^{\prime})}^{\mu}$.
Then $f_{(n^{\prime})}^{\prime}(x^{\prime},p^{\prime})$ in $n^{\prime}$
frame is given by \cite{Hidaka:2016yjf}
\begin{equation}
f_{(n^{\prime})}^{\prime}(x^{\prime},p^{\prime})=f_{(n)}(x,p)+\hbar N_{(nn^{\prime})}^{\mu}[\partial_{\mu}^{x}+F_{\nu\mu}\partial_{p}^{\nu}]f_{(n)}(x,p).
\end{equation}
To keep a relativistic vector $\mathscr{J}^{\mu}(x,p)$ in phase space
(vector component of the Wigner function for fermions with a fixed
chirality, see Secs \ref{sec:CCKE} and \ref{sec:solutions-wigner})
covariant, we should have 
\begin{equation}
\mathscr{J}^{\mu}(x,p)\sim p^{\mu}f+S_{(n)}^{\mu\nu}\partial_{\nu}f,
\end{equation}
in absence of background electromagnetic fields. In a three-vector
form, the above current becomes 
\begin{equation}
\mathbf{J}=\int\frac{d^{3}\mathbf{p}}{(2\pi)^{3}}\left(\widehat{\mathbf{p}}-\frac{\mathbf{p}}{2|\mathbf{p}|^{3}}\times\nabla_{\mathbf{x}}\right)f(t,\mathbf{x},\mathbf{p}),
\end{equation}
where the second term related to $S_{(n)}^{\mu\nu}$ corresponds to
the last term in the bracket in Eq.~(\ref{eq:current_03}). Such a
term can also be understood as the curl of the magnetization $\mathbf{M}$,
\begin{equation}
\mathbf{M}=\int\frac{d^{3}\mathbf{p}}{(2\pi)^{3}}\frac{\mathbf{p}}{2|\mathbf{p}|^{3}}f(t,\mathbf{x},\mathbf{p}).
\end{equation}
Such a contribution to the current from the magnetization is crucial
to reproduce the correct transport coefficient of the CVE \cite{Chen:2014cla}, also see Ref.~\cite{Kharzeev:2016sut}
for the discussion of similar terms for the CME
at finite frequencies. A systematic derivation of the current and
more discussion about the frame dependence are given in Sec.~\ref{sec:frame-decomposition}.

\subsection{Effective theories \label{subsec:Other-effective-theories}}

We briefly discuss in this subsection several other approaches to
the chiral kinetic equation from effective theories \cite{Son:2012zy, Lin:2019ytz, Manuel:2014dza,Carignano:2018gqt,Carignano:2019zsh,Manuel:2021oah}.
The starting point is to derive an effective Lagrangian through integrating
out irrelevant modes and obtain the equations of motion for the two-point
Green functions. One finally obtains the effective evolution equations
for distribution functions encoded in the two-point Green functions.

One convenient way to study chiral fermions is through the high-density
effective theory \cite{Son:2012zy} (also see the Refs. \cite{Hong:1998tn,Hong:1999ru,Schafer:2003jn} for the frameworks of the high-density effective theory). To extend the effective theories beyond the zero temperature limit, the authors in Refs. \cite{Manuel:2014dza,Carignano:2018gqt,Carignano:2019zsh,Manuel:2021oah} have 
developed the on-shell effective theory (OSEFT), which can recover the 
results derived from high-density effective theories in the zero temperature limits. 
Moreover, it has been pointed out by Ref. \cite{Carignano:2019zsh} that both high-density and on-shell effective theories are the counter-part of a Foldy-Wouthuysen diagonalization \cite{Foldy:1949wa},
in which the particles are antiparticles are disentangled after redefining.

For simplicity, 
in this subsection, we only concentrate on the high-density effective theory. 
In this theory, one concentrates on the physics near the Fermi surface.
The momentum of a particle can be written as 
\begin{equation}
p^{\mu}=\mu v^{\mu}+l^{\mu},\label{eq:p_high_density_01}
\end{equation}
where $\mu$ is the chemical potential for right-handed fermions,
$v^{\mu}=(1,\mathbf{v}),$ $\overline{v}^{\mu}=(1,-\mathbf{v})$ with
$\mathbf{v}$ being a unit three-vector, and $l^{\mu}=(l^{0},\mathbf{l})$
satisfying $|\mathbf{l}|\ll\mu$. The spinor can then be decomposed
as 
\begin{eqnarray}
\psi(x) & = & \int\frac{d^{4}p}{(2\pi)^{4}}e^{-ip\cdot x}\psi(p)=\sum_{v}e^{i\mu{\bf v}\cdot\mathbf{x}}\psi_{v}(x) 
=\sum_{v}e^{i\mu\mathbf{v}\cdot\mathbf{x}}[\psi_{+v}(x)+\psi_{-v}(x)],\label{eq:fast_slow_01}
\end{eqnarray}
where 
\begin{equation}
\psi_{\pm v}(x)\equiv P_{\pm}(\mathbf{v})\psi_{v}(x)=\int_{|l|<\mu}\frac{d^{4}l}{(2\pi)^{4}}e^{-il\cdot x}\psi_{\pm v}(l),\label{eq:wave_l_res_01}
\end{equation}
with the projector $P_{\pm}(\mathbf{v})=(1/2)(1\pm\boldsymbol{\sigma}\cdot\mathbf{v})$,
and $\psi_{\pm v}$ stands for the slow and fast modes, respectively.
Applying Eq.~(\ref{eq:fast_slow_01}) to the standard Lagrangian of
the right-handed fermions, 
\begin{equation}
\mathcal{L}=\psi^{\dagger}(i\sigma^{\mu}D_{\mu}+\mu)\psi,\label{eq:L_rhf_01}
\end{equation}
and expressing fast modes by slow modes via the equations of motion,
\begin{equation}
(2\mu+i\overline{v}\cdot D)\psi_{-v}+i\sigma\cdot D_{\perp}\psi_{+v}=0,
\end{equation}
we obtain the effective Lagrangian,
\begin{eqnarray}
\mathcal{L}_{\textrm{EFT}} & = & \sum_{v}\psi_{+v}^{\dagger}iv\cdot D\psi_{+v}(x)+\psi_{+v}^{\dagger}\sigma\cdot D_{\perp}\frac{1}{2\mu+i\overline{v}\cdot D}\sigma\cdot D_{\perp}\psi_{+v},\label{eq:effective_L_01}
\end{eqnarray}
where $D_{\mu}=\partial_{\mu}+iA_{\mu}$ and\textcolor{blue}{{} }$D_{\perp}^{\mu}=D^{\mu}-(v\cdot D)v^{\mu}$.
In the large $\mu$ limit, we can expand $\mathcal{L}_{\textrm{EFT}}$
in the power series of $1/\mu$, 
\begin{equation}
\mathcal{L}_{\textrm{EFT}}=\sum_{n}\mathcal{L}^{(n)},\;\mathcal{L}^{(n)}=\sum_{v}\psi_{+v}^{\dagger}\mathcal{D}^{(n)}\psi_{+v},
\end{equation}
where the effective operators are given by 
\begin{eqnarray}
\mathcal{D}^{(0)} & = & iv\cdot D,\nonumber \\
\mathcal{D}^{(1)} & = & \frac{(\sigma\cdot D_{\perp})^{2}}{2\mu},\nonumber \\
D^{(2)} & = & -\frac{i}{4\mu^{2}}(\sigma\cdot D^{\perp})(\overline{v}\cdot D)(\sigma\cdot D^{\perp}).
\end{eqnarray}
The explicit form of $\mathcal{L}^{(1)}$ at $O(1/\mu)$ is 
\begin{equation}
\mathcal{L}^{(1)}=\psi_{+v}^{\dagger}\left[\frac{D_{\perp}^{2}+\mathbf{B}\cdot\boldsymbol{\sigma}}{2\mu}\right]\psi_{+v}=\psi_{+v}^{\dagger}\left[\frac{D_{\perp}^{2}+\mathbf{B}\cdot\mathbf{v}}{2\mu}\right]\psi_{+v}.
\end{equation}

In order to derive the currents, we need to compute the effective
theory up to $O(1/\mu^{2})$. We introduce the two-point function
$G_{v}$, 
\begin{equation}
G_{v}(x_{1},x_{2})\equiv\left\langle \psi_{+v}(x_{1})\psi_{+v}^{\dagger}(x_{2})\right\rangle ,
\end{equation}
which satisfies the equations of motion,
\begin{equation}
\mathcal{D}_{x1}G_{v}(x_{1},x_{2})=0,\;G_{v}(x_{1},x_{2})\mathcal{D}_{x2}^{\dagger}=0,\label{eq:EOM_Gv_01}
\end{equation}
under the condition,
\begin{equation}
P_{-}G_{v}(x_{1},x_{2})=G_{v}(x_{1},x_{2})P_{-}=0,
\end{equation}
where $\mathcal{D}=\mathcal{D}^{(0)}+\mathcal{D}^{(1)}+\mathcal{D}^{(2)}+O(1/\mu^{3})$. 

We define 
\begin{equation}
x_{1}=x+\frac{y}{2},\;x_{2}=x-\frac{y}{2},
\end{equation}
and consider the gradient expansion under the following power counting,
\begin{eqnarray}
\partial_{x} & = & O(\epsilon_{1}),\;\partial_{s}=O(\epsilon_{2}),\;A_{\mu}=O(\epsilon_{3}),\nonumber \\
\epsilon_{1} & \ll & \epsilon_{2,3}\ll1,
\end{eqnarray}
where $\epsilon_{1,2,3}$ are all small and positive numbers playing
as expansion parameters. For simplicity, we assume that the system
is homogeneous, 
\begin{equation}
\partial_{x}^{\mu}F^{\alpha\beta}(x)=0,\;\partial_{i}^{x}f_v(x,l)=0,\;i=1,2,3,
\end{equation}
where $f_v(x,l)$ is the distribution function.

We now expand the equations of motion (\ref{eq:EOM_Gv_01}) order
by order and compute $G_{v}$. After a lengthy but straightforward
calculation, one finds \cite{Lin:2019ytz}
\begin{eqnarray}
G_{v} & = & 2\pi P_{+}(\mathbf{v})\delta\left(l_{0}-l_{\parallel}-\frac{l_{\perp}^{2}-\mathbf{B}\cdot\mathbf{v}}{2\mu}+\frac{l_{\parallel}[l_{\perp}^{2}-(\mathbf{B}\cdot\mathbf{v})]}{2\mu^{2}} \right. 
\left.+ \frac{\mathbf{B}\cdot\mathbf{l}_{\perp}+(\mathbf{E}\times \mathbf{l})\cdot \mathbf{v}}{4\mu^2}\right)f_v(x,l) +O(1/\mu^{3}),\label{eq:Gv_02}
\end{eqnarray}
and the particle's energy,
\begin{eqnarray}
\epsilon_{p} & \equiv & p^{0}=\mu+l^{0}=|\mathbf{p}|-\mathbf{B}\cdot\frac{\widehat{\mathbf{p}}}{2|\mathbf{p}|^{2}}+O(1/\mu^{3}),\label{eq:dispersion_01}
\end{eqnarray}
where 
\begin{eqnarray}
\mathbf{l}_{\parallel} & = & (\mathbf{v}\cdot\mathbf{l})\mathbf{v},\;\mathbf{l}_{\perp}=\mathbf{l}-\mathbf{l}_{\parallel},\nonumber \\
l_{\parallel} & \equiv & |\mathbf{l}_{\parallel}|=\mathbf{v}\cdot\mathbf{l},\;l\equiv|\mathbf{l}|=\sqrt{l_{\parallel}^{2}+l_{\perp}^{2}}.
\end{eqnarray}
Note that, the particle's energy $\epsilon_p$ also depends on $\mathbf{v}$. When one simply chooses $\mathbf{l}\parallel \mathbf{v}$, the dispersion relation reduces to a simple form in Eq.(\ref{eq:dispersion_01}) \cite{Lin:2019ytz}. 

Inserting Eq.~(\ref{eq:Gv_02}) into Eq.~(\ref{eq:EOM_Gv_01}) yields
the effective kinetic theory for the distribution function $f_v$.
We comment that the effective theory derived from the high density effective
theory looks different with Eq.~(\ref{eq:var_phase_space_01}). We emphasize that $f_v$ is the distribution function for the particles dressed with anti-particles, while the standard
distribution function $f$ describes distributions of the particles with positive energy only. 
Fortunately, one can prove that the kinetic theory for $f_v$ derived from the high density effective theory is equivalent to Eq.~(\ref{eq:var_phase_space_01}). More systematical discussion can be found in Ref.~\cite{Lin:2019ytz}.

Another method to derive the CKT is the on-shell effective field theory \cite{Manuel:2014dza,Carignano:2018gqt,Carignano:2019zsh,Manuel:2021oah},
in which the momentum of a fermion near its mass-shell can be written
as 
\begin{equation}
p^{\mu}=p^{0}v^{\mu}+l^{\mu},\qquad\mathrm{for}\;|l|\ll p^{0},
\end{equation}
similar to Eq.~(\ref{eq:p_high_density_01}) in the high-density effective
theory. Instead of integrating out slow modes, one can integrate out
off-shell modes. The advantage of on-shell effective theories is that it can
be implemented to the finite temperature cases. Recently the collision terms have been derived in the on-shell effective theory \cite{Carignano:2019zsh}.

Another important problem in effective theories is the reparametrization invariance. This problem for CKT was first discussed in the on-shell effective theory \cite{Carignano:2018gqt} and then in high-density effective theory \cite{Lin:2019ytz}. 
For example, the
high-density effective theory is invariant under re-parametrization
in general,
\begin{equation}
v^{\mu}\rightarrow v^{\prime\mu}=v^{\mu}+\delta v^{\mu},\qquad l^{\mu}\rightarrow l^{\prime\mu}=l^{\mu}-\mu\delta v^{\mu},\label{eq:reparameterization_01}
\end{equation}
where $p^{\mu}$ is invariant and $v\cdot\delta v=0$. However, it
is found that the delta function for $l_{0}$ in Eq.~(\ref{eq:Gv_02})
is not invariant under the transformation (\ref{eq:reparameterization_01}).
Although several works \cite{Carignano:2018gqt, Lin:2019ytz} have systematically discussed this problem, so far it is still an open question. 


It is also possible to derive the chiral kinetic theory from the world-line
formalism \cite{Mueller:2017lzw,Mueller:2017arw,Muller:2017rly}.
Note that the standard Schwinger proper time method or world-line
formalism provides the amplitude from the in-state to the out-state.
However, the evolution equations for particles require the expectation
value on the in-state \cite{Copinger:2018ftr,Copinger:2020nyx}. Although
the CME has been computed from the real time world-line
formalism \cite{Copinger:2018ftr,Copinger:2020nyx}, it is challenging
to derive the chiral kinetic theory in this formalism.
For the real-time simulations for the Schwinger mechanics from Wigner functions, one can see Refs. \cite{Hebenstreit:2010vz, Hebenstreit:2011pm, Kohlfurst:2015zxi, Kohlfurst:2019mag, Kohlfurst:2021skr} and the references therein. 

\subsection{Discussions}

A few remarks have to be made to end this section.

In the effective action (\ref{eq:action-11}), we have dropped off-diagonal
terms due to the assumption of adiabatic processes since they correspond
to transitions between states of different energies. However, the chiral
anomaly does not depend on this assumption. So a more rigorous way
free of such an assumption is to derive the chiral kinetic equation
from the quantum field theory, e.g., the covariant Wigner function
approach. The details of the covariant Wigner function approach to
the chiral kinetic equation are given in Sec.~\ref{sec:dwf-theorem}.

The collision terms in the chiral effective equation are neglected,
see, e.g., Eq.~(\ref{eq:CKT_non_interaction_01}). For transport phenomena,
particle collisions are essential and discussed in Sec.~\ref{sec:CKT_collisions} and \ref{sec:spin-Boltzmann}.

\newpage
\section{Kinetic equations for massless fermions with collisions}\label{sec:CKT_collisions}

In previous sections, we have not considered
interactions of massless fermions. This section will take them into
account and derive the Kadanoff-Baym equation with collisions for
massless fermions in background electromagnetic fields, from which
one can derive the chiral kinetic equation with collisions. Let us
start with the dressed propagator for right-handed fermions. In a
diagrammatic approach on the closed-time path in Fig.~\ref{fig:ctp},
the dressed propagator can be written as 
\begin{equation}
S=S_{0}-iS_{0}\Sigma S=S_{0}-iS\Sigma S_{0},\label{eq:SD equation}
\end{equation}
where $\Sigma$ is the self-energy, and $S_{0}$ is the free propagator.
For brevity of the notation, we omitted all indices such as the ones
for spinors and the CTP. Note that here $S$ does not necessarily
represent the lesser propagator taken in Sec.~\ref{sec:Master-equation}.
Explicitly, the product is defined as 
\begin{equation}
[AB]_{\alpha\beta}^{ab}(x_{1},x_{2})\equiv\sum_{\gamma}\int d^{4}z\left[A_{\alpha\gamma}^{a+}(x_{1},z)B_{\gamma\beta}^{+b}(z,x_{2})-A_{\alpha\gamma}^{a-}(x_{1},z)B_{\gamma\beta}^{-b}(z,x_{2})\right],\label{eq:product}
\end{equation}
where $a,b\in\{+,-\}$ denote the time branch on the CTP. Accordingly
the original integral is $\int_{C}=\int_{C^{+}}$ and $\int_{C^{-}}$,
where $C^{+}$ and $C^{-}$ are the forward and backward paths respectively.
The minus sign for the last term is due to the backward path in Fig.~\ref{fig:ctp}.
The lesser component of Eq.~\eqref{eq:SD equation} involve the following
terms,
\begin{eqnarray}
(S_{0}\Sigma S)^{<} & \equiv & -(S_{0}\Sigma S)^{+-}=-iS_{0}^{R}\Sigma^{R}S^{<}+iS_{0}^{R}\Sigma^{<}S^{A}-iS_{0}^{<}\Sigma^{A}S^{A},\nonumber \\
(S\Sigma S_{0})^{<} & \equiv & -(S\Sigma S_{0})^{+-}=-iS^{R}\Sigma^{R}S_{0}^{<}+iS^{R}\Sigma^{<}S_{0}^{A}-iS^{<}\Sigma^{A}S_{0}^{A}.\label{eq:S sigma S lessor}
\end{eqnarray}
Here, we define 
\begin{eqnarray}
\Sigma^{R} & = & \Sigma^{++}-\Sigma^{+-}=\Sigma^{-+}-\Sigma^{--},\nonumber \\
\Sigma^{A} & = & \Sigma^{++}-\Sigma^{-+}=\Sigma^{+-}-\Sigma^{--},\nonumber \\
i\Sigma^{+-} & = & -\Sigma^{<},\nonumber \\
i\Sigma^{-+} & = & \Sigma^{>},
\end{eqnarray}
and employ Eq.~(\ref{eq:def-green-func}) for fermionic propagators
to derive the equations above. In this expression, the integral and
spinor indices are omitted, while the indices for the time branch
on the CTP are explicitly shown. Substituting Eq.~\eqref{eq:S sigma S lessor}
into Eq.~\eqref{eq:SD equation}, we obtain 
\begin{eqnarray}
S^{<} & = & S_{0}^{<}-S_{0}^{R}\Sigma^{R}S^{<}+S_{0}^{R}\Sigma^{<}S^{A}-S_{0}^{<}\Sigma^{A}S^{A}\nonumber \\
 & = & S_{0}^{<}-S^{R}\Sigma^{R}S_{0}^{<}+S^{R}\Sigma^{<}S_{0}^{A}-S^{<}\Sigma^{A}S_{0}^{A}.\label{eq:Slessor}
\end{eqnarray}
The goal here is to rewrite Eq.~\eqref{eq:Slessor} into the form
of the quantum kinetic equation.

The free parts of retarded and advanced propagators satisfy 
\begin{eqnarray}
-i\sigma\cdot D_{x_{1}}S_{0}^{R}(x_{1},x_{2}) & = & \delta^{(4)}(x_{1}-x_{2}),\nonumber \\
S_{0}^{A}(x_{1},x_{2})i\sigma\cdot\overleftarrow{D}_{x_{2}}^{\dagger} & = & \delta^{(4)}(x_{1}-x_{2}).\label{eq:SR}
\end{eqnarray}
Acting $i\sigma\cdot D_{x_{1}}$($-i\sigma\cdot\overleftarrow{D}_{x_{2}}^{\dagger}$)
on the left (right) of both sides of the first (second) line of Eq.~\eqref{eq:Slessor}, respectively,
we obtain 
\begin{eqnarray}
i\sigma\cdot D_{x_{1}}S^{<} & = & \Sigma^{R}S^{<}-\Sigma^{<}S^{A},\label{eq:colliison1}\\
-S^{<}i\sigma\cdot\overleftarrow{D}_{x_{2}}^{\dagger} & = & -S^{R}\Sigma^{<}+S^{<}\Sigma^{A},\label{eq:colliison2}
\end{eqnarray}
where we employ Eqs.~\eqref{eq:Master_eq_pre_01} and \eqref{eq:SR}.
The retarded and advanced functions are further decomposed into 
\begin{align}
 & S^{R}=\overline{S}+\frac{i}{2}(S^{>}+S^{<}),\quad S^{A}=\overline{S}-\frac{i}{2}(S^{>}+S^{<}),\\
 & \Sigma^{R}=\overline{\Sigma}-\frac{i}{2}(\Sigma^{>}+\Sigma^{<}),\quad\Sigma^{A}=\overline{\Sigma}+\frac{i}{2}(\Sigma^{>}+\Sigma^{<}),
\end{align}
where $\overline{S}\equiv(S^{R}+S^{A})/2$ and $\overline{\Sigma}\equiv(\Sigma^{R}+\Sigma^{A})/2$.
After performing the Wigner transformation, $\overline{S}$ and $\overline{\Sigma}$
will turn to the real part of the retarded propagator and self-energy~\cite{Blaizot:2001nr}.
Substituting these equations into Eqs.~\eqref{eq:colliison1} and
\eqref{eq:colliison2}, we obtain the quantum kinetic equation with
collisions in coordinate space,
\begin{eqnarray}
(i\sigma\cdot D_{x_{1}}-\overline{\Sigma})S^{<}+\Sigma^{<}\overline{S} & = & -\frac{i}{2}(\Sigma^{>}S^{<}-\Sigma^{<}S^{>}),\label{eq:Kinetic equation 1}\\
S^{<}(-i\sigma\cdot\overleftarrow{D}_{x_{2}}^{\dagger}-\overline{\Sigma})+\overline{S}\Sigma^{<} & = & \frac{i}{2}(S^{<}\Sigma^{>}-S^{>}\Sigma^{<}).\label{eq:Kinetic equation 2}
\end{eqnarray}
Similarly, the greater propagators obey 
\begin{eqnarray}
(i\sigma\cdot D_{x_{1}}-\overline{\Sigma})S^{>}+\Sigma^{>}\overline{S} & = & -\frac{i}{2}(\Sigma^{<}S^{>}-\Sigma^{>}S^{<}),\label{eq:Kinetic equation greater 1}\\
S^{>}(-i\sigma\cdot\overleftarrow{D}_{x_{2}}^{\dagger}-\overline{\Sigma})+\overline{S}\Sigma^{>} & = & \frac{i}{2}(S^{>}\Sigma^{<}-S^{<}\Sigma^{>}).\label{eq:Kinetic equation greater 2}
\end{eqnarray}
As in Eq.~\eqref{eq:Master_eq_pre_01} for free massless fermions,
these expressions are not gauge invariant. To obtain the gauge-invariant
ones, we consider the following gauge-invariant product,
\begin{equation}
\int d^{4}zU(x_{2},x_{1})A(x_{1},z)B(z,x_{2})=\int d^{4}zU_{C}(x_{1},x_{2},z)\widetilde{A}(x_{1},z)\widetilde{B}(z,x_{2}),\label{eq:gauge invariant product}
\end{equation}
where $\widetilde{A}(x_{1},z)=A(x_{1},z)U(z,x_{1})$, $\widetilde{B}(z,x_{2})=B(z,x_{2})U(x_{2},z)$,
and we introduce 
\begin{equation}
U_{C}(x_{2},x_{1},z)=\exp\left(-\frac{i}{\hbar}\int_{L}dx^{\mu}A_{\mu}\right)=U(x_{2},x_{1})U(x_{1},z)U(z,x_{2}).
\end{equation}
Here $L$ represents the closed path: $z\to x_{1}\to x_{2}\to z$, which has
a triangle shape. Equation~\eqref{eq:gauge invariant product}
is manifestly gauge invariant. As a compensation for the gauge invariance,
$U_{C}(x_{1},x_{2},z)$ is added to the product.

To change variables in the coordinate space to the phase space, let
us consider the Wigner transformation of the product of Eq.~\eqref{eq:gauge invariant product},
\begin{eqnarray}
\widetilde{A}(x,p)\star\widetilde{B}(x,p) & \equiv & \int d^{4}y\exp\left(\frac{i}{\hbar}p\cdot y\right)\int d^{4}zU_{C}\left(x-\frac{y}{2},x+\frac{y}{2},z+x\right)\nonumber \\
 &  & \times\widetilde{A}\left(x+\frac{y}{2},z+x\right)\widetilde{B}\left(z+x,x-\frac{y}{2}\right),\label{eq:moyal-product-def}
\end{eqnarray}
where we have shifted the integral variable $z\to z+x$ for convenience
of later use, and the operator $\star$ is called the Moyal product,
whose explicit form is given above. For convenience, we express
$\widetilde{A}$ and $\widetilde{B}$ as 
\begin{eqnarray}
\widetilde{A}\left(x+\frac{y}{2},z+x\right) & = & \exp\left(\frac{1}{2}u\cdot\partial_{x}\right)\widetilde{A}\left(x-\frac{v}{2},x+\frac{v}{2}\right)\nonumber \\
 & = & \exp\left(\frac{1}{2}u\cdot\partial_{x}\right)\int\frac{d^{4}p}{(2\pi)^{4}}\exp\left(\frac{i}{\hbar}p\cdot v\right)\widetilde{A}(x,p),\nonumber \\
\widetilde{B}\left(z+x,x-\frac{y}{2}\right) & = & \exp\left(\frac{1}{2}v\cdot\partial_{x}\right)\widetilde{B}\left(x+\frac{u}{2},x-\frac{u}{2}\right)\nonumber \\
 & = & \exp\left(\frac{1}{2}v\cdot\partial_{x}\right)\int\frac{d^{4}p}{(2\pi)^{4}}\exp\left(-\frac{i}{\hbar}p\cdot u\right)\widetilde{B}(x,p),\label{eq:AB}
\end{eqnarray}
where we introduced $u=z+y/2$, $v=z-y/2$.

Let us focus on $U_{C}$ which, using the Stokes theorem, can be rewritten
as 
\begin{equation}
U_{C}\left(x-\frac{y}{2},x+\frac{y}{2},z+x\right)=\exp\left[-\frac{i}{\hbar}\int_{L}d\xi^{\mu}A_{\mu}(\xi)\right]=\exp\left[-\frac{i}{2\hbar}\int_{S}d\xi^{\mu}\wedge d\xi^{\nu}F_{\mu\nu}(\xi)\right],
\end{equation}
where $S$ is the surface area bounded by $L$, $\xi^{\mu}$ is the
parametrized coordinate to be defined below, and $\wedge$ is the
exterior product such that $d\xi^{\mu}\wedge d\xi^{\nu}=-d\xi^{\nu}\wedge d\xi^{\mu}$.
The parametrized coordinate is defined as 
\begin{equation}
\xi^{\mu}(s,t)=x^{\mu}+u^{\mu}s+v^{\mu}t,
\end{equation}
where $s,t\in[-1/2,1/2]$ and $s+t\geq0$. Since $dt\wedge dt=ds\wedge ds=0$
and $dt\wedge ds=-ds\wedge dt$, the surface element $d\xi^{\mu}\wedge d\xi^{\nu}$
can be expressed as 
\begin{equation}
d\xi^{\mu}\wedge d\xi^{\nu}=(u^{\mu}ds+v^{\mu}dt)\wedge(u^{\nu}ds+v^{\nu}dt)=2v^{\mu}u^{\nu}dt\wedge ds.
\end{equation}
Therefore, we have 
\begin{equation}
\int_{S}d\xi^{\mu}\wedge d\xi^{\nu}F_{\mu\nu}[\xi(s,t)]=2\int_{-1/2}^{1/2}dt\int_{-t}^{1/2}dsF_{\mu\nu}[\xi(s,t)]v^{\mu}u^{\nu}.
\end{equation}
Expanding $F_{\mu\nu}[\xi(s,t)]$ around $x$, we have 
\begin{eqnarray}
\int d\xi^{\mu}\wedge d\xi^{\nu} F_{\mu\nu}[\xi(s,t)] & = & 2\sum_{n=0}^{\infty}\frac{1}{n!}\int_{-1/2}^{1/2}dt\int_{-t}^{1/2}ds(su\cdot\partial+tv\cdot\partial)^{n}v^{\mu}u^{\nu}F_{\mu\nu}(x)\nonumber \\
 & \equiv & H(u\cdot\partial_{x},v\cdot\partial_{x})v^{\mu}u^{\nu}F_{\mu\nu}(x).\label{eq:H(ab)}
\end{eqnarray}
Here we used the notation $F_{\mu\nu}[x]\equiv F_{\mu\nu}(x)$ as
a function of space-time. With this formula, we can write $U_{C}$
as 
\begin{equation}
U_{C}\left(x-\frac{y}{2},x+\frac{y}{2},z+x\right)=\exp\left[-\frac{i}{2\hbar}H(u\cdot\partial_{x},v\cdot\partial_{x})v^{\mu}u^{\nu}F_{\mu\nu}(x)\right].\label{eq:U_C}
\end{equation}
The integral in Eq.~\eqref{eq:H(ab)} can be analytically performed as 
\begin{equation}
H(a,b)=\frac{2}{ab(a-b)}\left[b\exp\left(\frac{b-a}{2}\right)-a\exp\left(\frac{a-b}{2}\right)\right]+\frac{2}{ab}\exp\left(\frac{a+b}{2}\right).
\end{equation}
Although $a$ and $b$ are on the denominator, $H(a,b)$ is analytic
at the origin. For small $a$ and $b$, $H(a,b)$ is expanded as 
\begin{equation}
H(a,b)=1+\frac{a+b}{6}+\frac{1}{24}(a^{2}+b^{2})+\frac{1}{240}(a+b)(a^{2}+b^{2})+\cdots.
\end{equation}
In particular, if we set $b=0$, it reduces to a similar form to that
in Eq.~(\ref{eq:operator}), 
\begin{equation}
H(a,0)=\frac{2}{a^{2}}\exp\left(-\frac{a}{2}\right)\left(e^{a}a-e^{a}+1\right)=j_{0}\left(i\frac{a}{2}\right)-ij_{1}\left(i\frac{a}{2}\right).\label{eq:H(a,0)}
\end{equation}
From Eqs.~\eqref{eq:AB} and \eqref{eq:U_C}, the integrand in
Eq.~(\ref{eq:moyal-product-def}) can be put into the form,
\begin{eqnarray}
 &  & U_{C}\left(x-\frac{y}{2},x+\frac{y}{2},z+x\right)\widetilde{A}\left(x+\frac{y}{2},z+x\right)\widetilde{B}\left(z+x,x-\frac{y}{2}\right)\nonumber \\
 & = & \exp\left[-\frac{i}{2\hbar}H(u\cdot\partial_{x},v\cdot\partial_{x})v^{\mu}u^{\nu}F_{\mu\nu}(x)\right]\nonumber \\
 &  & \times\exp\left(\frac{1}{2}u\cdot\partial_{x^{\prime}}\right)\exp\left(\frac{1}{2}v\cdot\partial_{x^{\prime\prime}}\right)\nonumber \\
 &  & \times\left.\widetilde{A}\left(x^{\prime}-\frac{v}{2},x^{\prime}+\frac{v}{2}\right)\widetilde{B}\left(x^{\prime\prime}+\frac{u}{2},x^{\prime\prime}-\frac{u}{2}\right)\right|_{x^{\prime}=x^{\prime\prime}=x}.
\end{eqnarray}
Note the spacetime derivative $\partial_{x}$ in $H$ only acts on
$F_{\mu\nu}(x)$. The Wigner transformation replaces $u$ and $v$
with momentum derivatives. A straightforward calculation yields the
result of the Moyal product,
\begin{eqnarray}
\widetilde{A}(x,p)\star\widetilde{B}(x,p) & = & \exp\left[-\frac{i\hbar}{2}H(-i\hbar\partial_{p^{\prime\prime}}\cdot\partial_{x},i\hbar\partial_{p^{\prime}}\cdot\partial_{x})F_{\mu\nu}(x)\partial_{p^{\prime}}^{\mu}\partial_{p^{\prime\prime}}^{\nu}\right]\nonumber \\
 &  & \times\exp\left[\frac{i\hbar}{2}\left(\partial_{x^{\prime\prime}}\cdot\partial_{p^{\prime}}-\partial_{x^{\prime}}\cdot\partial_{p^{\prime\prime}}\right)\right] 
\times\left.\widetilde{A}(x^{\prime},p^{\prime})\widetilde{B}(x^{\prime\prime},p^{\prime\prime})\right|_{x^{\prime}=x^{\prime\prime}=x,p^{\prime}=p^{\prime\prime}=x},\label{eq:Moyal-product}
\end{eqnarray}
where $u$ and $v$ are replaced by $-i\hbar\partial_{p^{\prime\prime}}$
and $i\hbar\partial_{p^{\prime}}$ respectively. This product plays
an essential role in $\hbar$ expansion. Up to $O(\hbar)$, the Moyal
product is expanded as 
\begin{eqnarray}
\widetilde{A}(x,p)\star\widetilde{B}(x,p) & = & \widetilde{A}(x,p)\widetilde{B}(x,p)+\frac{i\hbar}{2}\left\{ \widetilde{A}(x,p),\widetilde{B}(x,p)\right\} _{{\rm P.B.}}\nonumber \\
 &  & -\frac{i\hbar}{2}F_{\mu\nu}\partial_{p}^{\mu}\widetilde{A}(x,p)\partial_{p}^{\nu}\widetilde{B}(x,p)+O(\hbar^{2}),
 \label{eq:Moyal_product}
\end{eqnarray}
where the Poisson bracket is defined as 
\begin{equation}
\left\{ \widetilde{A},\widetilde{B}\right\} _{{\rm P.B.}}\equiv(\partial_{p}^{\mu}\widetilde{A})(\partial_{\mu}^{x}\widetilde{B})-(\partial_{\mu}^{x}\widetilde{A})(\partial_{p}^{\mu}\widetilde{B}). \label{eq:def_PB}
\end{equation}


After performing the Wigner transformation of Eqs.~\eqref{eq:Kinetic equation 1}
and \eqref{eq:Kinetic equation 2} multiplied by $U(x_{2},x_{1})$,
the Kadanoff-Baym equation is put into the form,
\begin{eqnarray}
\sigma^{\mu}\left(\Pi_{\mu}+\frac{1}{2}i\hbar\nabla_{\mu}\right)S^{<}-\hbar\overline{\Sigma}\star S^{<}+\hbar\Sigma^{<}\star\overline{S} & = & \frac{i\hbar}{2}(\Sigma^{<}\star S^{>}-\Sigma^{>}\star S^{<}),\label{eq:kinetic collision 1}\\
\left(\Pi_{\mu}-\frac{1}{2}i\hbar\nabla_{\mu}\right)S^{<}\sigma^{\mu}-\hbar S^{<}\star\overline{\Sigma}+\hbar\overline{S}\star\Sigma^{<} & = & -\frac{i\hbar}{2}(S^{>}\star\Sigma^{<}-S^{<}\star\Sigma^{>}).\label{eq:kinetic collision 2}
\end{eqnarray}
Here we removed tildes from $\widetilde{S}$ and $\widetilde{\Sigma}$
for notational simplicity, and we employed the expression of Eqs.~\eqref{eq:master}
and \eqref{eq:master-conjugate} in first terms of Eqs.~\eqref{eq:kinetic collision 1}
and \eqref{eq:kinetic collision 2}, which are equivalent to $\sigma\cdot p\star S^{<}$
and $S^{<}\star\sigma\cdot p$. To prove these expressions, we note
the covariant derivative $iD_{x_{1}\mu}$ can be expressed as 
\begin{equation}
i\check{D}_{\mu}(x_{1},z)=\left[-i\hbar\partial_{\mu}^{z}-A_{\mu}\left(\frac{x_{1}+z}{2}\right)\right]\delta^{4}(x_{1}-z),
\end{equation}
in the product form~\eqref{eq:product} so that we have 
\begin{equation}
iD_{\mu}^{x_{1}}S^{<}(x_{1},x_{2})=\int d^{4}zi\check{D}_{\mu}(x_{1},z)S^{<}(z,x_{2}).
\end{equation}
The Wigner transformation of the gauge invariant covariant derivative
$U(z,x_{1})i\check{D}_{\mu}(x_{1},z)$ is evaluated as 
\begin{equation}
\int d^{4}\overline{y}\exp\left(\frac{i}{\hbar}p\cdot\overline{y}\right)U\left(\overline{x}-\frac{\overline{y}}{2},\overline{x}+\frac{\overline{y}}{2}\right)\left[i\hbar\partial_{\mu}^{\overline{y}}-A_{\mu}(\overline{x})\right]\delta(\overline{y})=p_{\mu},
\end{equation}
where $\overline{x}=(x_{1}+z)/2$, $\overline{y}=x_{1}-z$, and $\partial_\mu^z=-\partial_\mu^{\overline{y}}$. Therefore
the Wigner transformation of $\sigma^{\mu}iD_{\mu}^{x_{1}}S^{<}(x_{1},x_{2})$
multiplied by $U(x_{2},x_{1})$ turns to $\sigma\cdot p\star S^{<}$.
One can check $\sigma\cdot p\star S^{<}=\sigma^{\mu}\left[\Pi_{\mu}+(1/2)i\hbar\nabla_{\mu}\right]S^{<}$
by using Eq.~(\ref{eq:Moyal-product}) together with Eq.~\eqref{eq:H(a,0)}.
We assign the self-energy at $O(\hbar)$ such that it can reproduce
to the classical Boltzmann equation in the $\hbar\to0$ limit. Roughly
speaking, the right-hand sides in Eqs.~\eqref{eq:kinetic collision 1}
and \eqref{eq:kinetic collision 2} represent collisions and $\overline{\Sigma}\star S^{<}$
corresponds to the self-energy correction, while $\Sigma^{<}\star\overline{S}$
has no counter part in the classical kinetic theory. Since $\overline{S}$
represents the real part of the retarded propagator, it is negligible
in the quasi-particle approximation~\cite{Blaizot:2001nr}.


Parametrizing $S=\overline{\sigma}^{\mu}S_{\mu}$, $\Sigma=\sigma^{\mu}\Sigma_{\mu}$,
and using the same technique discussed in Sec.~\ref{sec:CCKE},
we find a set of equations,
\begin{eqnarray}
\nabla_{\mu}S^{<,\mu}+i\left[\overline{\Sigma}_{\mu},S^{<,\mu}\right]_{\star}-i\left[\Sigma_{\mu}^{<},\overline{S}^{\mu}\right]_{\star} & = & C_{\mu}^{\mu},\label{eq:CKT}\\
\Pi_{\mu}S^{<,\mu}-\frac{\hbar}{2}\left\{ \overline{\Sigma}_{\mu},S^{<,\mu}\right\} _{\star}+\frac{\hbar}{2}\left\{ \Sigma_{\mu}^{<},\overline{S}^{\mu}\right\} _{\star} & = & -\frac{\hbar^{2}}{4}D_{\mu}^{\mu},\\
\Pi^{[\nu}S^{<,\mu]}-\frac{\hbar}{2}\left\{ \overline{\Sigma}^{[\nu},S^{<,\mu]}\right\} _{\star}+\frac{\hbar}{2}\left\{ \Sigma^{<,[\nu},\overline{S}^{\mu]}\right\} _{\star} & = & \frac{\hbar}{2}\epsilon^{\mu\nu\rho\sigma}\left(\nabla_{\rho}S_{\sigma}^{<}-C_{\rho\sigma}\right)+\frac{\hbar^{2}}{4}D^{[\mu\nu]},\nonumber\\ \label{eq:CKT constraint 2}
\end{eqnarray}
where we introduced 
\begin{align}
C_{\mu\nu} & =\frac{1}{2}\left(\left\{ \Sigma_{\mu}^{<},S_{\nu}^{>}\right\} _{\star}-\left\{ \Sigma_{\mu}^{>},S_{\nu}^{<}\right\} _{\star}\right),\\
D_{\mu\nu} & =\frac{1}{i\hbar}\left(\left[ \Sigma_{\mu}^{<},S_{\nu}^{>}\right] _{\star}-\left[ \Sigma_{\mu}^{>},S_{\nu}^{<}\right] _{\star}\right),
\end{align}
with $\{A,B\}_{\star}=A\star B+B\star A$, $[A,B]_{\star}=A\star B-B\star A$,
and $A^{[\mu}B^{\nu]}\equiv A^{\mu}B^{\nu}-A^{\nu}B^{\mu}$. From
Eqs.~\eqref{eq:Kinetic equation greater 1} and \eqref{eq:Kinetic equation greater 2},
one can obtain the kinetic equations for $S^{>,\mu}$ by exchanging
$<$ and $>$ in Eqs.~\eqref{eq:CKT}-\eqref{eq:CKT constraint 2}.
In particular, Eq.~\eqref{eq:CKT} represents the kinetic equation
that describes the time evolution of the system, while other equations
are constraints. In Eqs.~\eqref{eq:CKT}-\eqref{eq:CKT constraint 2},
$\overline{\Sigma}_{\mu}$ gives a correction to the dispersion relation,
like a thermal mass. If $\overline{\Sigma}_{\mu}$ depends on $x$,
it plays a role of potential term in Eq.~\eqref{eq:CKT}. As mentioned
above, $\overline{S}^{\mu}$ is the real part of the retarded propagator,
which is negligible within the quasi-particle approximation~\cite{Blaizot:2001nr}.
These contributions are common in an ordinary quantum kinetic theory. In the
following, we focus on collision effects, so that we neglect terms
that contain $\overline{\Sigma}_{\mu}$ and $\overline{S}^{\mu}$.
Then, the chiral kinetic equation with collision terms (the transport
equation) has
\begin{equation}
\nabla_{\mu}S^{<,\mu}=C_{\mu}^{\mu}.\label{eq:CKT2}
\end{equation}

We can expand the propagator in powers of $\hbar$ and solve it order
by order,
\begin{equation}
S^{<}=S_{(0)}^{<}+\hbar S_{(1)}^{<}+\hbar^{2}S_{(2)}^{<}+\cdots.
\end{equation}
The leading order solution for $S_{(0)}^{<\mu}$ is just $\mathscr{J}_{(0)}^{\rho}(x,p)$
in Eq.~(\ref{eq:j-0}) for right-handed fermions, 
\begin{equation}
S_{(0)}^{<,\mu}=2\pi \mathrm{sgn}(p_0) p^{\mu}f_{(0)}\delta(p^{2}),\label{eq:leading S<}
\end{equation}
where $\mathrm{sgn}(p_0)=\Theta(p_0)-\Theta(-p_0)$ is the sign function, and 
we have pulled $2\pi\,\mathrm{sgn}(p_0)$ out of $f_{(0)}$ in Eq.~(\ref{eq:dist})
so that the new $f_{(0)}$ will reduce to the Fermi-Dirac distribution
at thermal equilibrium, $f_{(0)}=f_{\rm FD}(p\cdot u-\mu_{+})$. 
In the following, we focus on the positive energy state $p_0>0$, and 
drop $\mathrm{sgn}(p_0)$  throughout this section unless specified otherwise.
The greater component is then 
\begin{equation}
S_{(0)}^{>\mu}=2\pi p^{\mu}(1-f_{(0)})\delta(p^{2}).
\end{equation}
At this order, the distribution satisfies the ordinary Boltzmann equation,
\begin{equation}
p^{\mu}\nabla_{\mu}^{(0)}f_{(0)}=p^{\mu}C_{\mu},
\end{equation}
with 
\begin{equation}
C_{\mu}=\Sigma_{(0)\mu}^{<}(1-f_{(0)})-\Sigma_{(0)\mu}^{>}f_{(0)}.
\end{equation}
The constraint equation \eqref{eq:CKT constraint 2} at $O(\hbar)$
is modified by the self-energy, 
\begin{equation}
p^{\nu}S_{(1)}^{<,\mu}-p^{\mu}S_{(1)}^{<,\nu}-\frac{1}{2}\epsilon^{\mu\nu\rho\sigma}\left(\nabla_{\rho}^{(0)}S_{(0)\sigma}^{<}-\Sigma_{(0)\rho}^{<}S_{(0)\sigma}^{>}+\Sigma_{(0)\rho}^{>}S_{(0)\sigma}^{<}\right)=0.
\end{equation}
The solution can be obtained~\cite{Hidaka:2016yjf} as 
\begin{eqnarray}
S_{(1)}^{<,\mu} & = & 2\pi p^{\mu}f_{(1)}\delta(p^{2})+\pi\epsilon^{\mu\nu\rho\sigma}p_{\nu}F_{\rho\sigma}f_{(0)}\delta^{\prime}(p^{2})\nonumber \\
 &  & +2\pi S_{(n)}^{\mu\nu}(\nabla_{\nu}^{(0)}f_{(0)}-C_{\nu})\delta(p^{2}),\label{eq:next-leading-S<}
\end{eqnarray}
where $S_{(n)}^{\mu\nu}$ is defined by 
\begin{equation}
S_{(n)}^{\mu\nu}\equiv\frac{\epsilon^{\mu\nu\alpha\beta}p_{\alpha}n_{\beta}}{2n\cdot p},
\end{equation}
which is Eq.~\eqref{eq:spin} with $\lambda=+1$.
We see that the side jump term is shifted in presence of the collision
term,
\begin{equation}
S_{(n)}^{\mu\nu}\nabla_{\nu}^{(0)}f_{(0)}\to S_{(n)}^{\mu\nu}\left(\nabla_{\nu}^{(0)}f_{(0)}-C_{\nu}\right)\delta(p^{2}).
\end{equation}
The sum of Eqs.~\eqref{eq:leading S<} and \eqref{eq:next-leading-S<}
gives the solution up to $O(\hbar)$.
\begin{eqnarray}
S^{<,\mu} & = & 2\pi p^{\mu}f\delta(p^{2})+\pi\hbar\epsilon^{\mu\nu\rho\sigma}p_{\nu}F_{\rho\sigma}f\delta^{\prime}(p^{2}) 
+2\pi\hbar S_{(n)}^{\mu\nu}(\nabla_{\nu}f-C_{\nu})\delta(p^{2})+O(\hbar^{2}),
\end{eqnarray}
where $f=f_{(0)}+\hbar f_{(1)}$. For $S^{>,\mu}$, one simply needs to replace $f$ by $(1-f)$. When including both the contributions from fermions $(p\cdot n>0)$ and anti-fermions $(p\cdot n<0)$, one has to multiply $S^{<,\mu}$ or $S^{>,\mu}$ with an overall factor, ${\rm sgn}(p\cdot n)$ as in Eq.~\eqref{eq:leading S<}. 
The existence of the collision kernel
$C_{\nu}$ in $S^{<,\mu}$ is necessary to maintain the Lorentz covariance
of the current \cite{Chen:2015gta}, or equivalently, the frame independence
of $S^{<,\mu}$. One can show the frame independence by using $p^{\nu}(\nabla_{\nu}f-C_{\nu})=O(\hbar)$.
On the other hand, the CKT up to $O(\hbar)$ with collisions for a generic frame vector $n^{\mu}(x)$ is given by \cite{Hidaka:2016yjf,Hidaka:2017auj}
	\begin{eqnarray}\nonumber
		0&=&\delta\left(p^2-\hbar\frac{B_{(n)}\cdot p}{p\cdot n}\right)\Bigg[\bigg(p\cdot\hat{\mathcal{D}}+\frac{\hbar S^{\mu\nu}_{(n)}E_{(n)\mu}}{p\cdot n}\hat{\mathcal{D}}_{\nu}+\hbar S^{\mu\nu}_{(n)}\big(\partial_{\mu}F_{\rho\nu}\big)\partial^{\rho}_p+\hbar\big(\partial_{\mu}S^{\mu\nu}_{(n)}\big)\hat{\mathcal{D}}_{\nu}\bigg)f
		\\
		&&-\hbar S^{\mu\nu}_{(n)}
		\Big((1-f)\nabla_{\mu}\Sigma^{<}_{\nu}-f\nabla_{\mu}\Sigma^{>}_{\nu}\Big)	\Bigg]
		,
	\end{eqnarray}
where $\hat{\mathcal{D}}_{\mu}f=\nabla_{\mu} f-{C}_{\mu}$ and the nonlinear terms of self-energies are dropped
by assuming weak coupling.
	 $E_{(n)\mu}$ and $B_{(n)\mu}$ are defined by replacing $u^\mu$ by $n^\mu$ in Eq.~\eqref{eq:em-field-u} as
    \begin{equation}
         E_{(n)\sigma}=F_{\sigma\rho}n^{\rho},\;B_{(n)\sigma}=\frac{1}{2}\epsilon_{\sigma\mu\nu\rho}n^{\mu}F^{\nu\rho}.
    \end{equation}

So far, the collision terms are expressed in terms of self-energies
whose detailed forms depend on the theory. As an example for the self-energy
in the absence of the background field, let us consider the following
expression~\cite{Hidaka:2016yjf},
\begin{equation}
\Sigma_{\mu}^{<}=\int_{p',k,k'}P(p,k,k')S_{\mu}^{>}(p')S^{<}(k)\cdot S^{<}(k'),
\label{eq:self-energy}
\end{equation}
where $P(p,k,k')$ and $\int_{p',k,k'}$ are defined by 
\begin{equation}
P(p,k,k')\equiv4e^{2}\left[\frac{1}{(p-k)^{2}}+\frac{1}{(p-k')^{2}}\right]^{2},
\end{equation}
and 
\begin{equation}
\int_{p',k,k'}\equiv\int\frac{d^{4}p'}{(2\pi)^{4}}\int\frac{d^{4}k}{(2\pi)^{4}}\int\frac{d^{4}k'}{(2\pi)^{4}}(2\pi)^{4}\delta^{(4)}(p+p'-k-k').
\end{equation}
This self-energy represents the Coulomb scattering of right-handed
fermions. Similarly, one can obtain $\Sigma_{\mu}^{>}$ by interchanging
$<$ and $>$. The collision term in Eq.~\eqref{eq:CKT2} becomes
\begin{eqnarray}
C_{\mu}^{\mu} & = & \int_{p',k,k'}P(p,k,k')\left[S^{>}(p')\cdot S^{>}(p)S^{<}(k)\cdot S^{<}(k')\right.\nonumber \\
 &  & \left.-S^{<}(p')\cdot S^{<}(p)S^{>}(k)\cdot S^{>}(k')\right].
\end{eqnarray}
The explicit expression of the collision term is a little bit complicated
because of the existence of $S_{(n)}^{\mu\nu}$. However, if we choose
the center of mass frame called the no-jump frame~\cite{Chen:2015gta},
the terms proportional to $S_{(n)}^{\mu\nu}$ are canceled. Concretely,
we can choose 
\begin{equation}
n_{c,\mu}=\frac{1}{\sqrt{s}}(p_{\mu}+p'_{\mu})=\frac{1}{\sqrt{s}}(k_{\mu}+k'_{\mu}),
\end{equation}
with $s=(p+p')^{2}$, then the lessor propagator becomes
\begin{eqnarray}
S^{<,\mu}(x,p) & = & 2\pi p^{\mu}f^{(n_{c})}(x,p)\delta(p^{2})\nonumber \\
 &  & +\pi\hbar\frac{\epsilon^{\mu\nu\alpha\beta}p_{\alpha}p'_{\beta}}{p\cdot p'}\left[\nabla_{\nu}f^{(n_{c})}(x,p)-C_{\nu}\right]\delta(p^{2}).
\end{eqnarray}
 Therefore, the term of $O(\hbar)$ vanishes in $S^{<}(x,p)\cdot S^{<}(x,p')$
and we obtain 
\begin{eqnarray}
S^{<}(x,p)\cdot S^{<}(x,p') & = & (2\pi)^{2}(p\cdot p')f^{(n_{c})}(x,p')f^{(n_{c})}(x,p)\nonumber \\
 &  & \times\delta(p^{2})\delta(p'^{2})+O(\hbar^{2}).
\end{eqnarray}
In the no-jump frame, the collision term $C_{\mu}^{\mu}=2\pi p^{\mu}C_{\mu}\delta(p^{2})$
reduces to the form of standard one with $p^{\mu}C_{\mu}$ given
by 
\begin{eqnarray}
p^{\mu}C_{\mu} & = & \frac{1}{4}\int_{\mathbf{p}^{\prime},\mathbf{k},\mathbf{k}}|\mathcal{M}|^{2}\left\{ \left[1-f^{(n_{c})}(x,p)\right]\left[1-f^{(n_{c})}(x,p^{\prime})\right]f^{(n_{c})}(k)f^{(n_{c})}(x,k')\right.\nonumber \\
 &  & \left.-f^{(n_{c})}(x,p)f^{(n_{c})}(x,p')\left[1-f^{(n_{c})}(x,k)\right]\left[1-f^{(n_{c})}(x,k')\right]\right\} ,
\end{eqnarray}
where $|\mathcal{M}|^{2}=4P(p',k,k')(k\cdot k')(p\cdot p')$, all
momenta are now on-shell with positive energy~\footnote{{\color{blue}}
Here, we just focus on the Coulomb scattering process. In general, the self-energy~\eqref{eq:self-energy} also contains 
the fermion and anti-fermion annihilation process, when one takes into account the anti-fermion contribution.}, e.g., $k^{\mu}=(|\mathbf{k}|,\mathbf{k})$,
and the integral symbol $\int_{\mathbf{p}^{\prime},\mathbf{k},\mathbf{k}}$
is defined by 
\begin{equation}
\int_{\mathbf{p}^{\prime},\mathbf{k},\mathbf{k}}\equiv\int\frac{d^{3}\mathbf{p}'}{(2\pi)^{3}}\frac{1}{2|\mathbf{p}'|}\int\frac{d^{3}\mathbf{k}}{(2\pi)^{3}}\frac{1}{2|\mathbf{k}|}\int\frac{d^{3}\mathbf{k}'}{(2\pi)^{3}}\frac{1}{2|\mathbf{k}'|}(2\pi)^{4}\delta^{(4)}(p+p'-k-k').
\end{equation}
One can check that the collision term vanishes if the global equilibrium
solution obtained in Sec.~\ref{sec:solutions-wigner} is employed.

Moreover, when choosing $n^{\mu}(x)=u^{\mu}(x)$, a local-equilibrium solution for Wigner functions satisfying the vanishing collision term of the Coulomb scattering is found \cite{Hidaka:2017auj},
\begin{eqnarray}\label{L_equil_Wigner}\nonumber
	S_{\text{leq}}^{<,\mu}
	&=&2\pi\,{\rm sgn}(p\cdot u)\Bigg[\delta(p^2)\left(p^{\mu}+\frac{\hbar}{2}\big(u^{\mu}(p\cdot\omega)-\omega^{\mu}(p\cdot u)\big)\partial_{p\cdot u}-\hbar S^{\mu\nu}_{(u)}\tilde{E}_{\nu}
	\partial_{p\cdot u}
	\right)
	\\
	&&
	+\frac{\hbar\epsilon^{\mu\nu\alpha\beta}F_{\alpha\beta}}{4}\partial_{p\nu}\delta(p^2)
	\Bigg]f_{\rm FD}(p\cdot u-\mu_{+}),
\end{eqnarray}
where 
\begin{eqnarray}
	\tilde{E}_{\nu}=E_{\nu}+T \partial_{\nu}(\beta\mu_{+})+ \frac{(p\cdot u)}{T}\partial_{\nu}T
	-p^{\sigma}(\zeta_{\nu\sigma}+\kappa_{\nu\sigma})
\end{eqnarray}
with $\zeta_{\mu\nu}=(\partial_{\mu}u_{\nu}+\partial_{\nu}u_{\mu})/2$ and $\kappa_{\mu\nu}=(u_{\mu}u\cdot\partial u_{\nu}-u_{\nu}u\cdot\partial u_{\mu})/2$. Here we also incorporate the contribution from anti-fermions. Even though the $\hbar$ corrections in Eq.~(\ref{L_equil_Wigner}) only yield a non-vanishing current proportional to $\omega^{\mu}$ as the CVE after integrating over $p$, other terms could still affect the spectrum of spin polarization in momentum space. Nonetheless, the local-equilibrium solution does not satisfy the free-streaming part of the kinetic equation, and non-equilibrium corrections pertinent to interaction have to be included. Such contributions further engender nonlinear chiral transport (e.g., viscous corrections on CME and CVE) investigated in Refs.~\cite{Chen:2016xtg,Gorbar:2016qfh,Gorbar:2017toh,Rybalka:2018uzh, Hidaka:2017auj,Hidaka:2018ekt} by adopting the relaxation-time approximation.

We have derived in this section the Kadanoff-Baym equation with collision
terms for massless fermions. It is straightforward to extend the results
for massless fermions to the case of massive fermions, which will
be discussed in Sec.~\ref{sec:massive fermion with collisions} and \ref{sec:QKT_collision}.

\newpage
\section{Quantum kinetic theory for massive fermions without collisions}\label{sec:massive fermion with collisions}

To generalize the chiral
kinetic theory (CKT) for massless fermions to the quantum kinetic
theory (QKT) for massive fermions, we should identify several distinct
features between the massless and massive cases. Since chiral symmetry
is explicitly broken by the fermionic mass, it is no longer suitable
to decompose Wigner functions into chiral components. Furthermore,
the right-handed and left-handed components are now entangled even
in the absence of collisions. On the other hand, the orientation of spin
of massive particles is also no longer locked by the momentum direction
as for massless particles. One hence has to treat the spin degrees
of freedom as independent variables. In fact, the chirality mixing
has implied the intertwined dynamics of charge and spin transport
for massive fermions without collisions. Such a QKT, in the collisionless
case, has been recently derived from Wigner functions in Refs.~\cite{Gao:2019znl,Weickgenannt:2019dks,Hattori:2019ahi,Wang:2019moi}
with follow-up studies in, e.g., Refs.~\cite{Liu:2020flb,Guo:2020zpa,Sheng:2020oqs,Huang:2020wrr,Dayi:2020uwx,Sheng:2020oqs,Manuel:2021oah,Wang:2021owk,Dayi:2021yhf,Chen:2021rrl}.
In this section, we will mostly follow the approach in Ref.~\cite{Hattori:2019ahi},
which makes a smooth connection to the CKT in the massless limit and
covers the similar formalism constructed in Refs.~\cite{Gao:2019znl,Weickgenannt:2019dks}.
The inclusion of collisions into the QKT for massive fermions will
be done in the next section.

\subsection{Master equations and Wigner functions}

To work with suitable bases, it is more convenient to decompose the
(gauge-invariant) Wigner function for Dirac fermions in generators
of the Clifford algebra \cite{Vasak:1987um}, 
\begin{equation}
\widetilde{W}=\mathcal{F}+i\mathcal{P}\gamma^{5}+\mathcal{V}^{\mu}\gamma_{\mu}+\mathcal{A}^{\mu}\gamma^{5}\gamma_{\mu}+\frac{1}{2}\mathcal{S}^{\mu\nu}\sigma_{\mu\nu},\label{Clifford_decomp}
\end{equation}
where $\sigma_{\mu\nu}=i[\gamma_{\mu},\gamma_{\nu}]/2$ and $\gamma^{5}=i\gamma^{0}\gamma^{1}\gamma^{2}\gamma^{3}$.
For notational brevity, hereafter we will remove the tilde of $\widetilde{W}$
and write $W=\widetilde{W}$ as the gauge invariant Wigner function.
In this section, we will take $W=W^{<}$ and drop the superscript
$<$ for Clifford-algebra coefficients unless specified otherwise.
The coefficients $\mathcal{V}^{\mu}$ and $\mathcal{A}^{\mu}$ contribute
to the vector and axial charge currents, while $\mathcal{F}$ and
$\mathcal{P}$ are related to quark (scalar) and chiral (pseudo-scalar)
condensates, respectively. The antisymmetric $\mathcal{S}^{\mu\nu}$
is related to magnetization. Note that all these coefficients are
real functions in phase space since $W^{\dagger}=\gamma^{0}W\gamma^{0}$.
Nevertheless, these coefficients are not independent of each other.
Their relations are governed by the kinetic equation without collisions
and the Kadanoff-Baym equation with collisions. Eventually, the independent
components should encode four dynamical variables that lead to a $2\times2$
matrix in spin space constructed by two non-local fermionic fields
in the Wigner function. This fact will be manifested by the derivation
of the Wigner function explicitly from a mode expansion of Dirac fields
in the next subsection.

In light of the derivation for massless fermions in Sec.~\ref{sec:CKT_collisions},
the Kadanoff-Baym equation for massive fermions can be written as
\begin{equation}
(\gamma_{\mu}\Pi^{\mu}-m)W^{<}+\frac{1}{2}i\hbar\gamma_{\mu}\nabla^{\mu}W^{<}=\frac{1}{2}i\hbar\Big(\Sigma^{<}\star W^{>}-\Sigma^{>}\star W^{<}\Big),\label{eq:KBEq_massive}
\end{equation}
where $\Pi^{\mu}=p^{\mu}+O(\hbar^{2})$ with $\nabla^{\mu}=\partial_{x}^{\mu}-F^{\mu\nu}\partial_{\nu}^{p}+O(\hbar^{2})$,
$m$ is the mass of the fermion, $W^{\lessgtr}(x,p)$ denote the Wigner
transformation of $G^{\lessgtr}(x_{1},x_{2})$ in Eq.~(\ref{eq:def-green-func}),
and $\Sigma^{\lessgtr}$ demote the self-energies for fermions as
functions of $x$ and $p$. In the derivation of (\ref{eq:KBEq_massive}),
we retained necessary terms for obtaining the first-order quantum
corrections in $\hbar$ expansion and dropped all higher-order terms.
We then consider the collisionless case, $\Sigma^{\lessgtr}=0$, and
thus focus on just $W=W^{<}$. Inserting Eq.~(\ref{Clifford_decomp})
into Eq.~(\ref{eq:KBEq_massive}) without collisions and then separating
real and imaginary parts in terms of independent spinor bases yield
10 equations with 32 degrees of freedom ~\cite{Vasak:1987um,Hattori:2019ahi,Weickgenannt:2019dks}.
Three of them read 
\begin{eqnarray}
m\mathcal{F} & = & \Pi\cdot\mathcal{V},\nonumber \\
m\mathcal{P} & = & -\frac{\hbar}{2}\nabla_{\mu}\mathcal{A}^{\mu},\nonumber \\
m\mathcal{S}_{\mu\nu} & = & -\epsilon_{\mu\nu\rho\sigma}\Pi^{\rho}\mathcal{A}^{\sigma}+\frac{\hbar}{2}\nabla_{[\mu}\mathcal{V}_{\nu]},\label{replacement_eq-1}
\end{eqnarray}
where $A_{[\mu}B_{\nu]}\equiv A_{\mu}B_{\nu}-B_{\nu}A_{\mu}$. Accordingly,
one can choose either $\mathcal{F}$, $\mathcal{P}$, and $\mathcal{S}^{\mu\nu}$
as a set of independent variables, or the other components $\mathcal{V}^{\mu}$
and $\mathcal{A}^{\mu}$~\cite{Ochs:1998qj}, or $\mathcal{F}$ and
$\mathcal{S}^{\mu\nu}$~\cite{Weickgenannt:2019dks}, or $\mathcal{F}$
and $\mathcal{A}^{\mu}$~\cite{Gao:2019znl}, as independent variables.
All these choices are physically equivalent. In this article, we choose
$\mathcal{V}^{\mu}$ and $\mathcal{A}^{\mu}$ as an example and apply
the $\hbar$ expansion to the rest of equations, which results in
$6$ master equations \cite{Hattori:2019ahi}, 
\begin{eqnarray}
\nabla\cdot\mathcal{V} & = & 0,\label{eqv1}\\
(p^{2}-m^{2})\mathcal{V}_{\mu} & = & -\hbar\widetilde{F}_{\mu\nu}\mathcal{A}^{\nu},\label{eqv2}\\
p_{\nu}\mathcal{V}_{\mu}-p_{\mu}\mathcal{V}_{\nu} & = & \frac{\hbar}{2}\epsilon_{\mu\nu\rho\sigma}\nabla^{\rho}\mathcal{A}^{\sigma},\label{eqv3}\\
p\cdot\mathcal{A} & = & 0,\label{eqA1}\\
(p^{2}-m^{2})\mathcal{A}^{\mu} & = & \frac{\hbar}{2}\epsilon^{\mu\nu\rho\sigma}p_{\sigma}\nabla_{\nu}\mathcal{V}_{\rho},\label{eqA2}\\
p\cdot\nabla\mathcal{A}^{\mu}+F^{\nu\mu}\mathcal{A}_{\nu} & = & \frac{\hbar}{2}\epsilon^{\mu\nu\rho\sigma}(\partial_{\sigma}F_{\beta\nu})\partial_{p}^{\beta}\mathcal{V}_{\rho},\label{eqA3}
\end{eqnarray}
where one redundant equation that can be obtained from the above set
has been omitted. Notably, integrating over $p$ for Eq.~(\ref{eqv3})
corresponds to the conservation of angular momentum, where the right-hand
side lead to a non-vanishing anti-symmetric component of the canonical
energy-momentum tensor responsible for the angular-momentum transfer
from the orbital part to the spin part. A detailed analysis for the
massless case can be found in Ref.~\cite{Yang:2018lew}. Consequently,
solving master equations (\ref{eqv1})-(\ref{eqA3}) has implicitly
incorporated the quantum correction associated with the spin-orbit
interaction.

In the massless case, the two sets of master equations, Eqs.~(\ref{eqv1})-(\ref{eqv3})
and Eqs.~(\ref{eqA1})-(\ref{eqA3}), become degenerate due to the
decoupling of $\mathcal{F}$, $\mathcal{P}$, and $\mathcal{S}^{\mu\nu}$
as manifested by Eq.~(\ref{replacement_eq-1}); see also discussions
in Sec.~\ref{sec:CCKE}. Now, Eqs.~(\ref{eqv1})-(\ref{eqv3})
are similar to those in the massless case whereby one acquires the
on-shell condition from Eq.~(\ref{eqv2}) and perturbatively solves
for the extra $\hbar$ correction on $\mathcal{V}^{\mu}$ from Eq.~(\ref{eqv3})
and finally derives the kinetic equation from Eq.~(\ref{eqv1}).
However, as opposed to Eqs.~(\ref{eqv1})-(\ref{eqv3}), the extra
$\hbar$ correction for $\mathcal{A}^{\mu}$ cannot be fully solved
from Eq.~(\ref{eqA1}), while Eq.~(\ref{eqA2}) and Eq.~(\ref{eqA3})
still dictates the on-shell condition and yields the kinetic equation,
respectively. To determine the extra $\hbar$ correction on $\mathcal{A}^{\mu}$
with a smooth connection to the massless limit, we shall resort to
an alternative approach as presented in the next subsection. As mentioned
previously, $\mathcal{V}^{\mu}$ and $\mathcal{A}^{\mu}$ should contain
four dynamical variables.

We now seek perturbative solutions $\mathcal{V}^{\mu}=\mathcal{V}_{0}^{\mu}+\hbar\mathcal{V}_{1}^{\mu}$
and $\mathcal{A}^{\mu}=\mathcal{A}_{0}^{\mu}+\hbar\mathcal{A}_{1}^{\mu}$
up to $O(\hbar)$. The zeroth-order solutions are immediately obtained
from Eqs.~(\ref{eqv2})-(\ref{eqA2}) as 
\begin{eqnarray}
\mathcal{V}_{0}^{\mu} & = & 2\pi p^{\mu}\delta(p^{2}-m^{2})f_{V}(x,p),\nonumber \\
\mathcal{A}_{0}^{\mu} & = & 2\pi a^{\mu}\delta(p^{2}-m^{2})f_{A}(x,p),\label{LO_WF}
\end{eqnarray}
where $f_{V}(x,p)$ and $f_{A}(x,p)$ represent the vector and axial
distribution functions respectively, and $a^{\mu}(x,p)$ satisfies
$p\cdot a=p^{2}-m^{2}$ and corresponds to a non-normalized spin four-vector. Similar to the convention in Sec.~\ref{sec:CKT_collisions}, $f_{V}(x,p)$ and $f_{A}(x,p)$ only incorporate the contributions from fermions with positive energy, while those from anti-fermions are neglected in the present and next sections unless specified otherwise.
In the massless limit, we have $a^{\mu}=p^{\mu}$ because the spin
is enslaved by the momentum. However, $a^{\mu}$ is a dynamical variable
in the massive case, which should be determined by the kinetic theory.
Usually, one may simply introduce 
\begin{equation}
    \widetilde{a}^{\mu}\equiv a^{\mu}f_{A},
\end{equation}
without the decomposition adapted for making a comparison with the
massless result. Due to the constraint from Eq.~(\ref{eqA2}), $f_{V}$
and $\widetilde{a}^{\mu}$ in total delineate four dynamical variables
as expected.

Now, the solution of $\mathcal{V}^{\mu}$ from Eqs.~(\ref{eqv1})-(\ref{eqv3})
up to $O(\hbar)$ reads \cite{Hattori:2019ahi} 
\begin{eqnarray}
\mathcal{V}^{\mu} & = & 2\pi\left\{ \delta(p^{2}-m^{2})\left[p^{\mu}f_{V}+\frac{\hbar\epsilon^{\mu\nu\rho\sigma}n_{\nu}}{2p\cdot n}(\nabla_{\rho}\widetilde{a}_{\sigma}+F_{\rho\sigma}f_{A})\right]\right.\nonumber \\
 &  & \left.+\hbar\widetilde{F}^{\mu\nu}\widetilde{a}_{\nu}\delta^{\prime}(p^{2}-m^{2})\right\} .\label{vector_sol}
\end{eqnarray}
Here $n^{\mu}$ corresponds to a timelike frame vector analogous to
the massless case. The solution of $\mathcal{A}^{\mu}$ from Eqs.~(\ref{eqA1})-(\ref{eqA3})
up to $O(\hbar)$ is given by \cite{Hattori:2019ahi} 
\begin{eqnarray}
\mathcal{A}^{\mu} & = & 2\pi\left[\delta(p^{2}-m^{2})\left(\widetilde{a}^{\mu}+\hbar S_{m(n)}^{\mu\nu}\nabla_{\nu}f_{V}\right) +\hbar\widetilde{F}^{\mu\nu}p_{\nu}\delta^{\prime}(p^{2}-m^{2})f_{V}\right],\label{axial_sol}
\end{eqnarray}
where the tensor $S_{m(n)}^{\mu\nu}$ is defined as 
\begin{equation}
S_{m(n)}^{\mu\nu}=\frac{\epsilon^{\mu\nu\alpha\beta}p_{\alpha}n_{\beta}}{2a\cdot n}=\frac{\epsilon^{\mu\nu\alpha\beta}p_{\alpha}n_{\beta}}{2(p\cdot n+m)}.
\end{equation}
For completeness, we should, in principle, multiply $\mathcal{V}^{\mu}$
and $\mathcal{A}^{\mu}$ by the sign function for energy to include
antiparticle contributions that have been given explicitly in Refs.
\cite{Weickgenannt:2019dks,Gao:2019znl}. As already mentioned, for notational simplicity,
we will mostly focus on just the positive-energy solutions. Here only
the $\hbar$ correction proportional to $\delta^{\prime}(p^{2}-m^{2})$
in Eq.~(\ref{axial_sol}) is solved from Eq.~(\ref{eqA2}), whereas
the $S_{m(n)}^{\mu\nu}$ term cannot be determined by Eq.~(\ref{eqA1}),
which will be alternatively derived from a different approach. The
$\hbar$ terms containing the Levi-Civita symbols in Eqs.~(\ref{vector_sol}) and (\ref{axial_sol})
are dubbed as the magnetization-current terms associated with the
spin-orbit interaction, which correspond to the side-jump terms in
the massless limit. When taking $a^{\mu}=p^{\mu}$ and $m=0$, one
can show that Eqs.~(\ref{vector_sol}) and (\ref{axial_sol}) reduce to
the Wigner function of massless fermions presented in Sec.~\ref{sec:CCKE}.
Notably, unlike massless fermions, one can choose $n^{\mu}=n_{r}^{\mu}(p)=p^{\mu}/m$
as the rest frame of a massive fermion. One finds $S_{m(n_{r})}^{\mu\nu}=0$;
hence Eqs.~(\ref{vector_sol}) and (\ref{axial_sol}) and their corresponding
kinetic equations reduce to simpler forms. Such a frame choice has
been implicitly implemented in, e.g., Refs.~\cite{Weickgenannt:2019dks,Gao:2019znl}.
However, the price of such a frame choice is that a smooth connection
to the massless limit or more precisely to the small-mass regime for
$mW\ll|\gamma\cdot\Delta W|$ is lost.

Because of the explicit frame dependence of magnetization-current
terms, we anticipate also the frame dependence of $f_{V}$ and $\widetilde{a}^{\mu}$
as in the massless case. Based on the frame independence of $\mathcal{V}^{\mu}$
and $\mathcal{A}^{\mu}$, $f_{V}$ and $\widetilde{a}^{\mu}$ follow
the 
modified frame transformations between arbitrary frames
of $n^{\mu}$ and $n^{\prime\mu}$ \cite{Hattori:2019ahi}, 
\begin{equation}
f_{V}^{(n^{\prime})}-f_{V}^{(n)}=\frac{\hbar\epsilon^{\lambda\nu\rho\sigma}n_{\lambda}n_{\nu}^{\prime}}{2(p\cdot n)(p\cdot n^{\prime})}\big(\nabla_{\rho}a_{\sigma}+F_{\rho\sigma}\big)f_{A},
\end{equation}
and 
\begin{equation}
\widetilde{a}^{(n^{\prime})\mu}-\widetilde{a}^{(n)\mu}=\frac{\hbar\epsilon^{\mu\nu\alpha\beta}}{2}\left(\frac{n_{\beta}}{p\cdot n+m}-\frac{n_{\beta}^{\prime}}{p\cdot n^{\prime}+m}\right)p_{\alpha}\nabla_{\nu}f_{V}.
\end{equation}

\subsection{Magnetization-current terms from Dirac wave-functions}

In this subsection, we employ an alternative method to derive Wigner
functions without background fields up to $O(\hbar)$ presented in
Ref.~\cite{Hattori:2019ahi} (see also Ref.~\cite{Sheng:2021kfc}).
In particular, we will utilize the result to determine the magnetization-current
term in $\mathcal{A}_{\mu}$. Additionally, the origin of $f_{V}$
and $\widetilde{a}^{\mu}$ in quantum field theory will be manifested. We
now start with a free Dirac field \cite{Peskin:1995ev},
\begin{eqnarray}
\psi(x) & = & \int\frac{d^{3}{\bf k}}{(2\pi)^{3}}\frac{1}{\sqrt{2E_{k}}}\sum_{s}u_{s}(p)e^{-ik\cdot x}a_{s,{\bf k}},
\end{eqnarray}
where $k_{0}=k^{0}=E_{k}=\sqrt{|{\bf k}|^{2}+m^{2}}$ and we have
dropped the anti-fermion part for simplicity. We have the annihilation
(creation) operators $a_{s,{\bf k}}$ ($a_{s,{\bf k}}^{\dagger}$)
and the wave function $u_{s}(k)=(\sqrt{k\cdot\sigma}\xi_{s},\sqrt{k\cdot\overline{\sigma}}\xi_{s})^{T}$
with $\xi_{s}$ being a two-component spinor with a spin index $s=\pm1/2$~\cite{Peskin:1995ev}.
In $u_{s}(k)$, we introduced $\sigma^{\mu}=e_{a}^{\mu}\sigma^{a}$
from the local transformation of $\sigma^{a}=(1,\boldsymbol{\sigma})$
via the vierbein $e_{a}^{\mu}$ and similarly $\overline{\sigma}^{\mu}=e_{a}^{\mu}\overline{\sigma}^{a}$
from $\overline{\sigma}^{a}=(1,-\boldsymbol{\sigma})$ with $\boldsymbol{\sigma}=(\sigma_{1},\sigma_{2},\sigma_{3})$.
The frame vector mentioned in Sec.~\ref{subsec:WF_frame} can be
accordingly defined as $n^{\mu}\equiv e_{0}^{\mu}$ \cite{Hidaka:2018ekt}.
The Wigner function then takes the form,
\begin{eqnarray}
W(x,p) & = & \int\frac{d^{3}\mathbf{k}_{-}}{(2\pi)^{3}}\int\frac{d^{3}\mathbf{k}_{+}}{(2\pi)^{3}}e^{-ik_{-}\cdot x}\frac{1}{\sqrt{4E_{k_{+}+k_{-}/2}E_{k_{+}-k_{-}/2}}}(2\pi)^{4}\delta^{(4)}\left(p-k_{+}\right)\nonumber \\
 &  & \times\sum_{s,s^{\prime}}u_{s}\left(k_{+}+k_{-}/2\right)\overline{u}_{s^{\prime}}\left(k_{+}-k_{-}/2\right)\left\langle a_{s^{\prime},\mathbf{k}_{+}-\mathbf{k}_{-}/2}^{\dagger}a_{s,\mathbf{k}_{+}+\mathbf{k}_{-}/2}\right\rangle ,\label{WFs_mode_exp}
\end{eqnarray}
where $k_{+}^{\mu}=(k+k^{\prime})^{\mu}/2$ and $k_{-}^{\mu}=(k-k^{\prime})^{\mu}$
with $k$ and $k^{\prime}$ being on-shell momenta. One can integrate
over $\mathbf{k}_{+}$ to obtain 
\begin{eqnarray}
W(x,p) & = & \pi\int\frac{d^{3}\mathbf{k}_{-}}{(2\pi)^{3}}e^{-ik_{-}\cdot x}\frac{1}{\sqrt{E_{p+k_{-}/2}E_{p-k_{-}/2}}}\nonumber \\
 &  & \times\delta\left(p_{0}-\frac{1}{2}E_{p+k_{-}/2}-\frac{1}{2}E_{p-k_{-}/2}\right)\nonumber \\
 &  & \times\sum_{s,s^{\prime}}u_{s}\left(p+\frac{1}{2}k_{-}\right)\overline{u}_{s^{\prime}}\left(p-\frac{1}{2}k_{-}\right)\left\langle a_{s^{\prime},\mathbf{p}-\mathbf{k}_{-}/2}^{\dagger}a_{s,\mathbf{p}+\mathbf{k}_{-}/2}\right\rangle .
\end{eqnarray}
On the other hand, the expectation value of the density operator can
be written as 
\begin{equation}
\langle a_{s',{\bf k'}}^{\dagger}a_{s,{\bf k}}\rangle=\delta_{ss'}N_{V}({\bf k},{\bf k'})+\mathcal{A}_{ss'}({\bf k},{\bf k'}).
\end{equation}
Next, we parametrize the following spin sums as 
\begin{eqnarray}
\sum_{s}\xi_{s}\xi_{s}^{\dagger} & = & n\cdot\sigma=I,\nonumber \\
\sum_{s,s'}\xi_{s}\mathcal{A}_{ss'}\xi_{s'}^{\dagger} & = & S\cdot\sigma,\label{spin_sum}
\end{eqnarray}
where the four vectors $n^{\mu}$ and $S^{\mu}$ satisfy $S\cdot n=0$.
Here the frame vector $n^{\mu}$ also naturally appear as the choice
of a spin basis for the decomposition in $\sigma^{\mu}$. For simplicity,
we will work in the frame $n^{\mu}=(1,\bm{0})$ for the follow-up
computations in this subsection. After the Wigner transformation, we define 
\begin{eqnarray}
\widetilde{f}_{V}(x,p) & \equiv & \int\frac{d^{3}\bm{k}_{-}}{(2\pi)^{3}}N_{V}\left(\mathbf{p}+\frac{\mathbf{k}_{-}}{2},\mathbf{p}-\frac{\mathbf{k}_{-}}{2}\right)e^{-ik_{-}\cdot x},\nonumber \\
\widetilde{S}^{\mu}(x,p) & \equiv & \int\frac{d^{3}\bm{k}_{-}}{(2\pi)^{3}}S^{\mu}\left(\mathbf{p}+\frac{\mathbf{k}_{-}}{2},\mathbf{p}-\frac{\mathbf{k}_{-}}{2}\right)e^{-ik_{-}\cdot x},
\end{eqnarray}
where $\widetilde{f}_{V}(x,p)$ and $\widetilde{S}^{\mu}(x,p)$ are
related to the vector and spin-vector distribution functions, respectively.
From the construction above, it is clear to see $\widetilde{f}_{V}(x,p)$
and $\widetilde{S}_{\mu}(x,p)$ carry four degrees of freedom. We
will then reparameterize $\widetilde{f}_{V}(x,p)$ and $\widetilde{S}^{\mu}(x,p)$
in terms of $f_{V}(x,p)$ and $\widetilde{a}^{\mu}(x,p)$.

According to Eq.~(\ref{spin_sum}), one has 
\begin{eqnarray}
 &  & \sum_{s}u_{s}\left(p+\frac{k_{-}}{2}\right)\overline{u}_{s}\left(p-\frac{k_{-}}{2}\right)\nonumber \\
 & = & \left(\begin{array}{cc}
\sqrt{\sigma\cdot\left(p+\frac{k_{-}}{2}\right)\overline{\sigma}\cdot\left(p-\frac{k_{-}}{2}\right)} & \sqrt{\sigma\cdot\left(p+\frac{k_{-}}{2}\right)\sigma\cdot\left(p-\frac{k_{-}}{2}\right)}\\
\sqrt{\overline{\sigma}\cdot\left(p+\frac{k_{-}}{2}\right)\overline{\sigma}\cdot\left(p-\frac{k_{-}}{2}\right)} & \sqrt{\overline{\sigma}\cdot\left(p+\frac{k_{-}}{2}\right)\sigma\cdot\left(p-\frac{k_{-}}{2}\right)}
\end{array}\right),\label{vector_mtrx}
\end{eqnarray}
and 
\begin{eqnarray}
 &  & \sum_{s,s'}u_{s}\left(p+\frac{k_{-}}{2}\right)\mathcal{A}_{ss'}\overline{u}_{s'}\left(p-\frac{k_{-}}{2}\right)\nonumber \\
 & = & \left(\begin{array}{cc}
\sqrt{\sigma\cdot\left(p+\frac{k_{-}}{2}\right)}\sigma\cdot S\sqrt{\overline{\sigma}\cdot\left(p-\frac{k_{-}}{2}\right)} & \sqrt{\sigma\cdot\left(p+\frac{k_{-}}{2}\right)}\sigma\cdot S\sqrt{\sigma\cdot\left(p-\frac{k_{-}}{2}\right)}\\
\sqrt{\overline{\sigma}\cdot\left(p+\frac{k_{-}}{2}\right)}\sigma\cdot S\sqrt{\overline{\sigma}\cdot\left(p-\frac{k_{-}}{2}\right)} & \sqrt{\overline{\sigma}\cdot\left(p+\frac{k_{-}}{2}\right)}\sigma\cdot S\sqrt{\sigma\cdot\left(p-\frac{k_{-}}{2}\right)}
\end{array}\right).\nonumber \\
\label{axial_mtrx}
\end{eqnarray}
Note that the upper and lower off-diagonal terms in Eqs.~(\ref{vector_mtrx}) and (\ref{axial_mtrx})
are associated with $\mathcal{V}_{\mu}-\mathcal{A}_{\mu}$ and $\mathcal{V}_{\mu}+\mathcal{A}_{\mu}$,
respectively. The diagonal terms are in connection with other components
$\mathcal{F}$, $\mathcal{P}$, and $\mathcal{S}^{\mu\nu}$ in Wigner
functions. To evaluate the matrix elements above, we will employ the
following tricks for Pauli matrices. We may write 
\begin{eqnarray}
p\cdot\sigma & = & m\exp\left(\widetilde{p}_{\perp}\cdot\sigma\theta\right),\nonumber \\
\theta & = & \tanh^{-1}\left(\frac{|{\bf p}_{\perp}|}{E_{p}}\right),
\end{eqnarray}
where 
\begin{equation}
    p_{\perp}^{\mu}\equiv p^{\mu}-(n\cdot p)n^{\mu}\;,\;\;\;
    \widetilde{p}_{\perp}^{\mu}=p_{\perp}^{\mu}/\sqrt{-p_{\perp}^{2}}=p_{\perp}^{\mu}/|\bm{p}_{\perp}|\;,
\end{equation}
yielding
\begin{eqnarray}
\sqrt{p\cdot\sigma} & = & \sqrt{m}\left(\cosh\frac{\theta}{2}+\widetilde{p}_{\perp}\cdot\sigma\sinh\frac{\theta}{2}\right)\nonumber \\
 & = & \sqrt{\frac{1}{2(E_{p}+m)}}\left(E_{p}+m+p_{\perp}\cdot\sigma\right).
\end{eqnarray}
Hereafter we will use the subscripts $\perp$ to denote the component
of an arbitrary vector perpendicular to the frame vector $n^{\mu}$.

Subsequently, we further make the $\hbar$ expansion implicitly led
by the $k_{-}^{\mu}$ expansion for wave functions by analogy with
the derivation in Ref.~\cite{Hidaka:2016yjf} for Weyl fermions.
More precisely, the leading-order terms with $k_{-}^{\mu}=0$ in the
integrand of Eq.~(\ref{WFs_mode_exp}) contribute to the classical
contributions to Wigner functions, whereas $O(\hbar)$ corrections
arise from the terms linear to $k_{-}^{\mu}$. After a tedious yet
straightforward computation for Eq.~(\ref{WFs_mode_exp}) by using
above relations (see Ref.~\cite{Hattori:2019ahi} for details), up
to $O(k_{-})$, it is found 
\begin{eqnarray}
\sigma\cdot(\mathcal{V}-\mathcal{A}) & = & \pi\delta\left(p_{0}-E_{p}\right)\int\frac{d^{3}\mathbf{k}_{-}}{(2\pi)^{3}}\frac{e^{-ik_{-}\cdot x}}{2E_{p}(E_{p}+m)}\left(\mathscr{N}^{+}\cdot\sigma N_{V}+\mathscr{S}^{+}\cdot\sigma\right),\nonumber \\
\overline{\sigma}\cdot(\mathcal{V}+\mathcal{A}) & = & \pi\delta\left(p_{0}-E_{p}\right)\int\frac{d^{3}\mathbf{k}_{-}}{(2\pi)^{3}}\frac{e^{-ik_{-}\cdot x}}{2E_{p}(E_{p}+m)}\left(\mathscr{N}^{-}\cdot\sigma N_{V}+\mathscr{S}^{-}\cdot\sigma\right),
\end{eqnarray}
where $\mathscr{N}_{\mu}^{+}$ and $\mathscr{S}_{\mu}^{+}$ are given
by 
\begin{eqnarray}
\mathscr{N}^{+}\cdot n & = & 2E_{p}\left(E_{p}+m\right),\nonumber \\
\mathscr{N}_{\perp\mu}^{+} & = & 2\left(E_{p}+m\right)p_{\perp\mu}-i\epsilon_{\mu\nu\alpha\beta}n^{\alpha}p^{\beta}k_{-}^{\nu},\nonumber \\
\mathscr{S}^{+}\cdot n & = & -2\left(E_{p}+m\right)p_{\perp}\cdot S_{\perp}+i\epsilon^{\mu\nu\alpha\beta}p_{\mu}k_{-\nu}n_{\alpha}S_{\beta},\nonumber \\
\mathscr{S}_{\perp\mu}^{+} & = & 2\left[m\left(E_{p}+m\right)S_{\perp\mu}-\left(S_{\perp}\cdot p_{\perp}\right)p_{\perp\mu}\right]\nonumber \\
 &  & -i\epsilon_{\mu\nu\alpha\beta}n^{\alpha}S^{\beta}\left[\frac{p_{\perp}\cdot k_{-}}{E_{p}}p^{\nu}+\left(E_{p}+m\right)k_{-}^{\nu}\right],
\end{eqnarray}
Thus, one obtains 
\begin{eqnarray}
n\cdot\mathcal{V} & = & 2\pi\delta(p^{2}-m^{2})E_{p}f_{V},\nonumber \\
\mathcal{V}_{\perp\mu} & = & 2\pi\delta(p^{2}-m^{2})\left[p_{\perp\mu}f_{V}-\frac{1}{2E_{p}}\epsilon_{\mu\nu\alpha\beta}n^{\alpha}\partial^{\nu}\left(\frac{p\cdot\widetilde{S}^{\beta}}{E_{p}+m}-m\widetilde{S}^{\beta}\right)\right],\nonumber \\
n\cdot\mathcal{A} & = & 2\pi\delta(p^{2}-m^{2})p_{\perp}\cdot\widetilde{S}_{\perp},\nonumber \\
\mathcal{A}_{\perp\mu} & = & 2\pi\delta(p^{2}-m^{2})\left[\frac{\widetilde{S}_{\perp}\cdot p_{\perp}}{E_{p}+m}p_{\perp\mu}-m\widetilde{S}_{\perp\mu}-\frac{1}{2(E_{p}+m)}\epsilon_{\mu\nu\alpha\beta}n^{\alpha}p^{\beta}\partial^{\nu}\widetilde{f}_{V}\right],
\end{eqnarray}
where we have re-parameterized 
\begin{equation}
f_{V}(p,X)=\widetilde{f}_{V}(p,X)+\frac{1}{2E_{p}(E_{p}+m)}\epsilon^{\rho\nu\alpha\beta}n_{\alpha}p_{\beta}\partial_{\nu}\widetilde{S}_{\rho}(p,X).
\end{equation}
Finally, by further taking 
\begin{eqnarray}
\widetilde{S}\cdot p_{\perp} & = & a\cdot nf_{A},\nonumber \\
\frac{p\cdot\widetilde{S}}{E_{p}+m}p_{\perp\mu}-m\widetilde{S}_{\mu} & = & a_{\perp\mu}f_{A},\label{Arel_a_and_S}
\end{eqnarray}
and retrieving the $\hbar$ parameters, we arrive at 
\begin{eqnarray}
\mathcal{V}_{\mu} & = & 2\pi\delta(p^{2}-m^{2})\left[p_{\mu}f_{V}+\hbar\frac{\epsilon_{\mu\nu\alpha\beta}n^{\beta}}{2(p\cdot n)}\partial^{\nu}(a^{\alpha}f_{A})\right],\label{free_case_v}\\
\mathcal{A}_{\mu} & = & 2\pi\delta(p^{2}-m^{2})\left[a_{\mu}f_{A}+\hbar\frac{\epsilon_{\mu\nu\alpha\beta}p^{\alpha}n^{\beta}}{2(p\cdot n+m)}\partial^{\nu}f_{V}\right],\label{free_case_a}
\end{eqnarray}
where we have replaced $E_{p}$ by $p\cdot n$. We see that $\mathcal{V}^{\mu}$
in Eqs.~(\ref{free_case_v}) and (\ref{vector_sol}) agree with each
other when $F_{\mu\nu}=0$. Note that the previous constraint $p\cdot a=p^{2}-m^{2}$
is satisfied if we take $\widetilde{S}\cdot p=(p\cdot n+m)f_{A}$,
which implies $-\widetilde{S}^{\mu}=p_{\perp}^{\mu}f_{A}/(p\cdot n)$
when $m=0$. Thus, $(a\cdot n)/(2p\cdot n)$ is identified to be the
helicity in the massless limit.

One can now read out the magnetization-current term in $\mathcal{A}^{\mu}$
without background fields from Eq.~(\ref{free_case_v}), i.e., the
second term inside parentheses. We then generalize the derivative
operator to include a background field by analogy with the massless
case \cite{Hidaka:2016yjf} to obtain Eq.~(\ref{axial_sol}), and
find that $\mathcal{A}^{\mu}$ has a symmetric form with $\mathcal{V}^{\mu}$
under interchanges $p^{\mu}\leftrightarrow a^{\mu}$ and $f_{V}\leftrightarrow f_{A}$.
In fact, one could absorb this magnetization-current term into $\widetilde{a}^{\mu}$.
The freedom of such a redefinition reveals itself as the non-uniqueness
of the magnetization-current term as we saw when solving the master
equations (\ref{eqA1}) and (\ref{eqA2}) for $\mathcal{A}^{\mu}$,
and could occur in the massive case since $\widetilde{a}^{\mu}$ is
a dynamical variable to be determined by the kinetic theory. Nevertheless,
it is crucial to explicitly separate the magnetization-current term
from $\widetilde{a}^{\mu}$ in order to see a smooth reduction to
the CKT, where $\widetilde{a}^{\mu}$ is no longer an independent
dynamical variable and is locked by $p^{\mu}$. See also Refs.~\cite{Guo:2020zpa,Sheng:2020oqs}
for related discussions on the connection between massless and massive
fermions.

\subsection{Quantum kinetic equations for charge and spin transport}

Given the vector and axial-vector components of Wigner functions in
Eqs.~(\ref{vector_sol}) and (\ref{axial_sol}), we may derive the quantum
kinetic theory (or the so-called axial kinetic theory in Ref.~\cite{Hattori:2019ahi})
containing a scalar kinetic equation (SKE) and an axial-vector kinetic
equation (AKE) governing the dynamical degrees of freedoms $f_{V/A}$
and $a^{\mu}$. Plugging Eq.~(\ref{vector_sol}) into Eq.~(\ref{eqv1}),
with a complicated yet straightforward calculation (see Ref.~\cite{Hattori:2019ahi}
for details), one acquires the SKE with an arbitrary spacetime-dependent
frame vector up to $O(\hbar)$, 
\begin{equation}
\delta(p^{2}-m^{2})p\cdot\Delta f_{V}+\hbar\chi_{0}^{(n)}(a^{\mu},f_{A})+\hbar\chi_{m}^{(n)}(a^{\mu},f_{A})=0,\label{SKE_final}
\end{equation}
where the $O(\hbar)$ terms $\chi_{0}^{(n)}$ and $\chi_{m}^{(n)}$
as functions of $a^{\mu}$ and $f_{A}$ are given by 
\begin{eqnarray}
\chi_{0}^{(n)}(a^{\mu},f_{A}) & = & \delta(p^{2}-m^{2})\left[\frac{E_{\mu}S_{a(n)}^{\mu\nu}}{p\cdot n}\nabla_{\nu}+S_{a(n)}^{\mu\nu}\left(\partial_{\mu}F_{\rho\nu}\right)\partial_{p}^{\rho}+\left(\partial_{\mu}S_{a(n)}^{\mu\nu}\right)\nabla_{\nu}\right]f_{A}\nonumber \\
 &  & -\frac{\delta^{\prime}(q^{2}-m^{2})}{p\cdot n}B^{\mu}\Box_{\mu\nu}\widetilde{a}^{\nu},\nonumber \\
\chi_{m}^{(n)}(a^{\mu},f_{A}) & = & \frac{1}{2}\delta(q^{2}-m^{2})\epsilon^{\mu\nu\alpha\beta}\left\{ \nabla_{\mu}\left(\frac{n_{\beta}}{q\cdot n}\right)\left(\nabla_{\nu}a_{\alpha}+F_{\nu\alpha}\right)\right.\nonumber \\
 &  & \left.+\frac{n_{\beta}}{q\cdot n}\left[(\partial_{\mu}F_{\rho\nu})(\partial_{q}^{\rho}a_{\alpha})+\left(\nabla_{\nu}a_{\alpha}-F_{\rho\nu}(\partial_{p}^{\rho}a_{\alpha})\right)\nabla_{\mu}\right]\right\} f_{A},
\end{eqnarray}
with $E^{\mu}$and $B^{\mu}$ being defined in Eq.~(\ref{eq:em-field-u})
with the replacement of $u^{\mu}$ by $n^{\mu}$ and with $S_{a(n)}^{\mu\nu}$
and $\Box_{\mu\nu}$ being defined as 
\begin{eqnarray}
S_{a(n)}^{\mu\nu} & = & \frac{\epsilon^{\mu\nu\alpha\beta}a_{\alpha}n_{\beta}}{2p\cdot n},\nonumber \\
\Box_{\mu\nu}\widetilde{a}^{\nu} & = & p\cdot\nabla\widetilde{a}_{\mu}-F_{\mu\nu}\widetilde{a}^{\nu}.
\end{eqnarray}
Similarly, by inserting Eq.~(\ref{axial_sol}) into Eq.~(\ref{eqA3}),
one derives the AKE also with an arbitrary spacetime-dependent frame
vector up to $O(\hbar)$, 
\begin{equation}
\delta(p^{2}-m^{2})\Box^{\mu\nu}\widetilde{a}_{\nu}+\hbar\chi_{0}^{(n)\mu}(f_{V})+\hbar\chi_{m}^{(n)\mu}(f_{V})=0,\label{AKE_final}
\end{equation}
where the $O(\hbar)$ terms $\chi_{0}^{(n)\mu}$ and $\chi_{m}^{(n)\mu}$
as functions $f_{V}$ are given by 
\begin{eqnarray}
\chi_{0}^{(n)\mu}(f_{V}) & = & p^{\mu}\left\{ \delta(p^{2}-m^{2})\left[\left(\partial_{\alpha}S_{m(n)}^{\alpha\nu}\right)\nabla_{\nu}+\frac{S_{m(n)}^{\alpha\nu}E_{\alpha}\nabla_{\nu}}{p\cdot n+m}+S_{m(n)}^{\rho\nu}(\partial_{\rho}F_{\beta\nu})\partial_{p}^{\beta}\right]\right.\nonumber \\
 &  & \left.-\delta^{\prime}(p^{2}-m^{2})\frac{p\cdot B}{p\cdot n+m}p\cdot\nabla\right\} f_{V},\nonumber \\
\chi_{m}^{(n)\mu}(f_{V}) & = & m\left\{ \frac{\delta(p^{2}-m^{2})\epsilon^{\mu\nu\alpha\beta}}{2(p\cdot n+m)}\left[ \frac{}{} m(\partial_{\alpha}n_{\beta})\Delta_{\nu}\right.\right.\nonumber \\
 &  & \left.+(mn_{\beta}+p_{\beta})\left(\frac{E_{\alpha}-\partial_{\alpha}(p\cdot n)}{p\cdot n+m}\nabla_{\nu}-(\partial_{\nu}F_{\rho\alpha})\partial_{p}^{\rho}\right)\right]\nonumber \\
 &  & \left.+\delta^{\prime}(p^{2}-m^{2})\frac{mn_{\beta}+p_{\beta}}{p\cdot n+m}\tilde{F}^{\mu\beta}p\cdot\nabla\right\} f_{V}.
\end{eqnarray}

Taking $\hbar=0$, Eq.~(\ref{SKE_final}) reduces to the Vlasov equation,
while Eq.~(\ref{AKE_final}) corresponds to a relativistic version
of the Bargmann-Michel-Telegdi (BMT) equation \cite{BLT_spin}. In
the massless limit, we have $\chi_{m}^{(n)}[q^{\mu},f_{A}]=0$ and
$\chi_{0}^{(n)}[q^{\mu},f_{A}]$ reduces to the $O(\hbar)$ terms
in the CKT. On the other hand, for AKE, when $m=0$, we have $\chi_{m}^{(n)\mu}[f_{V}]=0$
and the rest part in Eq.~(\ref{AKE_final}) becomes the CKT multiplied
by $p^{\mu}$. As anticipated, Eqs.~(\ref{SKE_final}), and (\ref{AKE_final})
make a smooth connection between the massless case delineated by the
CKT and the massive one. As mentioned previously, when choosing the
rest frame of the particle, $n^{\mu}=n_{r}^{\mu}(p)=p^{\mu}/m$, the
magnetization-current term in $\mathcal{A}^{\mu}$ vanishes and the
AKE also becomes simpler \cite{Yang:2020hri}, which takes the form,
\begin{eqnarray}
0 & = & \delta(p^{2}-m^{2})\left[\Box_{\mu\nu}\widetilde{a}^{\nu}-\frac{\hbar}{2}\epsilon^{\mu\nu\rho\sigma}p_{\rho}(\partial_{\sigma}F_{\beta\nu})\partial_{p}^{\beta}f_{V}\right] 
+\hbar\widetilde{F}^{\mu\nu}p_{\nu}\delta^{\prime}(p^{2}-m^{2})p\cdot\nabla f_{V}.\label{AKE_free_nr}
\end{eqnarray}
One could also find the equivalent results in Refs.~\cite{Gao:2019znl,Weickgenannt:2019dks}.

When solving kinetic equations (\ref{SKE_final}) and (\ref{AKE_final}),
we need to handle the terms proportional to derivatives of delta functions
$\delta^{\prime}(p^{2}-m^{2})$. One may seek off-shell solutions
of $f_{V}$ and $\widetilde{a}^{\nu}$ independent of the explicit
expression of $p\cdot n$. In such a case, we can simply take $p\cdot\nabla f_{V}=O(\hbar)$
and $\Box^{\mu\nu}\widetilde{a}_{\nu}=O(\hbar)$. Accordingly, the
$\delta^{\prime}(p^{2}-m^{2})$ terms in Eqs.~(\ref{SKE_final}), and (\ref{AKE_final})
are at $O(\hbar^{2})$ and thus can be neglected. Nevertheless, when
handling on-shell kinetic equations and the on-shell $f_{V}$ and
$\widetilde{a}^{\mu}$, the $\delta^{\prime}(p^{2}-m^{2})$ terms
cannot simply be omitted. All these terms can be instead arranged
with the leading-order kinetic equations through 
\begin{eqnarray}
2\delta^{\prime}(p^{2}-m^{2})p^{\mu}p\cdot\nabla f_{V} & = & -\delta(p^{2}-m^{2})\partial_{p}^{\mu}(p\cdot\nabla f_{V}),\nonumber \\
2\delta^{\prime}(p^{2}-m^{2})\Box_{\mu\nu}\widetilde{a}^{\nu} & = & -\delta(p^{2}-m^{2})\partial_{p}^{\rho}\left(\frac{n_{\rho}\Box_{\mu\nu}\widetilde{a}^{\nu}}{p\cdot n}\right),
\end{eqnarray}
up to $O(\hbar)$. Then, all the delta functions can be factored out
for the SKE and AKE.

\subsection{Quantum transport in thermal equilibrium}

Given the solutions of quantum kinetic equations, one can obtain the
vector or axial vector current and the symmetric and anti-symmetric
part of the canonical energy-momentum tensor~\footnote{More precisely, one has to include both fermions and anti-fermions,
and put an overall sign function ${\rm sgn}(p\cdot n)$ in front of
the Wigner function.}, 
\begin{eqnarray}
 &  & J^{\mu}=4\int\frac{d^{4}p}{(2\pi)^{4}}\mathcal{V}^{\mu},\;\;J_{5}^{\mu}=4\int\frac{d^{4}p}{(2\pi)^{4}}\mathcal{A}^{\mu},\nonumber \\
 &  & T_{S}^{\mu\nu}=2\int\frac{d^{4}p}{(2\pi)^{4}}\left(\mathcal{V}^{\mu}p^{\nu}+\mathcal{V}^{\nu}p^{\mu}\right),\nonumber \\
 &  & T_{A}^{\mu\nu}=2\int\frac{d^{4}p}{(2\pi)^{4}}\left(\mathcal{V}^{\mu}p^{\nu}-\mathcal{V}^{\nu}p^{\mu}\right).\label{def_J_Tmunu}
\end{eqnarray}
Angular-momentum conservation arises from Eq.~(\ref{eqv3}) as discussed
in the massless case \cite{Yang:2018lew}, and $T_{A}^{\mu\nu}$ is
responsible for the angular-momentum transfer (see Ref.~\cite{Hattori:2019lfp}
and references therein).

While collisionless kinetic equations do not uniquely determine equilibrium
Wigner functions \cite{Hidaka:2017auj}, we may construct equilibrium
Wigner functions motivated by the following considerations. For the vector
charge, we may naturally take the Fermi distribution function,
\begin{equation}
f_{V}^{\text{eq}}=f_{\text{FD}}(p\cdot u-\mu)=\frac{1}{\exp[\beta(p\cdot u-\mu)]+1}.
\end{equation}
On the other hand, the axial
charge should be damped out as $t\rightarrow\infty$ when $m\neq0$
because of the scattering. But $f_{A}^{\text{eq}}$ may be at $O(\hbar)$
induced by the vorticity correction. Referring to the massless case
\cite{Chen:2015gta,Hidaka:2017auj}, we also expect that $\mathcal{A}_{\text{eq}}^{\mu}$
does not have an explicit dependence on $n^{\mu}$. Thus one may propose
an equilibrium Wigner function in constant magnetic fields and thermal
vorticity \cite{Fang:2016vpj,Hattori:2019ahi,Weickgenannt:2019dks},
\begin{eqnarray}
\mathcal{V}_{\text{eq}}^{\mu} & = & 2\pi\delta(p^{2}-m^{2})p^{\mu}f_{\text{FD}},\label{equ_V_mu}\\
\mathcal{A}_{\text{eq}}^{\mu} & = & 2\pi\hbar\left[\frac{1}{4}\delta(p^{2}-m^{2})p_{\nu}\epsilon^{\nu\mu\rho\sigma}\Omega_{\rho\sigma}\partial_{p\cdot\beta}+\widetilde{F}^{\mu\nu}p_{\nu}\delta^{\prime}(p^{2}-m^{2})\right]f_{\text{FD}}.\label{equ_A_mu}
\end{eqnarray}
Note that the leading-order SKE, merely the Vlasov equation, is satisfied under
the following conditions,
\begin{equation}
\partial_{(\mu}\beta_{\nu)}=0,\quad\partial_{\mu}\left(\frac{\mu}{T}\right)=F^{\mu\nu}\beta_{\nu},\quad\partial_{\lambda}\Omega_{\mu\nu}=0,
\end{equation}
where $A_{(\mu}B_{\nu)}=A_{\mu}B_{\nu}+A_{\nu}B_{\mu}$. In presence
of an electric field $E^{\mu}=F^{\mu\nu}u_{\nu}$ in the fluid rest
frame, the system may reach a steady state with an inhomogeneous local
chemical potential. Nonetheless, we here consider an equilibrium state
with $E^{\mu}=0$ and hence $\mu/T$ is also a constant. The equilibrium
$\mathcal{A}_{\text{eq}}^{\mu}$ takes the equivalent form as the
one for massless fermions at constant temperature except for the on-shell
condition \cite{Hidaka:2017auj}, and was also obtained from the statistical
field theory \cite{Becattini:2013fla}. See also Refs.~\cite{Prokhorov:2017atp,Prokhorov:2018qhq}
for Wigner functions beyond weak vorticity and with acceleration.

The equilibrium $\mathcal{A}^{\mu}$ in Eq.~(\ref{equ_A_mu}) now
leads to the CSE current $\hbar J_{(1)}^{\mu}$ induced by the vorticity
and magnetic field as in Eq.~(\ref{eq:2nd-porder-current-1}) but
with the conductivities given by 
\begin{eqnarray}
\xi_{5} & = & \frac{1}{2\pi^{2}}\int_{0}^{\infty}d|{\bf p}|\frac{2E_{p}^{2}-m^{2}}{E_{p}}\left[f_{\text{FD}}(E_{p}-\mu)+f_{\text{FD}}(E_{p}+\mu)\right],\nonumber \\
\xi_{B5} & = & \frac{1}{2\pi^{2}}\int_{0}^{\infty}d|{\bf p}|\left[f_{\text{FD}}(E_{p}-\mu)-f_{\text{FD}}(E_{p}+\mu)\right].
\end{eqnarray}
The above results agree with those derived from the Kubo formula with
thermal correlators \cite{Buzzegoli:2017cqy,Lin:2018aon}.

\newpage
\section{Quantum kinetic theory for massive fermions with collisions}\label{sec:QKT_collision}

In this section, we include collision terms into the QKT to describe
spin transport phenomena. Similar to the massless case \cite{Hidaka:2016yjf},
one may obtain a generic form of collision terms in self-energies
based on the Kadanoff-Baym equation \cite{Yang:2020hri,Sheng:2021kfc}.
Alternatively, one may apply a distinct formalism with generalized
distribution functions in extended phase space including spin degrees
of freedom and with non-local collisions \cite{Zhang:2019xya,Weickgenannt:2020aaf,Weickgenannt:2021cuo}.
A practical power counting scheme is adopted to treat spin degrees
of freedom in a subdominant order in the vector charge or energy transport
equations. There are several other studies of collisional effects
pertinent to spin polarization in heavy-ion collisions from different
or relevant approaches \cite{Sun:2017xhx,Li:2019qkf,Kapusta:2019sad,Carignano:2019zsh,Liu:2019krs,Bhadury:2020puc,Hou:2020mqp,Fauth:2021nwe,Lin:2021mvw, Carignano:2021zhu}.
Here we will mostly focus on the derivation from the Kadanoff-Baym
equation and that from non-local collisions.

\subsection{Effective quantum kinetic equations}

To include collisions into the QKT for massive fermions, we have to
incorporate non-vanishing self-energies in the Kadanoff-Baym equation
shown in Eq.~(\ref{eq:KBEq_massive}). In analogy with the massless
case, we only focus on scattering processes, so we drop the real parts
of retarded and advanced self-energies and of retarded propagators, see Eqs.~(\ref{eq:kinetic collision 1}), (\ref{eq:kinetic collision 2}), and (\ref{eq:CKT}) and accompanied discussions.
One then follows the same recipe as in the collisionless case to carry
out the decomposition of Wigner functions and derive master equations
with $\hbar$ expansion. For collision terms, we further decompose
the self-energies in the same way as for Wigner functions, 
\begin{equation}
\Sigma^{\lessgtr}=\Sigma_{F}^{\lessgtr}+i\Sigma_{P}^{\lessgtr}\gamma^{5}+\Sigma_{V\mu}^{\lessgtr}\gamma^{\mu}+\Sigma_{A\mu}^{\lessgtr}\gamma^{5}\gamma^{\mu}+\frac{1}{2}\Sigma_{T\mu\nu}^{\lessgtr}\sigma^{\mu\nu}.\label{Sigma_decomp}
\end{equation}
Inserting Eqs.~(\ref{Clifford_decomp}), and (\ref{Sigma_decomp}) into
Eq.~(\ref{eq:KBEq_massive}) with proper arrangements like the collisionless
case, we again acquire 10 coupled equations. Making similar replacements
as Eq.~(\ref{replacement_eq-1}) to rewrite $\mathcal{F}$, $\mathcal{P}$,
and $\mathcal{S}^{\mu\nu}$ in terms of $\mathcal{V}^{\mu}$ and $\mathcal{A}^{\mu}$,
we further obtain 6 independent master equations. Unlike the collisionless
case, these equations are more involved to find the perturbative solution
and derive the quantum kinetic equations. Nonetheless, for practical
purposes, we may propose a useful power-counting scheme to simplify
the problem.

In most of the practical situations such as in heavy-ion collisions, the
axial-charge current is usually smaller than the vector-charge current
since the spin polarization is generated by quantum effects. This
observation motivates us to introduce the power counting assumption
\cite{Yang:2020hri},
\begin{equation}
\mathcal{V}_{\mu}^{\lessgtr}\sim O(\hbar^{0})\quad{\rm and}\quad\mathcal{A}_{\mu}^{\lessgtr}\sim O(\hbar).
\end{equation}
Hereafter we attach the superscript $\lessgtr$ to specify the lesser
and greater components. The above power counting assumption also constrains
the orders of other components from master equations, 
\begin{equation}
\mathcal{F}^{\lessgtr}\sim O(\hbar^{0}),\quad\mathcal{S}_{\mu\nu}^{\lessgtr}\sim O(\hbar),\quad\mathcal{P}^{\lessgtr}\sim O(\hbar^{2}).
\end{equation}
On the other hand, since $\Sigma_{F}^{\lessgtr}$ and $\Sigma_{V,\mu}^{\lessgtr}$
are responsible for the collision terms of classical Boltzmann equations,
we should take 
\begin{equation}
\Sigma_{F}^{\lessgtr}\sim O(\hbar^{0})\quad{\rm and}\quad\Sigma_{V\mu}^{\lessgtr}\sim O(\hbar^{0}),
\end{equation}
which further results in 
\begin{equation}
\Sigma_{A\mu}^{\lessgtr}\sim O(\hbar^{1})\quad{\rm and}\quad\Sigma_{T\mu\nu}^{\lessgtr}\sim O(\hbar^{1}),
\end{equation}
by balancing the orders of coupled equations. However, there exists
no explicit restriction for $\Sigma_{P}^{\lessgtr}$ in generic master
equations, while it is expected that the presence of $\Sigma_{P}^{\lessgtr}$
has to be induced by nonzero pseudo-scalar condensate, which should
be at $O(\hbar^{2})$ from the consistency with the anomaly equation
(mass correction upon the chiral anomaly), see Ref.~\cite{Yang:2020hri}
for technical details. When adopting this power counting scheme, we
shall distinguish the explicit quantum correction dubbed as the explicit $\hbar$
term with $\hbar$ attached to it ($\hbar\mathcal{C}_{2}^{(n)\mu}$) and the implicit one without $\hbar$ attached ($\mathcal{C}_{1}^{(n)\mu}$) although they are now of the same order in $\hbar$ expansion. The latter further incorporates the "classical contribution" ($\mathcal{C}^{\mu}_{\text{cl}}$), which refers to the scattering with a classical medium without $\hbar$ corrections. Such a term is still of $O(\hbar)$ due to the involvement of $\tilde{a}^{\mu}$ from the probe fermion. Detailed expressions of $\mathcal{C}_{1}^{(n)\mu}$, $\mathcal{C}_{2}^{(n)\mu}$, and $\mathcal{C}^{\mu}_{\text{cl}}$ will be shown later.

By implementing the power counting, the master equations obtained
from the Kadanoff-Baym equations at the leading order are given by
\begin{eqnarray}
\mathcal{D}^{\mu}\mathcal{V}_{\mu}^{<} & = & -\frac{1}{m}p_{\mu}\widehat{\Sigma_{F}\mathcal{V}^{\mu}},\label{MA_1}\\
p^{[\mu}\mathcal{V}^{<\nu]} & = & 0,\label{MA_2}\\
(p^{2}-m^{2})\mathcal{V}_{\mu}^{<} & = & 0,\label{MA_3}\\
p\cdot\mathcal{A}^{<} & = & 0,\label{MA_4}\\
(p^{2}-m^{2})\mathcal{A}^{<\mu} & = & \frac{\hbar}{2}\epsilon^{\mu\alpha\beta\gamma}p_{\alpha}D_{\beta}\mathcal{V}_{\gamma}^{<},\label{MA_5}
\end{eqnarray}
\begin{eqnarray}
 &  & p\cdot\mathcal{D}\mathcal{A}_{\mu}^{<}-F_{\mu\nu}\mathcal{A}^{<\nu}-\frac{\hbar}{4}\epsilon_{\mu\nu\rho\sigma}[\mathcal{D}^{\nu},\mathcal{D}^{\rho}]\mathcal{V}^{<\sigma}\nonumber \\
 &  & =-m\Big(\widehat{\Sigma_{F}\mathcal{A}_{\mu}}-\frac{1}{2}\epsilon_{\mu\nu\rho\sigma}\widehat{\Sigma_{T}^{\nu\rho}\mathcal{V}^{\sigma}}\Big)-p_{\alpha}\widehat{\Sigma_{\mu}^{A}\mathcal{V}^{\alpha}}+p_{\mu}\widehat{\Sigma_{\alpha}^{A}\mathcal{V}^{\alpha}},\label{MA_6}
\end{eqnarray}
where we have defined $\mathcal{D}_{\mu}Z=\nabla_{\mu}Z+\widehat{\Sigma_{V\mu}Z}$
and $\widehat{XY}=X^{>}Y^{<}-X^{<}Y^{>}$ with $X$, $Y$, and $Z$
being the coefficients of the Clifford decomposition for propagators
and self-energies. Note that we have to keep the $O(\hbar)$ terms
linear to $\mathcal{V}_{\mu}^{\lessgtr}$ in Eqs.~(\ref{MA_5}) and
(\ref{MA_6}) since $\mathcal{A}_{\mu}^{\lessgtr}\sim O(\hbar)$.
In addition, we consider weakly coupled systems and thus drop the
$O(\Sigma^{2})$ terms. Since now we only focus on leading-order results,
we neglected quantum corrections of $\mathcal{V}_{\mu}^{<}$. Apparently,
$\mathcal{V}_{\mu}^{<}$ is led by the classical solution in Eq.~(\ref{LO_WF}),
whereas the SKE from Eq.~(\ref{MA_1}) now reads 
\begin{equation}
\delta(p^{2}-m^{2})\left(p\cdot\Delta f_{V}+p_{\mu}\widehat{\Sigma_{V}^{\mu}f_{V}}+m\widehat{\Sigma_{F}f_{V}}\right)=0,\label{SKE}
\end{equation}
as a standard Boltzmann equation, where $f_{V}^{<}=f_{V}$ and $f_{V}^{>}=1-f_{V}$.
On the other hand, similar to the collisionless case, the magnetization-current
term in $\mathcal{A}_{\mu}^{\lessgtr}$ cannot be determined by Eqs.~(\ref{MA_4})
and (\ref{MA_5}). We may, however propose a generalized solution in
connection with the massless solution. It is postulated \cite{Yang:2020hri},
\begin{eqnarray}
\mathcal{A}^{<\mu} & = & 2\pi\left[\delta(p^{2}-m^{2})\left(a^{\mu}f_{A}+\hbar S_{m(n)}^{\mu\nu}\mathcal{D}_{\nu}f_{V}\right)\right.\nonumber \\
 &  & \left.+\hbar\widetilde{F}^{\mu\nu}p_{\nu}\delta^{\prime}(p^{2}-m^{2})f_{V}\right],\label{axial_sol_collisions}
\end{eqnarray}
which is a generalization of $\mathcal{A}^{<\mu}$ in Eq.~(\ref{axial_sol})
with the replacement of $\nabla_{\nu}f_{V}$ by $\mathcal{D}_{\nu}f_{V}$.
The generalization of the magnetization-current term with collisions
here is rather natural since $\Sigma_{V\mu}^{\lessgtr}$ are the only
vectors at $O(\hbar^{0})$ in self-energies, which can then be coupled
to $S_{m(n)}^{\mu\nu}$. Furthermore, for the solution of $\mathcal{A}_{\mu}^{>}$,
we interchange the lesser and greater components therein, e.g., we
replace $f_{A}=f_{A}^{<}$ and $f_{V}=f_{V}^{<}$ in Eq.~(\ref{axial_sol_collisions})
by $f_{V}^{>}=1-f_{V}$ and $f_{A}^{>}=-f_{A}$ due to its origin
from the expectation value of the fermionic density operator in spinor
space.

Inserting Eq.~(\ref{axial_sol_collisions}) into Eq.~(\ref{MA_6}),
the AKE with collisional effects and the general frame vector $n^{\mu}=n^{\mu}(x)$
takes the form,
\begin{eqnarray}
 \delta(p^{2}-m^{2})\Box^{\mu\nu}\widetilde{a}_{\nu}+\hbar\chi_{0}^{(n)\mu}[f_{V}]+\hbar\chi_{m}^{(n)\mu}[f_{V}]
 =\mathcal{C}_{1}^{(n)\mu}+\hbar\mathcal{C}_{2}^{(n)\mu}.\label{AKE_collision_n}
\end{eqnarray}
The collisionless part is given in Eq.~(\ref{AKE_free_nr}). The collision
terms are 
\begin{eqnarray}
\mathcal{C}_{1}^{(n)\mu} & = & -\delta(p^{2}-m^{2})\left[a^{\mu}p_{\nu}\widehat{\Sigma_{V}^{\nu}f_{A}}+m^{2}\widehat{\Sigma_{A}^{\mu}f_{V}}-p^{\mu}p_{\nu}\widehat{\Sigma_{A}^{\nu}f_{V}}+\right.\nonumber \\
 &  & +m\left(a^{\mu}\widehat{\Sigma_{F}f_{A}}-\frac{1}{2}\epsilon^{\mu\nu\rho\sigma}p_{\nu}\widehat{\Sigma_{T\rho\sigma}f_{V}}\right),\label{C_n1}\\
\mathcal{C}_{2}^{(n)\mu} & = & \frac{1}{2}\delta(p^{2}-m^{2})\left\{ \epsilon^{\mu\nu\rho\sigma}p_{\nu}(\nabla_{\rho}\widehat{\Sigma_{V\sigma})f_{V}}\right.\nonumber \\
 &  & +2S_{m(n)}^{\mu\nu}\left[m(\nabla_{\nu}\widehat{\Sigma_{F})f_{V}}+F_{\,\,\,\nu}^{\rho}\widehat{\Sigma_{V\rho}f_{V}}-(p\cdot\Delta\widehat{\Sigma_{V\nu})f_{V}}\right]\nonumber \\
 &  & \left.+2\widehat{f_{V}\Sigma_{V\nu}}\left(p\cdot\nabla S_{m(n)}^{\mu\nu}-F_{\,\,\,\lambda}^{\mu}S_{m(n)}^{\lambda\nu}\right)\right\} \nonumber \\
 &  & -\left[\left(2S_{m(n)}^{\mu\nu}p^{\lambda}F_{\lambda\nu}-\widetilde{F}^{\mu\nu}p_{\nu}\right)\delta^{\prime}(p^{2}-m^{2})+\delta(p^{2}-m^{2})\left(\nabla_{\nu}S_{m(n)}^{\mu\nu}\right)\right]\nonumber \\
 &  & \times\left(C_{V}[f_{V}]+mC_{S}[f_{V}]\right),\label{C_nQ_2}
\end{eqnarray}
where $C_{V}[f_{V}]=-p^{\mu}\widehat{\Sigma_{V\mu}f_{V}}$ and $C_{S}[f_{V}]=-\widehat{\Sigma_{F}f_{V}}$.
Note that $\mathcal{C}_{1}^{(n)\mu}$ implicitly contains the $\hbar$
terms, from e.g., $\Sigma_{A}^{\mu}$ and $\Sigma_{T\rho\sigma}$,
and hence depends on the frame choice. In general, purely classical
contribution from $\mathcal{C}_{1}^{(n)\mu}$ only results in spin
relaxation and diffusion \footnote{In Ref.~\cite{Yang:2020hri}, both effects are less accurately dubbed as diffusion except for the case of heavy quarks, where only the diffusion term remains.}, whereas the implicit $\hbar$ terms therein and the explicit
ones from $\mathcal{C}_{2}^{(n)\mu}$ as quantum corrections could
yield dynamical spin polarization originating from, e.g., the spatial
inhomogeneity of $f_{V}$ like vorticity.

\subsection{Simplifications and analyses of collision terms}

Although $\mathcal{C}_{1}^{(n)\mu}+\hbar\mathcal{C}_{2}^{(n)\mu}$
given by Eqs.~(\ref{C_n1}) and (\ref{C_nQ_2}) serves as generic
collision terms in the presence of spacetime-dependent background electromagnetic
fields and an arbitrary spacetime-dependent frame vector $n^{\mu}(x)$,
we may consider simplified expressions for practical purposes \cite{Yang:2020hri}.
Based on the frame independence of Wigner functions and physical results,
it is more convenient to work with a constant frame vector such that
$\partial_{\mu}n^{\nu}=0$. More precisely, we could simply take $n^{\mu}=(1,{\bf 0})$,
which also corresponds to the frame choice for the CKT presented in
early works obtained from the Berry phase \cite{Stephanov:2012ki,Son:2012wh,Son:2012zy}.
For an application to the spin transport of quarks traveling in the
QGP, one may further neglect the influences from electromagnetic fields
that decay in late stages \cite{Deng:2012pc,McLerran:2013hla}. The
simplified AKE can be put into the form,
\begin{equation}
\delta(p^{2}-m^{2})p\cdot\partial\widetilde{a}^{\mu}=\mathcal{C}_{s1}^{(n)\mu}+\hbar\mathcal{C}_{s2}^{(n)\mu},\label{AKE_v1}
\end{equation}
where the collision terms $\mathcal{C}_{s1}^{(n)\mu}=\mathcal{C}_{1}^{(n)\mu}$
despite the implicit quantum corrections therein and $\mathcal{C}_{s2}^{(n)\mu}$
is given by 
\begin{eqnarray}
\mathcal{C}_{s2}^{(n)\mu} & = & \frac{1}{2}\delta(p^{2}-m^{2})\left\{ \epsilon^{\mu\nu\rho\sigma}p_{\nu}(\partial_{\rho}\widehat{\Sigma_{V\sigma})f_{V}}\right.\nonumber \\
 &  & +2S_{m(n)}^{\mu\nu}\left[m(\partial_{\nu}\widehat{\Sigma_{F})f_{V}}+p^{\rho}(\partial_{\nu}\widehat{\Sigma_{V\rho})f_{V}}-(p\cdot\partial\widehat{\Sigma_{V\nu})f_{V}}\right].\label{Cns2}
\end{eqnarray}
By rearranging the explicit $\hbar$ corrections in Eq.~(\ref{AKE_v1})
based on the decomposition for the terms proportional to $p^{\mu}$
and $m$, respectively, $\mathcal{C}_{s2}^{(n)\mu}$ can be equivalently
written as 
\begin{eqnarray}
\mathcal{C}_{s2}^{(n)\mu} & = & -\delta(p^{2}-m^{2})\left\{ p^{\mu}S_{m(n)}^{\rho\nu}\widehat{(\partial_{\rho}\Sigma_{V\nu})f_{V}}\right.\nonumber \\
 &  & \left.-m\left[S_{m(n)}^{\mu\nu}(\partial_{\nu}\widehat{\Sigma_{F})f_{V}}+\frac{\epsilon^{\mu\nu\rho\sigma}(p_{\rho}+mn_{\rho})}{2(p\cdot n+m)}(\partial_{\sigma}\widehat{\Sigma_{V\nu})f_{V}}\right]\right\} .\label{Cns2-1}
\end{eqnarray}
Although the collision term (\ref{Cns2}) and (\ref{Cns2-1}) are
mathematically equivalent, 
the collision term (\ref{Cns2})
is more useful in the large-mass regime while the term (\ref{Cns2-1})
is more suitable for light fermions.

In addition, when focusing on massive fermions with a mass much greater
than the gradient scale, we can set the frame vector at their rest
frame $n^{\mu}=n_{r}^{\mu}(p)=p^{\mu}/m$ to simplify both the Wigner
functions and AKE. Note that such a frame choice is also used in Refs.~\cite{Weickgenannt:2019dks,Gao:2019znl}.
In such a case, the magnetization-current term in $\mathcal{A}^{<\mu}$
vanishes and thus $\mathcal{A}^{<\mu}$ reduces to 
\begin{equation}
\mathcal{A}^{<\mu}=2\pi\left[\delta(p^{2}-m^{2})a^{\mu}f_{A}+\hbar\widetilde{F}^{\mu\nu}p_{\nu}\delta^{\prime}(p^{2}-m^{2})f_{V}\right],\label{axial_sol_nr}
\end{equation}
which is same as the form without collisions. From Eq.~(\ref{MA_6}),
we obtain the AKE in the form of Eq.~(\ref{AKE_collision_n}) with
$n^{\mu}$ replaced by $n_{r}^{\mu}(p)=p^{\mu}/m$ but with $\mathcal{C}_{2}^{(n_{r})\mu}$
defined as 
\begin{eqnarray}
\mathcal{C}_{2}^{(n_{r})\mu} & = & \frac{1}{2}\delta(p^{2}-m^{2})\epsilon^{\mu\nu\rho\sigma}p_{\nu}(\Delta_{\rho}\widehat{\Sigma_{V,\sigma})f_{V}}\nonumber \\
 &  & -\widetilde{F}^{\mu\nu}p_{\nu}\delta^{\prime}(p^{2}-m^{2})(p\cdot\widehat{\Sigma_{V}f_{V}}+m\widehat{\Sigma_{F}f_{V}}).\label{C_rQ}
\end{eqnarray}
Recall that the choice of a rest frame is valid only when the mass
is much larger than the gradient and electromagnetic scales in the
system. Here the quantum correction on collisions in Eq.~(\ref{C_rQ})
is also present in Eq.~(\ref{C_nQ_2}). More succinctly, Eqs.~(\ref{C_rQ})
and (\ref{Cns2}) share the same term $(1/2)\epsilon^{\mu\nu\rho\sigma}p_{\nu}(\partial_{\rho}\widehat{\Sigma_{V\sigma})f_{V}}$
when neglecting electric/magnetic fields though Eq.~(\ref{Cns2})
contains extra terms. We may regard these extra terms in Eq.~(\ref{C_nQ_2})
or Eq.~(\ref{Cns2}) as the $O(|{\bf p}|/m)$ corrections on top
of Eq.~(\ref{C_rQ}).

In summary, the leading-order SKE remains the same as a classical
Boltzmann equation, whereas the AKE governing the dynamics of spin
polarization. The AKE (\ref{AKE_collision_n}) delineates the evolution
of an axial-vector component $\mathcal{A}_{\mu}^{<}$ in Wigner functions
with collisional effects characterized by $\mathcal{C}_{1}^{(n)\mu}+\hbar\mathcal{C}_{2}^{(n)\mu}=\mathcal{C}_{\text{cl}}^{\mu}+\hbar\mathcal{C}_{\text{Q}}^{(n)\mu}$.
It is found $\mathcal{C}_{\text{cl}}^{\mu}$ and $\mathcal{C}_{\text{Q}}^{(n)\mu}$
are proportional to $f_{A}$ and $f_{V}$, respectively. Consequently,
$\mathcal{C}_{\text{cl}}^{\mu}$ serves as the term of spin relaxation and diffusion,
which vanishes when $f_{A}=0$. Such a fact will be explicitly shown
in the case for massive quarks traversing weakly coupled QGP in the
next subsection. On the contrary, the ``quantum'' correction $\mathcal{C}_{\text{Q}}^{(n)\mu}$,
which is led by the explicit $\hbar$ term such as $\mathcal{C}_{2}^{(n)\mu}$
and the implicit $\hbar$ corrections in $\mathcal{C}_{1}^{(n)\mu}$,
survives when $f_{A}=0$. It is dubbed as the spin-polarization term
and responsible for the polarization of particles via the intertwined
dynamics of vector-charge transport due to spin-orbit interaction.
Albeit not in QCD, a concrete example can be found in the application
of AKE to the Nambu-Jona-Lasinio model \cite{Wang:2020pej,Wang:2021qnt}, where
the equilibrium value of $\mathcal{A}^{\mu}$ as in Eq.~(\ref{equ_A_mu})
triggered by nonzero vorticity is acquired via the detailed balance
with a vanishing collision term. See also Refs.~\cite{Fang:2022ttm,Wang:2022yli} for follow-up studies in the gauge theory for massless and massive fermions. The so-called non-local collisions
also lead to equivalent results \cite{Weickgenannt:2020aaf,Weickgenannt:2021cuo}, which will be further elaborated in the next section.

\subsection{Spin relaxation and diffusion in weakly coupled QGP}\label{subsec:spin diffusion in QGP}

We now apply the formalism established in previous sections to investigate
the collision terms for massive quarks traversing weakly coupled QGP
in relativistic heavy-ion collisions. Nevertheless, to obtain explicit
expressions of $\hbar$ corrections in the collision terms for QGP
is rather challenging. Even though the implicit $\hbar$ corrections
encoded in self-energies from light quarks in equilibrium could be
derived from equilibrium Wigner functions shown in, e.g., Refs.~\cite{Hidaka:2017auj,Hattori:2019ahi},
how to include analogous corrections led by vorticity from polarized
gluons is currently unknown. For the future application on the dynamical
spin polarization of strange quarks traversing QGP, we will have to
work out the Wigner functions for polarized gluons up to $O(\hbar)$
with both classical and quantum contributions at least in equilibrium
as one of the essential ingredients. Recently, there have been related
studies for the Wigner functions and QKT of polarized photons in Refs.~\cite{Huang:2020kik,Hattori:2020gqh},
but the direct application to QCD is still not feasible. In this subsection,
we will only compute the classical contribution in the collision terms
responsible for spin relaxation and diffusion in weakly coupled QGP, which was first
derived in Ref.~\cite{Li:2019qkf} and reproduced in Ref.~\cite{Yang:2020hri}
with a more complete form.

To evaluate the classical contribution \textcolor{blue}{$\mathcal{C}^{\mu}_{\text{cl}}$} in $\mathcal{C}_{1}^{(n)\mu}$,
we have to first derive the fermionic self-energies therein. In the
following, we call fermions quarks and gauge bosons gluons interchangeably,
and include the color-group factors. Here, the color degrees of freedom
do not play a crucial role (like in the color conductivity), and the
same computation holds for QED with simple replacements of relevant
degrees of freedom. Furthermore, we consider the massive quarks with
their mass much greater than the scale of thermal mass in QGP and
accordingly neglect the Compton scattering with gluons as subleading
effects analogous to the study of heavy-quark transport in heavy-ion
collisions (see, e.g., Ref.~\cite{Svetitsky:1987gq}).

We shall consider a massive fermion, which may be regarded as a strange
quark, probing the weakly coupled QGP dominated by massless quarks
and gluons in thermal equilibrium. Having assumed the weakly coupled
system, we focus on the lowest-order contributions in the coupling
constant $g_{c}$, i.e., the 2-to-2 scatterings between a massive
quark and a massless quark/gluon. Since the computation is rather
complicated and lengthy, we will briefly mention the theoretical setup
with some critical steps of the derivation and highlight the primary
results. One may refer to Refs.~\cite{Li:2019qkf,Yang:2020hri} for
technical details.

The fermionic self-energies from gluon-exchange processes between
a massive fermion and the medium can be written as 
\begin{equation}
\Sigma^{\lessgtr}(x,p)=\lambda_{c}\int_{p^{\prime}}\gamma^{\mu}W^{\lessgtr}(x,p^{\prime})\gamma^{\nu}G_{\mu\nu}^{\gtrless}(x,p-p^{\prime}),\label{Sigma_in_Pi}
\end{equation}
where $\lambda_{c}$ denotes an overall coefficient including the
coupling and we dropped $O(\hbar^{2})$ and higher-order corrections
in our power counting. The gluon propagator $G_{\mu\nu}^{\lessgtr}$
in the Coulomb gauge and the fluid rest frame is given by 
\begin{equation}
G_{\mu\nu}^{\lessgtr}(x,q)=g_{\text{eq},q}^{\lessgtr}\Big[\rho_{L}(q)P_{\mu\nu}^{L}+\rho_{T}(q)P_{\mu\nu}^{T}\Big],\label{HTLA_Pi}
\end{equation}
where $q^{\mu}=p^{\mu}-p^{\prime\mu}$ denotes the four-momentum transfer
in scatterings, $g_{\text{eq},q}^{<}=g_{0q}=1/(e^{\beta q\cdot u}-1)$,
and $g_{\text{eq},q}^{>}=1+g_{0q}$. Here we introduce the longitudinal
and transverse projectors, 
\begin{equation}
P_{\mu\nu}^{L}\equiv u_{\mu}u_{\nu},\quad P_{\mu\nu}^{T}\equiv-\Theta_{\mu\alpha}\Theta_{\nu\beta}\left(\eta^{\alpha\beta}+\frac{q^{\alpha}q^{\beta}}{|{\bf q}|^{2}}\right)=-\left(\Theta_{\mu\nu}+\frac{q_{\perp\mu}q_{\perp\nu}}{|{\bf q}|^{2}}\right),
\end{equation}
where $\Theta^{\mu\nu}\equiv\eta^{\mu\nu}-u^{\mu}u^{\nu}$ with $u^{\mu}\approx(1,\bm{0})$.
We also adopt the hard-thermal-loop (HTL) approximation such that
$g_{c}T\ll|q^{\mu}|\ll T$, from which the HTL gluon spectral densities
take the form as (see, e.g., Ref.~\cite{Bellac:2011kqa}) 
\begin{equation}
\rho_{L}(q)\approx\frac{\pi m_{D}^{2}q_{0}}{|\bm{q}|^{5}},\quad\rho_{T}(q)\approx\frac{\pi m_{D}^{2}q_{0}}{2|\bm{q}|^{5}\left(1-q_{0}^{2}/|\mathbf{q}|^{2}\right)},
\end{equation}
where $m_{D}\sim g_{c}T$ corresponds to the Debye mass. The explicit
form of $m_{D}$ and $\lambda_{c}$ for gluons in $SU(N_{c})$ color
group and $N_{f}$-flavored massless quarks are given by 
\begin{equation}
m_{D}^{2}=\frac{g_{c}^{2}T^{2}(2N_{c}+N_{f})}{6},\quad\lambda_{c}=g_{c}^{2}C_{2}(F)=\frac{g_{c}^{2}(N_{c}^{2}-1)}{2N_{c}}.
\end{equation}

Given the explicit form of the gluon propagators $G_{\mu\nu}^{\lessgtr}(x,q)$
and fermionic Wigner functions $W^{\lessgtr}(x,p^{\prime})$, for
which only the classical contributions are included, one can derive
the fermionic self-energies $\Sigma^{\lessgtr}(x,p)$. Making the
decomposition of $\Sigma^{\lessgtr}(x,p)$ in light of Eq.~(\ref{Sigma_decomp})
and inserting all ingredients into Eq.~(\ref{SKE}) and $\mathcal{C}_{1}^{(n)\mu}$
in Eq.~(\ref{C_n1}) in the absence of background fields, one then
derives the classical collision terms for the SKE and AKE. Based on
the HTL approximation, in which one should conduct the small-$q$
expansion when evaluating the momentum integral in the collision kernel
with the infrared cutoff around $g_{c}T$, the leading-logarithmic
results in $g_{c}$ of collision terms are acquired \cite{Li:2019qkf,Yang:2020hri}.
The SKE takes the form,
\begin{eqnarray}
0 & = & \delta(p^{2}-m^{2})\left[p\cdot\partial-\kappa_{{\rm LL}}\left(2-2f_{V}+c_{(1)}\widetilde{p}_{\perp}^{\beta}\partial_{p_{\perp}^{\beta}}\right.\right.\nonumber \\
 &  & \left.\left.+c_{(2)}\widetilde{p}_{\perp}^{\alpha}\widetilde{p}_{\perp}^{\beta}\partial_{p_{\perp}^{\alpha}}\partial_{p_{\perp}^{\beta}}+c_{(3)}\eta^{\alpha\beta}\partial_{p_{\perp}^{\alpha}}\partial_{p_{\perp}^{\beta}}\right)\right]f_{V},\label{SKE_HTL_4_non}
\end{eqnarray}
where $\widetilde{p}_{\perp}^{\mu}\equiv p_{\perp}^{\mu}/|\mathbf{p}|$
as defined previously and we denote the coefficient of the leading
logarithmic result,
\begin{equation}
\kappa_{{\rm LL}}\equiv\frac{g_{c}^{2}C_{2}(F)m_{D}^{2}}{8\pi}\ln\frac{1}{g_{c}},
\end{equation}
We also introduce the quark four ``velocity'' $v^{\mu}=(v^{0},{\bf v}^{i})\equiv p^{\mu}/m$,
which has the normalization $v^{\mu}v_{\mu}=1$ under the on-shell
condition, and then the rapidity,
\begin{equation}
\eta_{p}\equiv\frac{1}{2}\ln\frac{E_{p}+|\mathbf{p}|}{E_{p}-|\mathbf{p}|}=\frac{1}{2}\ln\frac{v_{0}+|\mathbf{v}|}{v_{0}-|\mathbf{v}|}.
\end{equation}
The coefficients in the collision term of Eq.~(\ref{SKE_HTL_4_non}) are
given by 
\begin{eqnarray}
c_{(1)} & = & \frac{mv_{0}^{2}\theta_{-1}}{|{\bf v}|^{2}}(1-2f_{V}),\nonumber \\
c_{(2)} & = & \frac{mTv_{0}^{2}}{2|{\bf v}|^{3}}\left(\frac{|{\bf v}|^{2}\eta_{q}}{v_{0}^{2}}+\frac{3\theta_{1}}{v_{0}}\right),\nonumber \\
c_{(3)} & = & \frac{mT}{2}\left(\frac{v_{0}^{3}\theta_{-3}}{|{\bf v}|^{3}}-3v_{0}\right),
\end{eqnarray}
where $\theta_{n}\equiv|\mathbf{v}|-v_{0}^{n}\eta_{p}$. On the other
hand, the AKE reads 
\begin{eqnarray}
0 & = & \delta(p^{2}-m^{2})\Bigg[p\cdot\partial\widetilde{a}^{\mu}-\frac{\kappa_{{\rm LL}}T}{E_{p}}\left(\widetilde{a}^{\mu}\grave{\mathcal{Q}}_{\text{cl}}^{(1)}+u^{\mu}\grave{\mathcal{Q}}_{\text{cl}}^{(2)}+\widetilde{p}_{\perp}^{\mu}\grave{\mathcal{Q}}_{\text{cl}}^{(3)}+\grave{\mathcal{Q}}_{\text{cl}}^{(4)}\widetilde{p}_{\perp}^{\nu}\partial_{p_{\perp\mu}}\widetilde{a}_{\nu}\right.\nonumber \\
 &  & \left.+\grave{\mathcal{Q}}_{\text{cl}}^{(5)}\widetilde{p}^{\nu}\partial_{p_{\perp}^{\nu}}\widetilde{a}^{\mu}+\grave{\mathcal{Q}}_{\text{cl}}^{(6)}\eta^{\nu\rho}\partial_{p_{\perp}^{\nu}}\partial_{p_{\perp}^{\rho}}\widetilde{a}^{\mu}+\grave{\mathcal{Q}}_{\text{cl}}^{(7)}\widetilde{p}_{\perp}^{\nu}\widetilde{p}_{\perp}^{\rho}\partial_{p_{\perp}^{\nu}}\partial_{p_{\perp}^{\rho}}\widetilde{a}^{\mu}\right)\Bigg],\label{AKE_classical_QGP}
\end{eqnarray}
where the coefficients are given by 

\begin{eqnarray}
\grave{\mathcal{Q}}_{\text{cl}}^{(1)} & = &\frac{2m}{T}\left[v_{0}(1-2f_{V})-\frac{T}{m}-\frac{v_{0}^{3}}{|{\bf v}|^{2}}m\theta_{-1}\widetilde{p}_{\perp}^{\rho}\partial_{p_{\perp}^{\rho}}f_{V}\right],\label{grave_Qcl_1} \\
\grave{\mathcal{Q}}_{\text{cl}}^{(2)} & = & m\frac{v_{0}}{|{\bf v}|^{3}}\left[(\theta_{1}-|{\bf v}|^{3})\partial_{p_{\perp}^{\nu}}\widetilde{a}^{\nu}+(3\theta_{1}+|{\bf v}|^{3})\widetilde{p}_{\perp}^{\nu}\widetilde{p}_{\perp}^{\rho}\partial_{p^{\rho}}\widetilde{a}_{\nu}\right]\nonumber \\
 &  & +\frac{m}{v_{0}|{\bf v}|^{2}T}\left[v_{0}^{3}(1-2f_{V})\theta_{-1}-\frac{2T}{m}(\theta_{-1}+2|{\bf v}|^{3})\right]\widetilde{p}_{\perp}\cdot\widetilde{a}, \\
 \grave{\mathcal{Q}}_{\text{cl}}^{(3)} & = & \frac{m}{|{\bf v}|^{2}}\left[v_{0}^{2}\theta_{-1}\partial_{p_{\perp}^{\nu}}\widetilde{a}^{\nu}+(3\theta_{1}+|{\bf v}|^{3})\widetilde{p}_{\perp}^{\nu}\widetilde{p}_{\perp}^{\rho}\partial_{p^{\rho}}\widetilde{a}_{\nu}\right]\nonumber \\
 &  & +\frac{m}{|{\bf v}|T}\theta_{-1}\left[v_{0}(1-2f_{Vq})-\frac{2T}{m}\right]\widetilde{p}_{\perp}\cdot\widetilde{a},\label{grave_Qcl_3} \\
  \grave{\mathcal{Q}}_{\text{cl}}^{(4)} &=&-2m|{\bf v}|,\;\grave{\mathcal{Q}}_{\text{cl}}^{(5)}=\frac{m^{2}v_{0}^{3}}{|{\bf v}|^{2}T}\theta_{-1}(1-2f_{V}),\\
 \grave{\mathcal{Q}}_{\text{cl}}^{(6)}&=&-\frac{m^{2}v_{0}^{2}}{2|{\bf v}|^{3}}\left(3|{\bf v}|^{3}-v_{0}^{2}\theta_{-3}\right),\;\grave{\mathcal{Q}}_{\text{cl}}^{(7)}=\frac{m^{2}v_{0}}{2|{\bf v}|^{3}}\left(3v_{0}\theta_{1}+\eta_{q}|{\bf v}|^{2}\right).
\end{eqnarray}
It turns out that all components of the classical collision term of
AKE in Eq.~(\ref{AKE_classical_QGP}) depend on $\widetilde{a}^{\mu}$,
albeit the appearance of $f_{V}$ in some of them, and hence only
yield the spin relaxation and diffusion instead of dynamical polarization induced
by quantum corrections involving spacetime gradients upon $f_{V}$.
From Eq.~(\ref{SKE_HTL_4_non}), one can show the collision term
vanishes when $f_{V}$ takes the Fermi-Dirac distribution. Nevertheless,
the vanishing collision term in Eq.~(\ref{AKE_classical_QGP}) requires
$\widetilde{a}_{p}^{\mu}=0$. Consequently, the nonzero initial spin
polarization could be relaxed and diffused by the classical scattering in a purely
homogeneous medium. Notably, in the non-relativistic limit ($m\gg T,|{\bf p}|$),
the SKE and AKE reduce to 
\begin{eqnarray}
0 & = & \partial_{0}f_{V}+\frac{2}{3}\kappa_{{\rm LL}}T\eta^{\nu\rho}\partial_{p_{\perp}^{\nu}}\partial_{p_{\perp}^{\rho}}f_{V},\\
0 & = & \partial_{0}\widetilde{a}^{\mu}+\frac{2}{3}\kappa_{{\rm LL}}T\eta^{\nu\rho}\partial_{p_{\perp}^{\nu}}\partial_{p_{\perp}^{\rho}}\widetilde{a}^{\mu},
\end{eqnarray}
which share the same diffusion parameter. As anticipated, both the
vector and axial charges for heavy quarks should diffuse to zero at
late times. Also, the spin of heavy quarks serves as a conserved quantum
number with its orientation unchanged. Further study for the spin relaxation of heavy quarks can also be found from different methods \cite{Hongo:2022izs}.

\newpage
\section{Spin Boltzmann equations and nonlocal collisions in spin-dependent distributions}

\label{sec:spin-Boltzmann}An alternative way of describing spin transport
processes is through the Boltzmann equation in terms of spin-dependent
distribution functions, dubbed the spin Boltzmann equation. In
Ref.~\cite{Sheng:2021kfc}, Boltzmann equations for massive spin-1/2
fermions are derived with local and nonlocal collision terms. The
Boltzmann equation is expressed in terms of matrix-valued spin distribution
functions (MVSD), the building blocks for the quasi-classical part
of the Wigner function. An expansion in powers of $\hbar$ is made
for all quantities in spin Boltzmann equations. Nonlocal collision
terms appear at the next-to-leading order in $\hbar$ and are sources
for the polarization part of MVSD. Equivalently, the MVSD can be converted 
to the scalar spin-dependent distribution (SSD) through a continuous spin variable. 
The nonlocality of the collision manifests itself in a space shift in the SSD \cite{Weickgenannt:2020aaf,Weickgenannt:2021cuo}.  
The spin Boltzmann equation provides an effective way 
to simulate spin transport processes involving spin-vorticity couplings from the first principles.

\subsection{Spin Boltzmann equations}

The starting point for the derivation of the spin Boltzmann equation
is the Kadanoff-Baym equation (\ref{eq:KBEq_massive}) without background
electromagnetic fields, which can be re-written as 
\begin{equation}
\left(\gamma_{\mu}K^{\mu}-m\right)W^{<}(x,p)=I_{\mathrm{coll}},\label{eq:eom-Wigner-function}
\end{equation}
where the operator $K^{\mu}$ is defined as $K^{\mu}\equiv p^{\mu}+(i\hbar/2)\partial_{x}^{\mu}$,
and $I_{\mathrm{coll}}$ is given by the right-hand side of Eq.~(\ref{eq:KBEq_massive})
that has two parts: one without the Poisson bracket and one with it,
as shown in Eq.~(\ref{eq:Moyal_product}). Acting the operator $\gamma_{\nu}K^{\nu}+m$
on the Kadanoff-Baym equation (\ref{eq:eom-Wigner-function}), we can
obtain a Klein-Gordon-type equation,
\begin{equation}
\left(K^{2}-m^{2}\right)W^{<}(x,p)=\left(\gamma_{\nu}K^{\nu}+m\right)I_{\mathrm{coll}},\label{eq:kg-wigner}
\end{equation}
where we have used the property $\left[K_{\mu},K_{\nu}\right]=0$.
Taking the Hermitian conjugate of Eq.~(\ref{eq:eom-Wigner-function}),
we obtain a conjugate equation,
\begin{equation}
\left(K^{2*}-m^{2}\right)W^{<}(x,p)=\gamma^{0}\left[\left(\gamma_{\nu}K^{\nu}+m\right)I_{\mathrm{coll}}\right]^{\dagger}\gamma^{0},\label{eq:kg-wigner-conj}
\end{equation}
where we have used the Hermitian property of the Wigner function $[W^{<}(x,p)]^{\dagger}=\gamma^{0}W^{<}(x,p)\gamma^{0}$.
By taking the sum and difference of Eqs.~(\ref{eq:kg-wigner}) and
(\ref{eq:kg-wigner-conj}) we obtain the modified on-shell condition
and Boltzmann equation for the Wigner function,
\begin{eqnarray}
\left(p^{2}+\frac{\hbar^{2}}{4}\partial_{x}^{2}-m^{2}\right)W^{<}(x,p) & = & \frac{1}{2}(\gamma^{\mu}K_{\mu}+m)I_{\mathrm{coll}}+\frac{1}{2}\gamma^{0}\left[(\gamma^{\mu}K_{\mu}+m)I_{\mathrm{coll}}\right]^{\dagger}\gamma^{0},\nonumber \\
\hbar p\cdot\partial_{x}W^{<}(x,p) & = & -\frac{i}{2}(\gamma^{\mu}K_{\mu}+m)I_{\mathrm{coll}}+\frac{i}{2}\gamma^{0}\left[(\gamma^{\mu}K_{\mu}+m)I_{\mathrm{coll}}\right]^{\dagger}\gamma^{0}.
\label{eq:mass-sh-boltzmann}
\end{eqnarray}
Then we obtain the mass-shell equations and Boltzmann equations for
Clifford components 
\begin{eqnarray}
\left(p^{2}+\frac{\hbar^{2}}{4}\partial_{x}^{2}-m^{2}\right)\mathrm{Tr}\left(\Gamma_{a}W^{<}\right) & = & \mathrm{Re}\mathrm{Tr}\left[\Gamma_{a}(\gamma\cdot K+m)I_{\mathrm{coll}}\right],\nonumber \\
\hbar p\cdot\partial_{x}\mathrm{Tr}\left(\Gamma_{a}W^{<}\right) & = & \mathrm{Im}\mathrm{Tr}\left[\Gamma_{a}(\gamma\cdot K+m)I_{\mathrm{coll}}\right],\label{eq:on-shell-cond-boltzmann}
\end{eqnarray}
where we have used $\Gamma_{a}^{\dagger}=\gamma_{0}\Gamma_{a}\gamma_{0}$
with $\Gamma_{a}=\{1,\gamma^{\mu},i\gamma^{5},\gamma^{\mu}\gamma^{5},\sigma^{\mu\nu}\}$.

We make $\hbar$ expansion for all quantities in the Kadanoff-Baym
equation (\ref{eq:eom-Wigner-function}). At leading order, Eq.~(\ref{eq:eom-Wigner-function})
is just the kinetic equation without collisions since $I_{\mathrm{coll}}$
is at least of the next-to-leading order, so the Wigner function has
the free-streaming form 
\begin{eqnarray}
W_{\alpha\beta}^{<,(0)}(x,p) & = & 2\pi\hbar\,\theta(p_{0})\delta\left(p^{2}-m^{2}\right)\sum_{r,s}u_{r,\alpha}(p)\overline{u}_{s,\beta}(p)\,f_{sr}^{(+,0)}\left(x,p\right)\nonumber \\
 &  & +2\pi\hbar\,\theta(-p_{0})\delta\left(p^{2}-m^{2}\right)\sum_{r,s}v_{s,\alpha}(\overline{p})\overline{v}_{r,\beta}(\overline{p})\left[\delta_{sr}-f_{sr}^{(-,0)}\left(x,\overline{p}\right)\right]\;,
 \label{eq:g-less-0}
\end{eqnarray}
where we have used MVSD for fermions and antifermions in spin space
defined as 
\begin{eqnarray}
f_{sr}^{(+,0)}\left(x,p\right) & \equiv & \int\frac{d^{4}q}{2(2\pi\hbar)^{4}}\,2\pi\hbar\,\delta(p\cdot q)\,e^{-iq\cdot x/\hbar}\left\langle a_{s,\mathbf{p}-\mathbf{q}/2}^{\dagger}a_{r,\mathbf{p}+\mathbf{q}/2}\right\rangle \;,\label{eq:f_rs-distr_+}\\
f_{sr}^{(-,0)}\left(x,\overline{p}\right) & \equiv & \int\frac{d^{4}q}{2(2\pi\hbar)^{4}}\,2\pi\hbar\,\delta(\overline{p}\cdot q)\,e^{-iq\cdot x/\hbar}\left\langle b_{s,-\mathbf{p}-\mathbf{q}/2}^{\dagger}b_{r,-\mathbf{p}+\mathbf{q}/2}\right\rangle \;,\label{eq:f_rs-distr_-}
\end{eqnarray}
with $p\equiv(E_{\mathbf{p}},\mathbf{p})$ and $\overline{p}^{\mu}\equiv(E_{\mathbf{p}},-\mathbf{p})$.
In Eqs.~(\ref{eq:g-less-0})-(\ref{eq:f_rs-distr_-}), the index '$(n)$'
with $n=0,1,2,\cdots$ denotes the order of the expansion, so '(0)'
means the leading order and '(1)' means the next-to-leading order,
etc.. It can be verified that the components of $W^{<,(0)}$ satisfy
\begin{eqnarray}
\mathcal{P}^{(0)} & = & 0,\nonumber \\
\mathcal{V}_{\mu}^{(0)} & = & \frac{1}{m}p_{\mu}\mathcal{F}^{(0)},\nonumber \\
\mathcal{S}_{\mu\nu}^{(0)} & = & -\frac{1}{m}\epsilon_{\mu\nu\alpha\beta}p^{\alpha}\mathcal{A}^{(0)\beta}.\label{eq:wig-component-0}
\end{eqnarray}
Here we have chosen the scalar component $\mathcal{F}$ and the axial
vector component $\mathcal{A}^{\beta}$ as independent ones. As shown
in Eq.~(\ref{eq:g-less-0}), the leading order Wigner function is on-shell,
so are all of its Clifford components. Note that $W^{<}$ differs
by a sign from the corresponding equation in Ref.~\cite{Sheng:2021kfc}
due to the definition in Eq.~(\ref{eq:def-green-func}).

The next-to-leading order quantities $W^{\gtrless(1)}$ must satisfy
Eq.~(\ref{eq:eom-Wigner-function}) with non-vanishing $I_{\mathrm{coll}}$,
which contain additional terms other than those in the leading order
case. The components of $W^{<,(1)}$ satisfy 
\begin{eqnarray}
\mathcal{P}^{(1)} & = & -\frac{1}{2m}\partial_{x}^{\mu}\mathcal{A}_{\mu}^{(0)}+\frac{1}{m}\mathrm{Re\,Tr}\left(i\gamma^{5}I_{\mathrm{coll}}^{(1)}\right)\;,\label{eq:sol-p1}\\
\mathcal{V}_{\mu}^{(1)} & = & \frac{1}{m}p_{\mu}\mathcal{F}^{(1)}-\frac{1}{2m}\partial_{x}^{\nu}\mathcal{S}_{\nu\mu}^{(0)}-\frac{1}{m}\mathrm{Re\,Tr}\left(\gamma_{\mu}I_{\mathrm{coll}}^{(1)}\right)\;,\label{eq:sol-v1}\\
\mathcal{S}_{\mu\nu}^{(1)} & = & -\frac{1}{m}\epsilon_{\mu\nu\alpha\beta}p^{\alpha}\mathcal{A}^{(1)\beta}+\frac{1}{2m}\,\partial_{x[\mu}\mathcal{V}_{\nu]}^{(0)}-\frac{1}{m}\mathrm{Re\,Tr}\left(\sigma_{\mu\nu}I_{\mathrm{coll}}^{(1)}\right)\;,\label{eq:sol-a1}
\end{eqnarray}
where $I_{\mathrm{coll}}^{(1)}$ is the collision term at the leading
order without the Poisson bracket and it is on-shell. So it is helpful
to separate the full Wigner function at $O(\hbar)$ into different
parts as follows,
\begin{equation}
W^{(1)}=W_{\mathrm{qc}}^{(1)}+W_{\nabla}^{(1)}+W_{\mathrm{off}}^{(1)},\label{eq:wig-kin-nonkin-off}
\end{equation}
where $W^{(1)}\equiv W^{\lessgtr(1)}$, and $W_{\mathrm{qc}}^{(1)}$,
$W_{\nabla}^{(1)}$ and $W_{\mathrm{off}}^{(1)}$ denote the quasi-classical,
gradient and collision part and off-shell part of the Wigner function,
respectively. Note that $W_{\mathrm{qc}}^{(1)}$ and $W_{\nabla}^{(1)}$
are both on-shell. The quasi-classical part $W_{\mathrm{qc}}^{<(1)}$
is the one that its components satisfy 
\begin{eqnarray}
\mathcal{P}_{\mathrm{qc}}^{(1)} & = & 0,\nonumber \\
\mathcal{V}_{\mathrm{qc},\mu}^{(1)} & = & \frac{1}{m}p_{\mu}\mathcal{F}_{\mathrm{qc}}^{(1)},\nonumber \\
\mathcal{S}_{\mathrm{qc},\mu\nu}^{(1)} & = & -\frac{1}{m}\epsilon_{\mu\nu\alpha\beta}p^{\alpha}\mathcal{A}_{\mathrm{qc}}^{(1)\beta},\label{eq:1st-kin-comp}
\end{eqnarray}
which has the same form as Eq.~(\ref{eq:wig-component-0}) at the
leading order. The gradient and collision part $W_{\nabla}^{<(1)}$
is defined as 
\begin{eqnarray}
\mathcal{P}_{\nabla}^{(1)} & = & -\frac{1}{2m}\partial_{x}^{\mu}\mathcal{A}_{\mu}^{(0)}+\frac{1}{m}\mathrm{Re\,Tr}\left(i\gamma^{5}I_{\mathrm{coll}}^{(1)}\right),\nonumber \\
\mathcal{V}_{\nabla,\mu}^{(1)} & = & -\frac{1}{2m}\partial_{x}^{\nu}\mathcal{S}_{\nu\mu}^{(0)}-\frac{1}{m}\mathrm{Re\,Tr}\left(\gamma_{\mu}I_{\mathrm{coll}}^{(1)}\right),\nonumber \\
\mathcal{S}_{\nabla,\mu\nu}^{(1)} & = & \frac{1}{2m}\partial_{x[\mu}\mathcal{V}_{\nu]}^{(0)}-\frac{1}{m}\mathrm{Re\,Tr}\left(\sigma_{\mu\nu}I_{\mathrm{coll}}^{(1)}\right).\label{eq:1st-nonkin-comp}
\end{eqnarray}
The off-shell part comes from the interaction in vacuum and has nothing
to do with collisions in medium, so it has been neglected in Ref.~\cite{Sheng:2021kfc}.
With Eqs.~(\ref{eq:1st-kin-comp}) and (\ref{eq:1st-nonkin-comp}), we
see that the components of the Wigner function at the next-to-leading
order do satisfy Eqs.~(\ref{eq:sol-p1})-(\ref{eq:sol-a1}).

We further assume that the quasi-classical contribution to the Wigner
function at the next-to-leading order in $\hbar$ arises from $O(\hbar)$ corrections
of MVSD to Eq.~(\ref{eq:g-less-0}), 
\begin{eqnarray}
W_{\mathrm{qc},\alpha\beta}^{<,(1)}(x,p) & = & 2\pi\hbar\,\theta(p_{0})\delta\left(p^{2}-m^{2}\right)\sum_{r,s}u_{r,\alpha}(p)\overline{u}_{s,\beta}(p)f_{sr}^{(+,1)}\left(x,p\right)\nonumber \\
 &  & -2\pi\hbar\,\theta(-p_{0})\delta\left(p^{2}-m^{2}\right)\sum_{r,s}v_{s,\alpha}(\overline{p})\overline{v}_{r,\beta}(\overline{p})f_{sr}^{(-,1)}\left(x,\overline{p}\right),\label{eq:g-less-1}
\end{eqnarray}
Note that $f_{sr}^{(\pm,0)}$ and $f_{sr}^{(\pm,1)}$ are to be determined
by solving the Boltzmann equation at the leading and next-to-leading
order respectively. Therefore the quasi-classical part of the Wigner
function up to the next-to-leading order can be written as 
\begin{eqnarray}
W_{\mathrm{qc},\alpha\beta}^{<}(x,p) & = & W_{\alpha\beta}^{<,(0)}(x,p)+\hbar W_{\mathrm{qc},\alpha\beta}^{<,(1)}(x,p)+O(\hbar^{3})\nonumber \\
 & = & 2\pi\hbar\,\theta(p_{0})\delta\left(p^{2}-m^{2}\right)\sum_{r,s}u_{r,\alpha}(p)\overline{u}_{s,\beta}(p)f_{sr}^{(+)}\left(x,p\right)\nonumber \\
 &  & +2\pi\hbar\,\theta(-p_{0})\delta\left(p^{2}-m^{2}\right)\sum_{r,s}v_{s,\alpha}(\overline{p})\overline{v}_{r,\beta}(\overline{p})\left[\delta_{sr}-f_{sr}^{(-)}\left(x,\overline{p}\right)\right]+O(\hbar^{3}),\nonumber\\ \label{eq:g-less-all}
\end{eqnarray}
where $f_{sr}^{(+)}\left(x,p\right)$ is given by 
\begin{equation}
f_{sr}^{(+)}\left(x,p\right)=f_{sr}^{(+,0)}\left(x,p\right)+\hbar f_{sr}^{(+,1)}\left(x,p\right)+O(\hbar^{2}).\label{eq:mvsd-expand}
\end{equation}
Note that in Eq.~(\ref{eq:g-less-all}) we did not put the index
'qc' to $W_{\alpha\beta}^{<,(0)}$ since it is quasi-classical by
definition. Another Wigner function $W^{>,(0)}(x,p)$ at the leading
order can be obtained from $W^{<,(0)}(x,p)$ by replacements $f_{sr}^{(+,0)}\rightarrow(\delta_{sr}-f_{sr}^{(+,0)})$
and $\delta_{sr}-f_{sr}^{(-,0)}\rightarrow f_{sr}^{(-,0)}$ in Eq.
(\ref{eq:g-less-0}). But at the next-to-leading order we have $W_{\alpha\beta}^{>,(1)}(x,p)=W_{\alpha\beta}^{<,(1)}(x,p)$
all given by Eq.~(\ref{eq:g-less-1}).

We can express the components of the Wigner function in terms of spin
dependent distribution functions. Since we have chosen the scalar
and axial-vector components as independent ones, with $\mathcal{F}(x,p)=\mathrm{Tr}[W^{<}(x,p)]$
and Eqs.\ (\ref{eq:g-less-0}) and (\ref{eq:g-less-1}) we obtain the quasi-classical
and axial-vector components as independent ones, with $\mathcal{F}(x,p)=\mathrm{Tr}[W^{<}(x,p)]$,
\begin{eqnarray}
\mathcal{F}(x,p) & = & 2\pi\hbar\frac{m}{E_{p}}\left\{ \delta(p_{0}-E_{p})\,\mathrm{Tr}\left[f^{(+)}(x,p)\right]\right.\nonumber \\
 &  & \left.+\delta(p_{0}+E_{p})\,\mathrm{Tr}\left[f^{(-)}(x,\overline{p})-1\right]\right\} +O(\hbar^{3}),\label{eq:F_0}
\end{eqnarray}
where 'Tr' denotes the trace in the two-dimensional space of spin
indices and $f^{(+)}$ denotes the matrix form of MVSD in spin state
space. On the other hand, from $\mathcal{A}^{\mu}(x,p)=\mathrm{Tr}[\gamma^{\mu}\gamma^{5}G^{<}(x,p)]$
and Eqs.\ (\ref{eq:g-less-0}) and (\ref{eq:g-less-1}) we obtain the quasi-classical
part of the axial-vector component up to the next-to-leading order,
\begin{eqnarray}
\mathcal{A}^{\mu} & = & 2\pi\hbar\frac{m}{E_{p}}\left\{ \delta(p_{0}-E_{p})n_{j}^{(+)\mu}\mathrm{Tr}\left[\tau_{j}^{T}\,f^{(+)}(x,p)\right]\right.\nonumber \\
 &  & \left.+\delta(p_{0}+E_{p})\,n_{j}^{(-)\mu}\mathrm{Tr}\left[\tau_{j}^{T}f^{(-)}(x,\overline{p})\right]\right\} +O(\hbar^{3}),\label{eq:A_0}
\end{eqnarray}
where $n_{j}^{(\pm)\mu}$ with $j=1,2,3$ are polarization four-vectors
defined by 
\begin{equation}
n_{j}^{(\pm)\mu}\equiv n^{\mu}(\pm\mathbf{p},\mathbf{n}_{j})=\left(\pm\frac{\mathbf{n}_{j}\cdot\mathbf{p}}{m},\mathbf{n}_{j}+\frac{(\mathbf{n}_{j}\cdot\mathbf{p})\mathbf{p}}{m(E_{p}+m)}\right)^{T}.\label{eq:spin-direction-lab}
\end{equation}
Here $\mathbf{n}_{j}$ ($j=1,2,3$) are a set of orthonormal unit
vectors which form a right-handed basis in three spatial dimensions
with $\mathbf{n}_{3}$ being the spin quantization direction.

The Boltzmann equations for the scalar and axial-vector components
are given by Eq.~(\ref{eq:on-shell-cond-boltzmann}), from which
we can derive the Boltzmann equations for MVSD. At the leading order
the Boltzmann equation can be put into a compact form,
\begin{eqnarray}
\frac{1}{E_{p}}p\cdot\partial_{x}\mathrm{tr}\left[f^{(0)}(x,p)\right] & = & \mathscr{C}_{\mathrm{scalar}}\left[f^{(0)}\right],\nonumber \\
\frac{1}{E_{p}}p\cdot\partial_{x}\mathrm{tr}\left[n_{j}^{(+)\mu}\tau_{j}^{T}f^{(0)}(x,p)\right] & = & \mathscr{C}_{\mathrm{pol}}^{\mu}\left[f^{(0)}\right],\label{eq:main-1}
\end{eqnarray}
where $\mathscr{C}_{\mathrm{scalar}}\left[f^{(0)}\right]$ and $\mathscr{C}_{\mathrm{pol}}^{\mu}\left[f^{(0)}\right]$
are local collision terms which do not contain space-time derivatives
of distributions, i.e., collisions take place at the same space-time
point. The first equation describes how the scalar part of $f^{(0)}(x,p)$
evolves, while the second one describes how the polarization part
of $f^{(0)}(x,p)$ evolves. For the explicit form of Eq.\ (\ref{eq:main-1})
with an explanation of all variables, see Sec.~VI of Ref.~\cite{Sheng:2021kfc}.
At next-to-leading order, the Boltzmann equations describe how $f^{(1)}(x,p)$
evolves for given $f^{(0)}(x,p)$,
\begin{eqnarray}
\frac{1}{E_{p}}p\cdot\partial_{x}\mathrm{tr}\left[f^{(1)}(x,p)\right] & = & \mathscr{C}_{\mathrm{scalar}}\left[f^{(0)},\partial_{x}f^{(0)},f^{(1)}\right],\nonumber \\
\frac{1}{E_{p}}p\cdot\partial_{x}\mathrm{tr}\left[n_{j}^{(+)\mu}\tau_{j}^{T}f^{(1)}(x,p)\right] & = & \mathscr{C}_{\mathrm{pol}}^{\mu}\left[f^{(0)},\partial_{x}f^{(0)},f^{(1)}\right],\label{eq:main-2}
\end{eqnarray}
which contain space-time derivatives of $f^{(0)}(x,p)$ or nonlocal
collision terms, while $f^{(0)}(x,p)$ is determined by solving the
leading-order Boltzmann equations. The explicit form of Eq.~(\ref{eq:main-2})
as well as the physical meaning of all variables can be found in Sec.~VII 
of Ref.~\cite{Sheng:2021kfc}. The spin Boltzmann equations in
Eqs.~(\ref{eq:main-1}) and (\ref{eq:main-2}) can be solved numerically
and provide an effective tool to study spin transport phenomena.

\subsection{Scalar spin-dependent distributions}

In the previous subsection we have defined and used MVSD. In this
subsection we introduce an equivalent way of building the scalar spin
dependent distribution (SSD) through a continuous spin variable \cite{Weickgenannt:2020aaf,Weickgenannt:2021cuo}.
The ordinary phase space consisting of coordinate and momentum variables
can be extended by a continuous spin four-vector $\mathfrak{s}^{\mu}$.
Since this spin vector is not an observable, one has to integrate
it out in order to obtain physical quantities. This integration respects
the constraints $\mathfrak{s}\cdot p=0$ and $\mathfrak{s}^{2}=-3/c_{0}$,
with $c_{0}$ being a normalization constant.

Introducing the on-shell momentum four-vector $p^{\mu}=(E_{p},\mathbf{p})$
for fermions, the integration measure is defined as follows, 
\begin{equation}
[d\mathfrak{s}]\equiv\frac{1}{C}d^{4}\mathfrak{s}\delta(\mathfrak{s}\cdot p)\delta(\mathfrak{s}^{2}+3/c_{0}),
\end{equation}
with the normalization constant $C$ given by 
\begin{equation}
C=\int d^{4}\mathfrak{s}\delta(\mathfrak{s}\cdot p)\delta(\mathfrak{s}^{2}+3/c_{0})=\frac{2\pi}{m}\sqrt{\frac{3}{c_{0}}}.
\end{equation}
For antifermions, we define the on-shell momentum four-vector $\overline{p}^{\mu}=(E_{p},-\mathbf{p})$,
and the same relations hold with the replacement $p^{\mu}\rightarrow\overline{p}^{\mu}$.
We have the following properties for the integration over $\mathfrak{s}^{\mu}$,
\begin{align}
\int[d\mathfrak{s}] & =1,\label{eq:sint_1}\\
\int[d\mathfrak{s}]\mathfrak{s}^{\mu} & =0,\label{eq:sint_0}\\
\int[d\mathfrak{s}]\mathfrak{s}^{\mu}\mathfrak{s}^{\nu} & =-\frac{1}{c_{0}}\Theta^{(+)\mu\nu}\;\;\mathrm{for\:fermions},\label{eq:s-mu-s-nu_p}\\
\int[d\mathfrak{s}]\mathfrak{s}^{\mu}\mathfrak{s}^{\nu} & =-\frac{1}{c_{0}}\Theta^{(-)\mu\nu}\;\;\mathrm{for\:antifermions}.\label{eq:s-mu-s-nu_ap}
\end{align}
In the last two lines, we introduced the projection operators,
\begin{eqnarray}
\Theta_{\mu\nu}^{(+)} & \equiv & \eta_{\mu\nu}-\frac{p_{\mu}p_{\nu}}{m^{2}}=-n_{j,\mu}^{(+)}n_{j,\nu}^{(+)}\;,\label{eq:Delta+}\\
\Theta_{\mu\nu}^{(-)} & \equiv & \eta_{\mu\nu}-\frac{\overline{p}_{\mu}\overline{p}_{\nu}}{m^{2}}=-n_{j,\mu}^{(-)}n_{j,\nu}^{(-)}\;,\label{eq:Delta-}
\end{eqnarray}
where the last equality in each line is proven with the help of the
explicit expressions unit vectors $\mathbf{n}_{j}$ with $j=1,2,3$.

The scalar spin dependent distribution in the extended phase space
then depends on $x,p$ (the energy is on-shell for both fermions and
antifermions, $p_{0}=E_{p}$), and $\mathfrak{s}^{\mu}$, and can
be defined by MVSD introduced in the last subsection, 
\begin{eqnarray}
f_{+}\left(x,p,\mathfrak{s}\right) & = & \frac{1}{2}\mathrm{Tr}\left\{ \left(1-\mathfrak{s}\cdot n_{j}^{(+)}\tau_{j}^{T}\right)f^{(+)}\left(x,p\right)\right\} ,\label{eq:spin-depend-f_plus}\\
f_{-}\left(x,\overline{p},\mathfrak{s}\right) & = & \frac{1}{2}\mathrm{Tr}\left\{ \left(1-\mathfrak{s}\cdot n_{j}^{(-)}\tau_{j}^{T}\right)f^{(-)}\left(x,\overline{p}\right)\right\} ,\label{eq:spin-depend-f_minus}
\end{eqnarray}
where $f^{(\pm)}(x,p)$ denote the matrix form of MVSD in Eq.~(\ref{eq:mvsd-expand}),
and $n_{j}^{(\pm)\mu}$ are defined in Eq.\ (\ref{eq:spin-direction-lab}).

We now apply the well-known identities,
\begin{equation}
\sum_{r,s}u_{s}(p)\overline{u}_{r}(p)\delta_{rs}=p\cdot\gamma+m\;,\;\;\;\;\sum_{r,s}v_{r}(\overline{p})\overline{v}_{s}(\overline{p})\delta_{sr}=\overline{p}\cdot\gamma-m\;,
\end{equation}
as well as the identities,
\begin{eqnarray}
\sum_{r,s}u_{s}(p)\overline{u}_{r}(p)(\tau_{j}^{T})_{rs} & = & n_{j,\mu}^{(+)}\left(m\gamma^{5}\gamma^{\mu}+\frac{1}{2}\epsilon^{\mu\nu\alpha\beta}p_{\nu}\sigma_{\alpha\beta}\right),\nonumber \\
\sum_{r,s}v_{r}(\overline{p})\overline{v}_{s}(\overline{p})(\tau_{j})_{sr} & = & -n_{j,\mu}^{(-)}\left(m\gamma^{5}\gamma^{\mu}-\frac{1}{2}\epsilon^{\mu\nu\alpha\beta}\overline{p}_{\nu}\sigma_{\alpha\beta}\right),\label{sum_rs_uubar}
\end{eqnarray}
and use Eqs.\ (\ref{eq:Delta+}), and (\ref{eq:Delta-}) to obtain the
SSD up to the next-to-leading order,
\begin{eqnarray}
f_{+}\left(x,p,\mathfrak{s}\right) & = & \frac{E_{p}}{4\pi\hbar m}\int dp^{0}\theta(p^{0})\left[\mathcal{F}(x,p)-\mathfrak{s}^{\mu}\mathcal{A}_{\mu}(x,p)\right],\label{eq:spin-depend-f_plus-1}\\
f_{-}\left(x,\overline{p},\mathfrak{s}\right) & = & \frac{E_{p}}{4\pi\hbar m}\int dp^{0}\theta(-p^{0})\left[\mathcal{F}(x,p)-\mathfrak{s}^{\mu}\mathcal{A}_{\mu}(x,p)\right]+1.\label{eq:spin-depend-f_minus-1}
\end{eqnarray}
The quasi-classical parts of the Wigner functions up to the next-to-leading
order can be expressed in terms of SSD, 
\begin{eqnarray}
W_{\mathrm{qc}}^{<}(x,p) & = & 4\pi\hbar\delta(p^{2}-m^{2})\int[d\mathfrak{s}]\Pi(c_{0}\mathfrak{s})\bigg\{\theta(p^{0})f_{+}(x,p,\mathfrak{s})(p_{\mu}\gamma^{\mu}+m)\nonumber \\
 &  & +\theta(-p^{0})\left[1-f_{-}(x,\overline{p},\mathfrak{s})\right](\overline{p}_{\mu}\gamma^{\mu}-m)\bigg\},\label{eq:g-less-f-s}\\
W_{\mathrm{qc}}^{>}(x,p) & = & 4\pi\hbar\delta(p^{2}-m^{2})\int[d\mathfrak{s}]\Pi(c_{0}\mathfrak{s})\bigg\{\theta(p^{0})\left[1-f_{+}(x,p,\mathfrak{s})\right](p_{\mu}\gamma^{\mu}+m)\nonumber \\
 &  & +\theta(-p^{0})f_{-}(x,\overline{p},\mathfrak{s})(\overline{p}_{\mu}\gamma^{\mu}-m)\bigg\},\label{eq:g-larger-f-s}
\end{eqnarray}
where $\Pi(\mathfrak{s})$ denotes the spin projector,
\begin{equation}
\Pi(\mathfrak{s})=\frac{1}{2}(1+\gamma_{5}\mathfrak{s}\cdot\gamma).
\end{equation}
We see that $\Pi(c_{0}\mathfrak{s})$ instead of $\Pi(\mathfrak{s})$
appear in Eqs.~(\ref{eq:g-less-f-s}), and (\ref{eq:g-larger-f-s}) which
involves the normalization condition of $\mathfrak{s}$. Note that
there is a sign difference in $W^{<}$ as well as its components in
Eqs.~(\ref{eq:spin-depend-f_plus-1}), and (\ref{eq:spin-depend-f_minus-1})
from Ref.~\cite{Sheng:2021kfc} due to the definition of $G^{<}$
in Eq.~(\ref{eq:def-green-func}).

From Eq.~(\ref{eq:spin-depend-f_plus-1}), we see that the spin-projected
distribution has two components 
\begin{equation}
f(x,p,\mathfrak{s})=f_{\mathrm{scalar}}(x,p)-\mathfrak{s}_{\alpha}f_{\mathrm{polar}}^{\alpha}(x,p),\label{eq:f-scalar-polar}
\end{equation}
where $f_{\mathrm{scalar}}$ and $f_{\mathrm{polar}}^{\alpha}$ are
the scalar and polarization component of the distribution respectively.
The scalar component can be extracted by an integration of $f(x,p,\mathfrak{s})$
over $\mathfrak{s}$, while the polarization one can be extracted
by multiplying $\mathfrak{s}_{\rho}$ to $f(x,p,\mathfrak{s})$ and
taking an integration over $\mathfrak{s}$.

\subsection{Non-local collisions}

With the interaction Lagrangian $\mathcal{L}_{I}$ for Dirac fields,
the collision term in the Kadanoff-Baym equation (\ref{eq:eom-Wigner-function})
for the Wigner function can be written in the form,
\begin{equation}
\left[I_{\mathrm{coll}}\right]_{\alpha\beta}=\hbar\int\frac{d^{4}y}{(2\pi\hbar)^{4}}e^{-\frac{i}{\hbar}p\cdot y}\left\langle :\overline{\psi}_{\beta}\left(x+\frac{y}{2}\right)\rho_{\alpha}^{(I)}\left(x-\frac{y}{2}\right):\right\rangle ,\label{eq:collision-term}
\end{equation}
where the operator $\rho^{(I)}$ is defined as $\rho^{(I)}\equiv-(1/\hbar)\partial\mathcal{L}_{I}/\partial\overline{\psi}$.
By acting the operator $\gamma_{\nu}K^{\nu}+m$ on Eq.~(\ref{eq:eom-Wigner-function})
and separating the real and imaginary parts, one can derive a modified
mass-shell condition and a Boltzmann-like equation for the Wigner
function as in Eq.~(\ref{eq:mass-sh-boltzmann}).

For spin-1/2 particles, the Wigner function can be expressed by the
particle number density and the polarization density. Therefore we
find it convenient to define a single-particle distribution function
in phase space extended by an additional spin variable $\mathfrak{s}$
as in Eq.~(\ref{eq:f-scalar-polar}). The corresponding mass-shell
condition and Boltzmann equation are derived from Eq.~(\ref{eq:on-shell-cond-boltzmann}).
A detailed calculation shows that the off-shell part of the distribution
function cancels with that of the collision term in the Boltzmann
equation. Neglecting anti-particles, we obtain the on-shell Boltzmann
equation for particles,
\begin{equation}
p\cdot\partial f(x,p,\mathfrak{s})=C_{\text{on-shell}}[f],
\end{equation}
where $f(x,p,\mathfrak{s})\equiv f_{+}(x,p,\mathfrak{s})$. The derivation
of $C_{\text{on-shell}}[f]$ is based on the low-density (dilute)
approximation and the binary scattering assumption \cite{Weickgenannt:2020aaf,Weickgenannt:2021cuo}.
Then the ensemble average in Eq.~(\ref{eq:collision-term}) can be
performed over two-particle free states at the initial time. Correlations
between initial states can be neglected in a kinetic description.
Then the occupation number of a two-particle state can be written
as the product of the expectation values of two one-particle states,
which can be expressed in terms of the initial Wigner function. Note
that we adopt a low-density assumption, then the initial Wigner function
can be regarded as the interacting Wigner function up to the leading
order in the coupling constant. One can show that the collision term
has the form \cite{Weickgenannt:2020aaf,Weickgenannt:2021cuo},
\begin{eqnarray}
C_{\text{on-shell}}[f] & \sim & \prod_{j=1}^{2}\left[\int d^{4}x_{j}d^{4}p_{j}d^{4}q_{j}\exp\left(\frac{i}{\hbar}q_{i}\cdot x_{i}\right)\right.\nonumber \\
 &  & \times\overline{u}_{s_{j}}\left(p_{j}+\frac{q_{j}}{2}\right)W(x+x_{j},p_{j})u_{r_{j}}\left(p_{j}-\frac{q_{j}}{2}\right)u_{r_{j}}\left(p_{j}-\frac{q_{j}}{2}\right)\nonumber \\
 &  & \times\phi_{r_{1}r_{2}s_{1}s_{2}}(p;p_{1},p_{2},q_{1},q_{2})\,,
\end{eqnarray}
where the explicit expression of the collision kernel $\phi_{r_{1}r_{2}s_{1}s_{2}}(p;p_{1},p_{2},q_{1},q_{2})$
is given in Refs.~\cite{Weickgenannt:2020aaf,Weickgenannt:2021cuo}.
If the Wigner function varies slowly in space and time, we can take
a gradient expansion for $W(x+x_{j},p_{j})$ around $x$ and keep
terms up to the first order in gradients. By doing so, the collision
term can be separated into a local and a non-local contribution. We
find that the nonlocal term can be absorbed into the local one through
a space shift,
\begin{equation}
\Delta^{\mu}\equiv-\frac{\hbar}{2m(p\cdot\hat{t}+m)}\epsilon^{\mu\nu\alpha\beta}p_{\nu}\hat{t}_{\alpha}\mathfrak{s}_{\beta}\,,
\end{equation}
where $\hat{t}^{\mu}=(1,{\bf 0})$ is a time-like unit vector. The
collision term then have the following form,
\begin{eqnarray}
\mathfrak{C}_{\text{on-shell}}[f] & = & \int d\Gamma_{1}d\Gamma_{2}d\Gamma^{\prime}\widetilde{W}\left[f(x+\Delta_{1},p_{1},\mathfrak{s}_{1})\right.\nonumber \\
 &  & \left.\times f(x+\Delta_{2},p_{2},\mathfrak{s}_{2})-f(x+\Delta,p,\mathfrak{s})f(x+\Delta^{\prime},p^{\prime},\mathfrak{s}^{\prime})\right]\nonumber \\
 &  & +\int d\Gamma_{2}dS_{1}(p)\mathfrak{W}f(x+\Delta_{1},p,\mathfrak{s}_{1})f(x+\Delta_{2},p_{2},\mathfrak{s}_{2}), 
\label{eq:collision term-1}
\end{eqnarray}
where the phase-space measure is defined as $d\Gamma\equiv d^{4}p\delta(p^{2}-m^{2})[d\mathfrak{s}_{p}]$,
and the expressions of $\widetilde{W}$ and $\mathfrak{W}$ are given
in Refs.~\cite{Weickgenannt:2020aaf,Weickgenannt:2021cuo}. The first
term in Eq.~(\ref{eq:collision term-1}) arises from binary collisions
with both the momentum- and spin-exchange, while the second term describes
a pure spin-exchange without the momentum-exchange.

\section{Summary and outlook}
\label{sec:conclusion}

\begin{figure}[t]
    \centering
    \includegraphics[scale=0.4]{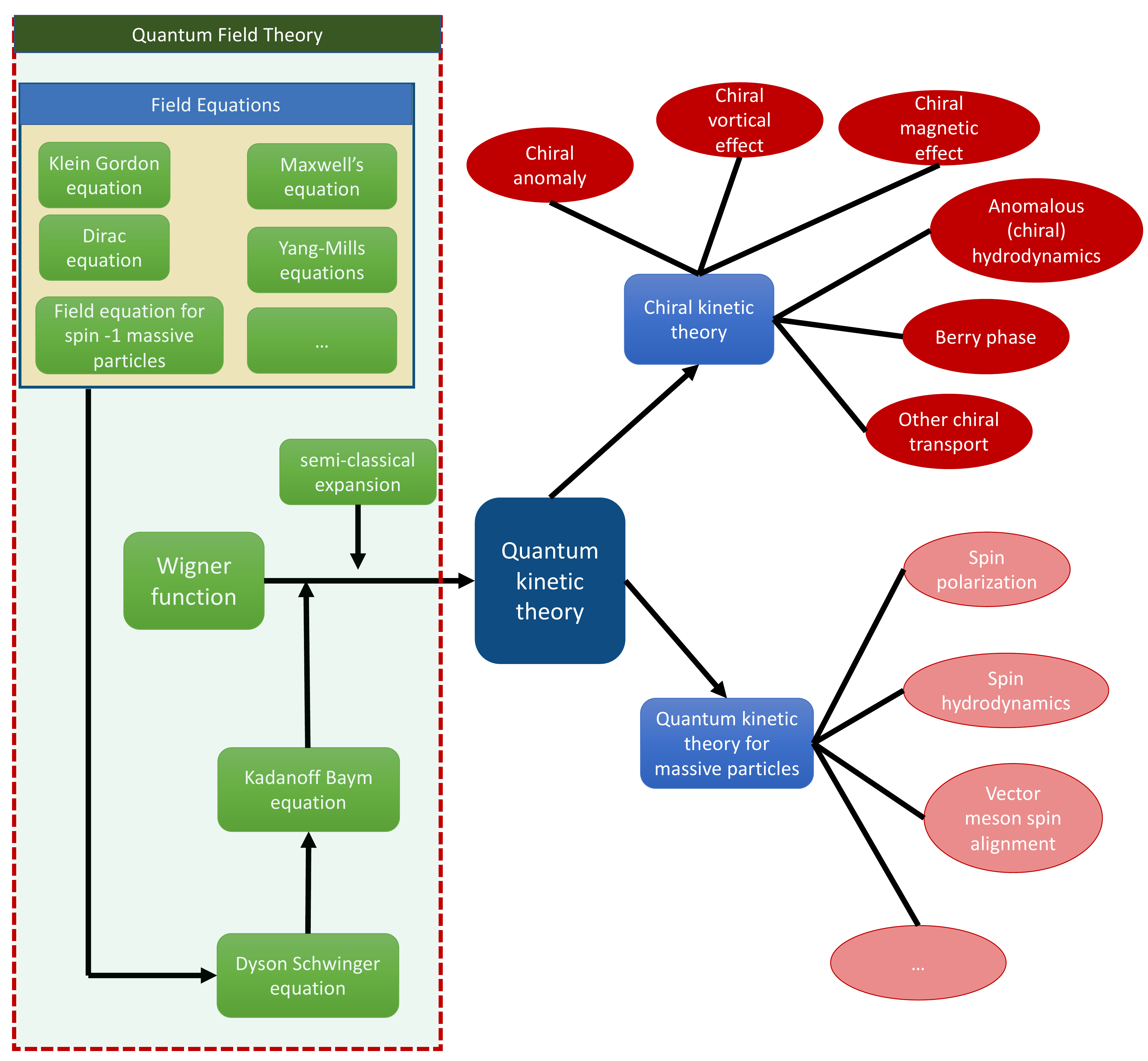}
    \caption{A schematic diagram for structure of the QKT.}
    \label{fig:QKT_structure}
\end{figure}

The quantum kinetic theory based on covariant Wigner functions is a powerful and systematic tool to describe quantum properties of particles in phase space. 
The structure of the quantum kinetic theory in connection with other theories or concepts is shown 
in Fig. \ref{fig:QKT_structure}.

The covariant Wigner function for chiral fermions can be decomposed into the left-handed and right-handed components dubbed the chiral Wigner function. A set of master equations for chiral Wigner functions in background electromagnetic fields can be derived from the Weyl equation.  The covariant Wigner function can be determined order by order in the Planck constant by solving the master equations which have an explicit Lorentz covariance. The vector (charge) and axial-vector (chiral charge) currents can be obtained from chiral Wigner functions by  integration over momenta. The CME and CVE currents emerge from chiral Wigner functions at the next-to-leading order in $\hbar$. 

For the left-handed Wigner function, which is a four-vector, there are four independent variables, while there are eight master equations about them (the same for the right-handed Wigner function). This indicates that there will be either redundancy or contradiction in these equations. By expressing space components by the time components of lower orders of the Planck constant, one can prove that these master equations can be reduced to one kinetic equation and one  constraint equation about the time components. Other equations are all consistent with these two equations, indicating that the extra number of equations are redundant. This is called the theorem of disentanglement of chiral Wigner function or DWF theorem for short. By solving chiral Wigner functions through expressing space components by time components, one introduces a frame automatically. Therefore the CME and CVE currents obtained in this way have two terms which contribute 1/3 and 2/3 of the total current, each term depends on the frame vector and has a clear physical meaning. This solves the '1/3' puzzle of the CME and CVE currents in the chiral kinetic theory for on-shell fermions in three-dimension. 

The second-order contributions in the Planck constant can be determined by solving master equations for chiral Wigner functions, which are in quadratic forms of the vorticity and electromagnetic field. These contributions include coupling terms of electromagnetic-field-electromagnetic-field (ee), vorticity-vorticity (vv), and vorticity-electromagnetic-field (ve).  All terms contribute to the charge current, while only 'ee' and 'vv' terms contribute to the chiral charge current or there is no contribution from the "ve" term. A Hall term $\epsilon ^{\mu\nu\rho\sigma}u_\nu E_\rho B_\sigma$ appears in the charge and chiral currents, while another Hall term $\epsilon ^{\mu\nu\rho\sigma}u_\nu E_\rho \omega_\sigma$ also appears in the charge current. The second-order energy-momentum tensor can also be obtained. The trace anomaly can be derived by including the quantum correction from electromagnetic fields to the renormalized energy-momentum tensor. 


We have also introduced the CKT from other effective theories such as the path integral of an effective Hamiltonian, the high-density effective theory, and other related methods. Then we showed the symplectic form for the phase space in EM fields with modified Poisson brackets. We argued that the distribution function is no longer a scalar but 
has quantum corrections under the Lorentz transformation. The CME and chiral anomaly in the context of 
the Berry phase are also discussed in these theories.

To include collisions, we have derived the Kadanoff-Baym equation with collision terms for massless fermions in EM fields, 
starting from the self-consistent equation for the dressed propagator in the CTP formalism. 
In the formal derivation, we have introduced the Moyal product for gauge-invariant operators~\eqref{eq:Moyal-product}, which includes the contribution from the EM fields.
This allows the equation to be systematically expanded with respect to $\hbar$.
The collision term plays an important role in thermalization of the system. 
In $O(\hbar)$, we have confirmed that the collision term vanishes 
for both local and global thermal equilibrium solutions, Eqs.~\eqref{eq:sol-first-order} and \eqref{L_equil_Wigner}. 
In general, the collision term has a complicated form; however, 
if one chooses the no-jump frame (the center of mass frame), 
the collision term has the form of the ordinary one. 

The generalization of the CKT for massless fermions to the QKT for massive fermions has been also achieved. By utilizing the Kadanoff-Baym equation and the expansion in the Planck constant, we derive the collisionless QKT and obtain the Wigner functions up to $O(\hbar)$ for massive fermions with arbitrary mass that can be smoothly connected to the CKT in the massless limit. Unlike the case of massless fermions, the spin four-vector has to be introduced as a new dynamical variable because the spin orientation of massive fermions is no longer locked by the momentum direction and chirality. The QKT for massive fermions thus incorporates coupled kinetic equations dubbed the SKE and AKE to delineate the intertwined charge and spin transport through the terms at $O(\hbar)$ associated with the spin-orbit  interaction. When applying a practical power-counting scheme such that the vector charge is dominant over the axial one due to the quantum origin of the latter in most physical systems, we further construct the collision term of the AKE from self-energies to track dynamical spin polarization up to $O(\hbar)$. It is found that the spin polarization can be produced by gradient terms resulting from space-time in-homogeneity of the scattered medium even without electromagnetic fields. The formalism is then applied to study the spin transport of massive quarks traversing a weakly coupled QGP. Nevertheless, due to the lack of the knowledge for quantum corrections from polarized gluons, only the classical collision term giving rise to spin relaxation and diffusion is obtained up to the leading logarithmic order in the QCD coupling constant. However, despite not being expatiated in this review, the quantum correction in the collision term of the AKE is shown to yield the spin polarization dictated by the thermal vorticity from the detailed balance in the Nambu-Jona-Lasinio model.  


Alternatively, the Wigner functions can be expressed in terms of matrix-valued spin distribution functions. From the Kadanoff-Baym equation, we can derive the Boltzmann equations for MVSD which can be decomposed into the scalar part and polarization part. The Boltzmann equation for the scalar part describes how the conventional distribution evolves through collisions, while that for the polarization part describes how the spin density in phase space for massless fermions evolves through collisions. The Boltzmann equation for the scalar and that for the polarization part are coupled together. The derivation of these Boltzmann equations is based on the property of Wigner functions in semi-classical expansion in the Planck constant that they can be decomposed into three sources of contributions: the quasi-classical term, the on-shell gradient and collision term, and the off-shell term (the off-shell term has been neglected in the derivation). At the leading order, only local collision terms appear in the Boltzmann equations without space-time derivatives, meaning that collisions take place at the same space-time point. At the next-to-leading order in $\hbar$,  the Boltzmann equations describe how the distribution function at next-to-leading order evolves under the influence of local as well as nonlocal collision terms with space-time derivatives. Therefore, the nonlocal collision terms can be regarded as sources for the polarization part. The MVSD can be converted to the scalar distribution function of a continuous spin variable (besides the position and momentum variable), with which one can rewrite the nonlocal collision term with a space-like shift of the collision point. This space-shift is related to the side jump in collisions for massless fermion. In this way, we demonstrate how the spin polarization arises from the vorticity in equilibrium through nonlocal collisions. 


Understanding dynamical spin polarization in relativistic heavy-ion collisions is one of the primary goals that motivate recent developments of the QKT. Most importantly, the QKT can be applied to investigate non-equilibrium spin transport. As briefly remarked in Sec.~\ref{sec:CKT_collisions}, the Wigner functions for Weyl fermions in local equilibrium \cite{Hidaka:2017auj} incorporate extra corrections on top of the thermal vorticity or kinetic vorticity at constant temperature in global equilibrium \cite{Becattini:2013fla,Fang:2016vpj}. In particular, the shear correction, which has been further derived from the linear-response theory and statistical quantum field theory for massive fermions \cite{Liu:2020dxg,Liu:2021uhn,Becattini:2021suc} (see also Ref.~\cite{Liu:2021nyg}), leads to a substantial contribution to the local spin polarization along the beam direction as shown by hydrodynamic simulations \cite{Fu:2021pok,Becattini:2021iol,Yi:2021ryh} (see also Ref.~\cite{Ryu:2021lnx} for simulations with the chemical-potential gradient). However, except for the global equilibrium condition, non-equilibrium corrections pertinent to interaction are inevitable to arise near local equilibrium. Further exploration of these corrections to spin polarization is necessary for phenomenological studies and the construction of spin hydrodynamics. Furthermore, even though the global-equilibrium solution of the Wigner functions for massive fermions can be acquired by the detailed balance of the QKT as shown in Refs.~\cite{Weickgenannt:2020aaf,Wang:2020pej}, the local-equilibrium solution in connection with the result obtained in the CKT \cite{Hidaka:2017auj,Fang:2022ttm} remains to be solved. 


On the other hand, in most studies of the QKT or alternative approaches to spin polarization in heavy-ion collisions, the role of gluons that can possibly affect the spin transport process of quarks has been overlooked despite a considerable number of previous developments in the kinetic theory of gluons \cite{Heinz:1984yq,Heinz:1985qe,Elze:1989un,Elze:1986hq,Elze:1989gm,Blaizot:1993zk,Blaizot:1993be, Blaizot:1999fq, Blaizot:1999xk,Blaizot:2001nr,Wang:2001dm}. Even in QED, the QKT for polarized photons (or more generally for bosons) and related quantum transport were less studied \cite{Avkhadiev:2017fxj,Yamamoto:2017uul,Yamamoto:2017gla,PhysRevA.96.043830,Huang:2018aly,Chernodub:2018era,Copetti:2018mxw,Prokhorov:2020okl,Hattori:2020gqh,Mameda:2022ojk}, compared to those in purely fermionic systems, and a direct generalization to gluon spin dynamics is not yet feasible as noted in Sec.~\ref{subsec:spin diffusion in QGP}. In addition, a recent extension of collisionless CKT and AKT to quarks coupled with non-Abelian background color fields has been made in Refs.~\cite{Luo:2021uog,Muller:2021hpe,Yang:2021fea} based on Refs.~\cite{Heinz:1983nx,Heinz:1984yq,Heinz:1985qe,Elze:1986qd,Elze:1989un}, from which quantum corrections upon spin polarization besides collision effects can be found \cite{Muller:2021hpe,Yang:2021fea} as an analogue of the anomalous diffusion led by turbulent color fields \cite{Asakawa:2006tc,Asakawa:2006jn}. A complete formalism to delineate entangled spin transport for both quarks and gluons in the QGP or pre-equilibrium phase is one of the next thread of research in future studies.

	
	\newpage
	\section*{Acknowledgements}
	We are grateful to thank  Koichi Hattori, Shu Lin, Cristina Manuel, Igor Shovkovy and Ho-Ung Yee for helpful discussions and comments. S.P. and Q.W. are partly supported by National Natural Science Foundation of China (NSFC) under Grants Nos. 12135011, 11890713 (a subgrant of No. 11890710) and 12075235. 
D.-L.Y. was supported by the Ministry of Science and Technology, Taiwan under Grant No. MOST 110-2112-M-001-070-MY3.
Y.H. was supported by JSPS KAKENHI Grant Numbers 17H06462 and 21H01084.

	\bibliography{qkt-ref}

\begin{thebibliography}{100}
\expandafter\ifx\csname url\endcsname\relax
  \def\url#1{\texttt{#1}}\fi
\expandafter\ifx\csname urlprefix\endcsname\relax\def\urlprefix{URL }\fi
\expandafter\ifx\csname href\endcsname\relax
  \def\href#1#2{#2} \def\path#1{#1}\fi

\bibitem{Rischke:2003mt}
D.~H. Rischke, {The Quark gluon plasma in equilibrium}, Prog. Part. Nucl. Phys.
  52 (2004) 197--296.
\newblock \href {http://arxiv.org/abs/nucl-th/0305030}
  {\path{arXiv:nucl-th/0305030}}, \href
  {https://doi.org/10.1016/j.ppnp.2003.09.002}
  {\path{doi:10.1016/j.ppnp.2003.09.002}}.

\bibitem{Gyulassy:2004vg}
M.~Gyulassy, {The QGP discovered at RHIC}, in: {NATO Advanced Study Institute:
  Structure and Dynamics of Elementary Matter}, 2004, pp. 159--182.
\newblock \href {http://arxiv.org/abs/nucl-th/0403032}
  {\path{arXiv:nucl-th/0403032}}.

\bibitem{Gyulassy:2004zy}
M.~Gyulassy, L.~McLerran, {New forms of QCD matter discovered at RHIC}, Nucl.
  Phys. A 750 (2005) 30--63.
\newblock \href {http://arxiv.org/abs/nucl-th/0405013}
  {\path{arXiv:nucl-th/0405013}}, \href
  {https://doi.org/10.1016/j.nuclphysa.2004.10.034}
  {\path{doi:10.1016/j.nuclphysa.2004.10.034}}.

\bibitem{Shuryak:2004cy}
E.~V. Shuryak, {What RHIC experiments and theory tell us about properties of
  quark-gluon plasma?}, Nucl. Phys. A 750 (2005) 64--83.
\newblock \href {http://arxiv.org/abs/hep-ph/0405066}
  {\path{arXiv:hep-ph/0405066}}, \href
  {https://doi.org/10.1016/j.nuclphysa.2004.10.022}
  {\path{doi:10.1016/j.nuclphysa.2004.10.022}}.

\bibitem{Csernai:2006zz}
L.~P. Csernai, J.~I. Kapusta, L.~D. McLerran, {On the Strongly-Interacting
  Low-Viscosity Matter Created in Relativistic Nuclear Collisions}, Phys. Rev.
  Lett. 97 (2006) 152303.
\newblock \href {http://arxiv.org/abs/nucl-th/0604032}
  {\path{arXiv:nucl-th/0604032}}, \href
  {https://doi.org/10.1103/PhysRevLett.97.152303}
  {\path{doi:10.1103/PhysRevLett.97.152303}}.

\bibitem{Akiba:2015jwa}
Y.~Akiba, et~al., {The Hot QCD White Paper: Exploring the Phases of QCD at RHIC
  and the LHC} (2 2015).
\newblock \href {http://arxiv.org/abs/1502.02730} {\path{arXiv:1502.02730}}.

\bibitem{Busza:2018rrf}
W.~Busza, K.~Rajagopal, W.~van~der Schee, {Heavy Ion Collisions: The Big
  Picture, and the Big Questions}, Ann. Rev. Nucl. Part. Sci. 68 (2018)
  339--376.
\newblock \href {http://arxiv.org/abs/1802.04801} {\path{arXiv:1802.04801}},
  \href {https://doi.org/10.1146/annurev-nucl-101917-020852}
  {\path{doi:10.1146/annurev-nucl-101917-020852}}.

\bibitem{STAR:2003wqp}
J.~Adams, et~al., {Particle type dependence of azimuthal anisotropy and nuclear
  modification of particle production in Au + Au collisions at s(NN)**(1/2) =
  200-GeV}, Phys. Rev. Lett. 92 (2004) 052302.
\newblock \href {http://arxiv.org/abs/nucl-ex/0306007}
  {\path{arXiv:nucl-ex/0306007}}, \href
  {https://doi.org/10.1103/PhysRevLett.92.052302}
  {\path{doi:10.1103/PhysRevLett.92.052302}}.

\bibitem{STAR:2003xyj}
J.~Adams, et~al., {Erratum: Azimuthal Anisotropy at the Relativistic Heavy Ion
  Collider: The First and Fourth Harmonics [Phys. Rev. Lett. 92, 062301
  (2004)]}, Phys. Rev. Lett. 92 (2004) 062301, [Erratum: Phys.Rev.Lett. 127,
  069901 (2021)].
\newblock \href {http://arxiv.org/abs/nucl-ex/0310029}
  {\path{arXiv:nucl-ex/0310029}}, \href
  {https://doi.org/10.1103/PhysRevLett.127.069901}
  {\path{doi:10.1103/PhysRevLett.127.069901}}.

\bibitem{STAR:2002hbo}
C.~Adler, et~al., {Elliptic flow from two and four particle correlations in
  Au+Au collisions at s(NN)**(1/2) = 130-GeV}, Phys. Rev. C 66 (2002) 034904.
\newblock \href {http://arxiv.org/abs/nucl-ex/0206001}
  {\path{arXiv:nucl-ex/0206001}}, \href
  {https://doi.org/10.1103/PhysRevC.66.034904}
  {\path{doi:10.1103/PhysRevC.66.034904}}.

\bibitem{Sorensen:2003kp}
P.~R. Sorensen, {Kaon and Lambda production at intermediate p(T): Insights into
  the hadronization of the bulk partonic matter created in Au+Au collisions at
  RHIC} (9 2003).
\newblock \href {http://arxiv.org/abs/nucl-ex/0309003}
  {\path{arXiv:nucl-ex/0309003}}.

\bibitem{Kolb:2000sd}
P.~F. Kolb, J.~Sollfrank, U.~W. Heinz, {Anisotropic transverse flow and the
  quark hadron phase transition}, Phys. Rev. C 62 (2000) 054909.
\newblock \href {http://arxiv.org/abs/hep-ph/0006129}
  {\path{arXiv:hep-ph/0006129}}, \href
  {https://doi.org/10.1103/PhysRevC.62.054909}
  {\path{doi:10.1103/PhysRevC.62.054909}}.

\bibitem{Kolb:2003dz}
P.~F. Kolb, U.~W. Heinz, {Hydrodynamic description of ultrarelativistic heavy
  ion collisions} (2003) 634--714\href {http://arxiv.org/abs/nucl-th/0305084}
  {\path{arXiv:nucl-th/0305084}}.

\bibitem{Hama:2004rr}
Y.~Hama, T.~Kodama, O.~Socolowski, Jr., {Topics on hydrodynamic model of
  nucleus-nucleus collisions}, Braz. J. Phys. 35 (2005) 24--51.
\newblock \href {http://arxiv.org/abs/hep-ph/0407264}
  {\path{arXiv:hep-ph/0407264}}, \href
  {https://doi.org/10.1590/S0103-97332005000100003}
  {\path{doi:10.1590/S0103-97332005000100003}}.

\bibitem{Huovinen:2006jp}
P.~Huovinen, P.~V. Ruuskanen, {Hydrodynamic Models for Heavy Ion Collisions},
  Ann. Rev. Nucl. Part. Sci. 56 (2006) 163--206.
\newblock \href {http://arxiv.org/abs/nucl-th/0605008}
  {\path{arXiv:nucl-th/0605008}}, \href
  {https://doi.org/10.1146/annurev.nucl.54.070103.181236}
  {\path{doi:10.1146/annurev.nucl.54.070103.181236}}.

\bibitem{Ollitrault:2007du}
J.-Y. Ollitrault, {Relativistic hydrodynamics for heavy-ion collisions}, Eur.
  J. Phys. 29 (2008) 275--302.
\newblock \href {http://arxiv.org/abs/0708.2433} {\path{arXiv:0708.2433}},
  \href {https://doi.org/10.1088/0143-0807/29/2/010}
  {\path{doi:10.1088/0143-0807/29/2/010}}.

\bibitem{Teaney:2003kp}
D.~Teaney, {The Effects of viscosity on spectra, elliptic flow, and HBT radii},
  Phys. Rev. C 68 (2003) 034913.
\newblock \href {http://arxiv.org/abs/nucl-th/0301099}
  {\path{arXiv:nucl-th/0301099}}, \href
  {https://doi.org/10.1103/PhysRevC.68.034913}
  {\path{doi:10.1103/PhysRevC.68.034913}}.

\bibitem{Lacey:2006bc}
R.~A. Lacey, N.~N. Ajitanand, J.~M. Alexander, P.~Chung, W.~G. Holzmann,
  M.~Issah, A.~Taranenko, P.~Danielewicz, H.~Stoecker, {Has the QCD Critical
  Point been Signaled by Observations at RHIC?}, Phys. Rev. Lett. 98 (2007)
  092301.
\newblock \href {http://arxiv.org/abs/nucl-ex/0609025}
  {\path{arXiv:nucl-ex/0609025}}, \href
  {https://doi.org/10.1103/PhysRevLett.98.092301}
  {\path{doi:10.1103/PhysRevLett.98.092301}}.

\bibitem{Gale:2013da}
C.~Gale, S.~Jeon, B.~Schenke, {Hydrodynamic Modeling of Heavy-Ion Collisions},
  Int. J. Mod. Phys. A 28 (2013) 1340011.
\newblock \href {http://arxiv.org/abs/1301.5893} {\path{arXiv:1301.5893}},
  \href {https://doi.org/10.1142/S0217751X13400113}
  {\path{doi:10.1142/S0217751X13400113}}.

\bibitem{Bloczynski:2012en}
J.~Bloczynski, X.-G. Huang, X.~Zhang, J.~Liao, {Azimuthally fluctuating
  magnetic field and its impacts on observables in heavy-ion collisions}, Phys.
  Lett. B718 (2013) 1529--1535.
\newblock \href {http://arxiv.org/abs/1209.6594} {\path{arXiv:1209.6594}},
  \href {https://doi.org/10.1016/j.physletb.2012.12.030}
  {\path{doi:10.1016/j.physletb.2012.12.030}}.

\bibitem{Deng:2012pc}
W.-T. Deng, X.-G. Huang, {Event-by-event generation of electromagnetic fields
  in heavy-ion collisions}, Phys. Rev. C 85 (2012) 044907.
\newblock \href {http://arxiv.org/abs/1201.5108} {\path{arXiv:1201.5108}},
  \href {https://doi.org/10.1103/PhysRevC.85.044907}
  {\path{doi:10.1103/PhysRevC.85.044907}}.

\bibitem{Tuchin:2013ie}
K.~Tuchin, {Particle production in strong electromagnetic fields in
  relativistic heavy-ion collisions}, Adv. High Energy Phys. 2013 (2013)
  490495.
\newblock \href {http://arxiv.org/abs/1301.0099} {\path{arXiv:1301.0099}},
  \href {https://doi.org/10.1155/2013/490495} {\path{doi:10.1155/2013/490495}}.

\bibitem{Tuchin:2013apa}
K.~Tuchin, {Time and space dependence of the electromagnetic field in
  relativistic heavy-ion collisions}, Phys. Rev. C 88~(2) (2013) 024911.
\newblock \href {http://arxiv.org/abs/1305.5806} {\path{arXiv:1305.5806}},
  \href {https://doi.org/10.1103/PhysRevC.88.024911}
  {\path{doi:10.1103/PhysRevC.88.024911}}.

\bibitem{Roy:2015coa}
V.~Roy, S.~Pu, {Event-by-event distribution of magnetic field energy over
  initial fluid energy density in $\sqrt{s_{\rm NN}}$= 200 GeV Au-Au
  collisions}, Phys. Rev. C92 (2015) 064902.
\newblock \href {http://arxiv.org/abs/1508.03761} {\path{arXiv:1508.03761}},
  \href {https://doi.org/10.1103/PhysRevC.92.064902}
  {\path{doi:10.1103/PhysRevC.92.064902}}.

\bibitem{Li:2016tel}
H.~Li, X.-l. Sheng, Q.~Wang, {Electromagnetic fields with electric and chiral
  magnetic conductivities in heavy ion collisions}, Phys. Rev. C94~(4) (2016)
  044903.
\newblock \href {http://arxiv.org/abs/1602.02223} {\path{arXiv:1602.02223}},
  \href {https://doi.org/10.1103/PhysRevC.94.044903}
  {\path{doi:10.1103/PhysRevC.94.044903}}.

\bibitem{Holliday:2016lbx}
R.~Holliday, R.~McCarty, B.~Peroutka, K.~Tuchin, {Classical Electromagnetic
  Fields from Quantum Sources in Heavy-Ion Collisions}, Nucl. Phys. A 957
  (2017) 406--415.
\newblock \href {http://arxiv.org/abs/1604.04572} {\path{arXiv:1604.04572}},
  \href {https://doi.org/10.1016/j.nuclphysa.2016.10.003}
  {\path{doi:10.1016/j.nuclphysa.2016.10.003}}.

\bibitem{Stewart:2017zsu}
E.~Stewart, K.~Tuchin, {Magnetic field in expanding quark-gluon plasma}, Phys.
  Rev. C 97~(4) (2018) 044906.
\newblock \href {http://arxiv.org/abs/1710.08793} {\path{arXiv:1710.08793}},
  \href {https://doi.org/10.1103/PhysRevC.97.044906}
  {\path{doi:10.1103/PhysRevC.97.044906}}.

\bibitem{Siddique:2021smf}
I.~Siddique, X.-L. Sheng, Q.~Wang, {Space-average electromagnetic fields and
  electromagnetic anomaly weighted by energy density in heavy-ion collisions},
  Phys. Rev. C 104~(3) (2021) 034907.
\newblock \href {http://arxiv.org/abs/2106.00478} {\path{arXiv:2106.00478}},
  \href {https://doi.org/10.1103/PhysRevC.104.034907}
  {\path{doi:10.1103/PhysRevC.104.034907}}.

\bibitem{Chen:2021nxs}
Y.~Chen, X.-L. Sheng, G.-L. Ma, {Electromagnetic fields from the extended
  Kharzeev-McLerran-Warringa model in relativistic heavy-ion collisions}, Nucl.
  Phys. A 1011 (2021) 122199.
\newblock \href {http://arxiv.org/abs/2101.09845} {\path{arXiv:2101.09845}},
  \href {https://doi.org/10.1016/j.nuclphysa.2021.122199}
  {\path{doi:10.1016/j.nuclphysa.2021.122199}}.

\bibitem{Kharzeev:2007jp}
D.~E. Kharzeev, L.~D. McLerran, H.~J. Warringa, {The Effects of topological
  charge change in heavy ion collisions: 'Event by event P and CP violation'},
  Nucl. Phys. A803 (2008) 227--253.
\newblock \href {http://arxiv.org/abs/0711.0950} {\path{arXiv:0711.0950}},
  \href {https://doi.org/10.1016/j.nuclphysa.2008.02.298}
  {\path{doi:10.1016/j.nuclphysa.2008.02.298}}.

\bibitem{Pu:2016ayh}
S.~Pu, V.~Roy, L.~Rezzolla, D.~H. Rischke, {Bjorken flow in one-dimensional
  relativistic magnetohydrodynamics with magnetization}, Phys. Rev. D93~(7)
  (2016) 074022.
\newblock \href {http://arxiv.org/abs/1602.04953} {\path{arXiv:1602.04953}},
  \href {https://doi.org/10.1103/PhysRevD.93.074022}
  {\path{doi:10.1103/PhysRevD.93.074022}}.

\bibitem{Roy:2015kma}
V.~Roy, S.~Pu, L.~Rezzolla, D.~Rischke, {Analytic Bjorken flow in
  one-dimensional relativistic magnetohydrodynamics}, Phys. Lett. B750 (2015)
  45--52.
\newblock \href {http://arxiv.org/abs/1506.06620} {\path{arXiv:1506.06620}},
  \href {https://doi.org/10.1016/j.physletb.2015.08.046}
  {\path{doi:10.1016/j.physletb.2015.08.046}}.

\bibitem{Pu:2016bxy}
S.~Pu, D.-L. Yang, {Transverse flow induced by inhomogeneous magnetic fields in
  the Bjorken expansion}, Phys. Rev. D93~(5) (2016) 054042.
\newblock \href {http://arxiv.org/abs/1602.04954} {\path{arXiv:1602.04954}},
  \href {https://doi.org/10.1103/PhysRevD.93.054042}
  {\path{doi:10.1103/PhysRevD.93.054042}}.

\bibitem{Pu:2016rdq}
S.~Pu, D.-L. Yang, {Analytic Solutions of Transverse Magneto-hydrodynamics
  under Bjorken Expansion}, EPJ Web Conf. 137 (2017) 13021.
\newblock \href {http://arxiv.org/abs/1611.04840} {\path{arXiv:1611.04840}},
  \href {https://doi.org/10.1051/epjconf/201713713021}
  {\path{doi:10.1051/epjconf/201713713021}}.

\bibitem{Roy:2017xtz}
V.~Roy, S.~Pu, L.~Rezzolla, D.~H. Rischke, {Effect of intense magnetic fields
  on reduced-MHD evolution in $\sqrt{s_{\rm NN}}$ = 200 GeV Au+Au collisions},
  DAE Symp. Nucl. Phys. 62 (2017) 926--927.

\bibitem{Siddique:2019gqh}
I.~Siddique, R.-j. Wang, S.~Pu, Q.~Wang, {Anomalous magnetohydrodynamics with
  longitudinal boost invariance and chiral magnetic effect}, Phys. Rev.
  D99~(11) (2019) 114029.
\newblock \href {http://arxiv.org/abs/1904.01807} {\path{arXiv:1904.01807}},
  \href {https://doi.org/10.1103/PhysRevD.99.114029}
  {\path{doi:10.1103/PhysRevD.99.114029}}.

\bibitem{Wang:2020qpx}
R.-j. Wang, P.~Copinger, S.~Pu, {Anomalous magnetohydrodynamics with constant
  anisotropic electric conductivities}, in: {28th International Conference on
  Ultrarelativistic Nucleus-Nucleus Collisions}, 2020.
\newblock \href {http://arxiv.org/abs/2004.06408} {\path{arXiv:2004.06408}}.

\bibitem{Inghirami:2016iru}
G.~Inghirami, L.~Del~Zanna, A.~Beraudo, M.~H. Moghaddam, F.~Becattini,
  M.~Bleicher, {Numerical magneto-hydrodynamics for relativistic nuclear
  collisions}, Eur. Phys. J. C76~(12) (2016) 659.
\newblock \href {http://arxiv.org/abs/1609.03042} {\path{arXiv:1609.03042}},
  \href {https://doi.org/10.1140/epjc/s10052-016-4516-8}
  {\path{doi:10.1140/epjc/s10052-016-4516-8}}.

\bibitem{Inghirami:2019mkc}
G.~Inghirami, M.~Mace, Y.~Hirono, L.~Del~Zanna, D.~E. Kharzeev, M.~Bleicher,
  {Magnetic fields in heavy ion collisions: flow and charge transport} (2019).
\newblock \href {http://arxiv.org/abs/1908.07605} {\path{arXiv:1908.07605}}.

\bibitem{Yan:2021zjc}
L.~Yan, X.-G. Huang, {Dynamical evolution of magnetic field in the
  pre-equilibrium quark-gluon plasma} (4 2021).
\newblock \href {http://arxiv.org/abs/2104.00831} {\path{arXiv:2104.00831}}.

\bibitem{Denicol:2018rbw}
G.~S. Denicol, X.-G. Huang, E.~Molnár, G.~M. Monteiro, H.~Niemi, J.~Noronha,
  D.~H. Rischke, Q.~Wang, {Nonresistive dissipative magnetohydrodynamics from
  the Boltzmann equation in the 14-moment approximation}, Phys. Rev. D98~(7)
  (2018) 076009.
\newblock \href {http://arxiv.org/abs/1804.05210} {\path{arXiv:1804.05210}},
  \href {https://doi.org/10.1103/PhysRevD.98.076009}
  {\path{doi:10.1103/PhysRevD.98.076009}}.

\bibitem{Denicol:2019iyh}
G.~S. Denicol, E.~Molnár, H.~Niemi, D.~H. Rischke, {Resistive dissipative
  magnetohydrodynamics from the Boltzmann-Vlasov equation}, Phys. Rev. D99~(5)
  (2019) 056017.
\newblock \href {http://arxiv.org/abs/1902.01699} {\path{arXiv:1902.01699}},
  \href {https://doi.org/10.1103/PhysRevD.99.056017}
  {\path{doi:10.1103/PhysRevD.99.056017}}.

\bibitem{Vilenkin:1980fu}
A.~Vilenkin, {EQUILIBRIUM PARITY VIOLATING CURRENT IN A MAGNETIC FIELD}, Phys.
  Rev. D22 (1980) 3080--3084.
\newblock \href {https://doi.org/10.1103/PhysRevD.22.3080}
  {\path{doi:10.1103/PhysRevD.22.3080}}.

\bibitem{Nielsen:1983rb}
H.~B. Nielsen, M.~Ninomiya, {ADLER-BELL-JACKIW ANOMALY AND WEYL FERMIONS IN
  CRYSTAL}, Phys. Lett. B 130 (1983) 389--396.
\newblock \href {https://doi.org/10.1016/0370-2693(83)91529-0}
  {\path{doi:10.1016/0370-2693(83)91529-0}}.

\bibitem{Alekseev:1998ds}
A.~Y. Alekseev, V.~V. Cheianov, J.~Frohlich, {Universality of transport
  properties in equilibrium, Goldstone theorem and chiral anomaly}, Phys. Rev.
  Lett. 81 (1998) 3503--3506.
\newblock \href {http://arxiv.org/abs/cond-mat/9803346}
  {\path{arXiv:cond-mat/9803346}}, \href
  {https://doi.org/10.1103/PhysRevLett.81.3503}
  {\path{doi:10.1103/PhysRevLett.81.3503}}.

\bibitem{Kharzeev:2004ey}
D.~Kharzeev, {Parity violation in hot QCD: Why it can happen, and how to look
  for it}, Phys. Lett. B633 (2006) 260--264.
\newblock \href {http://arxiv.org/abs/hep-ph/0406125}
  {\path{arXiv:hep-ph/0406125}}, \href
  {https://doi.org/10.1016/j.physletb.2005.11.075}
  {\path{doi:10.1016/j.physletb.2005.11.075}}.

\bibitem{Fukushima:2008xe}
K.~Fukushima, D.~E. Kharzeev, H.~J. Warringa, {The Chiral Magnetic Effect},
  Phys. Rev. D78 (2008) 074033.
\newblock \href {http://arxiv.org/abs/0808.3382} {\path{arXiv:0808.3382}},
  \href {https://doi.org/10.1103/PhysRevD.78.074033}
  {\path{doi:10.1103/PhysRevD.78.074033}}.

\bibitem{Manton:1983nd}
N.~S. Manton, {Topology in the Weinberg-Salam Theory}, Phys. Rev. D 28 (1983)
  2019.
\newblock \href {https://doi.org/10.1103/PhysRevD.28.2019}
  {\path{doi:10.1103/PhysRevD.28.2019}}.

\bibitem{Klinkhamer:1984di}
F.~R. Klinkhamer, N.~S. Manton, {A Saddle Point Solution in the Weinberg-Salam
  Theory}, Phys. Rev. D 30 (1984) 2212.
\newblock \href {https://doi.org/10.1103/PhysRevD.30.2212}
  {\path{doi:10.1103/PhysRevD.30.2212}}.

\bibitem{McLerran:1990de}
L.~D. McLerran, E.~Mottola, M.~E. Shaposhnikov, {Sphalerons and Axion Dynamics
  in High Temperature {QCD}}, Phys. Rev. D 43 (1991) 2027--2035.
\newblock \href {https://doi.org/10.1103/PhysRevD.43.2027}
  {\path{doi:10.1103/PhysRevD.43.2027}}.

\bibitem{Arnold:1996dy}
P.~B. Arnold, D.~Son, L.~G. Yaffe, {The Hot baryon violation rate is O
  (alpha-w**5 T**4)}, Phys. Rev. D 55 (1997) 6264--6273.
\newblock \href {http://arxiv.org/abs/hep-ph/9609481}
  {\path{arXiv:hep-ph/9609481}}, \href
  {https://doi.org/10.1103/PhysRevD.55.6264}
  {\path{doi:10.1103/PhysRevD.55.6264}}.

\bibitem{Moore:2010jd}
G.~D. Moore, M.~Tassler, {The Sphaleron Rate in SU(N) Gauge Theory}, JHEP 02
  (2011) 105.
\newblock \href {http://arxiv.org/abs/1011.1167} {\path{arXiv:1011.1167}},
  \href {https://doi.org/10.1007/JHEP02(2011)105}
  {\path{doi:10.1007/JHEP02(2011)105}}.

\bibitem{Mace:2016svc}
M.~Mace, S.~Schlichting, R.~Venugopalan, {Off-equilibrium sphaleron transitions
  in the Glasma}, Phys. Rev. D 93~(7) (2016) 074036.
\newblock \href {http://arxiv.org/abs/1601.07342} {\path{arXiv:1601.07342}},
  \href {https://doi.org/10.1103/PhysRevD.93.074036}
  {\path{doi:10.1103/PhysRevD.93.074036}}.

\bibitem{Tanji:2016dka}
N.~Tanji, N.~Mueller, J.~Berges, {Transient anomalous charge production in
  strong-field QCD}, Phys. Rev. D 93~(7) (2016) 074507.
\newblock \href {http://arxiv.org/abs/1603.03331} {\path{arXiv:1603.03331}},
  \href {https://doi.org/10.1103/PhysRevD.93.074507}
  {\path{doi:10.1103/PhysRevD.93.074507}}.

\bibitem{Joyce:1997uy}
M.~Joyce, M.~E. Shaposhnikov, {Primordial magnetic fields, right-handed
  electrons, and the Abelian anomaly}, Phys. Rev. Lett. 79 (1997) 1193--1196.
\newblock \href {http://arxiv.org/abs/astro-ph/9703005}
  {\path{arXiv:astro-ph/9703005}}, \href
  {https://doi.org/10.1103/PhysRevLett.79.1193}
  {\path{doi:10.1103/PhysRevLett.79.1193}}.

\bibitem{Akamatsu:2013pjd}
Y.~Akamatsu, N.~Yamamoto, {Chiral Plasma Instabilities}, Phys. Rev. Lett. 111
  (2013) 052002.
\newblock \href {http://arxiv.org/abs/1302.2125} {\path{arXiv:1302.2125}},
  \href {https://doi.org/10.1103/PhysRevLett.111.052002}
  {\path{doi:10.1103/PhysRevLett.111.052002}}.

\bibitem{Mace:2019cqo}
M.~Mace, N.~Mueller, S.~Schlichting, S.~Sharma, {Chiral Instabilities and the
  Onset of Chiral Turbulence in QED Plasmas}, Phys. Rev. Lett. 124~(19) (2020)
  191604.
\newblock \href {http://arxiv.org/abs/1910.01654} {\path{arXiv:1910.01654}},
  \href {https://doi.org/10.1103/PhysRevLett.124.191604}
  {\path{doi:10.1103/PhysRevLett.124.191604}}.

\bibitem{Son:2004tq}
D.~T. Son, A.~R. Zhitnitsky, {Quantum anomalies in dense matter}, Phys. Rev. D
  70 (2004) 074018.
\newblock \href {http://arxiv.org/abs/hep-ph/0405216}
  {\path{arXiv:hep-ph/0405216}}, \href
  {https://doi.org/10.1103/PhysRevD.70.074018}
  {\path{doi:10.1103/PhysRevD.70.074018}}.

\bibitem{Metlitski:2005pr}
M.~A. Metlitski, A.~R. Zhitnitsky, {Anomalous axion interactions and
  topological currents in dense matter}, Phys. Rev. D 72 (2005) 045011.
\newblock \href {http://arxiv.org/abs/hep-ph/0505072}
  {\path{arXiv:hep-ph/0505072}}, \href
  {https://doi.org/10.1103/PhysRevD.72.045011}
  {\path{doi:10.1103/PhysRevD.72.045011}}.

\bibitem{Kharzeev:2010gd}
D.~E. Kharzeev, H.-U. Yee, {Chiral Magnetic Wave}, Phys. Rev. D 83 (2011)
  085007.
\newblock \href {http://arxiv.org/abs/1012.6026} {\path{arXiv:1012.6026}},
  \href {https://doi.org/10.1103/PhysRevD.83.085007}
  {\path{doi:10.1103/PhysRevD.83.085007}}.

\bibitem{Burnier:2011bf}
Y.~Burnier, D.~E. Kharzeev, J.~Liao, H.-U. Yee, {Chiral magnetic wave at finite
  baryon density and the electric quadrupole moment of quark-gluon plasma in
  heavy ion collisions}, Phys. Rev. Lett. 107 (2011) 052303.
\newblock \href {http://arxiv.org/abs/1103.1307} {\path{arXiv:1103.1307}},
  \href {https://doi.org/10.1103/PhysRevLett.107.052303}
  {\path{doi:10.1103/PhysRevLett.107.052303}}.

\bibitem{Shovkovy:2018tks}
I.~A. Shovkovy, D.~O. Rybalka, E.~V. Gorbar, {The overdamped chiral magnetic
  wave}, PoS Confinement2018 (2018) 029.
\newblock \href {http://arxiv.org/abs/1811.10635} {\path{arXiv:1811.10635}},
  \href {https://doi.org/10.22323/1.336.0029} {\path{doi:10.22323/1.336.0029}}.

\bibitem{Vilenkin:1979ui}
A.~Vilenkin, {MACROSCOPIC PARITY VIOLATING EFFECTS: NEUTRINO FLUXES FROM
  ROTATING BLACK HOLES AND IN ROTATING THERMAL RADIATION}, Phys. Rev. D20
  (1979) 1807--1812.
\newblock \href {https://doi.org/10.1103/PhysRevD.20.1807}
  {\path{doi:10.1103/PhysRevD.20.1807}}.

\bibitem{Erdmenger:2008rm}
J.~Erdmenger, M.~Haack, M.~Kaminski, A.~Yarom, {Fluid dynamics of R-charged
  black holes}, JHEP 01 (2009) 055.
\newblock \href {http://arxiv.org/abs/0809.2488} {\path{arXiv:0809.2488}},
  \href {https://doi.org/10.1088/1126-6708/2009/01/055}
  {\path{doi:10.1088/1126-6708/2009/01/055}}.

\bibitem{Banerjee:2008th}
N.~Banerjee, J.~Bhattacharya, S.~Bhattacharyya, S.~Dutta, R.~Loganayagam,
  P.~Surowka, {Hydrodynamics from charged black branes}, JHEP 01 (2011) 094.
\newblock \href {http://arxiv.org/abs/0809.2596} {\path{arXiv:0809.2596}},
  \href {https://doi.org/10.1007/JHEP01(2011)094}
  {\path{doi:10.1007/JHEP01(2011)094}}.

\bibitem{Son:2009tf}
D.~T. Son, P.~Surowka, {Hydrodynamics with Triangle Anomalies}, Phys. Rev.
  Lett. 103 (2009) 191601.
\newblock \href {http://arxiv.org/abs/0906.5044} {\path{arXiv:0906.5044}},
  \href {https://doi.org/10.1103/PhysRevLett.103.191601}
  {\path{doi:10.1103/PhysRevLett.103.191601}}.

\bibitem{Landsteiner:2011cp}
K.~Landsteiner, E.~Megias, F.~Pena-Benitez, {Gravitational Anomaly and
  Transport}, Phys. Rev. Lett. 107 (2011) 021601.
\newblock \href {http://arxiv.org/abs/1103.5006} {\path{arXiv:1103.5006}},
  \href {https://doi.org/10.1103/PhysRevLett.107.021601}
  {\path{doi:10.1103/PhysRevLett.107.021601}}.

\bibitem{Pu:2010as}
S.~Pu, J.-h. Gao, Q.~Wang, {A consistent description of kinetic equation with
  triangle anomaly}, Phys. Rev. D 83 (2011) 094017.
\newblock \href {http://arxiv.org/abs/1008.2418} {\path{arXiv:1008.2418}},
  \href {https://doi.org/10.1103/PhysRevD.83.094017}
  {\path{doi:10.1103/PhysRevD.83.094017}}.

\bibitem{Sadofyev:2010is}
A.~V. Sadofyev, V.~I. Shevchenko, V.~I. Zakharov, {Notes on chiral
  hydrodynamics within effective theory approach}, Phys. Rev. D 83 (2011)
  105025.
\newblock \href {http://arxiv.org/abs/1012.1958} {\path{arXiv:1012.1958}},
  \href {https://doi.org/10.1103/PhysRevD.83.105025}
  {\path{doi:10.1103/PhysRevD.83.105025}}.

\bibitem{Gao:2012ix}
J.-H. Gao, Z.-T. Liang, S.~Pu, Q.~Wang, X.-N. Wang, {Chiral Anomaly and Local
  Polarization Effect from Quantum Kinetic Approach}, Phys. Rev. Lett. 109
  (2012) 232301.
\newblock \href {http://arxiv.org/abs/1203.0725} {\path{arXiv:1203.0725}},
  \href {https://doi.org/10.1103/PhysRevLett.109.232301}
  {\path{doi:10.1103/PhysRevLett.109.232301}}.

\bibitem{Jiang:2015cva}
Y.~Jiang, X.-G. Huang, J.~Liao, {Chiral vortical wave and induced flavor charge
  transport in a rotating quark-gluon plasma}, Phys. Rev. D 92~(7) (2015)
  071501.
\newblock \href {http://arxiv.org/abs/1504.03201} {\path{arXiv:1504.03201}},
  \href {https://doi.org/10.1103/PhysRevD.92.071501}
  {\path{doi:10.1103/PhysRevD.92.071501}}.

\bibitem{Huang:2013iia}
X.-G. Huang, J.~Liao, {Axial Current Generation from Electric Field: Chiral
  Electric Separation Effect}, Phys. Rev. Lett. 110~(23) (2013) 232302.
\newblock \href {http://arxiv.org/abs/1303.7192} {\path{arXiv:1303.7192}},
  \href {https://doi.org/10.1103/PhysRevLett.110.232302}
  {\path{doi:10.1103/PhysRevLett.110.232302}}.

\bibitem{Pu:2014cwa}
S.~Pu, S.-Y. Wu, D.-L. Yang, {Holographic Chiral Electric Separation Effect},
  Phys. Rev. D89~(8) (2014) 085024.
\newblock \href {http://arxiv.org/abs/1401.6972} {\path{arXiv:1401.6972}},
  \href {https://doi.org/10.1103/PhysRevD.89.085024}
  {\path{doi:10.1103/PhysRevD.89.085024}}.

\bibitem{Jiang:2014ura}
Y.~Jiang, X.-G. Huang, J.~Liao, {Chiral electric separation effect in the
  quark-gluon plasma}, Phys. Rev. D 91~(4) (2015) 045001.
\newblock \href {http://arxiv.org/abs/1409.6395} {\path{arXiv:1409.6395}},
  \href {https://doi.org/10.1103/PhysRevD.91.045001}
  {\path{doi:10.1103/PhysRevD.91.045001}}.

\bibitem{Pu:2014fva}
S.~Pu, S.-Y. Wu, D.-L. Yang, {Chiral Hall Effect and Chiral Electric Waves},
  Phys. Rev. D91~(2) (2015) 025011.
\newblock \href {http://arxiv.org/abs/1407.3168} {\path{arXiv:1407.3168}},
  \href {https://doi.org/10.1103/PhysRevD.91.025011}
  {\path{doi:10.1103/PhysRevD.91.025011}}.

\bibitem{Chen:2016xtg}
J.-W. Chen, T.~Ishii, S.~Pu, N.~Yamamoto, {Nonlinear Chiral Transport
  Phenomena}, Phys. Rev. D 93~(12) (2016) 125023.
\newblock \href {http://arxiv.org/abs/1603.03620} {\path{arXiv:1603.03620}},
  \href {https://doi.org/10.1103/PhysRevD.93.125023}
  {\path{doi:10.1103/PhysRevD.93.125023}}.

\bibitem{Gorbar:2016qfh}
E.~V. Gorbar, I.~A. Shovkovy, S.~Vilchinskii, I.~Rudenok, A.~Boyarsky,
  O.~Ruchayskiy, {Anomalous Maxwell equations for inhomogeneous chiral plasma},
  Phys. Rev. D 93~(10) (2016) 105028.
\newblock \href {http://arxiv.org/abs/1603.03442} {\path{arXiv:1603.03442}},
  \href {https://doi.org/10.1103/PhysRevD.93.105028}
  {\path{doi:10.1103/PhysRevD.93.105028}}.

\bibitem{Gorbar:2016sey}
E.~V. Gorbar, V.~A. Miransky, I.~A. Shovkovy, P.~O. Sukhachov, {Chiral magnetic
  plasmons in anomalous relativistic matter}, Phys. Rev. B 95~(11) (2017)
  115202.
\newblock \href {http://arxiv.org/abs/1611.05470} {\path{arXiv:1611.05470}},
  \href {https://doi.org/10.1103/PhysRevB.95.115202}
  {\path{doi:10.1103/PhysRevB.95.115202}}.

\bibitem{Gorbar:2016ygi}
E.~V. Gorbar, V.~A. Miransky, I.~A. Shovkovy, P.~O. Sukhachov, {Consistent
  Chiral Kinetic Theory in Weyl Materials: Chiral Magnetic Plasmons}, Phys.
  Rev. Lett. 118~(12) (2017) 127601.
\newblock \href {http://arxiv.org/abs/1610.07625} {\path{arXiv:1610.07625}},
  \href {https://doi.org/10.1103/PhysRevLett.118.127601}
  {\path{doi:10.1103/PhysRevLett.118.127601}}.

\bibitem{Gorbar:2017cwv}
E.~V. Gorbar, V.~A. Miransky, I.~A. Shovkovy, P.~O. Sukhachov, {Second-order
  chiral kinetic theory: Chiral magnetic and pseudomagnetic waves}, Phys. Rev.
  B 95~(20) (2017) 205141.
\newblock \href {http://arxiv.org/abs/1702.02950} {\path{arXiv:1702.02950}},
  \href {https://doi.org/10.1103/PhysRevB.95.205141}
  {\path{doi:10.1103/PhysRevB.95.205141}}.

\bibitem{Kharzeev:2012ph}
D.~E. Kharzeev, K.~Landsteiner, A.~Schmitt, H.-U. Yee, {'Strongly interacting
  matter in magnetic fields': an overview}, Lect. Notes Phys. 871 (2013) 1--11.
\newblock \href {http://arxiv.org/abs/1211.6245} {\path{arXiv:1211.6245}},
  \href {https://doi.org/10.1007/978-3-642-37305-3_1}
  {\path{doi:10.1007/978-3-642-37305-3_1}}.

\bibitem{Kharzeev:2015znc}
D.~E. Kharzeev, J.~Liao, S.~A. Voloshin, G.~Wang, {Chiral magnetic and vortical
  effects in high energy nuclear collisions: A status report}, Prog. Part.
  Nucl. Phys. 88 (2016) 1--28.
\newblock \href {http://arxiv.org/abs/1511.04050} {\path{arXiv:1511.04050}},
  \href {https://doi.org/10.1016/j.ppnp.2016.01.001}
  {\path{doi:10.1016/j.ppnp.2016.01.001}}.

\bibitem{Liao:2014ava}
J.~Liao, {Anomalous transport effects and possible environmental symmetry
  ‘violation’ in heavy-ion collisions}, Pramana 84~(5) (2015) 901--926.
\newblock \href {http://arxiv.org/abs/1401.2500} {\path{arXiv:1401.2500}},
  \href {https://doi.org/10.1007/s12043-015-0984-x}
  {\path{doi:10.1007/s12043-015-0984-x}}.

\bibitem{Miransky:2015ava}
V.~A. Miransky, I.~A. Shovkovy, {Quantum field theory in a magnetic field: From
  quantum chromodynamics to graphene and Dirac semimetals}, Phys. Rept. 576
  (2015) 1--209.
\newblock \href {http://arxiv.org/abs/1503.00732} {\path{arXiv:1503.00732}},
  \href {https://doi.org/10.1016/j.physrep.2015.02.003}
  {\path{doi:10.1016/j.physrep.2015.02.003}}.

\bibitem{Huang:2015oca}
X.-G. Huang, {Electromagnetic fields and anomalous transports in heavy-ion
  collisions --- A pedagogical review}, Rept. Prog. Phys. 79~(7) (2016) 076302.
\newblock \href {http://arxiv.org/abs/1509.04073} {\path{arXiv:1509.04073}},
  \href {https://doi.org/10.1088/0034-4885/79/7/076302}
  {\path{doi:10.1088/0034-4885/79/7/076302}}.

\bibitem{Fukushima:2018grm}
K.~Fukushima, {Extreme matter in electromagnetic fields and rotation}, Prog.
  Part. Nucl. Phys. 107 (2019) 167--199.
\newblock \href {http://arxiv.org/abs/1812.08886} {\path{arXiv:1812.08886}},
  \href {https://doi.org/10.1016/j.ppnp.2019.04.001}
  {\path{doi:10.1016/j.ppnp.2019.04.001}}.

\bibitem{Bzdak:2019pkr}
A.~Bzdak, S.~Esumi, V.~Koch, J.~Liao, M.~Stephanov, N.~Xu, {Mapping the Phases
  of Quantum Chromodynamics with Beam Energy Scan} (2019).
\newblock \href {http://arxiv.org/abs/1906.00936} {\path{arXiv:1906.00936}}.

\bibitem{Zhao:2019hta}
J.~Zhao, F.~Wang, {Experimental searches for the chiral magnetic effect in
  heavy-ion collisions}, Prog. Part. Nucl. Phys. 107 (2019) 200--236.
\newblock \href {http://arxiv.org/abs/1906.11413} {\path{arXiv:1906.11413}},
  \href {https://doi.org/10.1016/j.ppnp.2019.05.001}
  {\path{doi:10.1016/j.ppnp.2019.05.001}}.

\bibitem{Gao:2020vbh}
J.-H. Gao, G.-L. Ma, S.~Pu, Q.~Wang, {Recent developments in chiral and spin
  polarization effects in heavy-ion collisions}, Nucl. Sci. Tech. 31~(9) (2020)
  90.
\newblock \href {http://arxiv.org/abs/2005.10432} {\path{arXiv:2005.10432}},
  \href {https://doi.org/10.1007/s41365-020-00801-x}
  {\path{doi:10.1007/s41365-020-00801-x}}.

\bibitem{Shovkovy:2021yyw}
I.~A. Shovkovy, {Anomalous plasma: chiral magnetic effect and all that} (11
  2021).
\newblock \href {http://arxiv.org/abs/2111.11416} {\path{arXiv:2111.11416}}.

\bibitem{STAR:2009wot}
B.~I. Abelev, et~al., {Azimuthal Charged-Particle Correlations and Possible
  Local Strong Parity Violation}, Phys. Rev. Lett. 103 (2009) 251601.
\newblock \href {http://arxiv.org/abs/0909.1739} {\path{arXiv:0909.1739}},
  \href {https://doi.org/10.1103/PhysRevLett.103.251601}
  {\path{doi:10.1103/PhysRevLett.103.251601}}.

\bibitem{STAR:2009tro}
B.~I. Abelev, et~al., {Observation of charge-dependent azimuthal correlations
  and possible local strong parity violation in heavy ion collisions}, Phys.
  Rev. C 81 (2010) 054908.
\newblock \href {http://arxiv.org/abs/0909.1717} {\path{arXiv:0909.1717}},
  \href {https://doi.org/10.1103/PhysRevC.81.054908}
  {\path{doi:10.1103/PhysRevC.81.054908}}.

\bibitem{Wang:2012qs}
G.~Wang, {Search for Chiral Magnetic Effects in High-Energy Nuclear
  Collisions}, Nucl. Phys. A 904-905 (2013) 248c--255c.
\newblock \href {http://arxiv.org/abs/1210.5498} {\path{arXiv:1210.5498}},
  \href {https://doi.org/10.1016/j.nuclphysa.2013.01.069}
  {\path{doi:10.1016/j.nuclphysa.2013.01.069}}.

\bibitem{STAR:2013ksd}
L.~Adamczyk, et~al., {Fluctuations of charge separation perpendicular to the
  event plane and local parity violation in $\sqrt{s_{NN}}=200$ GeV Au+Au
  collisions at the BNL Relativistic Heavy Ion Collider}, Phys. Rev. C 88~(6)
  (2013) 064911.
\newblock \href {http://arxiv.org/abs/1302.3802} {\path{arXiv:1302.3802}},
  \href {https://doi.org/10.1103/PhysRevC.88.064911}
  {\path{doi:10.1103/PhysRevC.88.064911}}.

\bibitem{STAR:2013zgu}
L.~Adamczyk, et~al., {Measurement of charge multiplicity asymmetry correlations
  in high-energy nucleus-nucleus collisions at $\sqrt{{s}_{NN}} =$ 200 GeV},
  Phys. Rev. C 89~(4) (2014) 044908.
\newblock \href {http://arxiv.org/abs/1303.0901} {\path{arXiv:1303.0901}},
  \href {https://doi.org/10.1103/PhysRevC.89.044908}
  {\path{doi:10.1103/PhysRevC.89.044908}}.

\bibitem{STAR:2014uiw}
L.~Adamczyk, et~al., {Beam-energy dependence of charge separation along the
  magnetic field in Au+Au collisions at RHIC}, Phys. Rev. Lett. 113 (2014)
  052302.
\newblock \href {http://arxiv.org/abs/1404.1433} {\path{arXiv:1404.1433}},
  \href {https://doi.org/10.1103/PhysRevLett.113.052302}
  {\path{doi:10.1103/PhysRevLett.113.052302}}.

\bibitem{Tribedy:2017hwn}
P.~Tribedy, {Disentangling flow and signals of Chiral Magnetic Effect in U+U,
  Au+Au and p+Au collisions}, Nucl. Phys. A 967 (2017) 740--743.
\newblock \href {http://arxiv.org/abs/1704.03845} {\path{arXiv:1704.03845}},
  \href {https://doi.org/10.1016/j.nuclphysa.2017.05.078}
  {\path{doi:10.1016/j.nuclphysa.2017.05.078}}.

\bibitem{STAR:2019xzd}
J.~Adam, et~al., {Charge-dependent pair correlations relative to a third
  particle in $p$ + Au and $d$+ Au collisions at RHIC}, Phys. Lett. B 798
  (2019) 134975.
\newblock \href {http://arxiv.org/abs/1906.03373} {\path{arXiv:1906.03373}},
  \href {https://doi.org/10.1016/j.physletb.2019.134975}
  {\path{doi:10.1016/j.physletb.2019.134975}}.

\bibitem{ALICE:2012nhw}
B.~Abelev, et~al., {Charge separation relative to the reaction plane in Pb-Pb
  collisions at $\sqrt{s_{NN}}= 2.76$ TeV}, Phys. Rev. Lett. 110~(1) (2013)
  012301.
\newblock \href {http://arxiv.org/abs/1207.0900} {\path{arXiv:1207.0900}},
  \href {https://doi.org/10.1103/PhysRevLett.110.012301}
  {\path{doi:10.1103/PhysRevLett.110.012301}}.

\bibitem{CMS:2016wfo}
V.~Khachatryan, et~al., {Observation of charge-dependent azimuthal correlations
  in $p$-Pb collisions and its implication for the search for the chiral
  magnetic effect}, Phys. Rev. Lett. 118~(12) (2017) 122301.
\newblock \href {http://arxiv.org/abs/1610.00263} {\path{arXiv:1610.00263}},
  \href {https://doi.org/10.1103/PhysRevLett.118.122301}
  {\path{doi:10.1103/PhysRevLett.118.122301}}.

\bibitem{CMS:2017lrw}
A.~M. Sirunyan, et~al., {Constraints on the chiral magnetic effect using
  charge-dependent azimuthal correlations in $p\mathrm{Pb}$ and PbPb collisions
  at the CERN Large Hadron Collider}, Phys. Rev. C 97~(4) (2018) 044912.
\newblock \href {http://arxiv.org/abs/1708.01602} {\path{arXiv:1708.01602}},
  \href {https://doi.org/10.1103/PhysRevC.97.044912}
  {\path{doi:10.1103/PhysRevC.97.044912}}.

\bibitem{Li:2020dwr}
W.~Li, G.~Wang, {Chiral Magnetic Effects in Nuclear Collisions}, Ann. Rev.
  Nucl. Part. Sci. 70 (2020) 293--321.
\newblock \href {http://arxiv.org/abs/2002.10397} {\path{arXiv:2002.10397}},
  \href {https://doi.org/10.1146/annurev-nucl-030220-065203}
  {\path{doi:10.1146/annurev-nucl-030220-065203}}.

\bibitem{Voloshin:2004vk}
S.~A. Voloshin, {Parity violation in hot QCD: How to detect it}, Phys. Rev. C
  70 (2004) 057901.
\newblock \href {http://arxiv.org/abs/hep-ph/0406311}
  {\path{arXiv:hep-ph/0406311}}, \href
  {https://doi.org/10.1103/PhysRevC.70.057901}
  {\path{doi:10.1103/PhysRevC.70.057901}}.

\bibitem{Deng:2016knn}
W.-T. Deng, X.-G. Huang, G.-L. Ma, G.~Wang, {Test the chiral magnetic effect
  with isobaric collisions}, Phys. Rev. C 94 (2016) 041901.
\newblock \href {http://arxiv.org/abs/1607.04697} {\path{arXiv:1607.04697}},
  \href {https://doi.org/10.1103/PhysRevC.94.041901}
  {\path{doi:10.1103/PhysRevC.94.041901}}.

\bibitem{Zhao:2019crj}
X.-L. Zhao, G.-L. Ma, Y.-G. Ma, {Impact of magnetic-field fluctuations on
  measurements of the chiral magnetic effect in collisions of isobaric nuclei},
  Phys. Rev. C 99~(3) (2019) 034903.
\newblock \href {http://arxiv.org/abs/1901.04151} {\path{arXiv:1901.04151}},
  \href {https://doi.org/10.1103/PhysRevC.99.034903}
  {\path{doi:10.1103/PhysRevC.99.034903}}.

\bibitem{Pratt:2010zn}
S.~Pratt, S.~Schlichting, S.~Gavin, {Effects of Momentum Conservation and Flow
  on Angular Correlations at RHIC}, Phys. Rev. C 84 (2011) 024909.
\newblock \href {http://arxiv.org/abs/1011.6053} {\path{arXiv:1011.6053}},
  \href {https://doi.org/10.1103/PhysRevC.84.024909}
  {\path{doi:10.1103/PhysRevC.84.024909}}.

\bibitem{Bzdak:2012ia}
A.~Bzdak, V.~Koch, J.~Liao, {Charge-Dependent Correlations in Relativistic
  Heavy Ion Collisions and the Chiral Magnetic Effect}, Lect. Notes Phys. 871
  (2013) 503--536.
\newblock \href {http://arxiv.org/abs/1207.7327} {\path{arXiv:1207.7327}},
  \href {https://doi.org/10.1007/978-3-642-37305-3_19}
  {\path{doi:10.1007/978-3-642-37305-3_19}}.

\bibitem{Schlichting:2010qia}
S.~Schlichting, S.~Pratt, {Charge conservation at energies available at the BNL
  Relativistic Heavy Ion Collider and contributions to local parity violation
  observables}, Phys. Rev. C 83 (2011) 014913.
\newblock \href {http://arxiv.org/abs/1009.4283} {\path{arXiv:1009.4283}},
  \href {https://doi.org/10.1103/PhysRevC.83.014913}
  {\path{doi:10.1103/PhysRevC.83.014913}}.

\bibitem{Xu:2017zcn}
H.-J. Xu, X.~Wang, H.~Li, J.~Zhao, Z.-W. Lin, C.~Shen, F.~Wang, {Importance of
  isobar density distributions on the chiral magnetic effect search}, Phys.
  Rev. Lett. 121~(2) (2018) 022301.
\newblock \href {http://arxiv.org/abs/1710.03086} {\path{arXiv:1710.03086}},
  \href {https://doi.org/10.1103/PhysRevLett.121.022301}
  {\path{doi:10.1103/PhysRevLett.121.022301}}.

\bibitem{Xu:2020sln}
H.-j. Xu, J.~Zhao, Y.~Feng, F.~Wang, {Importance of non-flow background on the
  chiral magnetic wave search}, Nucl. Phys. A 1005 (2021) 121770.
\newblock \href {http://arxiv.org/abs/2002.05220} {\path{arXiv:2002.05220}},
  \href {https://doi.org/10.1016/j.nuclphysa.2020.121770}
  {\path{doi:10.1016/j.nuclphysa.2020.121770}}.

\bibitem{Ajitanand:2010rc}
N.~N. Ajitanand, R.~A. Lacey, A.~Taranenko, J.~M. Alexander, {A New method for
  the experimental study of topological effects in the quark-gluon plasma},
  Phys. Rev. C 83 (2011) 011901.
\newblock \href {http://arxiv.org/abs/1009.5624} {\path{arXiv:1009.5624}},
  \href {https://doi.org/10.1103/PhysRevC.83.011901}
  {\path{doi:10.1103/PhysRevC.83.011901}}.

\bibitem{Magdy:2017yje}
N.~Magdy, S.~Shi, J.~Liao, N.~Ajitanand, R.~A. Lacey, {New correlator to detect
  and characterize the chiral magnetic effect}, Phys. Rev. C 97~(6) (2018)
  061901.
\newblock \href {http://arxiv.org/abs/1710.01717} {\path{arXiv:1710.01717}},
  \href {https://doi.org/10.1103/PhysRevC.97.061901}
  {\path{doi:10.1103/PhysRevC.97.061901}}.

\bibitem{Tang:2019pbl}
A.~H. Tang, {Probe chiral magnetic effect with signed balance function}, Chin.
  Phys. C 44~(5) (2020) 054101.
\newblock \href {http://arxiv.org/abs/1903.04622} {\path{arXiv:1903.04622}},
  \href {https://doi.org/10.1088/1674-1137/44/5/054101}
  {\path{doi:10.1088/1674-1137/44/5/054101}}.

\bibitem{Voloshin:2010ut}
S.~A. Voloshin, {Testing the Chiral Magnetic Effect with Central U+U
  collisions}, Phys. Rev. Lett. 105 (2010) 172301.
\newblock \href {http://arxiv.org/abs/1006.1020} {\path{arXiv:1006.1020}},
  \href {https://doi.org/10.1103/PhysRevLett.105.172301}
  {\path{doi:10.1103/PhysRevLett.105.172301}}.

\bibitem{STAR:2021mii}
M.~Abdallah, et~al., {Search for the Chiral Magnetic Effect with Isobar
  Collisions at $\sqrt{s_{NN}}$ = 200 GeV by the STAR Collaboration at RHIC} (8
  2021).
\newblock \href {http://arxiv.org/abs/2109.00131} {\path{arXiv:2109.00131}}.

\bibitem{Li:2014bha}
Q.~Li, D.~E. Kharzeev, C.~Zhang, Y.~Huang, I.~Pletikosic, A.~V. Fedorov, R.~D.
  Zhong, J.~A. Schneeloch, G.~D. Gu, T.~Valla, {Observation of the chiral
  magnetic effect in ZrTe5}, Nature Phys. 12 (2016) 550--554.
\newblock \href {http://arxiv.org/abs/1412.6543} {\path{arXiv:1412.6543}},
  \href {https://doi.org/10.1038/nphys3648} {\path{doi:10.1038/nphys3648}}.

\bibitem{CMEScience}
J.~Xiong, S.~K. Kushwaha, T.~Liang, J.~W. Krizan, M.~Hirschberger, W.~Wang,
  R.~J. Cava, N.~P. Ong,
  \href{https://www.science.org/doi/abs/10.1126/science.aac6089}{Evidence for
  the chiral anomaly in the dirac semimetal na<sub>3</sub>bi}, Science
  350~(6259) (2015) 413--416.
\newblock \href
  {http://arxiv.org/abs/https://www.science.org/doi/pdf/10.1126/science.aac6089}
  {\path{arXiv:https://www.science.org/doi/pdf/10.1126/science.aac6089}}, \href
  {https://doi.org/10.1126/science.aac6089}
  {\path{doi:10.1126/science.aac6089}}.
\newline\urlprefix\url{https://www.science.org/doi/abs/10.1126/science.aac6089}

\bibitem{Feng:PhysRevB2015}
J.~Feng, Y.~Pang, D.~Wu, Z.~Wang, H.~Weng, J.~Li, X.~Dai, Z.~Fang, Y.~Shi,
  L.~Lu, \href{https://link.aps.org/doi/10.1103/PhysRevB.92.081306}{Large
  linear magnetoresistance in dirac semimetal
  ${\mathrm{cd}}_{3}{\mathrm{as}}_{2}$ with fermi surfaces close to the dirac
  points}, Phys. Rev. B 92 (2015) 081306.
\newblock \href {https://doi.org/10.1103/PhysRevB.92.081306}
  {\path{doi:10.1103/PhysRevB.92.081306}}.
\newline\urlprefix\url{https://link.aps.org/doi/10.1103/PhysRevB.92.081306}

\bibitem{Li:NC2015}
C.-Z. Li, L.-X. Wang, H.~Liu, J.~Wang, Z.-M. Liao, D.-P. Yu,
  \href{https://doi.org/10.1038/ncomms10137}{Giant negative magnetoresistance
  induced by the chiral anomaly in individual cd3as2 nanowires}, Nature
  Communications 6~(1) (2015) 10137.
\newblock \href {https://doi.org/10.1038/ncomms10137}
  {\path{doi:10.1038/ncomms10137}}.
\newline\urlprefix\url{https://doi.org/10.1038/ncomms10137}

\bibitem{Li:NC2016}
H.~Li, H.~He, H.-Z. Lu, H.~Zhang, H.~Liu, R.~Ma, Z.~Fan, S.-Q. Shen, J.~Wang,
  \href{https://doi.org/10.1038/ncomms10301}{Negative magnetoresistance in
  dirac semimetal cd3as2}, Nature Communications 7~(1) (2016) 10301.
\newblock \href {https://doi.org/10.1038/ncomms10301}
  {\path{doi:10.1038/ncomms10301}}.
\newline\urlprefix\url{https://doi.org/10.1038/ncomms10301}

\bibitem{Arnold:2015vvs}
F.~Arnold, et~al., {Negative magnetoresistance without well-defined chirality
  in the Weyl semimetal TaP}, Nature Commun. 7 (2016) 1615.
\newblock \href {http://arxiv.org/abs/1506.06577} {\path{arXiv:1506.06577}},
  \href {https://doi.org/10.1038/ncomms11615} {\path{doi:10.1038/ncomms11615}}.

\bibitem{Huang:2015eia}
X.~Huang, et~al., {Observation of the Chiral-Anomaly-Induced Negative
  Magnetoresistance in 3D Weyl Semimetal TaAs}, Phys. Rev. X 5~(3) (2015)
  031023.
\newblock \href {http://arxiv.org/abs/1503.01304} {\path{arXiv:1503.01304}},
  \href {https://doi.org/10.1103/PhysRevX.5.031023}
  {\path{doi:10.1103/PhysRevX.5.031023}}.

\bibitem{Zhang:NC2016}
C.-L. Zhang, S.-Y. Xu, I.~Belopolski, Z.~Yuan, Z.~Lin, B.~Tong, G.~Bian,
  N.~Alidoust, C.-C. Lee, S.-M. Huang, T.-R. Chang, G.~Chang, C.-H. Hsu, H.-T.
  Jeng, M.~Neupane, D.~S. Sanchez, H.~Zheng, J.~Wang, H.~Lin, C.~Zhang, H.-Z.
  Lu, S.-Q. Shen, T.~Neupert, M.~Zahid~Hasan, S.~Jia,
  \href{https://doi.org/10.1038/ncomms10735}{Signatures of the
  adler--bell--jackiw chiral anomaly in a weyl fermion semimetal}, Nature
  Communications 7~(1) (2016) 10735.
\newblock \href {https://doi.org/10.1038/ncomms10735}
  {\path{doi:10.1038/ncomms10735}}.
\newline\urlprefix\url{https://doi.org/10.1038/ncomms10735}

\bibitem{Hirschberger:NM2016}
M.~Hirschberger, S.~Kushwaha, Z.~Wang, Q.~Gibson, S.~Liang, C.~A. Belvin, B.~A.
  Bernevig, R.~J. Cava, N.~P. Ong, \href{https://doi.org/10.1038/nmat4684}{The
  chiral anomaly and thermopower of weyl fermions in the half-heusler gdptbi},
  Nature Materials 15~(11) (2016) 1161--1165.
\newblock \href {https://doi.org/10.1038/nmat4684}
  {\path{doi:10.1038/nmat4684}}.
\newline\urlprefix\url{https://doi.org/10.1038/nmat4684}

\bibitem{Wang:PhysRevB2016}
Z.~Wang, Y.~Zheng, Z.~Shen, Y.~Lu, H.~Fang, F.~Sheng, Y.~Zhou, X.~Yang, Y.~Li,
  C.~Feng, Z.-A. Xu,
  \href{https://link.aps.org/doi/10.1103/PhysRevB.93.121112}{Helicity-protected
  ultrahigh mobility weyl fermions in nbp}, Phys. Rev. B 93 (2016) 121112.
\newblock \href {https://doi.org/10.1103/PhysRevB.93.121112}
  {\path{doi:10.1103/PhysRevB.93.121112}}.
\newline\urlprefix\url{https://link.aps.org/doi/10.1103/PhysRevB.93.121112}

\bibitem{Du:SCP2016}
J.~Du, H.~Wang, Q.~Chen, Q.~Mao, R.~Khan, B.~Xu, Y.~Zhou, Y.~Zhang, J.~Yang,
  B.~Chen, C.~Feng, M.~Fang,
  \href{https://doi.org/10.1007/s11433-016-5798-4}{Large unsaturated positive
  and negative magnetoresistance in weyl semimetal tap}, Science China Physics,
  Mechanics \& Astronomy 59~(5) (2016) 657406.
\newblock \href {https://doi.org/10.1007/s11433-016-5798-4}
  {\path{doi:10.1007/s11433-016-5798-4}}.
\newline\urlprefix\url{https://doi.org/10.1007/s11433-016-5798-4}

\bibitem{Li:PNAS2018}
Y.~Li, Q.~Gu, C.~Chen, J.~Zhang, Q.~Liu, X.~Hu, J.~Liu, Y.~Liu, L.~Ling,
  M.~Tian, Y.~Wang, N.~Samarth, S.~Li, T.~Zhang, J.~Feng, J.~Wang,
  \href{https://www.pnas.org/content/115/38/9503}{Nontrivial superconductivity
  in topological mote2-xsx crystals}, Proceedings of the National Academy of
  Sciences 115~(38) (2018) 9503--9508.
\newblock \href
  {http://arxiv.org/abs/https://www.pnas.org/content/115/38/9503.full.pdf}
  {\path{arXiv:https://www.pnas.org/content/115/38/9503.full.pdf}}, \href
  {https://doi.org/10.1073/pnas.1801650115}
  {\path{doi:10.1073/pnas.1801650115}}.
\newline\urlprefix\url{https://www.pnas.org/content/115/38/9503}

\bibitem{Kharzeev:2019ceh}
D.~E. Kharzeev, Q.~Li, {The Chiral Qubit: quantum computing with chiral
  anomaly} (2019).
\newblock \href {http://arxiv.org/abs/1903.07133} {\path{arXiv:1903.07133}}.

\bibitem{Lublinsky:2009wr}
M.~Lublinsky, I.~Zahed, {Anomalous Chiral Superfluidity}, Phys. Lett. B 684
  (2010) 119--122.
\newblock \href {http://arxiv.org/abs/0910.1373} {\path{arXiv:0910.1373}},
  \href {https://doi.org/10.1016/j.physletb.2010.01.015}
  {\path{doi:10.1016/j.physletb.2010.01.015}}.

\bibitem{Sadofyev:2010pr}
A.~V. Sadofyev, M.~V. Isachenkov, {The Chiral magnetic effect in hydrodynamical
  approach}, Phys. Lett. B 697 (2011) 404--406.
\newblock \href {http://arxiv.org/abs/1010.1550} {\path{arXiv:1010.1550}},
  \href {https://doi.org/10.1016/j.physletb.2011.02.041}
  {\path{doi:10.1016/j.physletb.2011.02.041}}.

\bibitem{Kharzeev:2010gr}
D.~E. Kharzeev, D.~T. Son, {Testing the chiral magnetic and chiral vortical
  effects in heavy ion collisions}, Phys. Rev. Lett. 106 (2011) 062301.
\newblock \href {http://arxiv.org/abs/1010.0038} {\path{arXiv:1010.0038}},
  \href {https://doi.org/10.1103/PhysRevLett.106.062301}
  {\path{doi:10.1103/PhysRevLett.106.062301}}.

\bibitem{Jensen:2012jy}
K.~Jensen, {Triangle Anomalies, Thermodynamics, and Hydrodynamics}, Phys. Rev.
  D 85 (2012) 125017.
\newblock \href {http://arxiv.org/abs/1203.3599} {\path{arXiv:1203.3599}},
  \href {https://doi.org/10.1103/PhysRevD.85.125017}
  {\path{doi:10.1103/PhysRevD.85.125017}}.

\bibitem{Jensen:2012jh}
K.~Jensen, M.~Kaminski, P.~Kovtun, R.~Meyer, A.~Ritz, A.~Yarom, {Towards
  hydrodynamics without an entropy current}, Phys. Rev. Lett. 109 (2012)
  101601.
\newblock \href {http://arxiv.org/abs/1203.3556} {\path{arXiv:1203.3556}},
  \href {https://doi.org/10.1103/PhysRevLett.109.101601}
  {\path{doi:10.1103/PhysRevLett.109.101601}}.

\bibitem{Pu:2012wn}
S.~Pu, J.-h. Gao, {Induced anomalous current in relativistic hydrodynamics with
  chiral anomaly}, Central Eur. J. Phys. 10 (2012) 1258--1260.
\newblock \href {https://doi.org/10.2478/s11534-012-0142-8}
  {\path{doi:10.2478/s11534-012-0142-8}}.

\bibitem{Kalaydzhyan:2016dyr}
T.~Kalaydzhyan, E.~Murchikova, {Thermal chiral vortical and magnetic waves: new
  excitation modes in chiral fluids}, Nucl. Phys. B 919 (2017) 173--181.
\newblock \href {http://arxiv.org/abs/1609.00024} {\path{arXiv:1609.00024}},
  \href {https://doi.org/10.1016/j.nuclphysb.2017.03.019}
  {\path{doi:10.1016/j.nuclphysb.2017.03.019}}.

\bibitem{Hattori:2017usa}
K.~Hattori, Y.~Hirono, H.-U. Yee, Y.~Yin, {MagnetoHydrodynamics with chiral
  anomaly: phases of collective excitations and instabilities}, Phys. Rev. D
  100~(6) (2019) 065023.
\newblock \href {http://arxiv.org/abs/1711.08450} {\path{arXiv:1711.08450}},
  \href {https://doi.org/10.1103/PhysRevD.100.065023}
  {\path{doi:10.1103/PhysRevD.100.065023}}.

\bibitem{Ammon:2020rvg}
M.~Ammon, S.~Grieninger, J.~Hernandez, M.~Kaminski, R.~Koirala, J.~Leiber,
  J.~Wu, {Chiral hydrodynamics in strong external magnetic fields}, JHEP 04
  (2021) 078.
\newblock \href {http://arxiv.org/abs/2012.09183} {\path{arXiv:2012.09183}},
  \href {https://doi.org/10.1007/JHEP04(2021)078}
  {\path{doi:10.1007/JHEP04(2021)078}}.

\bibitem{Speranza:2021bxf}
E.~Speranza, F.~S. Bemfica, M.~M. Disconzi, J.~Noronha, {Challenges in Solving
  Chiral Hydrodynamics} (4 2021).
\newblock \href {http://arxiv.org/abs/2104.02110} {\path{arXiv:2104.02110}}.

\bibitem{Jiang:2016wve}
Y.~Jiang, S.~Shi, Y.~Yin, J.~Liao, {Quantifying the chiral magnetic effect from
  anomalous-viscous fluid dynamics}, Chin. Phys. C42~(1) (2018) 011001.
\newblock \href {http://arxiv.org/abs/1611.04586} {\path{arXiv:1611.04586}},
  \href {https://doi.org/10.1088/1674-1137/42/1/011001}
  {\path{doi:10.1088/1674-1137/42/1/011001}}.

\bibitem{Shi:2017cpu}
S.~Shi, Y.~Jiang, E.~Lilleskov, J.~Liao, {Anomalous Chiral Transport in Heavy
  Ion Collisions from Anomalous-Viscous Fluid Dynamics}, Annals Phys. 394
  (2018) 50--72.
\newblock \href {http://arxiv.org/abs/1711.02496} {\path{arXiv:1711.02496}},
  \href {https://doi.org/10.1016/j.aop.2018.04.026}
  {\path{doi:10.1016/j.aop.2018.04.026}}.

\bibitem{Shi:2017ucn}
S.~Shi, Y.~Jiang, E.~Lilleskov, J.~Liao, {Quantification of Chiral Magnetic
  Effect from Event-by-Event Anomalous-Viscous Fluid Mechanics}, PoS CPOD2017
  (2018) 021.
\newblock \href {http://arxiv.org/abs/1712.01386} {\path{arXiv:1712.01386}},
  \href {https://doi.org/10.22323/1.311.0021} {\path{doi:10.22323/1.311.0021}}.

\bibitem{Shi:2018sah}
S.~Shi, H.~Zhang, D.~Hou, J.~Liao, {Chiral Magnetic Effect in Isobaric
  Collisions from Anomalous-Viscous Fluid Dynamics (AVFD)}, Nucl. Phys. A982
  (2019) 539--542.
\newblock \href {http://arxiv.org/abs/1807.05604} {\path{arXiv:1807.05604}},
  \href {https://doi.org/10.1016/j.nuclphysa.2018.10.007}
  {\path{doi:10.1016/j.nuclphysa.2018.10.007}}.

\bibitem{Shi:2019wzi}
S.~Shi, H.~Zhang, D.~Hou, J.~Liao, {Signatures of Chiral Magnetic Effect in the
  Collisions of Isobars} (2019).
\newblock \href {http://arxiv.org/abs/1910.14010} {\path{arXiv:1910.14010}}.

\bibitem{Son:2012wh}
D.~T. Son, N.~Yamamoto, {Berry Curvature, Triangle Anomalies, and the Chiral
  Magnetic Effect in Fermi Liquids}, Phys. Rev. Lett. 109 (2012) 181602.
\newblock \href {http://arxiv.org/abs/1203.2697} {\path{arXiv:1203.2697}},
  \href {https://doi.org/10.1103/PhysRevLett.109.181602}
  {\path{doi:10.1103/PhysRevLett.109.181602}}.

\bibitem{Son:2012zy}
D.~T. Son, N.~Yamamoto, {Kinetic theory with Berry curvature from quantum field
  theories}, Phys. Rev. D 87~(8) (2013) 085016.
\newblock \href {http://arxiv.org/abs/1210.8158} {\path{arXiv:1210.8158}},
  \href {https://doi.org/10.1103/PhysRevD.87.085016}
  {\path{doi:10.1103/PhysRevD.87.085016}}.

\bibitem{Stephanov:2012ki}
M.~A. Stephanov, Y.~Yin, {Chiral Kinetic Theory}, Phys. Rev. Lett. 109 (2012)
  162001.
\newblock \href {http://arxiv.org/abs/1207.0747} {\path{arXiv:1207.0747}},
  \href {https://doi.org/10.1103/PhysRevLett.109.162001}
  {\path{doi:10.1103/PhysRevLett.109.162001}}.

\bibitem{Chen:2013iga}
J.-W. Chen, J.-y. Pang, S.~Pu, Q.~Wang, {Kinetic equations for massive Dirac
  fermions in electromagnetic field with non-Abelian Berry phase}, Phys. Rev. D
  89~(9) (2014) 094003.
\newblock \href {http://arxiv.org/abs/1312.2032} {\path{arXiv:1312.2032}},
  \href {https://doi.org/10.1103/PhysRevD.89.094003}
  {\path{doi:10.1103/PhysRevD.89.094003}}.

\bibitem{Chen:2014cla}
J.-Y. Chen, D.~T. Son, M.~A. Stephanov, H.-U. Yee, Y.~Yin, {Lorentz Invariance
  in Chiral Kinetic Theory}, Phys. Rev. Lett. 113~(18) (2014) 182302.
\newblock \href {http://arxiv.org/abs/1404.5963} {\path{arXiv:1404.5963}},
  \href {https://doi.org/10.1103/PhysRevLett.113.182302}
  {\path{doi:10.1103/PhysRevLett.113.182302}}.

\bibitem{Chen:2015gta}
J.-Y. Chen, D.~T. Son, M.~A. Stephanov, {Collisions in Chiral Kinetic Theory},
  Phys. Rev. Lett. 115~(2) (2015) 021601.
\newblock \href {http://arxiv.org/abs/1502.06966} {\path{arXiv:1502.06966}},
  \href {https://doi.org/10.1103/PhysRevLett.115.021601}
  {\path{doi:10.1103/PhysRevLett.115.021601}}.

\bibitem{Chen:2012ca}
J.-W. Chen, S.~Pu, Q.~Wang, X.-N. Wang, {Berry Curvature and Four-Dimensional
  Monopoles in the Relativistic Chiral Kinetic Equation}, Phys. Rev. Lett.
  110~(26) (2013) 262301.
\newblock \href {http://arxiv.org/abs/1210.8312} {\path{arXiv:1210.8312}},
  \href {https://doi.org/10.1103/PhysRevLett.110.262301}
  {\path{doi:10.1103/PhysRevLett.110.262301}}.

\bibitem{Hidaka:2016yjf}
Y.~Hidaka, S.~Pu, D.-L. Yang, {Relativistic Chiral Kinetic Theory from Quantum
  Field Theories}, Phys. Rev. D 95~(9) (2017) 091901.
\newblock \href {http://arxiv.org/abs/1612.04630} {\path{arXiv:1612.04630}},
  \href {https://doi.org/10.1103/PhysRevD.95.091901}
  {\path{doi:10.1103/PhysRevD.95.091901}}.

\bibitem{Huang:2018wdl}
A.~Huang, S.~Shi, Y.~Jiang, J.~Liao, P.~Zhuang, {Complete and Consistent Chiral
  Transport from Wigner Function Formalism}, Phys. Rev. D 98~(3) (2018) 036010.
\newblock \href {http://arxiv.org/abs/1801.03640} {\path{arXiv:1801.03640}},
  \href {https://doi.org/10.1103/PhysRevD.98.036010}
  {\path{doi:10.1103/PhysRevD.98.036010}}.

\bibitem{Mueller:2017arw}
N.~Mueller, R.~Venugopalan, {Worldline construction of a covariant chiral
  kinetic theory}, Phys. Rev. D 96~(1) (2017) 016023.
\newblock \href {http://arxiv.org/abs/1702.01233} {\path{arXiv:1702.01233}},
  \href {https://doi.org/10.1103/PhysRevD.96.016023}
  {\path{doi:10.1103/PhysRevD.96.016023}}.

\bibitem{Mueller:2017lzw}
N.~Mueller, R.~Venugopalan, {The chiral anomaly, Berry's phase and chiral
  kinetic theory, from world-lines in quantum field theory}, Phys. Rev. D
  97~(5) (2018) 051901.
\newblock \href {http://arxiv.org/abs/1701.03331} {\path{arXiv:1701.03331}},
  \href {https://doi.org/10.1103/PhysRevD.97.051901}
  {\path{doi:10.1103/PhysRevD.97.051901}}.

\bibitem{Manuel:2013zaa}
C.~Manuel, J.~M. Torres-Rincon, {Kinetic theory of chiral relativistic plasmas
  and energy density of their gauge collective excitations}, Phys. Rev. D
  89~(9) (2014) 096002.
\newblock \href {http://arxiv.org/abs/1312.1158} {\path{arXiv:1312.1158}},
  \href {https://doi.org/10.1103/PhysRevD.89.096002}
  {\path{doi:10.1103/PhysRevD.89.096002}}.

\bibitem{Manuel:2014dza}
C.~Manuel, J.~M. Torres-Rincon, {Chiral transport equation from the quantum
  Dirac Hamiltonian and the on-shell effective field theory}, Phys. Rev. D
  90~(7) (2014) 076007.
\newblock \href {http://arxiv.org/abs/1404.6409} {\path{arXiv:1404.6409}},
  \href {https://doi.org/10.1103/PhysRevD.90.076007}
  {\path{doi:10.1103/PhysRevD.90.076007}}.

\bibitem{Carignano:2018gqt}
S.~Carignano, C.~Manuel, J.~M. Torres-Rincon, {Consistent relativistic chiral
  kinetic theory: A derivation from on-shell effective field theory}, Phys.
  Rev. D 98~(7) (2018) 076005.
\newblock \href {http://arxiv.org/abs/1806.01684} {\path{arXiv:1806.01684}},
  \href {https://doi.org/10.1103/PhysRevD.98.076005}
  {\path{doi:10.1103/PhysRevD.98.076005}}.

\bibitem{Carignano:2019zsh}
S.~Carignano, C.~Manuel, J.~M. Torres-Rincon, {Chiral kinetic theory from the
  on-shell effective field theory: Derivation of collision terms}, Phys. Rev. D
  102~(1) (2020) 016003.
\newblock \href {http://arxiv.org/abs/1908.00561} {\path{arXiv:1908.00561}},
  \href {https://doi.org/10.1103/PhysRevD.102.016003}
  {\path{doi:10.1103/PhysRevD.102.016003}}.

\bibitem{Lin:2018aon}
S.~Lin, L.~Yang, {Mass correction to chiral vortical effect and chiral
  separation effect}, Phys. Rev. D 98~(11) (2018) 114022.
\newblock \href {http://arxiv.org/abs/1810.02979} {\path{arXiv:1810.02979}},
  \href {https://doi.org/10.1103/PhysRevD.98.114022}
  {\path{doi:10.1103/PhysRevD.98.114022}}.

\bibitem{Lin:2019ytz}
S.~Lin, A.~Shukla, {Chiral Kinetic Theory from Effective Field Theory
  Revisited}, JHEP 06 (2019) 060.
\newblock \href {http://arxiv.org/abs/1901.01528} {\path{arXiv:1901.01528}},
  \href {https://doi.org/10.1007/JHEP06(2019)060}
  {\path{doi:10.1007/JHEP06(2019)060}}.

\bibitem{Manuel:2021oah}
C.~Manuel, J.~M. Torres-Rincon, {Chiral kinetic theory with small mass
  corrections and quantum coherent states}, Phys. Rev. D 103~(9) (2021) 096022.
\newblock \href {http://arxiv.org/abs/2101.05832} {\path{arXiv:2101.05832}},
  \href {https://doi.org/10.1103/PhysRevD.103.096022}
  {\path{doi:10.1103/PhysRevD.103.096022}}.

\bibitem{Berry:1984jv}
M.~V. Berry, {Quantal phase factors accompanying adiabatic changes}, Proc. Roy.
  Soc. Lond. A 392 (1984) 45--57.
\newblock \href {https://doi.org/10.1098/rspa.1984.0023}
  {\path{doi:10.1098/rspa.1984.0023}}.

\bibitem{Xiao:2005qw}
D.~Xiao, J.-r. Shi, Q.~Niu, {Berry phase correction to electron density of
  states in solids}, Phys. Rev. Lett. 95 (2005) 137204.
\newblock \href {http://arxiv.org/abs/cond-mat/0502340}
  {\path{arXiv:cond-mat/0502340}}, \href
  {https://doi.org/10.1103/PhysRevLett.95.137204}
  {\path{doi:10.1103/PhysRevLett.95.137204}}.

\bibitem{Xiao:2009rm}
D.~Xiao, M.-C. Chang, Q.~Niu, {Berry Phase Effects on Electronic Properties},
  Rev. Mod. Phys. 82 (2010) 1959--2007.
\newblock \href {http://arxiv.org/abs/0907.2021} {\path{arXiv:0907.2021}},
  \href {https://doi.org/10.1103/RevModPhys.82.1959}
  {\path{doi:10.1103/RevModPhys.82.1959}}.

\bibitem{Stone:2013sga}
M.~Stone, V.~Dwivedi, {Classical version of the non-Abelian gauge anomaly},
  Phys. Rev. D 88~(4) (2013) 045012.
\newblock \href {http://arxiv.org/abs/1305.1955} {\path{arXiv:1305.1955}},
  \href {https://doi.org/10.1103/PhysRevD.88.045012}
  {\path{doi:10.1103/PhysRevD.88.045012}}.

\bibitem{Stone:2015kla}
M.~Stone, V.~Dwivedi, T.~Zhou, {Wigner Translations and the Observer Dependence
  of the Position of Massless Spinning Particles}, Phys. Rev. Lett. 114~(21)
  (2015) 210402.
\newblock \href {http://arxiv.org/abs/1501.04586} {\path{arXiv:1501.04586}},
  \href {https://doi.org/10.1103/PhysRevLett.114.210402}
  {\path{doi:10.1103/PhysRevLett.114.210402}}.

\bibitem{Gorbar:2017toh}
E.~V. Gorbar, D.~O. Rybalka, I.~A. Shovkovy, {Second-order dissipative
  hydrodynamics for plasma with chiral asymmetry and vorticity}, Phys. Rev. D
  95~(9) (2017) 096010.
\newblock \href {http://arxiv.org/abs/1702.07791} {\path{arXiv:1702.07791}},
  \href {https://doi.org/10.1103/PhysRevD.95.096010}
  {\path{doi:10.1103/PhysRevD.95.096010}}.

\bibitem{Hidaka:2017auj}
Y.~Hidaka, S.~Pu, D.-L. Yang, {Nonlinear Responses of Chiral Fluids from
  Kinetic Theory}, Phys. Rev. D 97~(1) (2018) 016004.
\newblock \href {http://arxiv.org/abs/1710.00278} {\path{arXiv:1710.00278}},
  \href {https://doi.org/10.1103/PhysRevD.97.016004}
  {\path{doi:10.1103/PhysRevD.97.016004}}.

\bibitem{Huang:2017tsq}
A.~Huang, Y.~Jiang, S.~Shi, J.~Liao, P.~Zhuang, {Out-of-equilibrium chiral
  magnetic effect from chiral kinetic theory}, Phys. Lett. B 777 (2018)
  177--183.
\newblock \href {http://arxiv.org/abs/1703.08856} {\path{arXiv:1703.08856}},
  \href {https://doi.org/10.1016/j.physletb.2017.12.025}
  {\path{doi:10.1016/j.physletb.2017.12.025}}.

\bibitem{Rybalka:2018uzh}
D.~O. Rybalka, E.~V. Gorbar, I.~A. Shovkovy, {Hydrodynamic modes in a
  magnetized chiral plasma with vorticity}, Phys. Rev. D 99~(1) (2019) 016017.
\newblock \href {http://arxiv.org/abs/1807.07608} {\path{arXiv:1807.07608}},
  \href {https://doi.org/10.1103/PhysRevD.99.016017}
  {\path{doi:10.1103/PhysRevD.99.016017}}.

\bibitem{Dayi:2017xrr}
O.~F. Dayi, E.~Kilin\c{c}arslan, {Nonlinear Chiral Plasma Transport in Rotating
  Coordinates}, Phys. Rev. D 96~(4) (2017) 043514.
\newblock \href {http://arxiv.org/abs/1705.01267} {\path{arXiv:1705.01267}},
  \href {https://doi.org/10.1103/PhysRevD.96.043514}
  {\path{doi:10.1103/PhysRevD.96.043514}}.

\bibitem{Ebihara:2017suq}
S.~Ebihara, K.~Fukushima, S.~Pu, {Boost invariant formulation of the chiral
  kinetic theory}, Phys. Rev. D 96~(1) (2017) 016016.
\newblock \href {http://arxiv.org/abs/1705.08611} {\path{arXiv:1705.08611}},
  \href {https://doi.org/10.1103/PhysRevD.96.016016}
  {\path{doi:10.1103/PhysRevD.96.016016}}.

\bibitem{Hidaka:2018ekt}
Y.~Hidaka, D.-L. Yang, {Nonequilibrium chiral magnetic/vortical effects in
  viscous fluids}, Phys. Rev. D 98~(1) (2018) 016012.
\newblock \href {http://arxiv.org/abs/1801.08253} {\path{arXiv:1801.08253}},
  \href {https://doi.org/10.1103/PhysRevD.98.016012}
  {\path{doi:10.1103/PhysRevD.98.016012}}.

\bibitem{Fukushima:2018osn}
K.~Fukushima, S.~Pu, Z.~Qiu, {Eddy magnetization from the chiral Barnett
  effect}, Phys. Rev. A99~(3) (2019) 032105.
\newblock \href {http://arxiv.org/abs/1808.08016} {\path{arXiv:1808.08016}},
  \href {https://doi.org/10.1103/PhysRevA.99.032105}
  {\path{doi:10.1103/PhysRevA.99.032105}}.

\bibitem{Yang:2018lew}
D.-L. Yang, {Side-Jump Induced Spin-Orbit Interaction of Chiral Fluids from
  Kinetic Theory}, Phys. Rev. D 98~(7) (2018) 076019.
\newblock \href {http://arxiv.org/abs/1807.02395} {\path{arXiv:1807.02395}},
  \href {https://doi.org/10.1103/PhysRevD.98.076019}
  {\path{doi:10.1103/PhysRevD.98.076019}}.

\bibitem{Dayi:2019hod}
O.~F. Dayi, E.~Kilin\c{c}arslan, {Some Features of Semiclassical Chiral
  Transport in Rotating Frames}, Phys. Rev. D 100~(4) (2019) 045012.
\newblock \href {http://arxiv.org/abs/1906.04504} {\path{arXiv:1906.04504}},
  \href {https://doi.org/10.1103/PhysRevD.100.045012}
  {\path{doi:10.1103/PhysRevD.100.045012}}.

\bibitem{Lin:2019fqo}
S.~Lin, L.~Yang, {Chiral kinetic theory from Landau level basis}, Phys. Rev. D
  101~(3) (2020) 034006.
\newblock \href {http://arxiv.org/abs/1909.11514} {\path{arXiv:1909.11514}},
  \href {https://doi.org/10.1103/PhysRevD.101.034006}
  {\path{doi:10.1103/PhysRevD.101.034006}}.

\bibitem{Lin:2021sjw}
S.~Lin, L.~Yang, {Magneto-vortical effect in strong magnetic field}, JHEP 06
  (2021) 054.
\newblock \href {http://arxiv.org/abs/2103.11577} {\path{arXiv:2103.11577}},
  \href {https://doi.org/10.1007/JHEP06(2021)054}
  {\path{doi:10.1007/JHEP06(2021)054}}.

\bibitem{Yang:2021eoz}
L.~Yang, {Two-point functions from chiral kinetic theory in magnetized plasma}
  (12 2021).
\newblock \href {http://arxiv.org/abs/2112.13351} {\path{arXiv:2112.13351}}.

\bibitem{Chen:2021azy}
Z.~Chen, S.~Lin, {Quantum Kinetic Theory with Vector and Axial Gauge Fields} (9
  2021).
\newblock \href {http://arxiv.org/abs/2109.08440} {\path{arXiv:2109.08440}}.

\bibitem{Fang:2022ttm}
S.~Fang, S.~Pu, D.-L. Yang, {Quantum kinetic theory for dynamical spin
  polarization from QED-type interaction} (4 2022).
\newblock \href {http://arxiv.org/abs/2204.11519} {\path{arXiv:2204.11519}}.

\bibitem{Sun:2016nig}
Y.~Sun, C.~M. Ko, F.~Li, {Anomalous transport model study of chiral magnetic
  effects in heavy ion collisions}, Phys. Rev. C94~(4) (2016) 045204.
\newblock \href {http://arxiv.org/abs/1606.05627} {\path{arXiv:1606.05627}},
  \href {https://doi.org/10.1103/PhysRevC.94.045204}
  {\path{doi:10.1103/PhysRevC.94.045204}}.

\bibitem{Sun:2016mvh}
Y.~Sun, C.~M. Ko, {Chiral vortical and magnetic effects in the anomalous
  transport model}, Phys. Rev. C95~(3) (2017) 034909.
\newblock \href {http://arxiv.org/abs/1612.02408} {\path{arXiv:1612.02408}},
  \href {https://doi.org/10.1103/PhysRevC.95.034909}
  {\path{doi:10.1103/PhysRevC.95.034909}}.

\bibitem{Sun:2017xhx}
Y.~Sun, C.~M. Ko,
  \href{https://link.aps.org/doi/10.1103/PhysRevC.96.024906}{{$\Lambda$ hyperon
  polarization in relativistic heavy ion collisions from a chiral kinetic
  approach}}, Phys. Rev. C96~(2) (2017) 024906.
\newblock \href {http://arxiv.org/abs/1706.09467} {\path{arXiv:1706.09467}},
  \href {https://doi.org/10.1103/PhysRevC.96.024906}
  {\path{doi:10.1103/PhysRevC.96.024906}}.
\newline\urlprefix\url{https://link.aps.org/doi/10.1103/PhysRevC.96.024906}

\bibitem{Sun:2018idn}
Y.~Sun, C.~M. Ko, {Chiral kinetic approach to the chiral magnetic effect in
  isobaric collisions}, Phys. Rev. C98~(1) (2018) 014911.
\newblock \href {http://arxiv.org/abs/1803.06043} {\path{arXiv:1803.06043}},
  \href {https://doi.org/10.1103/PhysRevC.98.014911}
  {\path{doi:10.1103/PhysRevC.98.014911}}.

\bibitem{Sun:2018bjl}
Y.~Sun, C.~M. Ko, {Azimuthal angle dependence of the longitudinal spin
  polarization in relativistic heavy ion collisions}, Phys. Rev. C99~(1) (2019)
  011903.
\newblock \href {http://arxiv.org/abs/1810.10359} {\path{arXiv:1810.10359}},
  \href {https://doi.org/10.1103/PhysRevC.99.011903}
  {\path{doi:10.1103/PhysRevC.99.011903}}.

\bibitem{Zhou:2018rkh}
W.-H. Zhou, J.~Xu, {Simulating the Chiral Magnetic Wave in a Box System}, Phys.
  Rev. C98~(4) (2018) 044904.
\newblock \href {http://arxiv.org/abs/1810.01030} {\path{arXiv:1810.01030}},
  \href {https://doi.org/10.1103/PhysRevC.98.044904}
  {\path{doi:10.1103/PhysRevC.98.044904}}.

\bibitem{Zhou:2019jag}
W.-H. Zhou, J.~Xu, {Simulating chiral anomalies with spin dynamics}, Phys.
  Lett. B798 (2019) 134932.
\newblock \href {http://arxiv.org/abs/1904.01834} {\path{arXiv:1904.01834}},
  \href {https://doi.org/10.1016/j.physletb.2019.134932}
  {\path{doi:10.1016/j.physletb.2019.134932}}.

\bibitem{Liu:2019krs}
S.~Y.~F. Liu, Y.~Sun, C.~M. Ko, {Spin Polarizations in a Covariant
  Angular-Momentum-Conserved Chiral Transport Model}, Phys. Rev. Lett. 125~(6)
  (2020) 062301.
\newblock \href {http://arxiv.org/abs/1910.06774} {\path{arXiv:1910.06774}},
  \href {https://doi.org/10.1103/PhysRevLett.125.062301}
  {\path{doi:10.1103/PhysRevLett.125.062301}}.

\bibitem{Charbonneau:2009ax}
J.~Charbonneau, A.~Zhitnitsky, {Topological Currents in Neutron Stars: Kicks,
  Precession, Toroidal Fields, and Magnetic Helicity}, JCAP 08 (2010) 010.
\newblock \href {http://arxiv.org/abs/0903.4450} {\path{arXiv:0903.4450}},
  \href {https://doi.org/10.1088/1475-7516/2010/08/010}
  {\path{doi:10.1088/1475-7516/2010/08/010}}.

\bibitem{Golkar:2012kb}
S.~Golkar, D.~T. Son, {(Non)-renormalization of the chiral vortical effect
  coefficient}, JHEP 02 (2015) 169.
\newblock \href {http://arxiv.org/abs/1207.5806} {\path{arXiv:1207.5806}},
  \href {https://doi.org/10.1007/JHEP02(2015)169}
  {\path{doi:10.1007/JHEP02(2015)169}}.

\bibitem{Hou:2012xg}
D.-F. Hou, H.~Liu, H.-c. Ren, {A Possible Higher Order Correction to the
  Vortical Conductivity in a Gauge Field Plasma}, Phys. Rev. D 86 (2012)
  121703.
\newblock \href {http://arxiv.org/abs/1210.0969} {\path{arXiv:1210.0969}},
  \href {https://doi.org/10.1103/PhysRevD.86.121703}
  {\path{doi:10.1103/PhysRevD.86.121703}}.

\bibitem{Jensen:2013vta}
K.~Jensen, P.~Kovtun, A.~Ritz, {Chiral conductivities and effective field
  theory}, JHEP 10 (2013) 186.
\newblock \href {http://arxiv.org/abs/1307.3234} {\path{arXiv:1307.3234}},
  \href {https://doi.org/10.1007/JHEP10(2013)186}
  {\path{doi:10.1007/JHEP10(2013)186}}.

\bibitem{Basar:2013qia}
G.~Basar, D.~E. Kharzeev, I.~Zahed, {Chiral and Gravitational Anomalies on
  Fermi Surfaces}, Phys. Rev. Lett. 111 (2013) 161601.
\newblock \href {http://arxiv.org/abs/1307.2234} {\path{arXiv:1307.2234}},
  \href {https://doi.org/10.1103/PhysRevLett.111.161601}
  {\path{doi:10.1103/PhysRevLett.111.161601}}.

\bibitem{Satow:2014lva}
D.~Satow, H.-U. Yee, {Chiral Magnetic Effect at Weak Coupling with Relaxation
  Dynamics}, Phys. Rev. D 90~(1) (2014) 014027.
\newblock \href {http://arxiv.org/abs/1406.1150} {\path{arXiv:1406.1150}},
  \href {https://doi.org/10.1103/PhysRevD.90.014027}
  {\path{doi:10.1103/PhysRevD.90.014027}}.

\bibitem{Wu:2016dam}
Y.~Wu, D.~Hou, H.-c. Ren, {Field theoretic perspectives of the Wigner function
  formulation of the chiral magnetic effect}, Phys. Rev. D 96~(9) (2017)
  096015.
\newblock \href {http://arxiv.org/abs/1601.06520} {\path{arXiv:1601.06520}},
  \href {https://doi.org/10.1103/PhysRevD.96.096015}
  {\path{doi:10.1103/PhysRevD.96.096015}}.

\bibitem{Feng:2017dom}
B.~Feng, D.-f. Hou, H.~Liu, H.-c. Ren, P.-p. Wu, Y.~Wu, {Chiral Magnetic Effect
  in a Lattice Model}, Phys. Rev. D 95~(11) (2017) 114023.
\newblock \href {http://arxiv.org/abs/1702.07980} {\path{arXiv:1702.07980}},
  \href {https://doi.org/10.1103/PhysRevD.95.114023}
  {\path{doi:10.1103/PhysRevD.95.114023}}.

\bibitem{Hou:2017szz}
D.-f. Hou, S.~Lin, {Fluctuation and Dissipation of Axial Charge from Massive
  Quarks}, Phys. Rev. D 98~(5) (2018) 054014.
\newblock \href {http://arxiv.org/abs/1712.08429} {\path{arXiv:1712.08429}},
  \href {https://doi.org/10.1103/PhysRevD.98.054014}
  {\path{doi:10.1103/PhysRevD.98.054014}}.

\bibitem{Feng:2018tpb}
B.~Feng, D.-F. Hou, H.-C. Ren, {QED radiative corrections to chiral magnetic
  effect}, Phys. Rev. D 99~(3) (2019) 036010.
\newblock \href {http://arxiv.org/abs/1810.05954} {\path{arXiv:1810.05954}},
  \href {https://doi.org/10.1103/PhysRevD.99.036010}
  {\path{doi:10.1103/PhysRevD.99.036010}}.

\bibitem{Lin:2018nxj}
S.~Lin, L.~Yan, G.-R. Liang, {Axial Charge Fluctuation and Chiral Magnetic
  Effect from Stochastic Hydrodynamics}, Phys. Rev. C 98~(1) (2018) 014903.
\newblock \href {http://arxiv.org/abs/1802.04941} {\path{arXiv:1802.04941}},
  \href {https://doi.org/10.1103/PhysRevC.98.014903}
  {\path{doi:10.1103/PhysRevC.98.014903}}.

\bibitem{Horvath:2019dvl}
M.~Horvath, D.~Hou, J.~Liao, H.-c. Ren, {Chiral magnetic response to arbitrary
  axial imbalance}, Phys. Rev. D 101~(7) (2020) 076026.
\newblock \href {http://arxiv.org/abs/1911.00933} {\path{arXiv:1911.00933}},
  \href {https://doi.org/10.1103/PhysRevD.101.076026}
  {\path{doi:10.1103/PhysRevD.101.076026}}.

\bibitem{Dong:2020zci}
R.-D. Dong, R.-H. Fang, D.-F. Hou, D.~She, {Chiral magnetic effect for chiral
  fermion system}, Chin. Phys. C 44~(7) (2020) 074106.
\newblock \href {http://arxiv.org/abs/2001.05801} {\path{arXiv:2001.05801}},
  \href {https://doi.org/10.1088/1674-1137/44/7/074106}
  {\path{doi:10.1088/1674-1137/44/7/074106}}.

\bibitem{Fukushima:2010vw}
K.~Fukushima, D.~E. Kharzeev, H.~J. Warringa, {Real-time dynamics of the Chiral
  Magnetic Effect}, Phys. Rev. Lett. 104 (2010) 212001.
\newblock \href {http://arxiv.org/abs/1002.2495} {\path{arXiv:1002.2495}},
  \href {https://doi.org/10.1103/PhysRevLett.104.212001}
  {\path{doi:10.1103/PhysRevLett.104.212001}}.

\bibitem{Copinger:2018ftr}
P.~Copinger, K.~Fukushima, S.~Pu, {Axial Ward identity and the Schwinger
  mechanism -- Applications to the real-time chiral magnetic effect and
  condensates}, Phys. Rev. Lett. 121~(26) (2018) 261602.
\newblock \href {http://arxiv.org/abs/1807.04416} {\path{arXiv:1807.04416}},
  \href {https://doi.org/10.1103/PhysRevLett.121.261602}
  {\path{doi:10.1103/PhysRevLett.121.261602}}.

\bibitem{Copinger:2020nyx}
P.~Copinger, S.~Pu, {Chirality production with mass effects \textemdash{}
  Schwinger pair production and the axial Ward identity}, Int. J. Mod. Phys. A
  35~(28) (2020) 2030015.
\newblock \href {http://arxiv.org/abs/2008.03635} {\path{arXiv:2008.03635}},
  \href {https://doi.org/10.1142/S0217751X2030015X}
  {\path{doi:10.1142/S0217751X2030015X}}.

\bibitem{Torabian:2009qk}
M.~Torabian, H.-U. Yee, {Holographic nonlinear hydrodynamics from AdS/CFT with
  multiple/non-Abelian symmetries}, JHEP 08 (2009) 020.
\newblock \href {http://arxiv.org/abs/0903.4894} {\path{arXiv:0903.4894}},
  \href {https://doi.org/10.1088/1126-6708/2009/08/020}
  {\path{doi:10.1088/1126-6708/2009/08/020}}.

\bibitem{Rebhan:2009vc}
A.~Rebhan, A.~Schmitt, S.~A. Stricker, {Anomalies and the chiral magnetic
  effect in the Sakai-Sugimoto model}, JHEP 01 (2010) 026.
\newblock \href {http://arxiv.org/abs/0909.4782} {\path{arXiv:0909.4782}},
  \href {https://doi.org/10.1007/JHEP01(2010)026}
  {\path{doi:10.1007/JHEP01(2010)026}}.

\bibitem{Sahoo:2009yq}
B.~Sahoo, H.-U. Yee, {Holographic chiral shear waves from anomaly}, Phys. Lett.
  B 689 (2010) 206--212.
\newblock \href {http://arxiv.org/abs/0910.5915} {\path{arXiv:0910.5915}},
  \href {https://doi.org/10.1016/j.physletb.2010.04.076}
  {\path{doi:10.1016/j.physletb.2010.04.076}}.

\bibitem{Yee:2009vw}
H.-U. Yee, {Holographic Chiral Magnetic Conductivity}, JHEP 11 (2009) 085.
\newblock \href {http://arxiv.org/abs/0908.4189} {\path{arXiv:0908.4189}},
  \href {https://doi.org/10.1088/1126-6708/2009/11/085}
  {\path{doi:10.1088/1126-6708/2009/11/085}}.

\bibitem{Gynther:2010ed}
A.~Gynther, K.~Landsteiner, F.~Pena-Benitez, A.~Rebhan, {Holographic Anomalous
  Conductivities and the Chiral Magnetic Effect}, JHEP 02 (2011) 110.
\newblock \href {http://arxiv.org/abs/1005.2587} {\path{arXiv:1005.2587}},
  \href {https://doi.org/10.1007/JHEP02(2011)110}
  {\path{doi:10.1007/JHEP02(2011)110}}.

\bibitem{Rebhan:2010ax}
A.~Rebhan, A.~Schmitt, S.~Stricker, {Holographic chiral currents in a magnetic
  field}, Prog. Theor. Phys. Suppl. 186 (2010) 463--470.
\newblock \href {http://arxiv.org/abs/1007.2494} {\path{arXiv:1007.2494}},
  \href {https://doi.org/10.1143/PTPS.186.463}
  {\path{doi:10.1143/PTPS.186.463}}.

\bibitem{Gorsky:2010xu}
A.~Gorsky, P.~N. Kopnin, A.~V. Zayakin, {On the Chiral Magnetic Effect in
  Soft-Wall AdS/QCD}, Phys. Rev. D 83 (2011) 014023.
\newblock \href {http://arxiv.org/abs/1003.2293} {\path{arXiv:1003.2293}},
  \href {https://doi.org/10.1103/PhysRevD.83.014023}
  {\path{doi:10.1103/PhysRevD.83.014023}}.

\bibitem{Kalaydzhyan:2011vx}
T.~Kalaydzhyan, I.~Kirsch, {Fluid/gravity model for the chiral magnetic
  effect}, Phys. Rev. Lett. 106 (2011) 211601.
\newblock \href {http://arxiv.org/abs/1102.4334} {\path{arXiv:1102.4334}},
  \href {https://doi.org/10.1103/PhysRevLett.106.211601}
  {\path{doi:10.1103/PhysRevLett.106.211601}}.

\bibitem{Hoyos:2011us}
C.~Hoyos, T.~Nishioka, A.~O'Bannon, {A Chiral Magnetic Effect from AdS/CFT with
  Flavor}, JHEP 10 (2011) 084.
\newblock \href {http://arxiv.org/abs/1106.4030} {\path{arXiv:1106.4030}},
  \href {https://doi.org/10.1007/JHEP10(2011)084}
  {\path{doi:10.1007/JHEP10(2011)084}}.

\bibitem{Kharzeev:2011rw}
D.~E. Kharzeev, H.-U. Yee, {Chiral helix in AdS/CFT with flavor}, Phys. Rev. D
  84 (2011) 125011.
\newblock \href {http://arxiv.org/abs/1109.0533} {\path{arXiv:1109.0533}},
  \href {https://doi.org/10.1103/PhysRevD.84.125011}
  {\path{doi:10.1103/PhysRevD.84.125011}}.

\bibitem{Landsteiner:2011iq}
K.~Landsteiner, E.~Megias, L.~Melgar, F.~Pena-Benitez, {Holographic
  Gravitational Anomaly and Chiral Vortical Effect}, JHEP 09 (2011) 121.
\newblock \href {http://arxiv.org/abs/1107.0368} {\path{arXiv:1107.0368}},
  \href {https://doi.org/10.1007/JHEP09(2011)121}
  {\path{doi:10.1007/JHEP09(2011)121}}.

\bibitem{Gahramanov:2012wz}
I.~Gahramanov, T.~Kalaydzhyan, I.~Kirsch, {Anisotropic hydrodynamics,
  holography and the chiral magnetic effect}, Phys. Rev. D 85 (2012) 126013.
\newblock \href {http://arxiv.org/abs/1203.4259} {\path{arXiv:1203.4259}},
  \href {https://doi.org/10.1103/PhysRevD.85.126013}
  {\path{doi:10.1103/PhysRevD.85.126013}}.

\bibitem{Ballon-Bayona:2012qnu}
A.~Ballon-Bayona, K.~Peeters, M.~Zamaklar, {A chiral magnetic spiral in the
  holographic Sakai-Sugimoto model}, JHEP 11 (2012) 164.
\newblock \href {http://arxiv.org/abs/1209.1953} {\path{arXiv:1209.1953}},
  \href {https://doi.org/10.1007/JHEP11(2012)164}
  {\path{doi:10.1007/JHEP11(2012)164}}.

\bibitem{Zakharov:2012vv}
V.~I. Zakharov, {Chiral Magnetic Effect in Hydrodynamic Approximation}, Lect.
  Notes Phys. 871 (2013) 295--330.
\newblock \href {http://arxiv.org/abs/1210.2186} {\path{arXiv:1210.2186}},
  \href {https://doi.org/10.1007/978-3-642-37305-3_11}
  {\path{doi:10.1007/978-3-642-37305-3_11}}.

\bibitem{Landsteiner:2012kd}
K.~Landsteiner, E.~Megias, F.~Pena-Benitez, {Anomalous Transport from Kubo
  Formulae}, Lect. Notes Phys. 871 (2013) 433--468.
\newblock \href {http://arxiv.org/abs/1207.5808} {\path{arXiv:1207.5808}},
  \href {https://doi.org/10.1007/978-3-642-37305-3_17}
  {\path{doi:10.1007/978-3-642-37305-3_17}}.

\bibitem{Jensen:2012kj}
K.~Jensen, R.~Loganayagam, A.~Yarom, {Thermodynamics, gravitational anomalies
  and cones}, JHEP 02 (2013) 088.
\newblock \href {http://arxiv.org/abs/1207.5824} {\path{arXiv:1207.5824}},
  \href {https://doi.org/10.1007/JHEP02(2013)088}
  {\path{doi:10.1007/JHEP02(2013)088}}.

\bibitem{Hoyos:2013qwa}
C.~Hoyos, T.~Nishioka, A.~O'Bannon, {A chiral magnetic effect from AdS/CFT with
  flavor}, Lect. Notes Phys. 871 (2013) 341--376.
\newblock \href {https://doi.org/10.1007/978-3-642-37305-3_13}
  {\path{doi:10.1007/978-3-642-37305-3_13}}.

\bibitem{Lin:2013sga}
S.~Lin, H.-U. Yee, {Out-of-Equilibrium Chiral Magnetic Effect at Strong
  Coupling}, Phys. Rev. D 88~(2) (2013) 025030.
\newblock \href {http://arxiv.org/abs/1305.3949} {\path{arXiv:1305.3949}},
  \href {https://doi.org/10.1103/PhysRevD.88.025030}
  {\path{doi:10.1103/PhysRevD.88.025030}}.

\bibitem{Landsteiner:2016led}
K.~Landsteiner, {Notes on Anomaly Induced Transport}, Acta Phys. Polon. B 47
  (2016) 2617.
\newblock \href {http://arxiv.org/abs/1610.04413} {\path{arXiv:1610.04413}},
  \href {https://doi.org/10.5506/APhysPolB.47.2617}
  {\path{doi:10.5506/APhysPolB.47.2617}}.

\bibitem{Bu:2016oba}
Y.~Bu, M.~Lublinsky, A.~Sharon, {Anomalous transport from holography: Part I},
  JHEP 11 (2016) 093.
\newblock \href {http://arxiv.org/abs/1608.08595} {\path{arXiv:1608.08595}},
  \href {https://doi.org/10.1007/JHEP11(2016)093}
  {\path{doi:10.1007/JHEP11(2016)093}}.

\bibitem{Bu:2016vum}
Y.~Bu, M.~Lublinsky, A.~Sharon, {Anomalous transport from holography: Part II},
  Eur. Phys. J. C 77~(3) (2017) 194.
\newblock \href {http://arxiv.org/abs/1609.09054} {\path{arXiv:1609.09054}},
  \href {https://doi.org/10.1140/epjc/s10052-017-4762-4}
  {\path{doi:10.1140/epjc/s10052-017-4762-4}}.

\bibitem{Ammon:2016fru}
M.~Ammon, S.~Grieninger, A.~Jimenez-Alba, R.~P. Macedo, L.~Melgar, {Holographic
  quenches and anomalous transport}, JHEP 09 (2016) 131.
\newblock \href {http://arxiv.org/abs/1607.06817} {\path{arXiv:1607.06817}},
  \href {https://doi.org/10.1007/JHEP09(2016)131}
  {\path{doi:10.1007/JHEP09(2016)131}}.

\bibitem{Landsteiner:2017lwm}
K.~Landsteiner, E.~Lopez, G.~Milans~del Bosch, {Quenching the Chiral Magnetic
  Effect via the Gravitational Anomaly and Holography}, Phys. Rev. Lett.
  120~(7) (2018) 071602.
\newblock \href {http://arxiv.org/abs/1709.08384} {\path{arXiv:1709.08384}},
  \href {https://doi.org/10.1103/PhysRevLett.120.071602}
  {\path{doi:10.1103/PhysRevLett.120.071602}}.

\bibitem{Bu:2018psl}
Y.~Bu, T.~Demircik, M.~Lublinsky, {Nonlinear chiral transport from holography},
  JHEP 01 (2019) 078.
\newblock \href {http://arxiv.org/abs/1807.08467} {\path{arXiv:1807.08467}},
  \href {https://doi.org/10.1007/JHEP01(2019)078}
  {\path{doi:10.1007/JHEP01(2019)078}}.

\bibitem{Bu:2019mow}
Y.~Bu, T.~Demircik, M.~Lublinsky, {Chiral transport in strong fields from
  holography}, JHEP 05 (2019) 071.
\newblock \href {http://arxiv.org/abs/1903.00896} {\path{arXiv:1903.00896}},
  \href {https://doi.org/10.1007/JHEP05(2019)071}
  {\path{doi:10.1007/JHEP05(2019)071}}.

\bibitem{Fernandez-Pendas:2019rkh}
J.~Fern\'andez-Pend\'as, K.~Landsteiner, {Out of equilibrium chiral magnetic
  effect and momentum relaxation in holography}, Phys. Rev. D 100~(12) (2019)
  126024.
\newblock \href {http://arxiv.org/abs/1907.09962} {\path{arXiv:1907.09962}},
  \href {https://doi.org/10.1103/PhysRevD.100.126024}
  {\path{doi:10.1103/PhysRevD.100.126024}}.

\bibitem{Morales-Tejera:2020xuv}
S.~Morales-Tejera, K.~Landsteiner, {Out of equilibrium chiral vortical effect
  in holography}, Phys. Rev. D 102~(10) (2020) 106020.
\newblock \href {http://arxiv.org/abs/2006.16031} {\path{arXiv:2006.16031}},
  \href {https://doi.org/10.1103/PhysRevD.102.106020}
  {\path{doi:10.1103/PhysRevD.102.106020}}.

\bibitem{Yin:2021zhs}
L.~Yin, D.~Hou, H.-c. Ren, {Chiral magnetic effect and three-point function
  from AdS/CFT correspondence}, JHEP 09 (2021) 117.
\newblock \href {http://arxiv.org/abs/2102.04851} {\path{arXiv:2102.04851}},
  \href {https://doi.org/10.1007/JHEP09(2021)117}
  {\path{doi:10.1007/JHEP09(2021)117}}.

\bibitem{Ghosh:2021naw}
J.~K. Ghosh, S.~Grieninger, K.~Landsteiner, S.~Morales-Tejera, {Is the chiral
  magnetic effect fast enough?}, Phys. Rev. D 104~(4) (2021) 046009.
\newblock \href {http://arxiv.org/abs/2105.05855} {\path{arXiv:2105.05855}},
  \href {https://doi.org/10.1103/PhysRevD.104.046009}
  {\path{doi:10.1103/PhysRevD.104.046009}}.

\bibitem{Grieninger:2021rxd}
S.~Grieninger, A.~Shukla, {Second order equilibrium transport in strongly
  coupled $ \mathcal{N} $ = 4 supersymmetric SU(N$_{c}$) Yang-Mills plasma via
  holography}, JHEP 08 (2021) 108.
\newblock \href {http://arxiv.org/abs/2105.08673} {\path{arXiv:2105.08673}},
  \href {https://doi.org/10.1007/JHEP08(2021)108}
  {\path{doi:10.1007/JHEP08(2021)108}}.

\bibitem{Yamamoto:2011gk}
A.~Yamamoto, {Chiral magnetic effect in lattice QCD with a chiral chemical
  potential}, Phys. Rev. Lett. 107 (2011) 031601.
\newblock \href {http://arxiv.org/abs/1105.0385} {\path{arXiv:1105.0385}},
  \href {https://doi.org/10.1103/PhysRevLett.107.031601}
  {\path{doi:10.1103/PhysRevLett.107.031601}}.

\bibitem{Muller:2016jod}
N.~M\"uller, S.~Schlichting, S.~Sharma, {Chiral magnetic effect and anomalous
  transport from real-time lattice simulations}, Phys. Rev. Lett. 117~(14)
  (2016) 142301.
\newblock \href {http://arxiv.org/abs/1606.00342} {\path{arXiv:1606.00342}},
  \href {https://doi.org/10.1103/PhysRevLett.117.142301}
  {\path{doi:10.1103/PhysRevLett.117.142301}}.

\bibitem{Mace:2016shq}
M.~Mace, N.~Mueller, S.~Schlichting, S.~Sharma, {Non-equilibrium study of the
  Chiral Magnetic Effect from real-time simulations with dynamical fermions},
  Phys. Rev. D 95~(3) (2017) 036023.
\newblock \href {http://arxiv.org/abs/1612.02477} {\path{arXiv:1612.02477}},
  \href {https://doi.org/10.1103/PhysRevD.95.036023}
  {\path{doi:10.1103/PhysRevD.95.036023}}.

\bibitem{Wan:2010fyf}
X.~Wan, A.~M. Turner, A.~Vishwanath, S.~Y. Savrasov, {Topological semimetal and
  Fermi-arc surface states in the electronic structure of pyrochlore iridates},
  Phys. Rev. B 83~(20) (2011) 205101.
\newblock \href {http://arxiv.org/abs/1007.0016} {\path{arXiv:1007.0016}},
  \href {https://doi.org/10.1103/PhysRevB.83.205101}
  {\path{doi:10.1103/PhysRevB.83.205101}}.

\bibitem{Burkov:2011ene}
A.~A. Burkov, L.~Balents, {Weyl Semimetal in a Topological Insulator
  Multilayer}, Phys. Rev. Lett. 107~(12) (2011) 127205.
\newblock \href {http://arxiv.org/abs/1105.5138} {\path{arXiv:1105.5138}},
  \href {https://doi.org/10.1103/PhysRevLett.107.127205}
  {\path{doi:10.1103/PhysRevLett.107.127205}}.

\bibitem{Xu:2011dn}
G.~Xu, H.~Weng, Z.~Wang, X.~Dai, Z.~Fang, {Chern semi-metal and Quantized
  Anomalous Hall Effect in $HgCr_2Se_4$}, Phys. Rev. Lett. 107 (2011) 186806.
\newblock \href {http://arxiv.org/abs/1106.3125} {\path{arXiv:1106.3125}},
  \href {https://doi.org/10.1103/PhysRevLett.107.186806}
  {\path{doi:10.1103/PhysRevLett.107.186806}}.

\bibitem{Son:2012bg}
D.~T. Son, B.~Z. Spivak, {Chiral Anomaly and Classical Negative
  Magnetoresistance of Weyl Metals}, Phys. Rev. B 88 (2013) 104412.
\newblock \href {http://arxiv.org/abs/1206.1627} {\path{arXiv:1206.1627}},
  \href {https://doi.org/10.1103/PhysRevB.88.104412}
  {\path{doi:10.1103/PhysRevB.88.104412}}.

\bibitem{Cortijo:2016wnf}
A.~Cortijo, D.~Kharzeev, K.~Landsteiner, M.~A.~H. Vozmediano, {Strain induced
  Chiral Magnetic Effect in Weyl semimetals}, Phys. Rev. B 94~(24) (2016)
  241405.
\newblock \href {http://arxiv.org/abs/1607.03491} {\path{arXiv:1607.03491}},
  \href {https://doi.org/10.1103/PhysRevB.94.241405}
  {\path{doi:10.1103/PhysRevB.94.241405}}.

\bibitem{Gorbar:2021ebc}
E.~V. Gorbar, V.~A. Miransky, I.~A. Shovkovy, P.~O. Sukhachov, {Electronic
  Properties of Dirac and Weyl Semimetals}, World Scientific, Singapore, 2021.
\newblock \href {https://doi.org/10.1142/11475} {\path{doi:10.1142/11475}}.

\bibitem{Yamamoto:2015gzz}
N.~Yamamoto, {Chiral transport of neutrinos in supernovae: Neutrino-induced
  fluid helicity and helical plasma instability}, Phys. Rev. D 93~(6) (2016)
  065017.
\newblock \href {http://arxiv.org/abs/1511.00933} {\path{arXiv:1511.00933}},
  \href {https://doi.org/10.1103/PhysRevD.93.065017}
  {\path{doi:10.1103/PhysRevD.93.065017}}.

\bibitem{Yamamoto:2016xtu}
N.~Yamamoto, {Scaling laws in chiral hydrodynamic turbulence}, Phys. Rev. D
  93~(12) (2016) 125016.
\newblock \href {http://arxiv.org/abs/1603.08864} {\path{arXiv:1603.08864}},
  \href {https://doi.org/10.1103/PhysRevD.93.125016}
  {\path{doi:10.1103/PhysRevD.93.125016}}.

\bibitem{Masada:2018swb}
Y.~Masada, K.~Kotake, T.~Takiwaki, N.~Yamamoto, {Chiral magnetohydrodynamic
  turbulence in core-collapse supernovae}, Phys. Rev. D 98~(8) (2018) 083018.
\newblock \href {http://arxiv.org/abs/1805.10419} {\path{arXiv:1805.10419}},
  \href {https://doi.org/10.1103/PhysRevD.98.083018}
  {\path{doi:10.1103/PhysRevD.98.083018}}.

\bibitem{Ohnishi:2014uea}
A.~Ohnishi, N.~Yamamoto, {Magnetars and the Chiral Plasma Instabilities} (2
  2014).
\newblock \href {http://arxiv.org/abs/1402.4760} {\path{arXiv:1402.4760}}.

\bibitem{Onishi:2020rqr}
N.~Onishi, T.~Maruyama, {Origin of the Strong Toroidal Magnetic Field in
  Magnetars} (2 2020).
\newblock \href {http://arxiv.org/abs/2002.02602} {\path{arXiv:2002.02602}}.

\bibitem{Kaminski:2014jda}
M.~Kaminski, C.~F. Uhlemann, M.~Bleicher, J.~Schaffner-Bielich, {Anomalous
  hydrodynamics kicks neutron stars}, Phys. Lett. B 760 (2016) 170--174.
\newblock \href {http://arxiv.org/abs/1410.3833} {\path{arXiv:1410.3833}},
  \href {https://doi.org/10.1016/j.physletb.2016.06.054}
  {\path{doi:10.1016/j.physletb.2016.06.054}}.

\bibitem{Yamamoto:2021hjs}
N.~Yamamoto, D.-L. Yang, {Magnetic field induced neutrino chiral transport near
  equilibrium}, Phys. Rev. D 104~(12) (2021) 123019.
\newblock \href {http://arxiv.org/abs/2103.13159} {\path{arXiv:2103.13159}},
  \href {https://doi.org/10.1103/PhysRevD.104.123019}
  {\path{doi:10.1103/PhysRevD.104.123019}}.

\bibitem{Yamamoto:2020zrs}
N.~Yamamoto, D.-L. Yang, {Chiral Radiation Transport Theory of Neutrinos},
  Astrophys. J. 895~(1) (2020) 56.
\newblock \href {http://arxiv.org/abs/2002.11348} {\path{arXiv:2002.11348}},
  \href {https://doi.org/10.3847/1538-4357/ab8468}
  {\path{doi:10.3847/1538-4357/ab8468}}.

\bibitem{Liu:2018xip}
Y.-C. Liu, L.-L. Gao, K.~Mameda, X.-G. Huang, {Chiral kinetic theory in curved
  spacetime}, Phys. Rev. D 99~(8) (2019) 085014.
\newblock \href {http://arxiv.org/abs/1812.10127} {\path{arXiv:1812.10127}},
  \href {https://doi.org/10.1103/PhysRevD.99.085014}
  {\path{doi:10.1103/PhysRevD.99.085014}}.

\bibitem{Hayata:2020sqz}
T.~Hayata, Y.~Hidaka, K.~Mameda, {Second order chiral kinetic theory under
  gravity and antiparallel charge-energy flow}, JHEP 05 (2021) 023.
\newblock \href {http://arxiv.org/abs/2012.12494} {\path{arXiv:2012.12494}},
  \href {https://doi.org/10.1007/JHEP05(2021)023}
  {\path{doi:10.1007/JHEP05(2021)023}}.

\bibitem{Liang:2004ph}
Z.-T. Liang, X.-N. Wang, {Globally polarized quark-gluon plasma in non-central
  A+A collisions}, Phys. Rev. Lett. 94 (2005) 102301, [Erratum: Phys.Rev.Lett.
  96, 039901 (2006)].
\newblock \href {http://arxiv.org/abs/nucl-th/0410079}
  {\path{arXiv:nucl-th/0410079}}, \href
  {https://doi.org/10.1103/PhysRevLett.94.102301}
  {\path{doi:10.1103/PhysRevLett.94.102301}}.

\bibitem{Liang:2004xn}
Z.-T. Liang, X.-N. Wang, {Spin alignment of vector mesons in non-central A+A
  collisions}, Phys. Lett. B 629 (2005) 20--26.
\newblock \href {http://arxiv.org/abs/nucl-th/0411101}
  {\path{arXiv:nucl-th/0411101}}, \href
  {https://doi.org/10.1016/j.physletb.2005.09.060}
  {\path{doi:10.1016/j.physletb.2005.09.060}}.

\bibitem{Gao:2007bc}
J.-H. Gao, S.-W. Chen, W.-t. Deng, Z.-T. Liang, Q.~Wang, X.-N. Wang, {Global
  quark polarization in non-central A+A collisions}, Phys. Rev. C 77 (2008)
  044902.
\newblock \href {http://arxiv.org/abs/0710.2943} {\path{arXiv:0710.2943}},
  \href {https://doi.org/10.1103/PhysRevC.77.044902}
  {\path{doi:10.1103/PhysRevC.77.044902}}.

\bibitem{Zhang:2019xya}
J.-j. Zhang, R.-h. Fang, Q.~Wang, X.-N. Wang, {A microscopic description for
  polarization in particle scatterings}, Phys. Rev. C 100~(6) (2019) 064904.
\newblock \href {http://arxiv.org/abs/1904.09152} {\path{arXiv:1904.09152}},
  \href {https://doi.org/10.1103/PhysRevC.100.064904}
  {\path{doi:10.1103/PhysRevC.100.064904}}.

\bibitem{STAR:2017ckg}
L.~Adamczyk, et~al., {Global $\Lambda$ hyperon polarization in nuclear
  collisions: evidence for the most vortical fluid}, Nature 548 (2017) 62--65.
\newblock \href {http://arxiv.org/abs/1701.06657} {\path{arXiv:1701.06657}},
  \href {https://doi.org/10.1038/nature23004} {\path{doi:10.1038/nature23004}}.

\bibitem{Betz:2007kg}
B.~Betz, M.~Gyulassy, G.~Torrieri, {Polarization probes of vorticity in heavy
  ion collisions}, Phys. Rev. C 76 (2007) 044901.
\newblock \href {http://arxiv.org/abs/0708.0035} {\path{arXiv:0708.0035}},
  \href {https://doi.org/10.1103/PhysRevC.76.044901}
  {\path{doi:10.1103/PhysRevC.76.044901}}.

\bibitem{Csernai:2013bqa}
L.~P. Csernai, V.~K. Magas, D.~J. Wang, {Flow Vorticity in Peripheral High
  Energy Heavy Ion Collisions}, Phys. Rev. C 87~(3) (2013) 034906.
\newblock \href {http://arxiv.org/abs/1302.5310} {\path{arXiv:1302.5310}},
  \href {https://doi.org/10.1103/PhysRevC.87.034906}
  {\path{doi:10.1103/PhysRevC.87.034906}}.

\bibitem{Becattini:2013vja}
F.~Becattini, L.~Csernai, D.~J. Wang, {$\Lambda$ polarization in peripheral
  heavy ion collisions}, Phys. Rev. C 88~(3) (2013) 034905, [Erratum:
  Phys.Rev.C 93, 069901 (2016)].
\newblock \href {http://arxiv.org/abs/1304.4427} {\path{arXiv:1304.4427}},
  \href {https://doi.org/10.1103/PhysRevC.88.034905}
  {\path{doi:10.1103/PhysRevC.88.034905}}.

\bibitem{Becattini:2015ska}
F.~Becattini, G.~Inghirami, V.~Rolando, A.~Beraudo, L.~Del~Zanna, A.~De~Pace,
  M.~Nardi, G.~Pagliara, V.~Chandra, {A study of vorticity formation in high
  energy nuclear collisions}, Eur. Phys. J. C 75~(9) (2015) 406, [Erratum:
  Eur.Phys.J.C 78, 354 (2018)].
\newblock \href {http://arxiv.org/abs/1501.04468} {\path{arXiv:1501.04468}},
  \href {https://doi.org/10.1140/epjc/s10052-015-3624-1}
  {\path{doi:10.1140/epjc/s10052-015-3624-1}}.

\bibitem{Pang:2016igs}
L.-G. Pang, H.~Petersen, Q.~Wang, X.-N. Wang, {Vortical Fluid and $\Lambda$
  Spin Correlations in High-Energy Heavy-Ion Collisions}, Phys. Rev. Lett.
  117~(19) (2016) 192301.
\newblock \href {http://arxiv.org/abs/1605.04024} {\path{arXiv:1605.04024}},
  \href {https://doi.org/10.1103/PhysRevLett.117.192301}
  {\path{doi:10.1103/PhysRevLett.117.192301}}.

\bibitem{Jiang:2016woz}
Y.~Jiang, Z.-W. Lin, J.~Liao, {Rotating quark-gluon plasma in relativistic
  heavy ion collisions}, Phys. Rev. C94~(4) (2016) 044910, [Erratum: Phys.
  Rev.C95,no.4,049904(2017)].
\newblock \href {http://arxiv.org/abs/1602.06580} {\path{arXiv:1602.06580}},
  \href {https://doi.org/10.1103/PhysRevC.94.044910,
  10.1103/PhysRevC.95.049904} {\path{doi:10.1103/PhysRevC.94.044910,
  10.1103/PhysRevC.95.049904}}.

\bibitem{Deng:2016gyh}
W.-T. Deng, X.-G. Huang, {Vorticity in Heavy-Ion Collisions}, Phys. Rev. C
  93~(6) (2016) 064907.
\newblock \href {http://arxiv.org/abs/1603.06117} {\path{arXiv:1603.06117}},
  \href {https://doi.org/10.1103/PhysRevC.93.064907}
  {\path{doi:10.1103/PhysRevC.93.064907}}.

\bibitem{Li:2017slc}
H.~Li, L.-G. Pang, Q.~Wang, X.-L. Xia, {Global $\Lambda$ polarization in
  heavy-ion collisions from a transport model}, Phys. Rev. C 96~(5) (2017)
  054908.
\newblock \href {http://arxiv.org/abs/1704.01507} {\path{arXiv:1704.01507}},
  \href {https://doi.org/10.1103/PhysRevC.96.054908}
  {\path{doi:10.1103/PhysRevC.96.054908}}.

\bibitem{Wei:2018zfb}
D.-X. Wei, W.-T. Deng, X.-G. Huang, {Thermal vorticity and spin polarization in
  heavy-ion collisions}, Phys. Rev. C 99~(1) (2019) 014905.
\newblock \href {http://arxiv.org/abs/1810.00151} {\path{arXiv:1810.00151}},
  \href {https://doi.org/10.1103/PhysRevC.99.014905}
  {\path{doi:10.1103/PhysRevC.99.014905}}.

\bibitem{Becattini:2007nd}
F.~Becattini, F.~Piccinini, {The Ideal relativistic spinning gas: Polarization
  and spectra}, Annals Phys. 323 (2008) 2452--2473.
\newblock \href {http://arxiv.org/abs/0710.5694} {\path{arXiv:0710.5694}},
  \href {https://doi.org/10.1016/j.aop.2008.01.001}
  {\path{doi:10.1016/j.aop.2008.01.001}}.

\bibitem{Becattini:2007sr}
F.~Becattini, F.~Piccinini, J.~Rizzo, {Angular momentum conservation in heavy
  ion collisions at very high energy}, Phys. Rev. C 77 (2008) 024906.
\newblock \href {http://arxiv.org/abs/0711.1253} {\path{arXiv:0711.1253}},
  \href {https://doi.org/10.1103/PhysRevC.77.024906}
  {\path{doi:10.1103/PhysRevC.77.024906}}.

\bibitem{Becattini:2013fla}
F.~Becattini, V.~Chandra, L.~Del~Zanna, E.~Grossi, {Relativistic distribution
  function for particles with spin at local thermodynamical equilibrium},
  Annals Phys. 338 (2013) 32--49.
\newblock \href {http://arxiv.org/abs/1303.3431} {\path{arXiv:1303.3431}},
  \href {https://doi.org/10.1016/j.aop.2013.07.004}
  {\path{doi:10.1016/j.aop.2013.07.004}}.

\bibitem{Becattini:2016gvu}
F.~Becattini, I.~Karpenko, M.~Lisa, I.~Upsal, S.~Voloshin, {Global hyperon
  polarization at local thermodynamic equilibrium with vorticity, magnetic
  field and feed-down}, Phys. Rev. C 95~(5) (2017) 054902.
\newblock \href {http://arxiv.org/abs/1610.02506} {\path{arXiv:1610.02506}},
  \href {https://doi.org/10.1103/PhysRevC.95.054902}
  {\path{doi:10.1103/PhysRevC.95.054902}}.

\bibitem{Wang:2017jpl}
Q.~Wang, {Global and local spin polarization in heavy ion collisions: a brief
  overview}, Nucl. Phys. A 967 (2017) 225--232.
\newblock \href {http://arxiv.org/abs/1704.04022} {\path{arXiv:1704.04022}},
  \href {https://doi.org/10.1016/j.nuclphysa.2017.06.053}
  {\path{doi:10.1016/j.nuclphysa.2017.06.053}}.

\bibitem{Becattini:2020ngo}
F.~Becattini, M.~A. Lisa, {Polarization and Vorticity in the
  Quark\textendash{}Gluon Plasma}, Ann. Rev. Nucl. Part. Sci. 70 (2020)
  395--423.
\newblock \href {http://arxiv.org/abs/2003.03640} {\path{arXiv:2003.03640}},
  \href {https://doi.org/10.1146/annurev-nucl-021920-095245}
  {\path{doi:10.1146/annurev-nucl-021920-095245}}.

\bibitem{Becattini:2020sww}
F.~Becattini, {Polarization in relativistic fluids: a quantum field theoretical
  derivation}, Lect. Notes Phys. 987 (2021) 15--52.
\newblock \href {http://arxiv.org/abs/2004.04050} {\path{arXiv:2004.04050}},
  \href {https://doi.org/10.1007/978-3-030-71427-7_2}
  {\path{doi:10.1007/978-3-030-71427-7_2}}.

\bibitem{Karpenko:2016jyx}
I.~Karpenko, F.~Becattini, {Study of $\Lambda $ polarization in relativistic
  nuclear collisions at $\sqrt{s_\mathrm {NN}}=7.7$ \textendash{}200 GeV}, Eur.
  Phys. J. C 77~(4) (2017) 213.
\newblock \href {http://arxiv.org/abs/1610.04717} {\path{arXiv:1610.04717}},
  \href {https://doi.org/10.1140/epjc/s10052-017-4765-1}
  {\path{doi:10.1140/epjc/s10052-017-4765-1}}.

\bibitem{Xie:2017upb}
Y.~Xie, D.~Wang, L.~P. Csernai, {Global \ensuremath{\Lambda} polarization in
  high energy collisions}, Phys. Rev. C 95~(3) (2017) 031901.
\newblock \href {http://arxiv.org/abs/1703.03770} {\path{arXiv:1703.03770}},
  \href {https://doi.org/10.1103/PhysRevC.95.031901}
  {\path{doi:10.1103/PhysRevC.95.031901}}.

\bibitem{Shi:2017wpk}
S.~Shi, K.~Li, J.~Liao, {Searching for the Subatomic Swirls in the CuCu and
  CuAu Collisions}, Phys. Lett. B 788 (2019) 409--413.
\newblock \href {http://arxiv.org/abs/1712.00878} {\path{arXiv:1712.00878}},
  \href {https://doi.org/10.1016/j.physletb.2018.09.066}
  {\path{doi:10.1016/j.physletb.2018.09.066}}.

\bibitem{Wu:2019eyi}
H.-Z. Wu, L.-G. Pang, X.-G. Huang, Q.~Wang, {Local spin polarization in high
  energy heavy ion collisions}, Phys. Rev. Research. 1 (2019) 033058.
\newblock \href {http://arxiv.org/abs/1906.09385} {\path{arXiv:1906.09385}},
  \href {https://doi.org/10.1103/PhysRevResearch.1.033058}
  {\path{doi:10.1103/PhysRevResearch.1.033058}}.

\bibitem{Wu:2020yiz}
H.-Z. Wu, L.-G. Pang, X.-G. Huang, Q.~Wang, {Local Spin Polarization in 200 GeV
  Au+Au and 2.76 TeV Pb+Pb Collisions}, Nucl. Phys. A 1005 (2021) 121831.
\newblock \href {http://arxiv.org/abs/2002.03360} {\path{arXiv:2002.03360}},
  \href {https://doi.org/10.1016/j.nuclphysa.2020.121831}
  {\path{doi:10.1016/j.nuclphysa.2020.121831}}.

\bibitem{Fu:2020oxj}
B.~Fu, K.~Xu, X.-G. Huang, H.~Song, {Hydrodynamic study of hyperon spin
  polarization in relativistic heavy ion collisions}, Phys. Rev. C 103~(2)
  (2021) 024903.
\newblock \href {http://arxiv.org/abs/2011.03740} {\path{arXiv:2011.03740}},
  \href {https://doi.org/10.1103/PhysRevC.103.024903}
  {\path{doi:10.1103/PhysRevC.103.024903}}.

\bibitem{Ryu:2021lnx}
S.~Ryu, V.~Jupic, C.~Shen, {Probing early-time longitudinal dynamics with the
  $\Lambda$ hyperon's spin polarization in relativistic heavy-ion collisions}
  (6 2021).
\newblock \href {http://arxiv.org/abs/2106.08125} {\path{arXiv:2106.08125}}.

\bibitem{ExHIC-P:2020tcv}
H.~Taya, et~al., {Signatures of the vortical quark-gluon plasma in hadron
  yields}, Phys. Rev. C 102~(2) (2020) 021901.
\newblock \href {http://arxiv.org/abs/2002.10082} {\path{arXiv:2002.10082}},
  \href {https://doi.org/10.1103/PhysRevC.102.021901}
  {\path{doi:10.1103/PhysRevC.102.021901}}.

\bibitem{Guo:2021udq}
Y.~Guo, J.~Liao, E.~Wang, H.~Xing, H.~Zhang, {Hyperon polarization from the
  vortical fluid in low-energy nuclear collisions}, Phys. Rev. C 104~(4) (2021)
  L041902.
\newblock \href {http://arxiv.org/abs/2105.13481} {\path{arXiv:2105.13481}},
  \href {https://doi.org/10.1103/PhysRevC.104.L041902}
  {\path{doi:10.1103/PhysRevC.104.L041902}}.

\bibitem{Ivanov:2020udj}
Y.~B. Ivanov, {Global $\Lambda$ polarization in moderately relativistic nuclear
  collisions}, Phys. Rev. C 103~(3) (2021) L031903.
\newblock \href {http://arxiv.org/abs/2012.07597} {\path{arXiv:2012.07597}},
  \href {https://doi.org/10.1103/PhysRevC.103.L031903}
  {\path{doi:10.1103/PhysRevC.103.L031903}}.

\bibitem{Deng:2020ygd}
X.-G. Deng, X.-G. Huang, Y.-G. Ma, S.~Zhang, {Vorticity in low-energy heavy-ion
  collisions}, Phys. Rev. C 101~(6) (2020) 064908.
\newblock \href {http://arxiv.org/abs/2001.01371} {\path{arXiv:2001.01371}},
  \href {https://doi.org/10.1103/PhysRevC.101.064908}
  {\path{doi:10.1103/PhysRevC.101.064908}}.

\bibitem{Deng:2021miw}
X.-G. Deng, X.-G. Huang, Y.-G. Ma, {Lambda polarization in $^{108}$Ag
  +$^{108}$Ag and $^{197}$Au +$^{197}$Au collisions around a few GeV} (9 2021).
\newblock \href {http://arxiv.org/abs/2109.09956} {\path{arXiv:2109.09956}}.

\bibitem{STAR:2021beb}
M.~S. Abdallah, et~al., {Global $\Lambda$-hyperon polarization in Au+Au
  collisions at $\sqrt{s_\mathrm{NN}}=3$ GeV} (7 2021).
\newblock \href {http://arxiv.org/abs/2108.00044} {\path{arXiv:2108.00044}}.

\bibitem{Kornas:2020qzi}
F.~J. Kornas, {$ \Lambda $ Polarization in Au+Au Collisions at $ \sqrt s_{NN} =
  2.4\,{\text{GeV}} $ Measured with HADES}, Springer Proc. Phys. 250 (2020)
  435--439.
\newblock \href {https://doi.org/10.1007/978-3-030-53448-6_68}
  {\path{doi:10.1007/978-3-030-53448-6_68}}.

\bibitem{STAR:2019erd}
J.~Adam, et~al., {Polarization of $\Lambda$ ($\bar{\Lambda}$) hyperons along
  the beam direction in Au+Au collisions at $\sqrt{s_{_{NN}}}$ = 200 GeV},
  Phys. Rev. Lett. 123~(13) (2019) 132301.
\newblock \href {http://arxiv.org/abs/1905.11917} {\path{arXiv:1905.11917}},
  \href {https://doi.org/10.1103/PhysRevLett.123.132301}
  {\path{doi:10.1103/PhysRevLett.123.132301}}.

\bibitem{Becattini:2017gcx}
F.~Becattini, I.~Karpenko, {Collective Longitudinal Polarization in
  Relativistic Heavy-Ion Collisions at Very High Energy}, Phys. Rev. Lett.
  120~(1) (2018) 012302.
\newblock \href {http://arxiv.org/abs/1707.07984} {\path{arXiv:1707.07984}},
  \href {https://doi.org/10.1103/PhysRevLett.120.012302}
  {\path{doi:10.1103/PhysRevLett.120.012302}}.

\bibitem{Xia:2018tes}
X.-L. Xia, H.~Li, Z.-B. Tang, Q.~Wang, {Probing vorticity structure in
  heavy-ion collisions by local $\Lambda$ polarization}, Phys. Rev. C 98 (2018)
  024905.
\newblock \href {http://arxiv.org/abs/1803.00867} {\path{arXiv:1803.00867}},
  \href {https://doi.org/10.1103/PhysRevC.98.024905}
  {\path{doi:10.1103/PhysRevC.98.024905}}.

\bibitem{Becattini:2019ntv}
F.~Becattini, G.~Cao, E.~Speranza, {Polarization transfer in hyperon decays and
  its effect in relativistic nuclear collisions}, Eur. Phys. J. C 79~(9) (2019)
  741.
\newblock \href {http://arxiv.org/abs/1905.03123} {\path{arXiv:1905.03123}},
  \href {https://doi.org/10.1140/epjc/s10052-019-7213-6}
  {\path{doi:10.1140/epjc/s10052-019-7213-6}}.

\bibitem{Xia:2019fjf}
X.-L. Xia, H.~Li, X.-G. Huang, H.~Z. Huang, {Feed-down effect on
  \ensuremath{\Lambda} spin polarization}, Phys. Rev. C 100~(1) (2019) 014913.
\newblock \href {http://arxiv.org/abs/1905.03120} {\path{arXiv:1905.03120}},
  \href {https://doi.org/10.1103/PhysRevC.100.014913}
  {\path{doi:10.1103/PhysRevC.100.014913}}.

\bibitem{Li:2021jvn}
H.~Li, X.-L. Xia, X.-G. Huang, H.~Z. Huang, {Global hyperon polarization and
  effects of decay feeddown}, in: {19th International Conference on Strangeness
  in Quark Matter}, 2021.
\newblock \href {http://arxiv.org/abs/2108.04111} {\path{arXiv:2108.04111}}.

\bibitem{Voloshin:2017kqp}
S.~A. Voloshin, {Vorticity and particle polarization in heavy ion collisions
  (experimental perspective)}, EPJ Web Conf. 171 (2018) 07002.
\newblock \href {http://arxiv.org/abs/1710.08934} {\path{arXiv:1710.08934}},
  \href {https://doi.org/10.1051/epjconf/201817107002}
  {\path{doi:10.1051/epjconf/201817107002}}.

\bibitem{Fu:2021pok}
B.~Fu, S.~Y.~F. Liu, L.~Pang, H.~Song, Y.~Yin, {Shear-induced spin polarization
  in heavy-ion collisions} (3 2021).
\newblock \href {http://arxiv.org/abs/2103.10403} {\path{arXiv:2103.10403}}.

\bibitem{Becattini:2021iol}
F.~Becattini, M.~Buzzegoli, A.~Palermo, G.~Inghirami, I.~Karpenko, {Local
  polarization and isothermal local equilibrium in relativistic heavy ion
  collisions} (3 2021).
\newblock \href {http://arxiv.org/abs/2103.14621} {\path{arXiv:2103.14621}}.

\bibitem{Liu:2020dxg}
S.~Y.~F. Liu, Y.~Yin, {Spin Hall effect in heavy-ion collisions}, Phys. Rev. D
  104~(5) (2021) 054043.
\newblock \href {http://arxiv.org/abs/2006.12421} {\path{arXiv:2006.12421}},
  \href {https://doi.org/10.1103/PhysRevD.104.054043}
  {\path{doi:10.1103/PhysRevD.104.054043}}.

\bibitem{Liu:2021uhn}
S.~Y.~F. Liu, Y.~Yin, {Spin polarization induced by the hydrodynamic
  gradients}, JHEP 07 (2021) 188.
\newblock \href {http://arxiv.org/abs/2103.09200} {\path{arXiv:2103.09200}},
  \href {https://doi.org/10.1007/JHEP07(2021)188}
  {\path{doi:10.1007/JHEP07(2021)188}}.

\bibitem{Becattini:2021suc}
F.~Becattini, M.~Buzzegoli, A.~Palermo, {Spin-thermal shear coupling in a
  relativistic fluid}, Phys. Lett. B 820 (2021) 136519.
\newblock \href {http://arxiv.org/abs/2103.10917} {\path{arXiv:2103.10917}},
  \href {https://doi.org/10.1016/j.physletb.2021.136519}
  {\path{doi:10.1016/j.physletb.2021.136519}}.

\bibitem{Liu:2021nyg}
Y.-C. Liu, X.-G. Huang, {Spin Polarization Formula for Dirac Fermions at Local
  Equilibrium} (9 2021).
\newblock \href {http://arxiv.org/abs/2109.15301} {\path{arXiv:2109.15301}}.

\bibitem{Florkowski:2021xvy}
W.~Florkowski, A.~Kumar, A.~Mazeliauskas, R.~Ryblewski, {Effect of thermal
  shear on longitudinal spin polarization in a thermal model} (12 2021).
\newblock \href {http://arxiv.org/abs/2112.02799} {\path{arXiv:2112.02799}}.

\bibitem{Yi:2021ryh}
C.~Yi, S.~Pu, D.-L. Yang, {Reexamination of local spin polarization beyond
  global equilibrium in relativistic heavy ion collisions}, Phys. Rev. C
  104~(6) (2021) 064901.
\newblock \href {http://arxiv.org/abs/2106.00238} {\path{arXiv:2106.00238}},
  \href {https://doi.org/10.1103/PhysRevC.104.064901}
  {\path{doi:10.1103/PhysRevC.104.064901}}.

\bibitem{Sun:2021nsg}
Y.~Sun, Z.~Zhang, C.~M. Ko, W.~Zhao, {Evolution of $\Lambda$ polarization in
  the hadronic phase of heavy-ion collisions} (12 2021).
\newblock \href {http://arxiv.org/abs/2112.14410} {\path{arXiv:2112.14410}}.

\bibitem{Becattini:2020xbh}
F.~Becattini, M.~Buzzegoli, A.~Palermo, G.~Prokhorov, {Polarization as a
  signature of local parity violation in hot QCD matter} (9 2020).
\newblock \href {http://arxiv.org/abs/2009.13449} {\path{arXiv:2009.13449}}.

\bibitem{Gao:2021rom}
J.-H. Gao, {Helicity polarization in relativistic heavy ion collisions}, Phys.
  Rev. D 104~(7) (2021) 076016.
\newblock \href {http://arxiv.org/abs/2105.08293} {\path{arXiv:2105.08293}},
  \href {https://doi.org/10.1103/PhysRevD.104.076016}
  {\path{doi:10.1103/PhysRevD.104.076016}}.

\bibitem{Yi:2021unq}
C.~Yi, S.~Pu, J.-H. Gao, D.-L. Yang, {Hydrodynamic helicity polarization in
  relativistic heavy ion collisions} (12 2021).
\newblock \href {http://arxiv.org/abs/2112.15531} {\path{arXiv:2112.15531}}.

\bibitem{Hattori:2019lfp}
K.~Hattori, M.~Hongo, X.-G. Huang, M.~Matsuo, H.~Taya, {Fate of spin
  polarization in a relativistic fluid: An entropy-current analysis}, Phys.
  Lett. B 795 (2019) 100--106.
\newblock \href {http://arxiv.org/abs/1901.06615} {\path{arXiv:1901.06615}},
  \href {https://doi.org/10.1016/j.physletb.2019.05.040}
  {\path{doi:10.1016/j.physletb.2019.05.040}}.

\bibitem{Fukushima:2020qta}
K.~Fukushima, S.~Pu, {Relativistic decomposition of the orbital and the spin
  angular momentum in chiral physics and Feynman's angular momentum paradox}
  (2020).
\newblock \href {http://arxiv.org/abs/2001.00359} {\path{arXiv:2001.00359}}.

\bibitem{Fukushima:2020ucl}
K.~Fukushima, S.~Pu, {Spin hydrodynamics and symmetric energy-momentum tensors
  \textendash{} A current induced by the spin vorticity \textendash{}}, Phys.
  Lett. B 817 (2021) 136346.
\newblock \href {http://arxiv.org/abs/2010.01608} {\path{arXiv:2010.01608}},
  \href {https://doi.org/10.1016/j.physletb.2021.136346}
  {\path{doi:10.1016/j.physletb.2021.136346}}.

\bibitem{Li:2020eon}
S.~Li, M.~A. Stephanov, H.-U. Yee, {Non-dissipative second-order transport,
  spin, and pseudo-gauge transformations in hydrodynamics} (11 2020).
\newblock \href {http://arxiv.org/abs/2011.12318} {\path{arXiv:2011.12318}}.

\bibitem{She:2021lhe}
D.~She, A.~Huang, D.~Hou, J.~Liao, {Relativistic Viscous Hydrodynamics with
  Angular Momentum} (5 2021).
\newblock \href {http://arxiv.org/abs/2105.04060} {\path{arXiv:2105.04060}}.

\bibitem{Montenegro:2017lvf}
D.~Montenegro, L.~Tinti, G.~Torrieri, {Sound waves and vortices in a polarized
  relativistic fluid}, Phys. Rev. D 96~(7) (2017) 076016.
\newblock \href {http://arxiv.org/abs/1703.03079} {\path{arXiv:1703.03079}},
  \href {https://doi.org/10.1103/PhysRevD.96.076016}
  {\path{doi:10.1103/PhysRevD.96.076016}}.

\bibitem{Montenegro:2017rbu}
D.~Montenegro, L.~Tinti, G.~Torrieri, {Ideal relativistic fluid limit for a
  medium with polarization}, Phys. Rev. D 96~(5) (2017) 056012, [Addendum:
  Phys.Rev.D 96, 079901 (2017)].
\newblock \href {http://arxiv.org/abs/1701.08263} {\path{arXiv:1701.08263}},
  \href {https://doi.org/10.1103/PhysRevD.96.056012}
  {\path{doi:10.1103/PhysRevD.96.056012}}.

\bibitem{Florkowski:2017ruc}
W.~Florkowski, B.~Friman, A.~Jaiswal, E.~Speranza, {Relativistic fluid dynamics
  with spin}, Phys. Rev. C 97~(4) (2018) 041901.
\newblock \href {http://arxiv.org/abs/1705.00587} {\path{arXiv:1705.00587}},
  \href {https://doi.org/10.1103/PhysRevC.97.041901}
  {\path{doi:10.1103/PhysRevC.97.041901}}.

\bibitem{Florkowski:2018myy}
W.~Florkowski, E.~Speranza, F.~Becattini, {Perfect-fluid hydrodynamics with
  constant acceleration along the stream lines and spin polarization}, Acta
  Phys. Polon. B 49 (2018) 1409.
\newblock \href {http://arxiv.org/abs/1803.11098} {\path{arXiv:1803.11098}},
  \href {https://doi.org/10.5506/APhysPolB.49.1409}
  {\path{doi:10.5506/APhysPolB.49.1409}}.

\bibitem{Florkowski:2018fap}
W.~Florkowski, A.~Kumar, R.~Ryblewski, {Relativistic hydrodynamics for
  spin-polarized fluids}, Prog. Part. Nucl. Phys. 108 (2019) 103709.
\newblock \href {http://arxiv.org/abs/1811.04409} {\path{arXiv:1811.04409}},
  \href {https://doi.org/10.1016/j.ppnp.2019.07.001}
  {\path{doi:10.1016/j.ppnp.2019.07.001}}.

\bibitem{Bhadury:2020puc}
S.~Bhadury, W.~Florkowski, A.~Jaiswal, A.~Kumar, R.~Ryblewski, {Relativistic
  dissipative spin dynamics in the relaxation time approximation}, Phys. Lett.
  B 814 (2021) 136096.
\newblock \href {http://arxiv.org/abs/2002.03937} {\path{arXiv:2002.03937}},
  \href {https://doi.org/10.1016/j.physletb.2021.136096}
  {\path{doi:10.1016/j.physletb.2021.136096}}.

\bibitem{Shi:2020qrx}
S.~Shi, C.~Gale, S.~Jeon, {From Chiral Kinetic Theory To Spin Hydrodynamics},
  Nucl. Phys. A 1005 (2021) 121949.
\newblock \href {http://arxiv.org/abs/2002.01911} {\path{arXiv:2002.01911}},
  \href {https://doi.org/10.1016/j.nuclphysa.2020.121949}
  {\path{doi:10.1016/j.nuclphysa.2020.121949}}.

\bibitem{Becattini:2018duy}
F.~Becattini, W.~Florkowski, E.~Speranza, {Spin tensor and its role in
  non-equilibrium thermodynamics}, Phys. Lett. B 789 (2019) 419--425.
\newblock \href {http://arxiv.org/abs/1807.10994} {\path{arXiv:1807.10994}},
  \href {https://doi.org/10.1016/j.physletb.2018.12.016}
  {\path{doi:10.1016/j.physletb.2018.12.016}}.

\bibitem{Gallegos:2021bzp}
A.~D. Gallegos, U.~G\"ursoy, A.~Yarom, {Hydrodynamics of spin currents},
  SciPost Phys. 11 (2021) 041.
\newblock \href {http://arxiv.org/abs/2101.04759} {\path{arXiv:2101.04759}},
  \href {https://doi.org/10.21468/SciPostPhys.11.2.041}
  {\path{doi:10.21468/SciPostPhys.11.2.041}}.

\bibitem{Hongo:2021ona}
M.~Hongo, X.-G. Huang, M.~Kaminski, M.~Stephanov, H.-U. Yee, {Relativistic spin
  hydrodynamics with torsion and linear response theory for spin relaxation},
  JHEP 11 (2021) 150.
\newblock \href {http://arxiv.org/abs/2107.14231} {\path{arXiv:2107.14231}},
  \href {https://doi.org/10.1007/JHEP11(2021)150}
  {\path{doi:10.1007/JHEP11(2021)150}}.

\bibitem{Florkowski:2017dyn}
W.~Florkowski, B.~Friman, A.~Jaiswal, R.~Ryblewski, E.~Speranza,
  {Spin-dependent distribution functions for relativistic hydrodynamics of
  spin-1/2 particles}, Phys. Rev. D 97~(11) (2018) 116017.
\newblock \href {http://arxiv.org/abs/1712.07676} {\path{arXiv:1712.07676}},
  \href {https://doi.org/10.1103/PhysRevD.97.116017}
  {\path{doi:10.1103/PhysRevD.97.116017}}.

\bibitem{Florkowski:2018ahw}
W.~Florkowski, A.~Kumar, R.~Ryblewski, {Thermodynamic versus kinetic approach
  to polarization-vorticity coupling}, Phys. Rev. C 98~(4) (2018) 044906.
\newblock \href {http://arxiv.org/abs/1806.02616} {\path{arXiv:1806.02616}},
  \href {https://doi.org/10.1103/PhysRevC.98.044906}
  {\path{doi:10.1103/PhysRevC.98.044906}}.

\bibitem{Florkowski:2019qdp}
W.~Florkowski, A.~Kumar, R.~Ryblewski, R.~Singh, {Spin polarization evolution
  in a boost invariant hydrodynamical background}, Phys. Rev. C 99~(4) (2019)
  044910.
\newblock \href {http://arxiv.org/abs/1901.09655} {\path{arXiv:1901.09655}},
  \href {https://doi.org/10.1103/PhysRevC.99.044910}
  {\path{doi:10.1103/PhysRevC.99.044910}}.

\bibitem{Florkowski:2019voj}
W.~Florkowski, A.~Kumar, R.~Ryblewski, A.~Mazeliauskas, {Longitudinal spin
  polarization in a thermal model}, Phys. Rev. C 100~(5) (2019) 054907.
\newblock \href {http://arxiv.org/abs/1904.00002} {\path{arXiv:1904.00002}},
  \href {https://doi.org/10.1103/PhysRevC.100.054907}
  {\path{doi:10.1103/PhysRevC.100.054907}}.

\bibitem{Bhadury:2020cop}
S.~Bhadury, W.~Florkowski, A.~Jaiswal, A.~Kumar, R.~Ryblewski, {Dissipative
  Spin Dynamics in Relativistic Matter}, Phys. Rev. D 103~(1) (2021) 014030.
\newblock \href {http://arxiv.org/abs/2008.10976} {\path{arXiv:2008.10976}},
  \href {https://doi.org/10.1103/PhysRevD.103.014030}
  {\path{doi:10.1103/PhysRevD.103.014030}}.

\bibitem{Shi:2020htn}
S.~Shi, C.~Gale, S.~Jeon, {From chiral kinetic theory to relativistic viscous
  spin hydrodynamics}, Phys. Rev. C 103~(4) (2021) 044906.
\newblock \href {http://arxiv.org/abs/2008.08618} {\path{arXiv:2008.08618}},
  \href {https://doi.org/10.1103/PhysRevC.103.044906}
  {\path{doi:10.1103/PhysRevC.103.044906}}.

\bibitem{Singh:2020rht}
R.~Singh, G.~Sophys, R.~Ryblewski, {Spin polarization dynamics in the
  Gubser-expanding background}, Phys. Rev. D 103~(7) (2021) 074024.
\newblock \href {http://arxiv.org/abs/2011.14907} {\path{arXiv:2011.14907}},
  \href {https://doi.org/10.1103/PhysRevD.103.074024}
  {\path{doi:10.1103/PhysRevD.103.074024}}.

\bibitem{Florkowski:2021wvk}
W.~Florkowski, R.~Ryblewski, R.~Singh, G.~Sophys, {Spin polarization dynamics
  in the non-boost-invariant background} (12 2021).
\newblock \href {http://arxiv.org/abs/2112.01856} {\path{arXiv:2112.01856}}.

\bibitem{Wang:2021ngp}
D.-L. Wang, S.~Fang, S.~Pu, {Analytic solutions of relativistic dissipative
  spin hydrodynamics with Bjorken expansion} (7 2021).
\newblock \href {http://arxiv.org/abs/2107.11726} {\path{arXiv:2107.11726}}.

\bibitem{Wang:2021wqq}
D.-L. Wang, X.-Q. Xie, S.~Fang, S.~Pu, {Analytic solutions of relativistic
  dissipative spin hydrodynamics with radial expansion in Gubser flow} (12
  2021).
\newblock \href {http://arxiv.org/abs/2112.15535} {\path{arXiv:2112.15535}}.

\bibitem{Liu:2020ymh}
Y.-C. Liu, X.-G. Huang, {Anomalous chiral transports and spin polarization in
  heavy-ion collisions}, Nucl. Sci. Tech. 31~(6) (2020) 56.
\newblock \href {http://arxiv.org/abs/2003.12482} {\path{arXiv:2003.12482}},
  \href {https://doi.org/10.1007/s41365-020-00764-z}
  {\path{doi:10.1007/s41365-020-00764-z}}.

\bibitem{Gao:2019znl}
J.-H. Gao, Z.-T. Liang, {Relativistic Quantum Kinetic Theory for Massive
  Fermions and Spin Effects}, Phys. Rev. D 100~(5) (2019) 056021.
\newblock \href {http://arxiv.org/abs/1902.06510} {\path{arXiv:1902.06510}},
  \href {https://doi.org/10.1103/PhysRevD.100.056021}
  {\path{doi:10.1103/PhysRevD.100.056021}}.

\bibitem{Weickgenannt:2019dks}
N.~Weickgenannt, X.-L. Sheng, E.~Speranza, Q.~Wang, D.~H. Rischke, {Kinetic
  theory for massive spin-1/2 particles from the Wigner-function formalism},
  Phys. Rev. D 100~(5) (2019) 056018.
\newblock \href {http://arxiv.org/abs/1902.06513} {\path{arXiv:1902.06513}},
  \href {https://doi.org/10.1103/PhysRevD.100.056018}
  {\path{doi:10.1103/PhysRevD.100.056018}}.

\bibitem{Weickgenannt:2020aaf}
N.~Weickgenannt, E.~Speranza, X.-l. Sheng, Q.~Wang, D.~H. Rischke, {Generating
  Spin Polarization from Vorticity through Nonlocal Collisions}, Phys. Rev.
  Lett. 127~(5) (2021) 052301.
\newblock \href {http://arxiv.org/abs/2005.01506} {\path{arXiv:2005.01506}},
  \href {https://doi.org/10.1103/PhysRevLett.127.052301}
  {\path{doi:10.1103/PhysRevLett.127.052301}}.

\bibitem{Hattori:2019ahi}
K.~Hattori, Y.~Hidaka, D.-L. Yang, {Axial Kinetic Theory and Spin Transport for
  Fermions with Arbitrary Mass}, Phys. Rev. D 100~(9) (2019) 096011.
\newblock \href {http://arxiv.org/abs/1903.01653} {\path{arXiv:1903.01653}},
  \href {https://doi.org/10.1103/PhysRevD.100.096011}
  {\path{doi:10.1103/PhysRevD.100.096011}}.

\bibitem{Yang:2020hri}
D.-L. Yang, K.~Hattori, Y.~Hidaka, {Effective quantum kinetic theory for spin
  transport of fermions with collsional effects}, JHEP 07 (2020) 070.
\newblock \href {http://arxiv.org/abs/2002.02612} {\path{arXiv:2002.02612}},
  \href {https://doi.org/10.1007/JHEP07(2020)070}
  {\path{doi:10.1007/JHEP07(2020)070}}.

\bibitem{Liu:2020flb}
Y.-C. Liu, K.~Mameda, X.-G. Huang, {Covariant Spin Kinetic Theory I:
  Collisionless Limit}, Chin. Phys. C 44~(9) (2020) 094101, [Erratum:
  Chin.Phys.C 45, 089001 (2021)].
\newblock \href {http://arxiv.org/abs/2002.03753} {\path{arXiv:2002.03753}},
  \href {https://doi.org/10.1088/1674-1137/ac009b}
  {\path{doi:10.1088/1674-1137/ac009b}}.

\bibitem{Weickgenannt:2021cuo}
N.~Weickgenannt, E.~Speranza, X.-l. Sheng, Q.~Wang, D.~H. Rischke, {Derivation
  of the nonlocal collision term in the relativistic Boltzmann equation for
  massive spin-1/2 particles from quantum field theory}, Phys. Rev. D 104~(1)
  (2021) 016022.
\newblock \href {http://arxiv.org/abs/2103.04896} {\path{arXiv:2103.04896}},
  \href {https://doi.org/10.1103/PhysRevD.104.016022}
  {\path{doi:10.1103/PhysRevD.104.016022}}.

\bibitem{Sheng:2021kfc}
X.-L. Sheng, N.~Weickgenannt, E.~Speranza, D.~H. Rischke, Q.~Wang, {From
  Kadanoff-Baym to Boltzmann equations for massive spin-1/2 fermions}, Phys.
  Rev. D 104~(1) (2021) 016029.
\newblock \href {http://arxiv.org/abs/2103.10636} {\path{arXiv:2103.10636}},
  \href {https://doi.org/10.1103/PhysRevD.104.016029}
  {\path{doi:10.1103/PhysRevD.104.016029}}.

\bibitem{Wang:2021qnt}
Z.~Wang, P.~Zhuang, {Damping and polarization rates in near equilibrium state}
  (5 2021).
\newblock \href {http://arxiv.org/abs/2105.00915} {\path{arXiv:2105.00915}}.

\bibitem{Huang:2020wrr}
A.~Huang, S.~Shi, X.~Zhu, L.~He, J.~Liao, P.~Zhuang, {Quantum kinetic equation
  and dynamical mass generation in 2+1 dimensions}, Phys. Rev. D 103~(5) (2021)
  056025.
\newblock \href {http://arxiv.org/abs/2007.02858} {\path{arXiv:2007.02858}},
  \href {https://doi.org/10.1103/PhysRevD.103.056025}
  {\path{doi:10.1103/PhysRevD.103.056025}}.

\bibitem{Wang:2020dws}
Z.~Wang, X.~Guo, S.~Shi, P.~Zhuang, {Mass Correction to Chiral Kinetic
  Equations}, Nucl. Phys. A 1005 (2021) 121976.
\newblock \href {http://arxiv.org/abs/2004.12174} {\path{arXiv:2004.12174}},
  \href {https://doi.org/10.1016/j.nuclphysa.2020.121976}
  {\path{doi:10.1016/j.nuclphysa.2020.121976}}.

\bibitem{Wang:2019moi}
Z.~Wang, X.~Guo, S.~Shi, P.~Zhuang, {Mass Correction to Chiral Kinetic
  Equations}, Phys. Rev. D 100~(1) (2019) 014015.
\newblock \href {http://arxiv.org/abs/1903.03461} {\path{arXiv:1903.03461}},
  \href {https://doi.org/10.1103/PhysRevD.100.014015}
  {\path{doi:10.1103/PhysRevD.100.014015}}.

\bibitem{Wang:2022yli}
Z.~Wang, {Spin evolution of massive fermion in QED plasma} (5 2022).
\newblock \href {http://arxiv.org/abs/2205.09334} {\path{arXiv:2205.09334}}.

\bibitem{Wigner:1932eb}
E.~P. Wigner, {On the quantum correction for thermodynamic equilibrium},
  Phys.Rev. 40 (1932) 749--760.
\newblock \href {https://doi.org/10.1103/PhysRev.40.749}
  {\path{doi:10.1103/PhysRev.40.749}}.

\bibitem{Groenewold:1946kp}
H.~Groenewold, {On the Principles of elementary quantum mechanics}, Physica 12
  (1946) 405--460.
\newblock \href {https://doi.org/10.1016/S0031-8914(46)80059-4}
  {\path{doi:10.1016/S0031-8914(46)80059-4}}.

\bibitem{Moyal:1949sk}
J.~Moyal, {Quantum mechanics as a statistical theory}, Proc.Cambridge Phil.Soc.
  45 (1949) 99--124.
\newblock \href {https://doi.org/10.1017/S0305004100000487}
  {\path{doi:10.1017/S0305004100000487}}.

\bibitem{Hillery:1983ms}
M.~Hillery, R.~O'Connell, M.~Scully, E.~P. Wigner, {Distribution functions in
  physics: Fundamentals}, Phys.Rept. 106 (1984) 121--167.
\newblock \href {https://doi.org/10.1016/0370-1573(84)90160-1}
  {\path{doi:10.1016/0370-1573(84)90160-1}}.

\bibitem{Zachos:2005gri}
C.~K. Zachos, D.~B. Fairlie, T.~L. Curtright (Eds.), {Quantum Mechanics in
  Phase Space}, World Scientific Publishing Co Pte Ltd, 2005.
\newblock \href {https://doi.org/10.1142/5287} {\path{doi:10.1142/5287}}.

\bibitem{Weyl:1927vd}
H.~Weyl, {Quantum mechanics and group theory}, Z. Phys. 46 (1927) 1.
\newblock \href {https://doi.org/10.1007/BF02055756}
  {\path{doi:10.1007/BF02055756}}.

\bibitem{Martin:1959jp}
P.~C. Martin, J.~S. Schwinger, {Theory of many particle systems. 1.}, Phys.
  Rev. 115 (1959) 1342--1373, [,427(1959)].
\newblock \href {https://doi.org/10.1103/PhysRev.115.1342}
  {\path{doi:10.1103/PhysRev.115.1342}}.

\bibitem{Keldysh:1964ud}
L.~V. Keldysh, {Diagram technique for nonequilibrium processes}, Zh. Eksp.
  Teor. Fiz. 47 (1964) 1515--1527, [Sov. Phys. JETP20,1018(1965)].

\bibitem{Chou:1984es}
K.-c. Chou, Z.-b. Su, B.-l. Hao, L.~Yu, {Equilibrium and Nonequilibrium
  Formalisms Made Unified}, Phys. Rept. 118 (1985) 1.
\newblock \href {https://doi.org/10.1016/0370-1573(85)90136-X}
  {\path{doi:10.1016/0370-1573(85)90136-X}}.

\bibitem{Blaizot:2001nr}
J.-P. Blaizot, E.~Iancu, {The Quark gluon plasma: Collective dynamics and hard
  thermal loops}, Phys. Rept. 359 (2002) 355--528.
\newblock \href {http://arxiv.org/abs/hep-ph/0101103}
  {\path{arXiv:hep-ph/0101103}}, \href
  {https://doi.org/10.1016/S0370-1573(01)00061-8}
  {\path{doi:10.1016/S0370-1573(01)00061-8}}.

\bibitem{Berges:2004yj}
J.~Berges, {Introduction to nonequilibrium quantum field theory}, AIP Conf.
  Proc. 739~(1) (2004) 3--62.
\newblock \href {http://arxiv.org/abs/hep-ph/0409233}
  {\path{arXiv:hep-ph/0409233}}, \href {https://doi.org/10.1063/1.1843591}
  {\path{doi:10.1063/1.1843591}}.

\bibitem{Schonhofen:1994zf}
M.~Schonhofen, M.~Cubero, B.~L. Friman, W.~Norenberg, G.~Wolf, {Covariant
  kinetic equations and relaxation processes in relativistic heavy ion
  collisions}, Nucl. Phys. A 572 (1994) 112--140.
\newblock \href {https://doi.org/10.1016/0375-9474(94)90424-3}
  {\path{doi:10.1016/0375-9474(94)90424-3}}.

\bibitem{Elze:1986ii}
H.~T. Elze, M.~Gyulassy, D.~Vasak, {The QCD quark Wigner operator and
  semiclassical transport equations}In *Santa Fe 1986, Proceedings, Hadronic
  matter in collision* 454-465. (see Conference Index).

\bibitem{Elze:1986qd}
H.~T. Elze, M.~Gyulassy, D.~Vasak, {TRANSPORT EQUATIONS FOR THE QCD QUARK
  WIGNER OPERATOR}, Nucl. Phys. B276 (1986) 706--728.
\newblock \href {https://doi.org/10.1016/0550-3213(86)90072-6}
  {\path{doi:10.1016/0550-3213(86)90072-6}}.

\bibitem{Vasak:1987um}
D.~Vasak, M.~Gyulassy, H.~T. Elze, {Quantum Transport Theory for Abelian
  Plasmas}, Annals Phys. 173 (1987) 462--492.
\newblock \href {https://doi.org/10.1016/0003-4916(87)90169-2}
  {\path{doi:10.1016/0003-4916(87)90169-2}}.

\bibitem{Gao:2015zka}
J.-h. Gao, Q.~Wang, {Magnetic moment, vorticity-spin coupling and parity-odd
  conductivity of chiral fermions in 4-dimensional Wigner functions}, Phys.
  Lett. B 749 (2015) 542--546.
\newblock \href {http://arxiv.org/abs/1504.07334} {\path{arXiv:1504.07334}},
  \href {https://doi.org/10.1016/j.physletb.2015.08.058}
  {\path{doi:10.1016/j.physletb.2015.08.058}}.

\bibitem{Gao:2018jsi}
J.-h. Gao, J.-Y. Pang, Q.~Wang, {Chiral vortical effect in Wigner function
  approach}, Phys. Rev. D 100~(1) (2019) 016008.
\newblock \href {http://arxiv.org/abs/1810.02028} {\path{arXiv:1810.02028}},
  \href {https://doi.org/10.1103/PhysRevD.100.016008}
  {\path{doi:10.1103/PhysRevD.100.016008}}.

\bibitem{Gao:2018wmr}
J.-H. Gao, Z.-T. Liang, Q.~Wang, X.-N. Wang, {Disentangling covariant Wigner
  functions for chiral fermions}, Phys. Rev. D 98~(3) (2018) 036019.
\newblock \href {http://arxiv.org/abs/1802.06216} {\path{arXiv:1802.06216}},
  \href {https://doi.org/10.1103/PhysRevD.98.036019}
  {\path{doi:10.1103/PhysRevD.98.036019}}.

\bibitem{Yang:2020mtz}
S.-Z. Yang, J.-H. Gao, Z.-T. Liang, Q.~Wang, {Second-order charge currents and
  stress tensor in a chiral system}, Phys. Rev. D 102~(11) (2020) 116024.
\newblock \href {http://arxiv.org/abs/2003.04517} {\path{arXiv:2003.04517}},
  \href {https://doi.org/10.1103/PhysRevD.102.116024}
  {\path{doi:10.1103/PhysRevD.102.116024}}.

\bibitem{Son:2002sd}
D.~T. Son, A.~O. Starinets, {Minkowski space correlators in AdS / CFT
  correspondence: Recipe and applications}, JHEP 09 (2002) 042.
\newblock \href {http://arxiv.org/abs/hep-th/0205051}
  {\path{arXiv:hep-th/0205051}}, \href
  {https://doi.org/10.1088/1126-6708/2002/09/042}
  {\path{doi:10.1088/1126-6708/2002/09/042}}.

\bibitem{Iatrakis:2015fma}
I.~Iatrakis, S.~Lin, Y.~Yin, {The anomalous transport of axial charge:
  topological vs non-topological fluctuations}, JHEP 09 (2015) 030.
\newblock \href {http://arxiv.org/abs/1506.01384} {\path{arXiv:1506.01384}},
  \href {https://doi.org/10.1007/JHEP09(2015)030}
  {\path{doi:10.1007/JHEP09(2015)030}}.

\bibitem{Blaizot:1999xk}
J.-P. Blaizot, E.~Iancu, {A Boltzmann equation for the QCD plasma}, Nucl. Phys.
  B 557 (1999) 183--236.
\newblock \href {http://arxiv.org/abs/hep-ph/9903389}
  {\path{arXiv:hep-ph/9903389}}, \href
  {https://doi.org/10.1016/S0550-3213(99)00341-7}
  {\path{doi:10.1016/S0550-3213(99)00341-7}}.

\bibitem{Peskin:1995ev}
M.~E. Peskin, D.~V. Schroeder, {An Introduction to quantum field theory},
  Addison-Wesley, Reading, USA, 1995.

\bibitem{Kharzeev:2016sut}
D.~E. Kharzeev, M.~A. Stephanov, H.-U. Yee, {Anatomy of chiral magnetic effect
  in and out of equilibrium}, Phys. Rev. D 95~(5) (2017) 051901.
\newblock \href {http://arxiv.org/abs/1612.01674} {\path{arXiv:1612.01674}},
  \href {https://doi.org/10.1103/PhysRevD.95.051901}
  {\path{doi:10.1103/PhysRevD.95.051901}}.

\bibitem{Gao:2019zhk}
J.-H. Gao, Z.-T. Liang, Q.~Wang, {Dirac sea and chiral anomaly in the quantum
  kinetic theory}, Phys. Rev. D 101~(9) (2020) 096015.
\newblock \href {http://arxiv.org/abs/1910.11060} {\path{arXiv:1910.11060}},
  \href {https://doi.org/10.1103/PhysRevD.101.096015}
  {\path{doi:10.1103/PhysRevD.101.096015}}.

\bibitem{Gao:2020pfu}
J.-H. Gao, Z.-T. Liang, Q.~Wang, {Quantum kinetic theory for spin-1/2 fermions
  in Wigner function formalism}, Int. J. Mod. Phys. A 36~(01) (2021) 2130001.
\newblock \href {http://arxiv.org/abs/2011.02629} {\path{arXiv:2011.02629}},
  \href {https://doi.org/10.1142/S0217751X21300015}
  {\path{doi:10.1142/S0217751X21300015}}.

\bibitem{Pu:2017apt}
S.~Pu, A.~Yamamoto, {Abelian and non-Abelian Berry curvatures in lattice QCD},
  Nucl. Phys. B933 (2018) 53--64.
\newblock \href {http://arxiv.org/abs/1712.02218} {\path{arXiv:1712.02218}},
  \href {https://doi.org/10.1016/j.nuclphysb.2018.06.005}
  {\path{doi:10.1016/j.nuclphysb.2018.06.005}}.

\bibitem{Duval:2005vn}
C.~Duval, Z.~Horvath, P.~A. Horvathy, L.~Martina, P.~Stichel, {Berry phase
  correction to electron density in solids and 'exotic' dynamics}, Mod. Phys.
  Lett. B 20 (2006) 373--378.
\newblock \href {http://arxiv.org/abs/cond-mat/0506051}
  {\path{arXiv:cond-mat/0506051}}, \href
  {https://doi.org/10.1142/S0217984906010573}
  {\path{doi:10.1142/S0217984906010573}}.

\bibitem{Skagerstam:1992er}
B.~S. Skagerstam, {Localization of massless spinning particles and the Berry
  phase} (4 1992).
\newblock \href {http://arxiv.org/abs/hep-th/9210054}
  {\path{arXiv:hep-th/9210054}}.

\bibitem{Duval:2014cfa}
C.~Duval, M.~Elbistan, P.~A. Horv\'athy, P.~M. Zhang,
  {Wigner\textendash{}Souriau translations and Lorentz symmetry of chiral
  fermions}, Phys. Lett. B 742 (2015) 322--326.
\newblock \href {http://arxiv.org/abs/1411.6541} {\path{arXiv:1411.6541}},
  \href {https://doi.org/10.1016/j.physletb.2015.01.048}
  {\path{doi:10.1016/j.physletb.2015.01.048}}.

\bibitem{Hong:1998tn}
D.~K. Hong, {An Effective field theory of QCD at high density}, Phys. Lett. B
  473 (2000) 118--125.
\newblock \href {http://arxiv.org/abs/hep-ph/9812510}
  {\path{arXiv:hep-ph/9812510}}, \href
  {https://doi.org/10.1016/S0370-2693(99)01472-0}
  {\path{doi:10.1016/S0370-2693(99)01472-0}}.

\bibitem{Hong:1999ru}
D.~K. Hong, {Aspects of high density effective theory in QCD}, Nucl. Phys. B
  582 (2000) 451--476.
\newblock \href {http://arxiv.org/abs/hep-ph/9905523}
  {\path{arXiv:hep-ph/9905523}}, \href
  {https://doi.org/10.1016/S0550-3213(00)00330-8}
  {\path{doi:10.1016/S0550-3213(00)00330-8}}.

\bibitem{Schafer:2003jn}
T.~Sch\"afer, {Hard loops, soft loops, and high density effective field
  theory}, Nucl. Phys. A 728 (2003) 251--271.
\newblock \href {http://arxiv.org/abs/hep-ph/0307074}
  {\path{arXiv:hep-ph/0307074}}, \href
  {https://doi.org/10.1016/j.nuclphysa.2003.08.028}
  {\path{doi:10.1016/j.nuclphysa.2003.08.028}}.

\bibitem{Foldy:1949wa}
L.~L. Foldy, S.~A. Wouthuysen, {On the Dirac theory of spin 1/2 particle and
  its nonrelativistic limit}, Phys. Rev. 78 (1950) 29--36.
\newblock \href {https://doi.org/10.1103/PhysRev.78.29}
  {\path{doi:10.1103/PhysRev.78.29}}.

\bibitem{Muller:2017rly}
N.~M\"uller, {Strong magnetic fields and non equilibrium dynamics in QCD},
  Ph.D. thesis, Heidelberg U. (2017).
\newblock \href {https://doi.org/10.11588/heidok.00023134}
  {\path{doi:10.11588/heidok.00023134}}.

\bibitem{Hebenstreit:2010vz}
F.~Hebenstreit, R.~Alkofer, H.~Gies, {Schwinger pair production in space and
  time-dependent electric fields: Relating the Wigner formalism to quantum
  kinetic theory}, Phys. Rev. D 82 (2010) 105026.
\newblock \href {http://arxiv.org/abs/1007.1099} {\path{arXiv:1007.1099}},
  \href {https://doi.org/10.1103/PhysRevD.82.105026}
  {\path{doi:10.1103/PhysRevD.82.105026}}.

\bibitem{Hebenstreit:2011pm}
F.~Hebenstreit, {Schwinger effect in inhomogeneous electric fields}, Ph.D.
  thesis, Graz U. (2011).
\newblock \href {http://arxiv.org/abs/1106.5965} {\path{arXiv:1106.5965}}.

\bibitem{Kohlfurst:2015zxi}
C.~Kohlf\"urst, {Electron-positron pair production in inhomogeneous
  electromagnetic fields}, Ph.D. thesis, Graz U. (2015).
\newblock \href {http://arxiv.org/abs/1512.06082} {\path{arXiv:1512.06082}}.

\bibitem{Kohlfurst:2019mag}
C.~Kohlf\"urst, {Effect of time-dependent inhomogeneous magnetic fields on the
  particle momentum spectrum in electron-positron pair production}, Phys. Rev.
  D 101~(9) (2020) 096003.
\newblock \href {http://arxiv.org/abs/1912.09359} {\path{arXiv:1912.09359}},
  \href {https://doi.org/10.1103/PhysRevD.101.096003}
  {\path{doi:10.1103/PhysRevD.101.096003}}.

\bibitem{Kohlfurst:2021skr}
C.~Kohlf\"urst, N.~Ahmadiniaz, J.~Oertel, R.~Sch\"utzhold, {Sauter-Schwinger
  effect for colliding laser pulses} (7 2021).
\newblock \href {http://arxiv.org/abs/2107.08741} {\path{arXiv:2107.08741}}.

\bibitem{Guo:2020zpa}
X.~Guo, {Massless Limit of Transport Theory for Massive Fermions}, Chin. Phys.
  C 44~(10) (2020) 104106.
\newblock \href {http://arxiv.org/abs/2005.00228} {\path{arXiv:2005.00228}},
  \href {https://doi.org/10.1088/1674-1137/ababf9}
  {\path{doi:10.1088/1674-1137/ababf9}}.

\bibitem{Sheng:2020oqs}
X.-L. Sheng, Q.~Wang, X.-G. Huang, {Kinetic theory with spin: From massive to
  massless fermions}, Phys. Rev. D 102~(2) (2020) 025019.
\newblock \href {http://arxiv.org/abs/2005.00204} {\path{arXiv:2005.00204}},
  \href {https://doi.org/10.1103/PhysRevD.102.025019}
  {\path{doi:10.1103/PhysRevD.102.025019}}.

\bibitem{Dayi:2020uwx}
O.~F. Dayi, E.~Kilin\c{c}arslan, {Semiclassical transport equations of Dirac
  particles in rotating frames}, Phys. Rev. D 102~(4) (2020) 045015.
\newblock \href {http://arxiv.org/abs/2004.07510} {\path{arXiv:2004.07510}},
  \href {https://doi.org/10.1103/PhysRevD.102.045015}
  {\path{doi:10.1103/PhysRevD.102.045015}}.

\bibitem{Wang:2021owk}
Z.~Wang, P.~Zhuang, {Spin Polarization Induced by Inhomogeneous Dynamical
  Condensate} (1 2021).
\newblock \href {http://arxiv.org/abs/2101.00586} {\path{arXiv:2101.00586}}.

\bibitem{Dayi:2021yhf}
O.~F. Dayi, E.~Kilin\c{c}arslan, {Quantum kinetic equation for fluids of
  spin-1/2 fermions}, JHEP 11 (2021) 086.
\newblock \href {http://arxiv.org/abs/2106.12780} {\path{arXiv:2106.12780}},
  \href {https://doi.org/10.1007/JHEP11(2021)086}
  {\path{doi:10.1007/JHEP11(2021)086}}.

\bibitem{Chen:2021rrl}
S.~Chen, Z.~Wang, P.~Zhuang, {Equal-time kinetic equations in a rotational
  field} (1 2021).
\newblock \href {http://arxiv.org/abs/2101.07596} {\path{arXiv:2101.07596}}.

\bibitem{Ochs:1998qj}
S.~Ochs, U.~W. Heinz, {Wigner functions in covariant and single time
  formulations}, Annals Phys. 266 (1998) 351--416.
\newblock \href {http://arxiv.org/abs/hep-th/9806118}
  {\path{arXiv:hep-th/9806118}}, \href {https://doi.org/10.1006/aphy.1998.5796}
  {\path{doi:10.1006/aphy.1998.5796}}.

\bibitem{BLT_spin}
V.~Bargmann, L.~Michel, V.~L. Telegdi,
  \href{https://link.aps.org/doi/10.1103/PhysRevLett.2.435}{Precession of the
  polarization of particles moving in a homogeneous electromagnetic field},
  Phys. Rev. Lett. 2 (1959) 435--436.
\newblock \href {https://doi.org/10.1103/PhysRevLett.2.435}
  {\path{doi:10.1103/PhysRevLett.2.435}}.
\newline\urlprefix\url{https://link.aps.org/doi/10.1103/PhysRevLett.2.435}

\bibitem{Fang:2016vpj}
R.-h. Fang, L.-g. Pang, Q.~Wang, X.-n. Wang, {Polarization of massive fermions
  in a vortical fluid}, Phys. Rev. C 94~(2) (2016) 024904.
\newblock \href {http://arxiv.org/abs/1604.04036} {\path{arXiv:1604.04036}},
  \href {https://doi.org/10.1103/PhysRevC.94.024904}
  {\path{doi:10.1103/PhysRevC.94.024904}}.

\bibitem{Prokhorov:2017atp}
G.~Prokhorov, O.~Teryaev, {Anomalous current from the covariant Wigner
  function}, Phys. Rev. D 97~(7) (2018) 076013.
\newblock \href {http://arxiv.org/abs/1707.02491} {\path{arXiv:1707.02491}},
  \href {https://doi.org/10.1103/PhysRevD.97.076013}
  {\path{doi:10.1103/PhysRevD.97.076013}}.

\bibitem{Prokhorov:2018qhq}
G.~Prokhorov, O.~Teryaev, V.~Zakharov, {Axial current in rotating and
  accelerating medium}, Phys. Rev. D 98~(7) (2018) 071901.
\newblock \href {http://arxiv.org/abs/1805.12029} {\path{arXiv:1805.12029}},
  \href {https://doi.org/10.1103/PhysRevD.98.071901}
  {\path{doi:10.1103/PhysRevD.98.071901}}.

\bibitem{Buzzegoli:2017cqy}
M.~Buzzegoli, E.~Grossi, F.~Becattini, {General equilibrium second-order
  hydrodynamic coefficients for free quantum fields}, JHEP 10 (2017) 091,
  [Erratum: JHEP 07, 119 (2018)].
\newblock \href {http://arxiv.org/abs/1704.02808} {\path{arXiv:1704.02808}},
  \href {https://doi.org/10.1007/JHEP10(2017)091}
  {\path{doi:10.1007/JHEP10(2017)091}}.

\bibitem{Li:2019qkf}
S.~Li, H.-U. Yee, {Quantum Kinetic Theory of Spin Polarization of Massive
  Quarks in Perturbative QCD: Leading Log}, Phys. Rev. D 100~(5) (2019) 056022.
\newblock \href {http://arxiv.org/abs/1905.10463} {\path{arXiv:1905.10463}},
  \href {https://doi.org/10.1103/PhysRevD.100.056022}
  {\path{doi:10.1103/PhysRevD.100.056022}}.

\bibitem{Kapusta:2019sad}
J.~I. Kapusta, E.~Rrapaj, S.~Rudaz, {Relaxation Time for Strange Quark Spin in
  Rotating Quark-Gluon Plasma}, Phys. Rev. C 101~(2) (2020) 024907.
\newblock \href {http://arxiv.org/abs/1907.10750} {\path{arXiv:1907.10750}},
  \href {https://doi.org/10.1103/PhysRevC.101.024907}
  {\path{doi:10.1103/PhysRevC.101.024907}}.

\bibitem{Hou:2020mqp}
D.~Hou, S.~Lin, {Polarization Rotation of Chiral Fermions in Vortical Fluid},
  Phys. Lett. B 818 (2021) 136386.
\newblock \href {http://arxiv.org/abs/2008.03862} {\path{arXiv:2008.03862}},
  \href {https://doi.org/10.1016/j.physletb.2021.136386}
  {\path{doi:10.1016/j.physletb.2021.136386}}.

\bibitem{Fauth:2021nwe}
G.~Fauth, J.~Berges, A.~Di~Piazza, {Collisional strong-field QED kinetic
  equations from first principles}, Phys. Rev. D 104~(3) (2021) 036007.
\newblock \href {http://arxiv.org/abs/2103.13437} {\path{arXiv:2103.13437}},
  \href {https://doi.org/10.1103/PhysRevD.104.036007}
  {\path{doi:10.1103/PhysRevD.104.036007}}.

\bibitem{Lin:2021mvw}
S.~Lin, {Quantum Kinetic Theory for Quantum Electrodynamics} (9 2021).
\newblock \href {http://arxiv.org/abs/2109.00184} {\path{arXiv:2109.00184}}.

\bibitem{Carignano:2021zhu}
S.~Carignano, C.~Manuel, {Power corrections and gradient expansion in QED
  transport theory}, Phys. Rev. D 104~(5) (2021) 056031.
\newblock \href {http://arxiv.org/abs/2107.03655} {\path{arXiv:2107.03655}},
  \href {https://doi.org/10.1103/PhysRevD.104.056031}
  {\path{doi:10.1103/PhysRevD.104.056031}}.

\bibitem{McLerran:2013hla}
L.~McLerran, V.~Skokov, {Comments About the Electromagnetic Field in Heavy-Ion
  Collisions}, Nucl. Phys. A 929 (2014) 184--190.
\newblock \href {http://arxiv.org/abs/1305.0774} {\path{arXiv:1305.0774}},
  \href {https://doi.org/10.1016/j.nuclphysa.2014.05.008}
  {\path{doi:10.1016/j.nuclphysa.2014.05.008}}.

\bibitem{Wang:2020pej}
Z.~Wang, X.~Guo, P.~Zhuang, {Equilibrium Spin Distribution From Detailed
  Balance}, Eur. Phys. J. C 81~(9) (2021) 799.
\newblock \href {http://arxiv.org/abs/2009.10930} {\path{arXiv:2009.10930}},
  \href {https://doi.org/10.1140/epjc/s10052-021-09586-8}
  {\path{doi:10.1140/epjc/s10052-021-09586-8}}.

\bibitem{Huang:2020kik}
X.-G. Huang, P.~Mitkin, A.~V. Sadofyev, E.~Speranza, {Zilch Vortical Effect,
  Berry Phase, and Kinetic Theory}, JHEP 10 (2020) 117.
\newblock \href {http://arxiv.org/abs/2006.03591} {\path{arXiv:2006.03591}},
  \href {https://doi.org/10.1007/JHEP10(2020)117}
  {\path{doi:10.1007/JHEP10(2020)117}}.

\bibitem{Hattori:2020gqh}
K.~Hattori, Y.~Hidaka, N.~Yamamoto, D.-L. Yang, {Wigner functions and quantum
  kinetic theory of polarized photons}, JHEP 02 (2021) 001.
\newblock \href {http://arxiv.org/abs/2010.13368} {\path{arXiv:2010.13368}},
  \href {https://doi.org/10.1007/JHEP02(2021)001}
  {\path{doi:10.1007/JHEP02(2021)001}}.

\bibitem{Svetitsky:1987gq}
B.~Svetitsky, {Diffusion of charmed quarks in the quark-gluon plasma}, Phys.
  Rev. D 37 (1988) 2484--2491.
\newblock \href {https://doi.org/10.1103/PhysRevD.37.2484}
  {\path{doi:10.1103/PhysRevD.37.2484}}.

\bibitem{Bellac:2011kqa}
M.~L. Bellac, {Thermal Field Theory}, Cambridge Monographs on Mathematical
  Physics, Cambridge University Press, 2011.
\newblock \href {https://doi.org/10.1017/CBO9780511721700}
  {\path{doi:10.1017/CBO9780511721700}}.

\bibitem{Hongo:2022izs}
M.~Hongo, X.-G. Huang, M.~Kaminski, M.~Stephanov, H.-U. Yee, {Spin relaxation
  rate for heavy quarks in weakly coupled QCD plasma} (1 2022).
\newblock \href {http://arxiv.org/abs/2201.12390} {\path{arXiv:2201.12390}}.

\bibitem{Heinz:1984yq}
U.~W. Heinz, {Quark - Gluon Transport Theory. Part 1. the Classical Theory},
  Annals Phys. 161 (1985) 48.
\newblock \href {https://doi.org/10.1016/0003-4916(85)90336-7}
  {\path{doi:10.1016/0003-4916(85)90336-7}}.

\bibitem{Heinz:1985qe}
U.~W. Heinz, {Quark - Gluon Transport Theory. Part 2. Color Response and Color
  Correlations in a Quark - Gluon Plasma}, Annals Phys. 168 (1986) 148.
\newblock \href {https://doi.org/10.1016/0003-4916(86)90114-4}
  {\path{doi:10.1016/0003-4916(86)90114-4}}.

\bibitem{Elze:1989un}
H.-T. Elze, U.~W. Heinz, {Quark - Gluon Transport Theory}, Phys. Rept. 183
  (1989) 81--135.
\newblock \href {https://doi.org/10.1016/0370-1573(89)90059-8}
  {\path{doi:10.1016/0370-1573(89)90059-8}}.

\bibitem{Elze:1986hq}
H.~T. Elze, M.~Gyulassy, D.~Vasak, {Transport Equations for the {QCD} Gluon
  Wigner Operator}, Phys. Lett. B 177 (1986) 402--408.
\newblock \href {https://doi.org/10.1016/0370-2693(86)90778-1}
  {\path{doi:10.1016/0370-2693(86)90778-1}}.

\bibitem{Elze:1989gm}
H.-T. Elze, {Transport Equations and {QCD} Collective Modes in a Selfconsistent
  Covariant Background Gauge}, Z. Phys. C 47 (1990) 647--662.
\newblock \href {https://doi.org/10.1007/BF01552332}
  {\path{doi:10.1007/BF01552332}}.

\bibitem{Blaizot:1993zk}
J.~P. Blaizot, E.~Iancu, {Kinetic equations for long wavelength excitations of
  the quark - gluon plasma}, Phys. Rev. Lett. 70 (1993) 3376--3379.
\newblock \href {http://arxiv.org/abs/hep-ph/9301236}
  {\path{arXiv:hep-ph/9301236}}, \href
  {https://doi.org/10.1103/PhysRevLett.70.3376}
  {\path{doi:10.1103/PhysRevLett.70.3376}}.

\bibitem{Blaizot:1993be}
J.~P. Blaizot, E.~Iancu, {Soft collective excitations in hot gauge theories},
  Nucl. Phys. B 417 (1994) 608--673.
\newblock \href {http://arxiv.org/abs/hep-ph/9306294}
  {\path{arXiv:hep-ph/9306294}}, \href
  {https://doi.org/10.1016/0550-3213(94)90486-3}
  {\path{doi:10.1016/0550-3213(94)90486-3}}.

\bibitem{Blaizot:1999fq}
J.-P. Blaizot, E.~Iancu, {Ultrasoft amplitudes in hot QCD}, Nucl. Phys. B 570
  (2000) 326--358.
\newblock \href {http://arxiv.org/abs/hep-ph/9906485}
  {\path{arXiv:hep-ph/9906485}}, \href
  {https://doi.org/10.1016/S0550-3213(99)00783-X}
  {\path{doi:10.1016/S0550-3213(99)00783-X}}.

\bibitem{Wang:2001dm}
Q.~Wang, K.~Redlich, H.~Stoecker, W.~Greiner, {Kinetic equation for gluons in
  the background gauge of QCD}, Phys. Rev. Lett. 88 (2002) 132303.
\newblock \href {http://arxiv.org/abs/nucl-th/0111040}
  {\path{arXiv:nucl-th/0111040}}, \href
  {https://doi.org/10.1103/PhysRevLett.88.132303}
  {\path{doi:10.1103/PhysRevLett.88.132303}}.

\bibitem{Avkhadiev:2017fxj}
A.~Avkhadiev, A.~V. Sadofyev, {Chiral Vortical Effect for Bosons}, Phys. Rev. D
  96~(4) (2017) 045015.
\newblock \href {http://arxiv.org/abs/1702.07340} {\path{arXiv:1702.07340}},
  \href {https://doi.org/10.1103/PhysRevD.96.045015}
  {\path{doi:10.1103/PhysRevD.96.045015}}.

\bibitem{Yamamoto:2017uul}
N.~Yamamoto, {Photonic chiral vortical effect}, Phys. Rev. D 96~(5) (2017)
  051902.
\newblock \href {http://arxiv.org/abs/1702.08886} {\path{arXiv:1702.08886}},
  \href {https://doi.org/10.1103/PhysRevD.96.051902}
  {\path{doi:10.1103/PhysRevD.96.051902}}.

\bibitem{Yamamoto:2017gla}
N.~Yamamoto, {Spin Hall effect of gravitational waves}, Phys. Rev. D 98~(6)
  (2018) 061701.
\newblock \href {http://arxiv.org/abs/1708.03113} {\path{arXiv:1708.03113}},
  \href {https://doi.org/10.1103/PhysRevD.98.061701}
  {\path{doi:10.1103/PhysRevD.98.061701}}.

\bibitem{PhysRevA.96.043830}
V.~A. Zyuzin, \href{https://link.aps.org/doi/10.1103/PhysRevA.96.043830}{Landau
  levels for an electromagnetic wave}, Phys. Rev. A 96 (2017) 043830.
\newblock \href {https://doi.org/10.1103/PhysRevA.96.043830}
  {\path{doi:10.1103/PhysRevA.96.043830}}.
\newline\urlprefix\url{https://link.aps.org/doi/10.1103/PhysRevA.96.043830}

\bibitem{Huang:2018aly}
X.-G. Huang, A.~V. Sadofyev, {Chiral Vortical Effect For An Arbitrary Spin},
  JHEP 03 (2019) 084.
\newblock \href {http://arxiv.org/abs/1805.08779} {\path{arXiv:1805.08779}},
  \href {https://doi.org/10.1007/JHEP03(2019)084}
  {\path{doi:10.1007/JHEP03(2019)084}}.

\bibitem{Chernodub:2018era}
M.~N. Chernodub, A.~Cortijo, K.~Landsteiner, {Zilch vortical effect}, Phys.
  Rev. D 98~(6) (2018) 065016.
\newblock \href {http://arxiv.org/abs/1807.10705} {\path{arXiv:1807.10705}},
  \href {https://doi.org/10.1103/PhysRevD.98.065016}
  {\path{doi:10.1103/PhysRevD.98.065016}}.

\bibitem{Copetti:2018mxw}
C.~Copetti, J.~Fern\'andez-Pend\'as, {Higher spin vortical Zilches from Kubo
  formulae}, Phys. Rev. D 98~(10) (2018) 105008.
\newblock \href {http://arxiv.org/abs/1809.08255} {\path{arXiv:1809.08255}},
  \href {https://doi.org/10.1103/PhysRevD.98.105008}
  {\path{doi:10.1103/PhysRevD.98.105008}}.

\bibitem{Prokhorov:2020okl}
G.~Y. Prokhorov, O.~V. Teryaev, V.~I. Zakharov, {Chiral vortical effect:
  Black-hole versus flat-space derivation}, Phys. Rev. D 102~(12) (2020)
  121702.
\newblock \href {http://arxiv.org/abs/2003.11119} {\path{arXiv:2003.11119}},
  \href {https://doi.org/10.1103/PhysRevD.102.121702}
  {\path{doi:10.1103/PhysRevD.102.121702}}.

\bibitem{Mameda:2022ojk}
K.~Mameda, N.~Yamamoto, D.-L. Yang, {Photonic spin Hall effect from quantum
  kinetic theory in curved spacetime}, Phys. Rev. D 105 (2022) 096019.
\newblock \href {http://arxiv.org/abs/2203.08449} {\path{arXiv:2203.08449}},
  \href {https://doi.org/10.1103/PhysRevD.105.096019}
  {\path{doi:10.1103/PhysRevD.105.096019}}.

\bibitem{Luo:2021uog}
X.-L. Luo, J.-H. Gao, {Covariant chiral kinetic equation in non-Abelian gauge
  field from \textquotedblleft{}covariant gradient
  expansion\textquotedblright{}}, JHEP 11 (2021) 115.
\newblock \href {http://arxiv.org/abs/2107.11709} {\path{arXiv:2107.11709}},
  \href {https://doi.org/10.1007/JHEP11(2021)115}
  {\path{doi:10.1007/JHEP11(2021)115}}.

\bibitem{Muller:2021hpe}
B.~M\"uller, D.-L. Yang, {Anomalous spin polarization from turbulent color
  fields} (10 2021).
\newblock \href {http://arxiv.org/abs/2110.15630} {\path{arXiv:2110.15630}}.

\bibitem{Yang:2021fea}
D.-L. Yang, {Quantum kinetic theory for spin transport of quarks with
  background chromo-electromagnetic fields} (12 2021).
\newblock \href {http://arxiv.org/abs/2112.14392} {\path{arXiv:2112.14392}}.

\bibitem{Heinz:1983nx}
U.~W. Heinz, {Kinetic Theory for Nonabelian Plasmas}, Phys. Rev. Lett. 51
  (1983) 351.
\newblock \href {https://doi.org/10.1103/PhysRevLett.51.351}
  {\path{doi:10.1103/PhysRevLett.51.351}}.

\bibitem{Asakawa:2006tc}
M.~Asakawa, S.~A. Bass, B.~Muller, {Anomalous viscosity of an expanding
  quark-gluon plasma}, Phys. Rev. Lett. 96 (2006) 252301.
\newblock \href {http://arxiv.org/abs/hep-ph/0603092}
  {\path{arXiv:hep-ph/0603092}}, \href
  {https://doi.org/10.1103/PhysRevLett.96.252301}
  {\path{doi:10.1103/PhysRevLett.96.252301}}.

\bibitem{Asakawa:2006jn}
M.~Asakawa, S.~A. Bass, B.~Muller, {Anomalous transport processes in
  anisotropically expanding quark-gluon plasmas}, Prog. Theor. Phys. 116 (2007)
  725--755.
\newblock \href {http://arxiv.org/abs/hep-ph/0608270}
  {\path{arXiv:hep-ph/0608270}}, \href {https://doi.org/10.1143/PTP.116.725}
  {\path{doi:10.1143/PTP.116.725}}.

\end{thebibliography}
	


	
	\newpage
	\appendix
	\renewcommand*{\thesection}{\Alph{section}}
	
	\section{List of acronyms and symbols}

\begin{longtable}{|c|l|}
\caption{List of acronyms and their meanings.}
\label{Tab:Acronyms}\\
\hline 
AdS/CFT & Anti-de Sitter/Conformal Field Theory\tabularnewline
\hline 
AKE & axial-vector kinetic equation\tabularnewline
\hline 
AMPT & A Multi-Phase Transport model\tabularnewline
\hline 
BMT & Bargmann-Michel-Telegdi equation\tabularnewline
\hline 
CKE & chiral kinetic equation\tabularnewline
\hline 
CKT & chiral kinetic theory\tabularnewline
\hline 
CME & chiral magnetic effect\tabularnewline
\hline 
CSE & chiral separation effect\tabularnewline
\hline 
CTP & Closed-Time-Path\tabularnewline
\hline 
CVE & chiral vortical effect\tabularnewline
\hline 
DWF & Disentanglement of chiral Wigner function\tabularnewline
\hline 
EM & electromagnetic\tabularnewline
\hline 
HIJING & Heavy Ion Jet Interaction Generator\tabularnewline
\hline 
LHC & Large Hadron Collider\tabularnewline
\hline 
MVSD & matrix-valued spin distribution function\tabularnewline
\hline 
OAM & orbital angular momentum\tabularnewline
\hline 
OSEFT & on-shell effective theory\tabularnewline
\hline 
QCD & quantum chromodynamics\tabularnewline
\hline 
QGP & quark gluon plasma\tabularnewline
\hline 
QKT & quantum kinetic theory\tabularnewline
\hline 
RHIC & relativistic heavy ion collider\tabularnewline
\hline 
SKE & scalar kinetic equation\tabularnewline
\hline 
sQGP & strongly coupled QGP\tabularnewline
\hline 
SSD & spin-dependent distribution\tabularnewline
\hline 
UrQMD & Ultra-relativistic Quantum Molecular Dynamics model\tabularnewline
\hline 
WF & Wigner function\tabularnewline
\hline 
	\end{longtable}

\newpage



\begin{longtable}{|c|l|}
\caption{List of symbols and their meanings.}
\label{Tab:Symbols}\\
\hline
\multicolumn{1}{|l}{} & \tabularnewline
\multicolumn{1}{|l}{} & Sec. \ref{sec:Master-equation} \tabularnewline
\multicolumn{1}{|l}{} & \tabularnewline
\hline
$G(x_{1},x_{2})$ & Generic two-point Green function on CTP\tabularnewline
\hline
$G^{F}(x_{1},x_{2})$ & Feynman-type Green function with normal time-order\tabularnewline
\hline
$G^{\overline{F}}(x_{1},x_{2})$ & Feynman-type Green function with anti-time-order\tabularnewline
\hline
$G^{\lessgtr}(x_{1},x_{2})$ & Two-point Green function on CTP with $t_{1}=x_{1}^{0}$ and $t_{2}=x_{2}^{0}$\tabularnewline
 & on positive/negative and negative/positive time-branch respectively\tabularnewline
\hline
$G^{R/A}(x_{1},x_{2})$ & Retarded/Advanced Green function\tabularnewline
\hline
$\chi_{R/L}$ & Pauli spinor for massless fermions with right/left chirality\tabularnewline
\hline
$D_{\mu}$ & Covariant derivative\tabularnewline
\hline
$S^{\lessgtr}(x_{1},x_{2})$ & Two-point Green function on CTP for chiral fermions constructed\tabularnewline
 & from Pauli spinors for a specific chirality R or L (suppressed)\tabularnewline
\hline
$S(x_{1},x_{2})$ & synonymous symbol for $S^{<}(x_{1},x_{2})$ for a specific chirality
R or L\tabularnewline
\hline
$\widetilde{S}(x,y)$ & Gauge invariant two-point Green function for $S(x_{1},x_{2})$ expressed\tabularnewline
 & in terms of $x=(x_{1}+x_{2})/2$ and $y=x_{1}-x_{2}$\tabularnewline
\hline
$\widetilde{S}(x,p)$ & Fourier transform of $\widetilde{S}(x,y)$ with respect to $y$,\tabularnewline
 & also denoted as $S(x,p)$\tabularnewline
\hline
$\nabla^{\mu},\Pi^{\mu}$ & Differential operators defined in Eq. (\ref{eq:operator})\tabularnewline
\hline
\multicolumn{1}{|l}{} & \tabularnewline
\multicolumn{1}{|c}{} & Sec. \ref{sec:CCKE} \tabularnewline
\multicolumn{1}{|l}{} & \tabularnewline
\hline
$S_{R,L}(x,p)$ & Wigner functions for right-handed and left-handed fermions,\tabularnewline
 & $2\times2$ matrices\tabularnewline
\hline
$\mathscr{J}_{\mu}^{s}(x,p)$ & The right-handed ($s=+$) and left-handed ($s=-$) component of \tabularnewline
 & the Wigner function for massless fermions, sometimes simply \tabularnewline
 & denoted as $\mathscr{J}_{\mu}(x,p)$ (the chirality $s=\pm$ is suppressed)\tabularnewline
\hline
$\mathcal{V}_{\mu}(x,p)$ & Vector component of the Wigner function for massless fermions\tabularnewline
\hline
$\mathcal{A}_{\mu}(x,p)$ & Axial vector component of the Wigner function for massless fermions\tabularnewline
\hline
$J_{\mu}^{s}(x),J_{s}^{\mu}(x)$ & Current from $\mathscr{J}_{\mu}^{s}(x,p)$ and $\mathscr{J}_{s}^{\mu}(x,p)$,
sometimes simply denoted\tabularnewline
 & as $J_{\mu}(x)$ or $J^{\mu}(x)$ with the chirality $s=\pm$ being
suppressed in case \tabularnewline
 & of no ambiguity\tabularnewline
\hline
$J^{\mu}(x)$ & Vector (charge) current defined as $J_{+}^{\mu}(x)+J_{-}^{\mu}(x)$\tabularnewline
\hline
$J_{5}^{\mu}(x)$ & Axial vector (Chiral charge) current defined as $J_{+}^{\mu}(x)-J_{-}^{\mu}(x)$\tabularnewline
\hline
$\mathscr{J}_{\mu}^{(n)}$,$\mathscr{J}_{(n)}^{\mu}$ & The $n$-th order terms of $\mathscr{J}_{\mu}(x,p)$ and $\mathscr{J}^{\mu}(x,p)$
in $\hbar$-expansion\tabularnewline
\hline
$X_{\mu}^{(n)},X_{(n)}^{\mu}$ & Residue terms in $\mathscr{J}_{\mu}^{(n)}$ and $\mathscr{J}_{(n)}^{\mu}$\tabularnewline
\hline
$f_{(0)}(x,p)$ & Phase space distribution of the massless fermion\tabularnewline
\hline 
 &  at the zeroth order \tabularnewline
\hline
$f_{(0)}(x,p)$ & Phase space distribution of the chiral fermion at the zeroth\tabularnewline
 &   order, with the chirality $s=\pm$ being suppressed\tabularnewline
\hline
$f_{\mathrm{FD}}(p_{0}-\mu_{s})$ & Fermi-Dirac distribution function where $p_{0}\equiv u\cdot p$ and
$\mu_{s}=\mu+s\mu_{5}$\tabularnewline
 & being the chemical potential for the fermion with the chirality $s$\tabularnewline
\hline
$\beta^{\rho}$ & Thermal velocity defined as $\beta u^{\rho}=u^{\rho}/T$\tabularnewline
\hline
$\overline{\mu}_{s},\overline{\mu},\overline{\mu}_{5}$ & Dimensionless chemical potentials defined as $\beta\mu_{s}$, $\beta\mu$,
$\beta\mu_{5}$\tabularnewline
\hline
$\Theta(p_{0})$ & Step function with $\Theta(p_{0}>0)=1$ and $\Theta(p_{0}<0)=0$\tabularnewline
\hline
$\Theta^{\mu\nu}$ & Projector orthorgonal to $u^{\mu}$ defined as $\eta^{\mu\nu}-u^{\mu\nu}$\tabularnewline
\hline
$\epsilon^{\mu\nu\alpha\beta},\epsilon_{\mu\nu\alpha\beta}$ & Anti-symmetric tensor with $\epsilon^{0123}=1$ and $\epsilon_{0123}=-1$\tabularnewline
\hline
$\Omega^{\mu\nu}$ & Thermal vorticity tensor\tabularnewline
\hline
$\widetilde{\Omega}^{\mu\nu}$ & Dual tensor of the thermal vorticity\tabularnewline
\hline
$\varepsilon^{\mu},\omega^{\mu}$ & Electric and magnetic component of $\Omega^{\mu\nu}$\tabularnewline
\hline
$F^{\mu\nu}$ & Strength tensor of the electromagnetic field\tabularnewline
\hline
$\widetilde{F}^{\mu\nu}$ & Dual strength tensor of the electromagnetic field\tabularnewline
\hline
$E^{\mu},B^{\mu}$ & Electric and magnetic field component of $F^{\mu\nu}$\tabularnewline
\hline
$\rho$ & Charge density from $J_{(0)}^{\mu}(x)$\tabularnewline
\hline
$\rho_{5}$ & Chiral charge density from $J_{5,(0)}^{\mu}(x)$\tabularnewline
\hline
$\xi$ & Coefficient of $\omega^{\mu}$ term in $J_{(1)}^{\mu}(x)$, CVE coefficient\tabularnewline
\hline
$\xi_{B}$ & Coefficient of $B^{\mu}$ term in $J_{(1)}^{\mu}(x)$, CME conductivity\tabularnewline
\hline
$\xi_{5}$ & Coefficient of $\omega^{\mu}$ term in $J_{5,(1)}^{\mu}(x)$\tabularnewline
\hline
$\xi_{B5}$ & Coefficient of $B^{\mu}$ term in $J_{5,(1)}^{\mu}(x)$\tabularnewline
\hline
$T^{\mu\nu}$ & Energy-momentum tensor\tabularnewline
\hline
$T_{(n)}^{\mu\nu}$ & The $n$-th order term of $T^{\mu\nu}$ in $\hbar$-expansion\tabularnewline
\hline
$\epsilon$ & Energy density from $T_{(0)}^{\mu\nu}$\tabularnewline
\hline
\multicolumn{1}{|l}{} & \tabularnewline
\multicolumn{1}{|c}{} & Sec.  \ref{sec:dwf-theorem} 
\tabularnewline
\multicolumn{1}{|l}{} & \tabularnewline
\hline
$\mathscr{J}^{0}$ & The time component of $\mathscr{J}^{\mu}(x,p)$\tabularnewline
\hline
$\pmb{\mathscr{J}}$ & The time component of $\mathscr{J}^{\mu}(x,p)$\tabularnewline
\hline
$\boldsymbol{\Omega}_{p}$ & Berry curvature in momentum space, also denoted as $\boldsymbol{\Omega}_{\mathbf{p}}$\tabularnewline
\hline
$n_{\mu},n_{\mu}^{\prime}$ & Frame vectors\tabularnewline
\hline
\multicolumn{1}{|l}{} & \tabularnewline
\multicolumn{1}{|c}{} & Sec.  \ref{sec:CKT_effective} \tabularnewline
\multicolumn{1}{|l}{} & \tabularnewline
\hline 
$\boldsymbol{\Omega}$ or $\boldsymbol{\Omega}_{\mathbf{p}}$ & Berry curvature\tabularnewline
\hline 
$\mathbf{a}_{\mathbf{p}}$ or $\mathcal{A}(\mathbf{p})$ & Berry connection \tabularnewline
\hline 
$\mathbf{p}_{c}$ & Canonical momentum\tabularnewline
\hline 
$\mathbf{p}$ & Kinetic momentum\tabularnewline
\hline
$\xi^a$  & Parameters in phase space used in action (\ref{eq:action_03}) \tabularnewline
$(a=1,2,...,6)$ & \tabularnewline
\hline
$w_a$ & Coefficients for $\xi^a$ in the action (\ref{eq:action_03}) \tabularnewline
$(a=1,2,...,6)$ & \tabularnewline
\hline 
$\omega_{ab}$ & Symplectic matrix\tabularnewline
\hline 
$\{,\}_{\omega}$ & Poisson bracket in symplectic form\tabularnewline
\hline 
$f(t,\mathbf{x},\mathbf{p})$ & Distribution function\tabularnewline
\hline 
$n_{\mathbf{p}}(\mathbf{x})$ & Occupation number density operator\tabularnewline
\hline 
$S^{\mu\nu}$ & Rank-two canonical spin tensor\tabularnewline
\hline 
$n^{\mu}$ & Frame vector\tabularnewline
\hline 
$v^{\mu}$ & Velocity at Fermi surface\tabularnewline
\hline 
$G_{v}$ & Two-point Green functions in effective theory\tabularnewline
\hline 
\multicolumn{1}{|l}{} & \tabularnewline
\multicolumn{1}{|c}{} & Sec.  \ref{sec:CKT_collisions}\tabularnewline
\multicolumn{1}{|l}{} & \tabularnewline
\hline 
$S^{\lessgtr}, S^{R/A}$ & Lessor, greater, retarded, and advanced propagators \tabularnewline
& for right-handed Weyl fermions \tabularnewline
\hline 
$S_0^{\lessgtr}, S_0^{R/A}$  &  Free part of lessor, greater, retarded, and advance propagators \tabularnewline
& for right-handed Weyl fermions \tabularnewline
\hline 
$\Sigma^\lessgtr$, $\Sigma^{R/A}$ &  Lessor, greater, retarded, and advanced self-energies \tabularnewline
& for right-handed Weyl fermions\tabularnewline
\hline 
 $\overline{S}$ & Average value of retarded and advanced propagators, $(S^{R}+S^{A})/2$ \tabularnewline
 \hline 
$\overline{\Sigma}$ & Average value of retarded and advanced self-energies, $(\Sigma^{R}+\Sigma^{A})/2$ \tabularnewline
\hline 
$S^{\lessgtr,\mu}$ & 
Vector decomposition of
lessor and greater propagators,
$S^{\lessgtr}=S^{\lessgtr,\mu}\bar{\sigma}_\mu$
\tabularnewline
\hline 
$S_{\text{leq}}^{<,\mu}$ & $S^{<,\mu}$ in a local thermal equilibrium state\tabularnewline
\hline 
$\star$  & Moyal product defined in Eq.~\eqref{eq:Moyal-product} \tabularnewline
\hline 
$\{A,B\}_{\star}$ & Anti-commutator with Moyal product,
$A\star B+B\star A$
\tabularnewline
\hline
$[A,B]_{\star}$ &
Commutator with Moyal product, $A\star B-B\star A$
\tabularnewline
\hline
$C_\mu^\mu$  & Collision term of the chiral kinetic equation
\tabularnewline
\hline 
$\{\}_{\mathrm{P.B.}}$  & Poisson bracket for the relativistic phase space $(x^\mu, p^\mu)$ \tabularnewline
\hline 
$f(x,p)$ & Distribution function  \tabularnewline
\hline 
$S^{\mu\nu}_{(n)}$  & Spin tensor with a frame vector $n_\mu$ \tabularnewline
\hline
$E_{(n)}^\mu, B_{(n)}^\mu$ & 
Electric and magnetic field components of $F_{\mu\nu}$
with a frame vector $n_\mu$
\tabularnewline
\hline 
\multicolumn{1}{|l}{} & \tabularnewline
\multicolumn{1}{|c}{} & Sec.  \ref{sec:massive fermion with collisions} and Sec. \ref{sec:QKT_collision} \tabularnewline
\multicolumn{1}{|l}{} & \tabularnewline
\hline
$\mathcal{V}^{\lessgtr}_{\mu}$ & Vector components of the lesser and greater propagators \tabularnewline
& for massive fermions \tabularnewline
\hline
$\mathcal{A}^{\lessgtr}_{\mu}$ & Axial-vector components of the lesser and greater propagators \tabularnewline
& for massive fermions \tabularnewline
\hline
$\Sigma^{\lessgtr}$ & Lesser and greater self-energies for massive fermions \tabularnewline
\hline
$\mathcal{V}_{\mu}$ and $\mathcal{A}_{\mu}$ & Vector and axial-vector components of the lesser propagator for \tabularnewline & massive fermions \tabularnewline
\hline
$f_V(x,p)$& Vector distribution functions \tabularnewline
\hline
$f_A(x,p)$ & Axial distribution functions \tabularnewline
\hline
$a^{\mu}(x,p)$ & Non-normalized spin four vector and  $\widetilde{a}^{\mu}\equiv a^{\mu}f_A$
\tabularnewline
\hline
$S^{\mu\nu}_{a(n)}$ and & Anti-symmetric tensors coupled to the magnetization-current  \tabularnewline
$S^{\mu\nu}_{m(n)}$ & terms, which reduce to $S^{\mu\nu}_{(n)}$ in the massless limit.
\tabularnewline
\hline
$\chi_0^{(n)}(a^{\mu},f_A)$ & Quantum corrections in the collisionless SKE, where $\chi_m^{(n)}(a^{\mu},f_A)\rightarrow 0$  \tabularnewline 
 and & and $\chi_0^{(n)}(a^{\mu},f_A)$ reduces to the quantum corrections in the CKT
  \tabularnewline  $\chi_m^{(n)}(a^{\mu},f_A)$ & when $m\rightarrow 0$. 
\tabularnewline
\hline
$\chi_0^{(n)\mu}(f_V)$ and & Quantum corrections in the collisionless AKE, where $\chi_m^{(n)\mu}(f_V)\rightarrow 0$ \tabularnewline 
$\chi_m^{(n)\mu}(f_V)$ & and $\chi_0^{(n)\mu}(f_V)$ reduces to the quantum corrections multiplied by $p^{\mu}$
\tabularnewline  & in the CKT when $m\rightarrow 0$
\tabularnewline
\hline
$n^{\mu}_{r}(p)$& The particle rest frame for massive fermions
\tabularnewline
\hline
$\Omega_{\mu\nu}$& Thermal vorticity
\tabularnewline
\hline
$\Sigma^{\lessgtr}_{F}$, $\Sigma^{\lessgtr}_{V\mu}$, & The Clifford-algebra components of the lesser and greater \tabularnewline
$\Sigma^{\lessgtr}_{A\mu}$, etc. & self-energies
\tabularnewline
\hline
$\mathcal{C}^{(n)\mu}_1$& The explicit classical collision term with implicit quantum corrections \tabularnewline & for a general frame vector $n^{\mu}(x)$ in the AKE
\tabularnewline
\hline
$\mathcal{C}^{(n)\mu}_2$& The explicit quantum collision term in the AKE for a general frame \tabularnewline & vector $n^{\mu}(x)$
\tabularnewline
\hline
$\mathcal{C}^{(n)\mu}_{\rm s2}$& The explicit quantum collision term in the AKE with a constant \tabularnewline & frame vector and zero electromagnetic fields
\tabularnewline
\hline
$\mathcal{C}^{(n_r)\mu}_{\rm 2}$& The explicit quantum collision term in the AKE with the particle \tabularnewline & rest frame
\tabularnewline
\hline
$\kappa_{\rm LL}$& A leading-logarithmic order coefficient for the classical collision \tabularnewline 
& term of QCD in the HTL approximation.
\tabularnewline
\hline
\multicolumn{1}{|l}{} & \tabularnewline
\multicolumn{1}{|c}{} & Sec.  \ref{sec:spin-Boltzmann}\tabularnewline
\multicolumn{1}{|l}{} & \tabularnewline
\hline
$K^{\mu}$ & Operator defined as $p^{\mu}+(i\hbar/2)\partial_{x}^{\mu}$\tabularnewline
\hline
$W^{<}(x,p)$ & Wigner function for the massive fermion corresponding to\tabularnewline
 & two-point Green function $G^{<}(x_{1},x_{2})$\tabularnewline
\hline
$I_{\mathrm{coll}}$ & Collision term in the equation of motion for $W^{<}(x,p)$\tabularnewline
\hline
$\Gamma_{a}$ & Generators of Clifford algebra\tabularnewline
\hline
$u_{r,\alpha}(p),v_{s,\lambda}(p)$ & Dirac spinors for fermions and anti-fermions with spin and\tabularnewline
 & Dirac indices\tabularnewline
\hline
$f_{rs}^{(e,n)}$ & Matrix-valued spin distribution function for massive\tabularnewline
 & fermions in the $n$-th order, where $n=0,1$ and $e=\pm$  \tabularnewline
 & stands for fermions and anti-fermions\tabularnewline
\hline
$\overline{p}^{\mu}$ & On-shell momentum for massive fermions defined as $(E_{p},-\mathbf{p})$\tabularnewline
\hline
$W^{(1)}$ & The first order Wigner function $W^{\lessgtr(1)}$\tabularnewline
\hline
$W_{\mathrm{qc}}^{(1)}$ & Quasi-classical part of $W^{(1)}$\tabularnewline
\hline
$W_{\nabla}^{(1)}$ & Gradient and collision part of $W^{(1)}$\tabularnewline
\hline
$W_{\mathrm{off}}^{(1)}$ & Off-shell part of $W^{(1)}$\tabularnewline
\hline
$\mathbf{n}_{i}$ & A set of orthonormal unit vectors which form a right-handed\tabularnewline
($i=1,2,3$) & basis in three spatial dimensions with $\mathbf{n}_{3}$ being the
spin\tabularnewline
 & quantization direction in the rest frame of the fermion\tabularnewline
\hline
$n_{i}^{(e)\mu}(\mathbf{p})$ & Polarization four-vector for fermions ($e=+$) and\tabularnewline
 & anti-fermions ($e=-$) boosted from $\mathbf{n}_{i}$ in the rest
frame to  \tabularnewline
 & a frame in which the partcle has the momentum $\mathbf{p}$\tabularnewline
\hline
$\tau_{i}$ & Pauli matrices in the spin space with $i=1,2,3$\tabularnewline
\hline
$\mathscr{C}_{\mathrm{scalar}}$ & Scalar part of the collision term in the spin\tabularnewline
 & Boltzmann equation\tabularnewline
\hline
$\mathscr{C}_{\mathrm{pol}}$ & Polarization part of the collision term in the spin\tabularnewline
 & Boltzmann equation\tabularnewline
\hline
$\mathfrak{s}^{\mu}$ & Continuous spin four-vector\tabularnewline
\hline
$\Theta_{\mu\nu}^{(+)}$ & Projector operator orthorgonal to $p^{\mu}$ for massive fermions \tabularnewline
\hline
$\Theta_{\mu\nu}^{(-)}$ & Projector operator orthorgonal to $\overline{p}^{\mu}$ for massive fermions \tabularnewline
\hline
$f_{\pm}(x,p,\mathfrak{s})$ & Scalar spin dependent distribution in the extended\tabularnewline
 & phase space built from $f_{rs}^{(\pm,n)}$\tabularnewline
\hline
$\Pi(\mathfrak{s})$ & Spin projector on $\mathfrak{s}^{\mu}$\tabularnewline
\hline
$\Delta^{\mu}$ & Space shift in non-local collisions\tabularnewline
\hline
\end{longtable}

\end{document}